\documentclass[a4paper,usenames,dvipsnames,11pt]{article}
\pdfoutput=1

\usepackage{jheppub}
\usepackage{slashed}
\usepackage{mathrsfs,booktabs,multirow,tabularx}
\usepackage{stmaryrd}
\usepackage{xspace}
\usepackage{fancyvrb}
\usepackage[makeroom]{cancel}
\usepackage{amsmath}    
\usepackage{amssymb}    %
\usepackage{graphicx}   
\usepackage{verbatim}   
\usepackage{lscape}
\usepackage{subfig}
\usepackage{listings}

\def\beq{\begin{equation}}
\def\eeq{\end{equation}}
\def\beqn{\begin{eqnarray}}
\def\eeqn{\end{eqnarray}}
\def\abs#1{\left|#1\right|}

\def\xidistr#1{\left(\frac{1}{\xi}\right)_{\!#1}}
\def\pdistr#1#2{\left(\frac{1}{#1}\right)_{\!#2}}
\def\lxidistr#1{\left(\frac{\log\xi}{\xi}\right)_{\!#1}}
\def\lpdistr#1#2{\left(\frac{\log#1}{#1}\right)_{\!#2}}
\def\lppdistr#1#2{\left(\frac{\log\left(#1\right)}{#1}\right)_{\!#2}}

\def\FKSeq#1{eq.~({\bf II}.#1)}
\def\MadFKSeq#1{eq.~({\bf I}.#1)}
\def\set#1#2#3{\left\{#1\right\}_{#2,#3}}
\def\setS#1#2#3#4{\left\{#1\right\}_{#2,#3}^{[#4]}}

\newcommand{\bqa}{\begin{eqnarray}}
\newcommand{\eqa}{\end{eqnarray}}
\chardef\MyArticleWithColor=\pdfcolorstackinit page direct{0 g}

\newcommand{\plaat}[3]{\raisebox{#3pt}{\epsfig{figure=#1.pdf,width=#2cm}}}
\def\cCode#1{\begin{lstlisting}[mathescape,basicstyle=\small
\ttfamily,frame=leftline,aboveskip=4mm,belowskip=4mm,xleftmargin=20pt,framexleftmargin=10pt,
numbers=none,framerule=2pt,abovecaptionskip=0.0mm,belowcaptionskip=3.5mm #1]}

\newcommand\sss{\scriptscriptstyle}
\newcommand\mydot{\!\cdot\!}
\newcommand\ep{\epsilon}
\newcommand\half{\frac{1}{2}}

\newcommand\as{\alpha_{\sss S}}
\newcommand\gs{g_{\sss S}}

\newcommand\aem{\alpha}
\newcommand\alo{\as}
\newcommand\alt{\aem}
\newcommand{\gev}{\,\textrm{GeV}}
\newcommand{\mev}{\,\textrm{MeV}}

\newcommand\bt{\bar{t}}
\newcommand\aNLO{{\sc\small MadGraph5\_aMC@NLO}}
\newcommand\aNLOs{{\sc\small MG5\_aMC}}
\newcommand\MadGraph{{\sc\small MadGraph}}

\newcommand\UFO{{\sc\small UFO}}
\newcommand\mf{{\sc\small MadFKS}}
\newcommand\ml{{\sc\small MadLoop}}
\newcommand\ct{{\sc\small CutTools}}
\newcommand\nin{{\sc\small Ninja}}
\newcommand\coll{{\sc\small Collier}}
\newcommand\IREGI{{\sc\small IREGI}}

\newcommand\HWsv{{\sc\small Herwig7}}
\newcommand\PYe{{\sc\small Pythia8}}

\newcommand\FJ{{\sc\small FastJet}}

\newcommand\recola{{\sc\small RECOLA}}
\newcommand\sherpa{{\sc\small Sherpa}}
\newcommand\gosam{{\sc\small GoSam}}
\newcommand\maddip{{\sc\small MadDipole}}
\newcommand\OL{{\sc\small OpenLoops}}
\newcommand{\FeynRules}{{\sc \small FeynRules}}

\newcommand{\nloct}{{\sc \small NloCt}}
\newcommand\prompt{{\tt MG5\_aMC>}}
\newcommand\pt{p_{\sss T}}
\newcommand\pti{p_{{\sss T},i}}
\newcommand\kt{k_{\sss T}}
\newcommand{\Ht}{H_{\sss T}}
\newcommand{\LO}{{\rm LO}}
\newcommand{\LOi}{{\rm LO}_i}
\newcommand{\LOipo}{{\rm LO}_{i+1}}
\newcommand{\LOo}{{\rm LO}_1}
\newcommand{\LOt}{{\rm LO}_2}
\newcommand{\LOth}{{\rm LO}_3}
\newcommand{\LOf}{{\rm LO}_4}
\newcommand{\NLO}{{\rm NLO}}
\newcommand{\NLOi}{{\rm NLO}_i}
\newcommand{\NpLOi}{{\rm N}^p{\rm LO}_i}
\newcommand{\NLOipo}{{\rm NLO}_{i+1}}
\newcommand{\NLOo}{{\rm NLO}_1}
\newcommand{\NLOt}{{\rm NLO}_2}
\newcommand{\NLOth}{{\rm NLO}_3}
\newcommand{\NLOf}{{\rm NLO}_4}
\newcommand{\NLOfv}{{\rm NLO}_5}

\newcommand{\MSb}{\overline{\rm MS}}

\newcommand{\BW}{{\rm BW}}
\newcommand{\stepf}{\Theta}
\newcommand\muF{\mu_{\sss F}}
\newcommand\muR{\mu_{\sss R}}
\newcommand{\ttv}{t\bar{t}V}
\newcommand{\ttV}{\ttv}
\newcommand{\nmax}{{\tt n_{\tt max}}}
\newcommand{\mmax}{{\tt m_{\tt max}}}
\newcommand\FKSpairs{{\cal P}_{\sss\rm FKS}}
\newcommand\ren{{\rm R}}
\newcommand\unren{{\rm U}}
\newcommand\oG{\overline{G}}
\newcommand\vertmuCM{{\cal V}^{(0)\mu}_{\rm CM}}
\newcommand\vertnuCM{{\cal V}^{(0)\nu}_{\rm CM}}
\newcommand\vertmuZW{{\cal V}^{(0)\mu}_{\rm ZW}}
\newcommand\vertnuZW{{\cal V}^{(0)\nu}_{\rm ZW}}
\newcommand\bM{\bar{M}}
\newcommand\bGa{\bar{\Gamma}}
\newcommand\bga{\bar{\gamma}}
\newcommand\bMW{\bM_W}
\newcommand\bGaW{\bGa_W}
\newcommand\bMZ{\bM_Z}
\newcommand\bGaZ{\bGa_Z}
\newcommand\ZW{{\rm ZW}}
\newcommand\CM{{\rm CM}}

\newcommand{\bq}{\bar{q}}
\newcommand{\epem}{e^+e^-}
\newcommand{\mpmm}{\mu^+\mu^-}
\newcommand{\ord}{{\cal O}}
\newcommand{\Sfun}{{\cal S}}
\newcommand{\Sfunij}{\Sfun_{ij}}
\newcommand\asotpi{\frac{\as}{2\pi}}
\newcommand\aotpi{\frac{\aem}{2\pi}}
\newcommand\muoQ{\frac{\mu^2}{Q^2}}
\newcommand\muoQep{\left(\frac{\mu^2}{Q^2}\right)^\ep}
\newcommand\Dz{{\cal D}^{(0)}}
\newcommand\Do{{\cal D}^{(1)}}

\newcommand\Tt{{\rm T}}
\newcommand\QCD{{\rm QCD}}
\newcommand\QED{{\rm QED}}
\newcommand\proc{r}
\newcommand\nini{n_{\sss I}}
\newcommand\nlight{n_{\sss L}}
\newcommand\nlightB{\nlight^{\sss (B)}}

\newcommand\nlightBorR{\nlight^{\sss (B/R)}}
\newcommand\nheavy{n_{\sss H}}
\newcommand\nzero{n_\emptyset}
\newcommand\ident{{\cal I}}
\newcommand\Ione{\ident_1}
\newcommand\Itwo{\ident_2}
\newcommand\amp{{\cal A}}
\newcommand\ampmt{\amp^{(m,0)}}
\newcommand\ampnt{\amp^{(n,0)}}

\newcommand\ampnl{\amp^{(n,1)}}
\newcommand\ampsq{{\cal M}}
\newcommand\ampsqmt{\ampsq^{(m,0)}}
\newcommand\ampsqnt{\ampsq^{(n,0)}}

\newcommand\ampsqnl{\ampsq^{(n,1)}}
\newcommand\vampsqnl{{\cal V}^{(n,1)}}
\newcommand\hvampsqnl{\hat{\cal V}^{(n,1)}}

\newcommand\Qop{\vec{Q}}
\newcommand\Qops{Q}
\newcommand\JetsB{J^{\nlightB}}

\newcommand\velkl{v_{kl}}
\newcommand\avg{{\cal N}}
\newcommand\xicut{\xi_{cut}}
\newcommand\deltaO{\delta_{\sss O}}

\newcommand\NC{N_{\sss c}}

\newcommand\nC{n_{\sss c}}
\newcommand\NF{N_{\sss F}}
\newcommand\Nl{N_l}
\newcommand\eikint{{\cal E}}
\newcommand\phsp{d\phi}
\newcommand\phspn{\phsp_{n}}

\newcommand\mua{\mu_\alpha}

\newcommand\ampzCM{\amp^{(0)}_{\rm CM}}
\newcommand\ampzZW{\amp^{(0)}_{\rm ZW}}

\newcommand{\bb}{\bar{b}}

\newcommand\epb{\overline{\epsilon}}
\newcommand\xii{\xi_i}
\newcommand\xij{\xi_j}
\newcommand\xik{\xi_k}
\newcommand\xia{\xi_\alpha}
\newcommand\xib{\xi_\beta}
\newcommand\xinpf{\xi_{n+4}}
\newcommand\xic{\xi_c}

\newcommand\yi{y_i}
\newcommand\yj{y_j}
\newcommand\ya{y_\alpha}
\newcommand\yb{y_\beta}

\newcommand\ynpf{y_{n+4}}
\newcommand\vol{{\cal V}}

\newcommand{\LOone}{\ensuremath{\LOo}\xspace}
\newcommand{\LOtwo}{\ensuremath{\LOt}\xspace}
\newcommand{\LOthree}{\ensuremath{\LOth}\xspace}
\newcommand{\LOfour}{\ensuremath{\LOf}\xspace}
\newcommand{\NLOone}{\ensuremath{\NLOo}\xspace}
\newcommand{\NLOtwo}{\ensuremath{\NLOt}\xspace}
\newcommand{\NLOthree}{\ensuremath{\NLOth}\xspace}
\newcommand{\NLOfour}{\ensuremath{\NLOf}\xspace}
\newcommand{\NLOfive}{\ensuremath{\NLOfv}\xspace}

\newcommand{\mz}{m_Z}

\newcommand{\SMWidth}{{\sc\small SMWidth}}
\newcommand{\FeynArts}{{\sc\small FeynArts}}
\newcommand{\HDecay}{{\sc\small HDecay}}
\newcommand{\OneLoop}{{\sc\small OneLoop}}

\newcommand{\MadLoop}{\ml}

\title{The automation of next-to-leading order electroweak calculations}

\author[a]{R. Frederix,}
\author[b]{S. Frixione,}
\author[c]{V. Hirschi,}
\author[a]{D. Pagani,}
\author[d]{H.-S. Shao,}
\author[e]{M. Zaro}
\affiliation[a]{Physik Department T31, Technische Universit\"at M\"unchen, 
James-Franck-Str. 1,\\ D-85748 Garching, Germany}
\affiliation[b]{INFN, Sezione di Genova, Via Dodecaneso 33, I-16146, 
Genoa, Italy}
\affiliation[c]{Institute for Theoretical Physics, ETH Z\"urich, 
8093 Z\"urich, Switzerland}
\affiliation[d]{Laboratoire de Physique Th\'eorique et Hautes Energies
(LPTHE), UMR 7589,\\ Sorbonne Universit\'e et CNRS, 4 place Jussieu, 
75252 Paris Cedex 05, France}
\affiliation[e]{Nikhef, Science Park 105, NL-1098 XG Amsterdam, The Netherlands}

\emailAdd{rikkert.frederix@tum.de}
\emailAdd{Stefano.Frixione@cern.ch}
\emailAdd{hirschva@itp.phys.ethz.ch}
\emailAdd{davide.pagani@tum.de}
\emailAdd{huasheng.shao@lpthe.jussieu.fr}
\emailAdd{m.zaro@nikhef.nl}

\abstract{We present the key features relevant to the automated
computation of all the leading- and next-to-leading order contributions
to short-distance cross sections in a mixed-coupling expansion, with 
special emphasis on the first subleading NLO term in the QCD+EW scenario, 
commonly referred to as NLO EW corrections. We discuss, in particular, 
the FKS subtraction in the context of a mixed-coupling expansion; the 
extension of the FKS subtraction to processes that include final-state
tagged particles, defined by means of fragmentation functions; and
some properties of the complex mass scheme. We combine the present 
paper with the release of a new version of {\sc\small MadGraph5\_aMC@NLO},
capable of dealing with mixed-coupling expansions. We use the code to 
obtain illustrative inclusive and differential results for the 13-TeV LHC.
}

\keywords{NLO computations, Electroweak theory, Automation}

\preprint{
\begin{flushright}
Nikhef/2018-015\\
TUM-HEP-1138/18\\
\today
\end{flushright}
}

\begin{document}
\maketitle
\flushbottom

\section{Introduction\label{sec:intro}}
With more than two-thirds of the LHC Run II completed, and a collected 
integrated luminosity in excess of 85 fb$^{-1}$ at 13 TeV, it appears 
that physics beyond the Standard Model is not eager to be discovered.
This could be either because its characteristic scale is larger than 
usually assumed, and thus beyond the LHC reach; or because its signals are 
particularly difficult to identify by means of final-state direct searches.
While we have no way at present to tell which of these two scenarios is
the one realised by Nature, we know that they entail different long-term
strategies. In the former instance, we must be able to explore larger
scales, chiefly by increasing collider c.m.~energies; in the latter
instance, we must collect more statistics and fight against systematics. 
In both cases, indirect searches might turn out to play a vital role.

In order to cope with this situation, phenomenologists must continue in 
the trend which has now been established for a few years, namely to progress 
in the direction of increasing both the flexibility and the precision of
their calculations. In terms of flexibility, the key aspects are
the ability of computer codes to readily deal with the matrix elements
relevant to both new-physics models and the SM, and that of embedding
these matrix element results in fully realistic final-state simulations,
such as those provided by parton shower Monte Carlos. As far as precision
is concerned, in the vast majority of applications this is a synonym for
the computation of higher orders in a fixed-order coupling-constant 
perturbative expansion. Which specific perturbative orders depends 
obviously on the theory one considers. Owing to both the large numerical
values assumed by $\as$ and the paramount importance of hadron-collision 
physics in the LHC era, QCD has played a particularly prominent role
in recent years -- fully-differential NLO results are by now standard, 
including the cases of processes with very involved final states,
while more and more NNLO and even N$^3$LO predictions are becoming
available, for low-multiplicity reactions and with different degrees
of inclusiveness (see e.g.~ref.~\cite{Bendavid:2018nar} for a recent
review).

Currently, there is therefore a compelling case for considering in full
generality the calculation of NLO contributions in the EW theory, and this
for at least three reasons. Firstly, based on the values of $\as$ and $\aem$,
one expects NNLO QCD and NLO EW effects to be numerically comparable.
Secondly, this naive scaling behaviour could actually be violated
in those regions of the phase space associated with large mass scales,
since there the coefficients of the EW series might grow faster than 
their QCD counterparts, owing to the presence of large Sudakov
logarithms (see e.g.~refs.~\cite{Kuroda:1990wn,Degrassi:1992ue,
Ciafaloni:1998xg,Ciafaloni:2000df,Denner:2000jv,Denner:2001gw}
for discussions about their origin and universal nature) -- 
large transverse momenta are a particularly important example, 
given the relevance of the high-$\pt$ regions in new-physics searches. 
Thirdly, it is likely that among future colliders there will be at 
least one $\epem$ machine, and possibly an $eH$ one, for whose physics
simulations EW computations will be crucial.

When combining the necessity of the calculation of EW NLO corrections
with the requirement that they be flexible and available for arbitrary
processes, one is naturally led to their automation, not least because
the case for automation is strongly supported by the striking success that 
this strategy has enjoyed in the case of QCD, where NLO results are now 
mass-produced and constitute the backbone of ATLAS and CMS $pp$ simulations.
Indeed, although not as comprehensive as for QCD computations, there has been
steady progress in the automation of EW NLO corrections, for both
one-loop and real-emission contributions, by collaborations such as
\recola~\cite{Actis:2016mpe,Actis:2012qn} with \sherpa~\cite{Gleisberg:2008ta,
Schonherr:2017qcj}, \OL~\cite{Cascioli:2011va} with \sherpa, 
\gosam~\cite{Cullen:2011ac,Cullen:2014yla} with either 
\maddip~\cite{Frederix:2008hu,Gehrmann:2010ry} or \sherpa, 
and \aNLO~\cite{Alwall:2014hca}. Recent results obtained with these
tools~\cite{Denner:2014ina,Denner:2014wka,Denner:2015yca,Kallweit:2014xda,
Frixione:2014qaa,Chiesa:2015mya,Kallweit:2015dum,Frixione:2015zaa,
Biedermann:2016guo,Biedermann:2016yvs,Denner:2016jyo,Biedermann:2016yds,
Biedermann:2016lvg,Denner:2016wet,Frederix:2016ost,Pagani:2016caq,
Biedermann:2017yoi,Kallweit:2017khh,Biedermann:2017bss,Biedermann:2017oae,
Chiesa:2017gqx,Czakon:2017wor,Frederix:2017wme,Gutschow:2018tuk} clearly 
demonstrate how automation is allowing us to attack problems whose complexity 
is too great to justify their solutions through traditional approaches.

The goal of this paper is that of presenting the most important features
that underpin NLO calculations performed by \aNLO\ (\aNLOs\ henceforth)
in a mixed-coupling scenario, i.e.~when two coupling constants are 
simultaneously treated as small parameters in a perturbative expansion.
While the QCD+EW case, in the context of which \aNLOs\ is able to compute
the NLO QCD and the (so-called) NLO EW corrections among other things, is the 
most prominent in today phenomenology, the code is not necessarily restricted 
to that, and is structured to handle analogous situations in user-defined
theories (including, in particular, EFTs). We also combine the present paper 
with a new release of the \aNLOs\ code, which will be the first {\em public} 
one compatible with a mixed-coupling expansion -- previous applications that 
included non-QCD effects~\cite{Frixione:2014qaa,Frixione:2015zaa,
Frederix:2016ost,Czakon:2017wor,Frederix:2017wme,Pagani:2016caq,
Czakon:2017lgo,Maltoni:2017ims,Plehn:2015cta} had been obtained 
with preliminary and still-private versions. Conversely, we stress that 
EW-loop-induced processes could already be automatically generated by
\aNLOs, owing to the work of ref.~\cite{Hirschi:2015iia}.

This paper is organised as follows. In sect.~\ref{sec:gen} we lay out
the scope of our work, discuss the strategy upon which \aNLOs\ is based,
and point out the limitations of the current version of the code. In 
sect.~\ref{sec:Xsec} we extend the short-distance FKS~\cite{Frixione:1995ms,
Frixione:1997np} formulae to the mixed-coupling case. Sect.~\ref{sec:Frag} 
shows how one can apply the FKS subtraction formalism to those cases in 
which the use of fragmentation functions is required. 
In sect.~\ref{sec:cmscheme} we discuss some features 
of the complex-mass scheme~\cite{Denner:1999gp,Denner:2005fg} that are 
particularly relevant to its automation in \aNLOs. 
In sect.~\ref{sec:Res} we present some illustrative results,
both inclusive and at the differential level, that have been obtained
in a 13-TeV LHC configuration. Finally, in sect.~\ref{sec:conc} we 
draw our conclusions. Some extra technical material is reported in
the appendices.

\vfill

\section{General considerations\label{sec:gen}}
We start by writing an observable $\Sigma(\alo,\alt)$ to all orders
in $\as$ and $\aem$ by adopting the notation introduced in 
ref.~\cite{Alwall:2014hca}:
\beqn
\Sigma(\alo,\alt)&=&\alo^{c_s(k_0)}\alt^{c(k_0)}\sum_{p=0}^\infty
\sum_{q=0}^{\Delta(k_0)+p}\Sigma_{k_0+p,q}\,
\alo^{\Delta(k_0)+p-q}\alt^q
\label{taylor40}
\\*
&=&
\Sigma^{\rm (LO)}(\alo,\alt)+\Sigma^{\rm (NLO)}(\alo,\alt)+\ldots\,,
\label{taylor4}
\eeqn
where we have identified the LO ($\Sigma^{\rm (LO)}$) and NLO 
($\Sigma^{\rm (NLO)}$) contributions in eq.~(\ref{taylor4}) with the 
$p=0$ and $p=1$ terms in eq.~(\ref{taylor40}). The integer numbers 
$k_0$, $c_s(k_0)$, $c(k_0)$, and $\Delta(k_0)$ are process-specific
quantities (of which we shall give an explicit example below) that
obey the constraint \mbox{$k_0=c_s(k_0)+c(k_0)+\Delta(k_0)$}. We can further 
decompose these contributions by defining terms that factorise a single 
coupling-constant combination $\as^n\aem^m$; each of these terms corresponds
to a single value of $q$ in the second sum in eq.~(\ref{taylor40}).
Explicitly:
\beqn
\Sigma^{\rm (LO)}(\alo,\alt)&=&
\alo^{c_s(k_0)}\alt^{c(k_0)}
\sum_{q=0}^{\Delta(k_0)}\Sigma_{k_0,q}\,
\alo^{\Delta(k_0)-q}\alt^q
\nonumber
\\*&=&
\Sigma_{\LOo}+\ldots+\Sigma_{{\rm LO}_{\Delta(k_0)+1}}\,,
\label{SigB}
\\*
\Sigma^{\rm (NLO)}(\alo,\alt)&=&
\alo^{c_s(k_0)}\alt^{c(k_0)}
\sum_{q=0}^{\Delta(k_0)+1}\Sigma_{k_0+1,q}\,
\alo^{\Delta(k_0)+1-q}\alt^q
\nonumber
\\*&=&
\Sigma_{\NLOo}+\ldots+\Sigma_{{\rm NLO}_{\Delta(k_0)+2}}\,.
\label{SigNLO}
\eeqn
In a well-behaved perturbative series, and given that $\aem\ll\as$,
eqs.~(\ref{SigB}) and~(\ref{SigNLO}) imply:
\beq
\Sigma_{\LOi}\gg\Sigma_{\LOipo}\,,
\;\;\;\;\;\;\;\;
\Sigma_{\NLOi}\gg\Sigma_{\NLOipo}\,,
\;\;\;\;\;\;\;\;\forall\,i\,.
\eeq
This hierarchy suggests (see ref.~\cite{Frixione:2014qaa}) to call
$\Sigma_{\NpLOi}$ the leading ($i=1$) or $i^{th}$-leading ($i>1$: 
second-leading, third-leading, and so forth) term of the N$^p$LO 
contribution to the cross section. It is customary to identify
$\Sigma_{\NLOo}$ and $\Sigma_{\NLOt}$ with the NLO QCD and the NLO EW
corrections, respectively. While this is unambiguous in the case of 
$\Sigma_{\NLOo}$, it is somehow misleading in the case of $\Sigma_{\NLOt}$, 
for two reasons. Firstly, at one loop there is no clear-cut way to define
pure-EW contributions on a diagrammatic basis. Secondly, $\Sigma_{\NLOt}$ 
may receive contributions from the so-called heavy-boson radiation (HBR), 
namely from diagrams that feature the emission of a real $W$, $Z$, or $H$ 
boson (see e.g.~eq.~(2.3) of ref.~\cite{Frixione:2015zaa}) -- these
are typically not included in what are conventionally denoted as NLO EW
corrections. With these caveats in mind, in what follows we shall also 
often refer to $\Sigma_{\NLOt}$ as the NLO EW contribution, in those cases
where no ambiguity is possible.

In order to give an explicit example, let us consider the case of
$t\bt$ production in association with a heavy EW boson. 
Equations~(\ref{taylor40})--(\ref{SigNLO}) read:
\beqn
\Sigma_{\ttV}^{\rm (LO)}(\as,\aem)&=&
\as^2\aem\,\Sigma_{3,0}
+\as\aem^2\,\Sigma_{3,1}
+\aem^3\,\Sigma_{3,2}
\nonumber\\*
&\equiv&
\Sigma_{\LOo}+\Sigma_{\LOt}+\Sigma_{\LOth}\,.
\label{SigBttv}
\\
\Sigma_{\ttV}^{\rm (NLO)}(\as,\aem)&=&
\as^3\aem\,\Sigma_{4,0}
+\as^2\aem^2\,\Sigma_{4,1}
+\as\aem^3\,\Sigma_{4,2}
+\aem^4\,\Sigma_{4,3}\,,
\nonumber\\*
&\equiv&
\Sigma_{\NLOo}+\Sigma_{\NLOt}+\Sigma_{\NLOth}+\Sigma_{\NLOf}\,.
\label{SigNLOttv}
\eeqn
These imply that, for this process, we have $k_0=3$, $c_s(k_0)=0$,
$c(k_0)=1$, and $\Delta(k_0)=2$.

Finally, we point out that eqs.~(\ref{taylor40})--(\ref{SigNLO}),
although written here for the QCD+EW case, actually apply to the 
perturbative expansion in any two couplings\footnote{In fact, the 
internal bookkeeping in \aNLOs\ also works for expansions in more
than two couplings, although the number of IR-subtraction types is
limited to two (assumed to be QCD- and QED-like).} $\alpha_1$ and $\alpha_2$.
The formal replacements $\as\to\alpha_1$ and $\aem\to\alpha_2$ suffice
to obtain the general formulae.

As it has been anticipated in sect.~\ref{sec:intro}, our final goal is 
the automation of the computation of eqs.~(\ref{SigB}) and~(\ref{SigNLO})
in \aNLOs. We briefly recall here the basic building blocks that form the 
core of the code. \aNLOs\ makes use of the FKS method~\cite{Frixione:1995ms,
Frixione:1997np} (automated in the module \mf~\cite{Frederix:2009yq,
Frederix:2016rdc}) for dealing with IR singularities. The computations 
of one-loop amplitudes are carried out by switching dynamically between 
two integral-reduction techniques, OPP~\cite{Ossola:2006us} or 
Laurent-series expansion~\cite{Mastrolia:2012bu}, and 
TIR~\cite{Passarino:1978jh,Davydychev:1991va,Denner:2005nn}. These have 
been automated in the module \ml~\cite{Hirschi:2011pa,Alwall:2014hca}, 
which in turn exploits \ct~\cite{Ossola:2007ax}, \nin~\cite{Peraro:2014cba,
Hirschi:2016mdz}, \IREGI~\cite{ShaoIREGI}, or \coll~\cite{Denner:2016kdg},
together with an in-house implementation of the {\sc OpenLoops} 
optimisation~\cite{Cascioli:2011va}. Finally, in the case of matching 
with parton showers, the MC@NLO formalism~\cite{Frixione:2002ik} is employed.

We also remind the reader that the building blocks in \aNLOs\ parametrise
the cross sections in a theory- and process-independent manner. Theory-
and process-specific information are given in input at runtime, and are
fully  under the user's control. In particular, the information on the
theory (such as the spectrum of the particles and their interactions)
are collectively referred to as a {\em model}. Nowadays, most models can
be constructed automatically given the Lagrangian by employing tools
such as \FeynRules~\cite{Christensen:2008py,Christensen:2009jx,
Christensen:2010wz,Duhr:2011se,Alloul:2013bka,Alloul:2013fw} and
\nloct~\cite{Degrande:2014vpa} -- the latter is essential for embedding
in the model the quantities necessary for one-loop computations, namely 
the UV and $R_2$ counterterms. Further details on these topics are given 
in sect.~2.1 of ref.~\cite{Alwall:2014hca}.

The implications of what has been said above are the following.
The building blocks of the \aNLOs\ code have been made compatible 
with the structure of the generic mixed-coupling expansion, 
eqs.~(\ref{SigB}) and~(\ref{SigNLO}). This implies, in particular, 
upgrading the handling of the inputs, and implementing a much 
more involved bookkeeping, both of which are due to the necessity of 
retaining independent control on the different $\Sigma_{k,q}$ contributions. 
In terms of matrix-element computations, the code is aware of the 
possible presence of more than one type of interactions (e.g.~QCD and EW),
whose details are assumed to be inherited from the model. In particular,
inspection of the model allows \aNLOs\ to construct all of the real-emission 
and one-loop diagrams that contribute to the process under study at the
desired perturbative order. Crucially, it also allows the code to figure
out the structure of the IR singularities, by considering the set of all
possible $1\to 2$ branchings with at least one outgoing massless particle,
and so to set up the appropriate FKS subtractions. More details on the
calculation strategies underlying \mf\ and \ml\ can be found in sect.~2.4.1 
of ref.~\cite{Alwall:2014hca}, and sects.~2.4.2 and~4.3 of 
ref.~\cite{Alwall:2014hca}, respectively. Mixed-coupling expansion 
capabilities have been gradually added to \mf\ and \ml, in different stages 
for the two modules, during the course of the work relevant to 
refs.~\cite{Frixione:2014qaa,Frixione:2015zaa,Frederix:2016ost}. 
They have been completed and validated for the present paper.
The bottom line is that the typical \aNLOs\ usage in the case e.g.~of
a QCD+EW expansion may read as follows\footnote{Here, {\tt QED} is a 
conventional keyword that stands for both electromagnetic and weak effects. 
See footnote~\ref{ft:QED} for more details.}:

\vskip 0.25truecm
\noindent
~~\prompt\ {\tt ~import~model~myNLOmodel\_w\_qcd\_qed}

\noindent
~~\prompt\ {\tt ~generate~p$_1$ p$_2$ > p$_3$ p$_4$ p$_5$ p$_6$ 
QCD=$\nmax$ QED=$\mmax$ [QCD QED]}

\vskip 0.25truecm
\noindent
with {\tt p$_i$} (multi)particles that belong to the spectrum of
the model {\tt myNLOmodel\_w\_qcd\_qed}. The syntax above implies
computing the following LO and NLO contributions:
\beqn
&&{\rm LO}:\phantom{Naaa}
\as^n\aem^m\,,\;\;\;\;\;n\le\nmax\,,\;\;\phantom{+1,}\;\,
m\le\mmax\,,\;\;\phantom{+1,}\;\,\,n+m=k_0\,,
\label{LOsynt}
\\*
&&{\rm NLO}:\phantom{aaa}\,
\as^n\aem^m\,,\;\;\;\;\;n\le\nmax+1\,,\;\;\;
m\le\mmax+1\,,\;\;\;\,n+m=k_0+1\,.
\label{NLOsynt}
\eeqn
We point out that the largest power of $\as$ in eq.~(\ref{NLOsynt}) is 
exactly one unity larger than its LO counterpart in eq.~(\ref{LOsynt})
because of the presence of the keyword {\tt [QCD]} in the process-generation
command. This instructs the code to consider all diagrams that feature
two extra QCD vertices\footnote{Here ``diagrams'' is meant in the 
Cutkowsky sense: the vertices will be on the same side (opposite sides) 
of the Cutkowsky cut for one-loop (real-emission) contributions.}
w.r.t.~those present at the Born level. Likewise, it is the 
keyword {\tt [QED]} that sets the largest power of $\aem$ that appears
at the NLO level equal to \mbox{$\mmax+1$}. Either keyword 
can be omitted -- we shall give explicit examples in sect.~\ref{sec:Res}.
More details on the syntax above, relevant to the inner workings of
\aNLOs, are given in appendix~\ref{sec:tech}.

We now turn to underscore an issue which is specific to QED, and
that stems from the fact that, in such a theory, photons and leptons 
can be regarded both as particles that enter the short-distance process 
and as observable (taggable) objects. This point has been already 
discussed in ref.~\cite{Frederix:2016ost}, and we shall limit ourselves
here to summarise the conclusions of that paper. The key point is that
short-distance photons and massless leptons can be identified 
with the corresponding taggable objects only up to a certain 
$\Sigma_{{\rm NLO}_{i_0}}$ term, beyond which (i.e.~for $\Sigma_{\NLOi}$,
$i>i_0$) this identification leads to IR-unsafe observables. The value 
of $i_0$ is process dependent; typically, IR unsafety manifests itself 
in the third- or even the second-leading NLO contribution\footnote{In 
the case of the second-leading NLO term one can work around this issue
by means of the $\aem(0)$ scheme, which is the reason why it has 
never been prominent in the context of NLO EW computations.}. The solution 
proposed in ref.~\cite{Frederix:2016ost} is as follows:
\begin{enumerate}
\item Short-distance photons and massless leptons are not taggable objects.
\item A taggable photon is a photon that emerges from a fragmentation process.
\item A taggable massless lepton is either a lepton that emerges from a 
fragmentation process or a dressed lepton, i.e.~an object whose 
four-momentum is equal to that of a very narrow jet that contains
the short-distance lepton.
\item These rules imply that photons and massless leptons must be treated
on the same footing as gluons and quarks in short-distance computations
(democratic approach).
\item Short-distance computations should be performed in $\MSb$-like 
EW renormalisation schemes, such as the $G_\mu$ or $\aem(m_Z)$ ones,
regardless of the initial- and final-state particle contents of the
process of interest.
\end{enumerate}
We point out that, since the $\ord(\aem^0)$ term of both the photon and 
the lepton fragmentation functions is equal to $\delta(1-z)$, the above 
prescriptions exactly coincide with the usual procedure followed in
the calculation of NLO QCD corrections, where no distinction is made
between short-distance and taggable EW objects. 

Furthermore, we would like to stress that, as was already mentioned in 
ref.~\cite{Frederix:2016ost}, a direct consequence of the evolution
equations for fragmentation functions at the first non-trivial order
in QED is that the $z\to 1$ dominant term in the photon-to-photon
fragmentation is equal to \mbox{$\aem_0/\aem(Q^2)\,\delta(1-z)$},
with $\aem_0$ an $\ord(\aem)$ small-scale constant, and $Q^2$ a scale
of the order of the hardness of the process. This shows that, by
retaining only such a term, a fragmentation-function approach allows
one to recover naturally and in a straightforward manner the results 
obtained in the standard EW approach that makes use of the $\aem(0)$-scheme.
On top of being fully general and backward-compatible with established
procedures, working with fragmentation functions has also the appealing
feature of putting QCD and QED on a similar footing, and of rendering
conceptually alike the treatment of initial- and final-state
photons\footnote{Obviously, differences remain. In particular, the
different scales associated with the evolution of the photon-to-photon
fragmentation function and the photon-in-photon PDF imply that the 
issues (see e.g.~ref.~\cite{Harland-Lang:2016lhw}) relevant to 
photon-initiated processes are simply not present here.}.
The case of massless leptons has not been discussed in
ref.~\cite{Frederix:2016ost}, and we also leave it to future work; however,
we point out that universal logarithmic mass effects can be included in 
lepton fragmentation functions through the computation of its perturbative 
part, for example by working in analogy to what has been done in QCD
for the case of massive quarks~\cite{Mele:1990cw}.
In conclusion, there is a compelling motivation for a fully general
treatment of fragmentation functions in the context of NLO computations,
which is what will be done in sect.~\ref{sec:Frag}.

We also note that items~4 and~5 in the list above put the MC@NLO-type
matching to QED showers on the same footing as its QCD counterpart,
at least for those QED showers implemented in a QCD-like manner.

We conclude this section by enumerating the limitations of the new version
of the \aNLOs\ code whose release is associated with the present paper.
Firstly, the implementation of eqs.~(\ref{SigB}) and~(\ref{SigNLO}) implies
that the {\em automated} calculation of NLO QCD+EW effects is always carried
out in an additive scheme. If one is interested in a multiplicative approach
(where QCD and EW terms are assumed to factorise), one must compute the
leading and second-leading contributions separately, and manually combine 
them afterwards\footnote{We point out that all user-selected cross
section contributions are separately available during the course of
a single run. Therefore, their combination can be straightforwardly
achieved at the analysis level.}. Secondly, the implementation of the 
convolution of short-distance cross sections with fragmentation functions is 
not made publicly available (thus, observables with tagged photons and/or
leptons cannot be constructed). This stems from phenomenological considerations,
given that the perturbative terms for which those functions are a necessity 
are strongly suppressed, and the functions are not easy to extract from 
current datasets\footnote{Having said that, the possibility of using
purely-theoretical, leading-behaviour photon and lepton fragmentation
functions will be made available in the \aNLOs\ code in the near
future. This will allow one to recover standard results 
for processes where such particles are tagged.}.
Thirdly, although the implementation of the MC counterterms for QCD-like
QED showers (which are available in both \PYe~\cite{Sjostrand:2014zea} 
and \HWsv~\cite{Bellm:2015jjp}) has been completed, some further validation
will be necessary in order to bring the MC@NLO matching in the QED sector
on par with the QCD one. Thus, in such a sector the present release
of \aNLOs\ is restricted to performing fixed-order computations.
Finally, ISR and beamstrahlung effects are not yet implemented
in \aNLOs, and therefore we refrain from presenting $\epem$-collisions
results in this paper.

\section{FKS subtraction in mixed-coupling expansions\label{sec:Xsec}}
In this section we introduce the basic mechanisms and the notation that 
underpin the implementation of a mixed-coupling scenario in \mf; in the 
process, we shall thus extend the FKS subtraction to cover such a case.
In keeping with what has been done so far and in order to be definite, 
we shall deal explicitly with the QCD+QED case, which has the further
advantage of being essentially a worst-case scenario as far as IR 
subtractions are concerned\footnote{As the formulae in this section
show, a double copy of QCD (possibly with different colour factors), or 
any typical BSM model, do not pose any additional complications.}. For 
backward compatibility, the notation is modified in a manner as minimal as is 
possible w.r.t.~the one introduced in ref.~\cite{Frederix:2009yq}; eq.~(n.m) 
of that paper will be denoted by \MadFKSeq{n.m} here.

\subsection{Matrix elements\label{sec:ME}}
What is done in sect.~3.1 of ref.~\cite{Frederix:2009yq}, and in particular
\MadFKSeq{3.1}--\MadFKSeq{3.18} is unchanged, bar for the following
amendments:
\begin{itemize}
\item $\nlightBorR$ is the total number of light quarks, gluons, massless
charged leptons, and photons, at the Born ($B$) and real-emission ($R$) level;
\item $\nheavy$ is the number of massive particles that are either
strongly interacting, or electrically charged, or both;
\item $\nzero$ is the number of particles that do not belong to either
of the two previous categories;
\item $\nini=1$ regardless of the type of incoming particles (see
\MadFKSeq{3.11}--\MadFKSeq{3.13});
\item the symbol $\oplus$ as is used in \MadFKSeq{3.18} understands either
a QCD or a QED vertex.
\end{itemize}
As far as the matrix elements are concerned, the notation used so
far, introduced in sect.~2.4 of ref.~\cite{Alwall:2014hca}, emphasises
the role of the hierarchy among coupling combinations $\as^n\aem^m$
with constant $n+m$, which is convenient from a physics viewpoint.
However, for internal code usage a different notation is more apt,
that tells one immediately which powers of the coupling constants are 
relevant to individual contributions to short-distance cross sections.
Thus, for tree-level $m$-body matrix elements we shall write:
\beq
\ampsqmt_{(p,q)}\propto \as^p\aem^q\,,
\label{ampsq0}
\eeq
with
\beqn
p+q&=&k_0\,,\phantom{+1a}\;\;\;\;\;\;\;\;{\rm LO}\,,
\label{ppqlo}
\\
p+q&=&k_0+1\,,\;\;\;\;\;\;\;\;{\rm NLO}\,.
\label{ppqnlo}
\eeqn
For some given $p$ and $q$, the generalisation of \MadFKSeq{3.22}
and \MadFKSeq{3.23} reads as follows ($m=n,n+1$):
\beqn
\ampsqmt_{(p,q)}(\proc)&=&\frac{1}{2s}\frac{1}{\omega(\Ione)\omega(\Itwo)}
\mathop{\sum_{\rm colour}}_{\rm spin}
\sum_{p_1p_2}\sum_{q_1q_2}\delta_{p_1+p_2,p}\delta_{q_1+q_2,q}
\nonumber
\\&\times&
\left(2-\delta_{p_1p_2}\delta_{q_1q_2}\right)
\Re\left\{\ampmt_{(p_1,q_1)}(\proc)\ampmt_{(p_2,q_2)}(\proc)^{\star}\right\},
\label{Mtreenpo}
\eeqn
with amplitudes following similar conventions as in eq.~(\ref{ampsq0}):
\beq
\ampmt_{(p,q)}\propto \gs^p e^q\,,
\label{amp0}
\eeq
and $\omega(\ident)$ the product of spin and colour (if relevant) degrees 
of freedom for particle $\ident$.
Colour-linked Born's are defined as in \MadFKSeq{3.24}, by taking into
account the factors introduced in eq.~(\ref{Mtreenpo}):
\beqn
\ampsqnt_{\QCD(p,q)kl}(\proc)&=&-\frac{1}{2s}
\frac{2-\delta_{kl}}{\omega(\Ione)\omega(\Itwo)}
\mathop{\sum_{\rm colour}}_{\rm spin}
\sum_{p_1p_2}\sum_{q_1q_2}\delta_{p_1+p_2,p}\delta_{q_1+q_2,q}
\label{MQCDlinked}
\\&\times&
\left(2-\delta_{p_1p_2}\delta_{q_1q_2}\right)
\Re\left\{\ampnt_{(p_1,q_1)}(\proc) \Qop(\ident_k)\mydot\Qop(\ident_l)
{\ampnt_{(p_2,q_2)}(\proc)}^{\star}\right\}.
\nonumber
\eeqn
The colour operators are defined in \MadFKSeq{3.26}; that definition 
must be supplemented with:
\beq
\Qop(\ident)=0\,,\;\;\;\;\;\;\;\;
{\rm if}~\ident~{\rm is~not~strongly~interacting}\,.
\eeq
In a mixed-coupling scenario there will be a QED analogue of
eq.~(\ref{MQCDlinked}), i.e.~the charge-linked Born's that stem from 
the insertion of a soft-photon line. We have:
\beqn
\ampsqnt_{\QED(p,q)kl}(\proc)&=&-\frac{1}{2s}
\frac{2-\delta_{kl}}{\omega(\Ione)\omega(\Itwo)}
\mathop{\sum_{\rm colour}}_{\rm spin}
\sum_{p_1p_2}\sum_{q_1q_2}\delta_{p_1+p_2,p}\delta_{q_1+q_2,q}
\label{MQEDlinked}
\\&\times&
\left(2-\delta_{p_1p_2}\delta_{q_1q_2}\right)
\Re\left\{\ampnt_{(p_1,q_1)}(\proc) \Qops(\ident_k)\Qops(\ident_l)
{\ampnt_{(p_2,q_2)}(\proc)}^{\star}\right\},
\nonumber
\eeqn
with $Q(\ident)$ the charge operator, defined as follows:
\beq
Q(\ident)=(-)^{s(\ident)}e(\ident)\,,
\eeq
where $e(\ident)$ is the electric charge of particle $\ident$ in unit of
the positron charge, and:
\beqn
s(\ident)&=&\left\{
\begin{array}{c}
2\phantom{aaa}{\rm outgoing~(anti)particle}\,,\\
1\phantom{aaa}{\rm incoming~(anti)particle}\,.\\
\end{array}
\right.
\eeqn
This leads directly to the analogue of \MadFKSeq{3.31}, i.e.~to
charge conservation:
\beq
\sum_{k=\nini}^{\nlightB+\nheavy+2}\Qops(\ident_k)=0\,.
\eeq
The trivial nature of the charge operator implies an immediate simplification
of the form of the charge-linked Borns:
\beq
\ampsqnt_{\QED(p,q)kl}(\proc)=
(-)^{1+s(\ident_k)+s(\ident_l)}\left(2-\delta_{kl}\right)\,
e(\ident_k)\,e(\ident_l)\,
\ampsqnt_{(p,q)}(\proc)\,.
\eeq
Both the colour- and charge-linked Born's satisfy 
\MadFKSeq{3.32}--\MadFKSeq{3.34}, which we re-write as follows
(${\rm T}=\QCD,\QED$):
\beqn
\ampsqnt_{{\rm T}(p,q)kl} &=& \ampsqnt_{{\rm T}(p,q)lk}\,,
\label{Mklident0}
\\
\mathop{\sum_{k\ne l}}_{k=\nini}^{\nlightB+\nheavy+2}
\ampsqnt_{{\rm T}(p,q)kl} &=& 2C_{\rm T}(\ident_l)\,\ampsqnt_{(p,q)}\,,
\label{Mklident1}
\\
\ampsqnt_{{\rm T}(p,q)kk} &=& - C_{\rm T}(\ident_k)\,\ampsqnt{(p,q)}\,.
\label{Mklident2}
\eeqn
The QCD Casimirs $C_{\QCD}(\ident)$ are defined in the usual way
(see \MadFKSeq{3.29}--\MadFKSeq{3.30}) as the squares of the colour
operators; their QED analogues are the squares of the charge operator, 
and thus:
\beq
C_{\QED}(\ident)=e(\ident)^2\,.
\label{CcasQED}
\eeq
This result is identical to that which one obtains directly from the 
Altarelli-Parisi QED kernels, as it must as a consequence of the commutation 
of the soft and collinear limits (see e.g.~ref.~\cite{Frixione:2011kh}).
Such kernels are reported in appendix~\ref{sec:QEDkern}, together with 
the results for all of the QED-charge factors that are relevant to FKS 
subtraction, and that will appear in sect.~\ref{sec:xsec}.
By construction, a given colour-linked (charge-linked) Born will be non-null
only if both $k$ and $l$ correspond to strongly-interacting (electrically
charged) lines. This implies that, for any given $l$ and ${\rm T}$,
several terms in the sum on the l.h.s.~of eq.~(\ref{Mklident1}) will
be identically equal to zero. Finally, the one-loop matrix elements
that generalise those of \MadFKSeq{3.25} are:
\beqn
\ampsqnl_{(p,q)}(\proc)&=&\frac{1}{2s}\frac{1}{\omega(\Ione)\omega(\Itwo)}
\mathop{\sum_{\rm colour}}_{\rm spin}
\sum_{p_1p_2}\sum_{q_1q_2}\delta_{p_1+p_2,p}\delta_{q_1+q_2,q}
\nonumber
\\&\times&
2\Re\left\{\ampnt_{(p_1,q_1)}(\proc)\ampnl_{(p_2,q_2)}(\proc)^{\star}\right\}.
\label{Moneloop}
\eeqn
Note that the factor $\delta_{p_1p_2}\delta_{q_1q_2}$ that appears in
eqs.~(\ref{Mtreenpo}), (\ref{MQCDlinked}), and~(\ref{MQEDlinked})
vanishes identically here, since otherwise the tree-level and one-loop
amplitudes that enter eq.~(\ref{Moneloop}) would have the same number
of QCD and QED vertices.

\subsection{Short-distance cross sections\label{sec:xsec}}
The short-distance cross sections emerge from applying the FKS
subtraction procedure to all possible IR-divergent configurations,
be them QCD- or QED-induced. The key observation is that such 
configurations are ``squared'' in nature (i.e.~they never lead to
relative ${\cal O}(\gs e)$ corrections). In other words, one either 
inserts a soft-gluon line or a soft-photon line; likewise, a collinear
splitting is of either QCD or QED type. Hence, provided that the
$\Sfun$ functions take into account all IR-divergent sectors regardless
of their origins, the subtraction proceeds as before. In turn, this is
equivalent (see sect.~5.2 of ref.~\cite{Frederix:2009yq}) to constructing 
the $\FKSpairs$ set so that all singularities are represented there.
This can be done by following the procedure advocated in sect.~2.4.1 
of ref.~\cite{Alwall:2014hca}, which has been implemented in that paper
for the QCD case, and extended in the course of the present work to cover 
mixed-coupling scenarios. Basically, such a procedure simply requires one 
to consider, within the adopted theory model, all possible $1\to 2$ branchings 
with at least one outgoing massless particle. In so doing, one treats
all massless particles in a democratic manner (which is consistent
with item~4 in sect.~\ref{sec:gen}) which, among other things, implies
that any two-particle combination in the initial state is allowed.
This is the main reason that motivates the amendments listed at the 
beginning of sect.~\ref{sec:ME}. 

The hadron-level result is obtained by extending \MadFKSeq{C.1} (for 
simplicity, we start by setting the renormalisation and factorisation 
scales equal to each other):
\beq
d\sigma_{\sss H_1H_2}=f^{(H_1)}\star f^{(H_2)}\star\sum_{pq}\left(
d\sigma_{(p,q)}^{(n+1)}+d\bar{\sigma}_{(p,q)}^{(n+1)}+
d\sigma_{(p,q)}^{(n)}\right),
\label{factTH3}
\eeq
where the sums are constrained by eqs.~(\ref{ppqlo}) and~(\ref{ppqnlo}),
which are always understood. The sums over incoming-parton species are
also left implicit. The symbol $\star$ in eq.~(\ref{factTH3}) 
denotes the usual initial-state convolution of short-distance cross 
sections and PDFs, i.e.~for a given initial-state leg associated with 
hadron momentum $P$:
\beq
f\star d\sigma = \int dx\,f(x)\,d\sigma(xP)\,.
\label{convISR}
\eeq
With the caveat on the $\Sfun$ functions mentioned above, 
the $(n+1)$-body contribution has the same form as in 
\MadFKSeq{4.29}, with the matrix elements taken from eq.~(\ref{Mtreenpo}).
The degenerate $(n+1)$-body contributions can be derived from \MadFKSeq{4.41}
and \MadFKSeq{4.42}. We adopt a short-hand notation to re-write the sum of 
those contributions for a given initial-state parton configuration $(a,b)$ 
as follows:
\beq
d\bar{\sigma}_{ab}^{(n+1)}=
\asotpi\,{\cal K}_{da}^{\QCD}\star d\sigma_{db}^{(B,n)}+
\asotpi\,{\cal K}_{db}^{\QCD}\star d\sigma_{ad}^{(B,n)}\,,
\eeq
where
\beqn
{\cal K}_{ab}^{\QCD}&=&\xi P_{ab}^{\QCD<}(1-\xi)
\left[\xidistr{c}\log\frac{s\delta_I}{2\mu^2}+2\lxidistr{c}\right]
\nonumber
\\
&-&\xi P_{ab}^{\QCD\prime<}(1-\xi)\xidistr{c}-K_{ab}^{\QCD}(1-\xi)\,.
\label{Kdef}
\eeqn
The result for the mixed-coupling case then reads as follows:
\beqn
d\bar{\sigma}_{(p,q)ab}^{(n+1)}&=&
\asotpi\,{\cal K}_{da}^{\QCD}\star d\sigma_{(p-1,q)db}^{(B,n)}+
\asotpi\,{\cal K}_{db}^{\QCD}\star d\sigma_{(p-1,q)ad}^{(B,n)}
\nonumber\\
&+&
\aotpi\,{\cal K}_{da}^{\QED}\star d\sigma_{(p,q-1)db}^{(B,n)}+
\aotpi\,{\cal K}_{db}^{\QED}\star d\sigma_{(p,q-1)ad}^{(B,n)}\,,
\label{degnpo}
\eeqn
with
\beq
{\cal K}^{\QED}={\cal K}^{\QCD}\left(P^{\QCD<}\to P^{\QED<},
P^{\QCD\prime<}\to P^{\QED\prime<},
K^{\QCD<}\to K^{\QED<}\right),
\eeq
where we have admitted the possibility of different PDF schemes
relevant to QCD and QED evolution. It is immediate to see that the
first (last) two terms on the r.h.s.~of eq.~(\ref{degnpo}) correspond
to the blue right-to-left (red left-to-right) arrows of fig.~1 of 
ref.~\cite{Alwall:2014hca}. Thus, they will identically vanish when
$(p,q)$ are associated with the pure QED (QCD) NLO term.

The $n$-body cross section in eq.~(\ref{factTH3}) can be read from
\MadFKSeq{4.3}:
\beq
d\sigma_{(p,q)}^{(n)}=d\sigma_{(p,q)}^{(B,n)}+d\sigma_{(p,q)}^{(C,n)}+
d\sigma_{(p,q)}^{(S,n)}+d\sigma_{(p,q)}^{(V,n)}\,.
\label{nbody}
\eeq
We point out that, for a given $(p,q)$, either the Born or the other
three terms on the r.h.s.~vanish, owing to eqs.~(\ref{ppqlo}) 
and~(\ref{ppqnlo}); the final result is correct because $p$ and $q$ 
are summed over in eq.~(\ref{factTH3}). The Born term differs from 
that of \MadFKSeq{4.4} only by notation:
\beq
d\sigma_{(p,q)}^{(B,n)}(\proc)=
\ampsqnt_{(p,q)}(\proc)\,\frac{\JetsB}{\avg(\proc)}\,\phspn\,,
\label{dsignB}
\eeq
The term $d\sigma_{(p,q)}^{(C,n)}$ collects the Born-like remainders 
of the final- and initial-state collinear subtractions. Thus, in full
analogy with eq.~(\ref{degnpo}) and by taking \MadFKSeq{4.5} and
\MadFKSeq{4.6} into account, we have:
\beq
d\sigma_{(p,q)}^{(C,n)}(\proc)=
\asotpi\,{\cal Q}^{\QCD}(\proc)\,d\sigma_{(p-1,q)}^{(B,n)}(\proc) +
\aotpi\,{\cal Q}^{\QED}(\proc)\,d\sigma_{(p,q-1)}^{(B,n)}(\proc)
\eeq
with (${\rm T}=\QCD,\QED$):
\beqn
{\cal Q}^{\Tt}(\proc)\!\!&=&\!\!-\log\frac{\mu^2}{Q^2}\,
\Bigg(\gamma_{\Tt}(\ident_1)+2C_{\Tt}(\ident_1)\log\xicut
+\gamma_{\Tt}(\ident_2)+2C_{\Tt}(\ident_2)\log\xicut\Bigg)
\nonumber \\*\!\!&+&\!\!
\sum_{k=3}^{\nlightB+2}\Bigg[\gamma_{\Tt}^\prime(\ident_k)
-\log\frac{s\deltaO}{2Q^2}\left(\gamma_{\Tt}(\ident_k)
-2C_{\Tt}(\ident_k)\log\frac{2E_k}{\xicut\sqrt{s}}\right)
\nonumber \\*&&\phantom{\sum_{k=3}^{\nlightB+2}}\!\!
+2C_{\Tt}(\ident_k)\left(\log^2\frac{2E_k}{\sqrt{s}}-\log^2\xicut\right)
-2\gamma_{\Tt}(\ident_k)\log\frac{2E_k}{\sqrt{s}}\Bigg].
\label{Qdef}
\eeqn
The colour and charge factors that appear in eq.~(\ref{Qdef})
are reported in appendix~\ref{sec:QEDkern}.
Analogously, from \MadFKSeq{4.12} one obtains the soft term:
\beqn
d\sigma_{(p,q)}^{(S,n)}(\proc)&=&
\sum_{k=\nini}^{\nlightB+\nheavy+2}\,\,\sum_{l=k}^{\nlightB+\nheavy+2}
\eikint_{kl}^{(m_k,m_l)}\frac{\JetsB}{\avg(\proc)}
\label{dsignS}
\\&\times&
\left(\asotpi\ampsqnt_{\QCD(p-1,q)kl}(\proc)+
\aotpi\ampsqnt_{\QED(p,q-1)kl}(\proc)\right)\phspn\,.
\nonumber
\eeqn
Note that the integrated eikonal has the same expression in the QCD- 
and QED-induced terms, being of kinematical origin. Finally, the virtual
contribution is analogous to that of \MadFKSeq{4.14}:
\beq
d\sigma_{(p,q)}^{(V,n)}(\proc)=
\frac{1}{2\pi}\,
\vampsqnl_{(p,q){\sss FIN}}(\proc)\frac{\JetsB}{\avg(\proc)}\phspn\,.
\label{dsignV}
\eeq
The finite part of the virtual contribution that appears in this expression
depends on what is included in the divergent part of the one-loop matrix
elements. We use the same conventions as in ref.~\cite{Frederix:2009yq}, 
and write the latter as in \MadFKSeq{B.1} and \MadFKSeq{B.2}:
\beqn
\ampsqnl_{(p,q)}(\proc)&=&\frac{1}{2\pi}\frac{(4\pi)^\ep}{\Gamma(1-\ep)}
\left(\frac{\mu^2}{Q^2}\right)^\ep {\cal V}_{(p,q)}(\proc)\,,
\label{Virt1}
\\
{\cal V}_{(p,q)}(\proc)&=&
\as {\cal V}_{(p-1,q){\sss DIV}}^{(n,1)\QCD}(\proc)+
\aem {\cal V}_{(p,q-1){\sss DIV}}^{(n,1)\QED}(\proc)+
\vampsqnl_{(p,q){\sss FIN}}(\proc)\,,
\eeqn
where (${\rm T}=\QCD,\QED$):
\beqn
{\cal V}_{(p,q){\sss DIV}}^{(n,1)\Tt}&=&-\Bigg(
\frac{1}{\ep^2}\sum_{k=\nini}^{\nlightB+2}C_{\Tt}(\ident_k)
+\frac{1}{\ep}\sum_{k=\nini}^{\nlightB+2}\gamma_{\Tt}(\ident_k)
\nonumber\\*&&\phantom{aaaaaa}
+\frac{1}{\ep}\sum_{k=\nlightB+3}^{\nlightB+\nheavy+2}C_{\Tt}(\ident_k)
\Bigg)\ampsqnt_{(p,q)}
\nonumber\\*&&
+\frac{1}{\ep}\sum_{k=\nini}^{\nlightB+2}
\sum_{l=k+1}^{\nlightB+\nheavy+2}\log\frac{2k_k\mydot k_l}{Q^2}
\ampsqnt_{\Tt(p,q)kl}
\nonumber\\*&&
+\frac{1}{2\ep}\sum_{k=\nlightB+3}^{\nlightB+\nheavy+1}
\sum_{l=k+1}^{\nlightB+\nheavy+2}
\frac{1}{\velkl}\log\frac{1+\velkl}{1-\velkl}
\ampsqnt_{\Tt(p,q)kl}
\nonumber\\*&&
-\frac{1}{2\ep}\sum_{k=\nlightB+3}^{\nlightB+\nheavy+2}
\log\frac{m_k^2}{Q^2}
\sum_{l=\nini}^{\nlightB+2}\ampsqnt_{\Tt(p,q)kl}\,.
\label{Virt2}
\eeqn
Finally, the separate dependence on the renormalisation and factorisation 
scales can be re-instated by following the procedure outlined in
appendix C of ref.~\cite{Frederix:2009yq}. We report here in
appendix~\ref{sec:RGE} its extension to the mixed-coupling case.
We remind the reader that \aNLOs\ allows the evaluation of the 
hard-scale and PDF uncertainties at no extra computational costs,
by following the procedure introduced in ref.~\cite{Frederix:2011ss}.
This feature is also supported in mixed-coupling applications.
Note, however, that the scale dependence of $\aem$ is ignored,
and that, in keeping with what is done in current PDF fits, the
QCD- and QED-factorisation scales are set equal to each other.

\section{FKS subtraction with fragmentation\label{sec:Frag}}
As was discussed in sect.~\ref{sec:gen}, the computation of
sufficiently subleading NLO contributions might require that some
physical objects be defined through fragmentation functions (FFs
henceforth). This motivates the extension of the FKS subtraction 
procedure to cover the cases where FFs would also be present -- in 
other words, measurable quantities may be obtained by means of FFs, 
or with a jet-finding algorithm, or as stable-particle taggable massive 
objects (e.g.~a Higgs boson); any combination of the objects thus 
constructed is allowed. This must be done by respecting the chief FKS 
subtraction characteristics, namely that it be universal, and observable- 
and process-independent.

The aim of this section is that of achieving such an extension
of the FKS method. To the best of our knowledge, this is the
first time that a general treatment of collinear fragmentation has
been included in a universal subtraction formalism.

In order to proceed, we observe that in NLO computations at most two
final-state partons are not well separated. Since hard and isolated
partons can be fragmented in a trivial manner, this implies that the
only non-trivial case we shall have to deal with is that which features a 
single FF. Furthermore, sect.~\ref{sec:Xsec} shows that the FKS subtraction 
in a mixed-coupling scenario can readily be obtained from its QCD counterpart,
and from the knowledge of the relevant kernels and colour or charge factors.
Therefore, in order to simplify the notation of the formal proof that 
follows, we limit ourselves to presenting explicitly only the QCD case; the 
mixed-coupling results can then be readily obtained by following the same 
procedure as in sect.~\ref{sec:Xsec}. For the same reason, we assume that
all final-state particles are massless.

Following what has been done in the original paper~\cite{Frixione:1995ms},
such a proof is essentially that of the cancellation of the IR divergences
that arise in the intermediate steps of the calculation of a generic
final-state observable. Its by-products are the main results we are
interested in, namely the IR-finite short-distance cross sections 
that can be numerically integrated.

\subsection{Fragmentation in perturbative QCD\label{sec:fragm}}
The case we are considering, that of a single FF, is by definition
equivalent to the computation of a single-hadron cross section in QCD.
That is typically written as follows (see e.g.~refs.~\cite{Ellis:1979sj,
Aversa:1988vb}):
\beq
K_H^0\frac{d\sigma_H}{d^3 K_H}=\sum_p\int\frac{d\zeta}{\zeta^2}\,
D^{(a_p)}_H(\zeta)\,k_p^0\left.\frac{d\sigma}{d^3 k_p}
\right|_{\vec{k}_p=\vec{K}_H/\zeta}\,.
\label{FFxsec}
\eeq
Here, $d\sigma$ is the short-distance partonic cross section (for a given
partonic process), already convoluted with the PDFs and integrated over the 
degrees of freedom not explicitly indicated in eq.~(\ref{FFxsec}); 
$D^{(a_p)}_H(\zeta)$ is the FF of parton $p$ (with momentum $k_p$ and 
flavour $a_p$) which fragments into hadron $H$ (with momentum $K_H$); 
the latter carries a fraction $\zeta$ of the three-momentum of the former 
(in leading-twist QCD factorisation, fragmentation is strictly 
collinear); the index $p$ runs over all final-state partons. 
Equation~(\ref{FFxsec}) is not suited to the definition of a parton-level
generator through the FKS (or any equivalent) procedure; to that end,
a fully differential form is required. Such a form is:
\beq
d\sigma_H\equiv\sum_p d\sigma_{p\to H}=
\sum_p D^{(a_p)}_H(\zeta)d\sigma d\zeta\,,
\label{FFxsecd}
\eeq
where in the r.h.s.~one understands that the momenta of hadron $H$ and 
parton $p$ are related by \mbox{$\vec{K}_H=\zeta\vec{k}_p$}. 

Equation~(\ref{FFxsecd}) has an unusual form, because the variables
are seemingly separated (i.e.~at variance with eq.~(\ref{FFxsec}) there is no 
dependence on $\zeta$ in the partonic cross section $d\sigma$ on the r.h.s.).
However, this is purely 
an artifact that stems from the fully differential nature of 
eq.~(\ref{FFxsecd}). One must keep in mind that formulae of 
this kind (including those of FKS) are never meaningful if they do not 
understand the inclusion of an IR-safe observable. In other words,
eq.~(\ref{FFxsecd}) formally expresses the computation of the 
expectation value:
\beq
\langle O\rangle=
\sum_p\int d\zeta d\sigma(k_p)\,D^{(a_p)}_H(\zeta)\,O(\zeta k_p)\,,
\label{Oavg}
\eeq
for any observable $O(K_H)$ constructed with the momentum of the
hadron emerging from the fragmentation process (the dependence on any
other momentum, if present, is understood here). Eq.~(\ref{Oavg})
manifestly shows that the variables are not separated due to the
presence of the observable.

In order to see that eqs.~(\ref{FFxsec}) and~(\ref{FFxsecd}) are mutually
consistent, we demonstrate that the former equation can be derived from the 
latter one. By writing the partonic cross section in terms of a matrix 
element and the phase space:
\beq
d\sigma=N\ampsq^{(n)}\,\delta^4\left(Q-\sum_{i=1}^n k_i\right)
\prod_{i=1}^n\frac{d^3k_i}{k_i^0}\,,
\eeq
with $N$ a normalisation factor, one obtains:
\beq
\frac{d\sigma_{p\to H}}{d^3 K_H}=D^{(a_p)}_H(\zeta)
N\ampsq^{(n)}\,\delta^4\left(Q-\sum_{i=1}^n k_i\right)
\prod_{i=1}^n\frac{d^3k_i}{k_i^0}\,
\delta\!\left(\vec{K}_H-\zeta\vec{k}_p\right) d\zeta\,.
\eeq
With
\beq
\delta\!\left(\vec{K}_H-\zeta\vec{k}_p\right)=
\frac{1}{\zeta^3}
\delta\!\left(\vec{k}_p-\vec{K}_H/\zeta\right)\,,
\eeq
and keeping in mind that parton $p$ is massless ($k_p^0=|\vec{k_p}|$)
we have:
\beq
\frac{d\sigma_{p\to H}}{d^3 K_H}=D^{(a_p)}_H(\zeta)
N\ampsq^{(n)}\,\delta^4\left(Q-\sum_{\stackrel{i\ne p}{i=1}}^n k_i
-K_H/\zeta\right)\prod_{\stackrel{i\ne p}{i=1}}^n\frac{d^3k_i}{k_i^0}
\frac{1}{\abs{\vec{K}_H}/\zeta}\,\frac{d\zeta}{\zeta^3}\,,
\eeq
whence:
\beq
\abs{\vec{K}_H}\frac{d\sigma_{p\to H}}{d^3 K_H}=
\frac{d\zeta}{\zeta^2}D^{(a_p)}_H(\zeta)
\left.k_p^0\frac{d\sigma}{d^3 k_p}\right|_{\vec{k}_p=\vec{K}_H/\zeta,
k_p^0=\abs{\vec{K}_H}/\zeta}\,,
\label{FFxsec2}
\eeq
since by construction
\beq
\left.k_p^0\frac{d\sigma}{d^3 k_p}\right|_{k_p=\bar{k}_p}=
N\ampsq^{(n)}\,\delta^4\left(Q-\sum_{\stackrel{i\ne p}{i=1}}^n k_i
-\bar{k}_p\right)\prod_{\stackrel{i\ne p}{i=1}}^n\frac{d^3k_i}{k_i^0}\,,
\eeq
for any $\bar{k}_p$. It is apparent that, after summing over $p$,
eq.~(\ref{FFxsec2}) is identical to eq.~(\ref{FFxsec}), if one assumes that 
hadron $H$ is massless, i.e.~\mbox{$K_H^0=|\vec{K}_H|$}. This is of course 
not the case for any physical hadron, but it appears to be a reasonable
approximation in the present context; note, also, that by forcing the
fragmented parton to have a non-zero mass one is obliged to perform
an arbitrary kinematic reshuffling. Therefore, in what follows hadrons
will always be treated as massless.

Finally, note that by inserting the identity
\mbox{$d^3 K_H\,\delta(\vec{K}_H-\zeta \vec{k}_p)$} into the r.h.s.~of
eq.~(\ref{Oavg}), and by using eq.~(\ref{FFxsec}), one readily obtains:
\beq
\langle O\rangle=\int d^3 K_H\,\frac{d\sigma_H}{d^3 K_H}\,O(K_H)\,.
\label{Oavg2}
\eeq
The rightmost side of eq.~(\ref{Oavg2}) is by definition the expectation
value of $O$, which shows again the correctness of eq.~(\ref{FFxsecd}),
and the way it must be understood.

In summary, we shall use eq.~(\ref{FFxsecd}) as our master formula
for the fully-differential implementation of single-FF cross sections.
We remark that, at variance with eq.~(\ref{FFxsec}) whose l.h.s.~is a
function of a given four-momentum $K_H$, eq.~(\ref{FFxsecd}) is naturally
associated with $n$ different $n$-body final-state kinematic configurations,
each of which corresponds to a different parton $p$ undergoing fragmentation.

\subsection{FKS: notation\label{sec:not}}
While the \aNLOs\ implementation of the FKS subtraction is done
according to ref.~\cite{Frederix:2009yq}, the notation used in that
paper is not ideal to carry out formal manipulations such as those
required to prove the cancellation of IR singularities, and certain
aspects of the original FKS paper, ref.~\cite{Frixione:1995ms}, have
to be preferred. In particular, ref.~\cite{Frixione:1995ms} employs
partonic cross sections that are highly symmetric in the kinematic and
flavour spaces, which is a key feature we want to use here.
Conversely, the $\Sfun$ functions of ref.~\cite{Frixione:1995ms} are
cumbersome -- they include the definition of the observable jets and
use it to partition the phase space, which was shown later in
ref.~\cite{Frixione:1997np} to be unnecessary; thus, we shall use here
the $\Sfun$ functions as defined in ref.~\cite{Frederix:2009yq}.  The
merging of the two notations is straightforward and we shall not
explicitly spell it out here.  We shall denote by \FKSeq{n.m}
eq.~(n.m) of ref.~\cite{Frixione:1995ms}.  We remind the reader (see
sect.~\ref{sec:Xsec}) that by \MadFKSeq{x.y} we denote eq.~(x.y) of
ref.~\cite{Frederix:2009yq}.

When one of the final-state partons is fragmented into a hadron, the
short-distance cross sections have to be modified as discussed in
sect.~\ref{sec:fragm}. Hence, we write the unsubtracted $(n-1)$-jet
$(m+2)$-parton $L$-loop partonic cross section with one parton
fragmented into a hadron as follows:
\begin{multline}
d\sigma^{(m,L;p)}\left(\set{a_l}{1}{m+2};\set{k_l}{1}{m+2}\right)=
\frac{1}{m!}\ampsq^{(m,L)}\left(\set{a_l}{1}{m+2};\set{k_l}{1}{m+2}\right)\\
\times
F^{(p)}_H(\zeta k_p) J_m^{n-1}\left(\setS{k_l}{3}{m+2}{p};H\right)
d\phi_m\left(k_1,k_2\to \set{k_l}{3}{m+2}\right).
\label{mFxsec}
\end{multline}
Here, $F^{(p)}_H$ denotes a purely kinematical factor, responsible
for enforcing final-state cuts on hadron $H$. While there will be
no need to specify the functional form of this factor, its key property 
is that it must vanish when the transverse momentum of hadron $H$ is below
a given threshold. This implies that:
\beq
\lim_{k^0\to 0}F^{(p)}_H(k)=
\lim_{k\parallel k_1}F^{(p)}_H(k)=
\lim_{k\parallel k_2}F^{(p)}_H(k)=0\,.
\label{SClimF}
\eeq
The factor $J$ reconstructs $n-1$ jets (the momentum of parton $p$
being excluded from the reconstruction); as the notation suggests, an
isolation condition (of the jets w.r.t.~hadron $H$) is enforced as
well. These two features are not mandatory, and are given here merely
as examples; in particular, it is possible to include hadron $H$ in
the jet-finding procedure, and to obtain $n$ final-state jets. The
hadron-level contribution induced by eq.~(\ref{mFxsec}) is:
\beq
d\sigma_{had}^{(m,L)}=\sum_{\set{a_l}{1}{m+2}}\sum_{p=3}^{m+2}
f_{a_1}^{(H_1)}\star f_{a_2}^{(H_2)}\star 
d\sigma^{(m,L;p)}\left(\set{a_l}{1}{m+2}\right)\star D^{(a_p)}_H\,,
\label{HmFxsec}
\eeq
where the sum over flavours runs over all possible QCD partons
(i.e.~the gluon and all the light quarks and antiquarks). We stress that
the flavour of the fragmenting parton $p$ is also summed over; this is
obviously right from a physics viewpoint, and amends the notation used
in sect.~\ref{sec:fragm}, where such a sum has been neglected since
that case was relevant to a single partonic process. Loosely speaking,
this sum restores the complete symmetry over final-state partons,
which is originally broken by the fragmentation of one of them,
and is thus ultimately responsible for the fact that the symmetry
factor in eq.~(\ref{mFxsec}) is equal to $1/m!$; more details on
this point are given in appendix~\ref{sec:symm}. Since no confusion
is possible, in eq.~(\ref{HmFxsec}) we have employed the symbol $\star$
to denote the convolution of the short-distance cross section with 
both the PDFs and the FFs. The former convolution has already
been introduced in eq.~(\ref{convISR}), while the latter one is 
defined as follows:
\beq
d\sigma\star D = \int d\zeta \,d\sigma(k)\,D(\zeta)\,,
\label{convFSR}
\eeq
which has to be interpreted as explained in eq.~(\ref{Oavg}).
Note that by construction, the condition of hadron-level 
four-momentum conservation is:
\beq
k_1+k_2=\sum_{i\ne p}k_i+K_H+(1-\zeta)k_p\,,
\eeq
where the last term on the r.h.s.~of this equation is the momentum
associated with the fragmentation remnants. As is customary, we shall 
neglect to indicate explicitly any dependence on this momentum.

\subsection{Final-state collinear counterterms\label{sec:cnts}}
By means of the FKS procedure one arrives at subtracted partonic cross 
sections. In the case dealt with in ref.~\cite{Frixione:1995ms},
the combination of all subtracted cross sections is still IR divergent,
owing to the fact that the complete integration over initial-state
degrees of freedom cannot be performed at the short-distance level
(the incoming-parton momenta are constrained by their hadronic counterparts).
In order to obtain IR-finite cross sections one must thus add the so-called
(initial-state) collinear counterterms, whose forms can be derived from
first principles following the procedure outlined at the beginning of
sect.~2.1 of ref.~\cite{Frixione:1995ms} (see \FKSeq{2.2}--\FKSeq{2.5}).

In the fragmentation case one of the final-state momenta is constrained
as well, and therefore final-state collinear counterterms (one for 
each fragmenting parton) must also be envisaged. To work out their forms,
we start from eq.~(\ref{FFxsecd}) and proceed by analogy to sect.~2.1 
of ref.~\cite{Frixione:1995ms}. This implies that we need to replace
hadron $H$ with parton $a_q$, and interpret the cross section on the
r.h.s.~of that equation as a subtracted one. As shown in eq.~(\ref{HmFxsec}),
we also need to consider explicitly the sum over the flavour of the
fragmenting parton. Thus, the $p$-parton contribution to eq.~(\ref{FFxsecd}) 
becomes:
\beq
d\sigma_{a_q}(k_q)=
\sum_{a_p} D^{(a_p)}_{a_q}(\zeta)d\hat{\sigma}_{a_p}(k_q/\zeta) d\zeta\,,
\label{xscnt}
\eeq
where a shortened notation is used. Now we expand the cross sections in
eq.~(\ref{xscnt}) by introducing their LO and NLO contributions,
as is done in \FKSeq{2.3}:
\beq
d\sigma_a=d\sigma_a^{(0)}+d\sigma_a^{(1)}\,,\;\;\;\;
d\hat{\sigma}_b=d\hat{\sigma}_b^{(0)}+d\hat{\sigma}_b^{(1)}\,,\;\;\;\;
\label{exps}
\eeq
and we use the analogue of \FKSeq{2.2}:
\beq
D^{(b)}_a(\zeta)=\delta_{ab}\delta(1-\zeta)-\frac{\as}{2\pi}
\left(\frac{1}{\epb}P_{ab}(\zeta)-K_{ab}^{\rm FF}(\zeta)\right)+
{\cal O}(\as^2)\,.
\label{partFF}
\eeq
By replacing eqs.~(\ref{exps}) and~(\ref{partFF}) into eq.~(\ref{xscnt})
one obtains:
\beqn
d\hat{\sigma}_a^{(0)}(k)&=&d\sigma_a^{(0)}(k)\,,
\label{cnt0}
\\
d\hat{\sigma}_a^{(1)}(k)&=&d\sigma_a^{(1)}(k)+
\frac{\as}{2\pi}\sum_b
\left(\frac{1}{\epb}P_{ab}(\zeta)-K_{ab}^{\rm FF}(\zeta)\right)
d\sigma_b^{(0)}(k/\zeta) d\zeta\,.
\label{cnt1}
\eeqn
Eqs.~(\ref{cnt0}) and~(\ref{cnt1}) are the analogues of \FKSeq{2.4}
and \FKSeq{2.5}, respectively. Note, however, that the role of the 
indices of the Altarelli-Parisi kernels is opposite in eq.~(\ref{cnt1}) 
w.r.t.~that in \FKSeq{2.5}, consistently with the fact that the former
is a timelike branching, while the latter is a spacelike one. The 
second term on the r.h.s.~of eq.~(\ref{cnt1}) defines the final-state
collinear counterterm we were seeking; as was said before, there will
be one such counterterm per fragmenting parton.

\subsection{Born and virtual contributions\label{sec:nbody}}
As in the original FKS formulation, the Born and virtual cross
sections follow trivially from the master equation, eq.~(\ref{mFxsec}) -- 
we simply need to replace there the values $m=n$ and $L=0\,$:
\beq
d\sigma^{(n,0;p)}=\frac{1}{n!}\ampsq^{(n,0)}
F^{(p)}_H J_n^{n-1} d\phi_n\,,
\label{BornFF}
\eeq
to obtain the Born cross section, and the values $m=n$ and $L=1\,$:
\beq
d\sigma^{(n,1;p)}=\frac{1}{n!}\ampsq^{(n,1)}F^{(p)}_H
J_n^{n-1}d\phi_n\,,
\label{virtFF}
\eeq
to obtain the virtual cross section. Equations~(\ref{BornFF})
and~(\ref{virtFF}) are, as expected, extremely similar to their 
standard-FKS counterparts -- \FKSeq{2.6} and \FKSeq{2.15} respectively,
the only formal differences being due to the fragmenting parton and 
its associated factor. In both of these equations \mbox{$3\le p\le n+2$},
owing to eq.~(\ref{HmFxsec}).

Equation~(\ref{virtFF}) implies that the IR divergences of virtual origin in 
the fragmentation case are given by \FKSeq{3.2}, times the fragmenting-parton 
factors. Therefore, the cancellation of the singularities is proven if one 
shows that all of the contributions of real-emission origin have the same 
forms as those given in ref.~\cite{Frixione:1995ms}, times the fragmentation
factors. It is obvious that this is a sufficient but not necessary condition
(since the {\em individual} real-emission contributions might have 
different forms w.r.t.~the standard-FKS ones, with only their sum being equal
to the sum of the latter, up to the fragmentation factors); but it will
turn out that by and large this is indeed the case, which is a 
significant simplification.

\subsection{Real-emission contribution\label{sec:reals}}
We shall follow here as closely as possible the procedure outlined in
sect.~4 of ref.~\cite{Frixione:1995ms}. We split the real-emission
contribution into a soft and a non-soft part, as is done in \FKSeq{4.11} 
and \FKSeq{4.12}. For the latter, the initial- and final-state collinear 
contributions are dealt with separately, by using a decomposition identical 
to that which leads to \FKSeq{4.14} and \FKSeq{4.15}. By exploiting
eq.~(\ref{SClimF}), the complete symmetrisation over final states, and 
the invariance over parton relabeling (see appendix~\ref{sec:symm}), the 
results of ref.~\cite{Frixione:1995ms} can straightforwardly be extended 
to the fragmentation case for the soft and the initial-state collinear
contribution. In particular, the analogue of \FKSeq{4.27} reads:
\beqn
d\sigma^{(n+1,0;p)}_s&=&\asotpi\frac{(4\pi)^\ep}{\Gamma(1-\ep)}\muoQep
\Bigg[\frac{1}{\ep^2}\sum_{k=1}^{n+2}C(a_k)
+\frac{2}{\ep}\sum_{k=3}^{n+2}C(a_k)\log\frac{2E_k}{\xic\sqrt{S}}
\nonumber
\\*&&\phantom{aaaaaa}
-\frac{2}{\ep}\left(C(a_1)+C(a_2)\right)\log\xic\Bigg]
d\sigma^{(n,0;p)}
\nonumber
\\*&-&
\asotpi\frac{(4\pi)^\ep}{\Gamma(1-\ep)}\muoQep
\frac{1}{2\ep}\sum_{k,l=1}^{n+2}\log\frac{k_k\mydot k_l}{Q^2}\,
d\sigma_{kl}^{(n,0;p)}
\nonumber
\\*&+&
\asotpi\half\sum_{k,l=1}^{n+2}{\cal I}_{kl}^{(reg)}d\sigma_{kl}^{(n,0;p)}\,,
\label{realsF}
\eeqn
with \mbox{$3\le p\le n+2$}. In eq.~(\ref{realsF}) we have defined the 
colour-linked Borns, in analogy with \FKSeq{4.26} and consistently with 
eq.~(\ref{BornFF}):
\beq
d\sigma_{kl}^{(n,0;p)}=\frac{1}{n!}\ampsq_{kl}^{(n,0)}
F^{(p)}_H J_n^{n-1} d\phi_n\,.
\label{realsF2}
\eeq
For the initial-state collinear contributions we find that the analogues 
of \FKSeq{4.37}, \FKSeq{4.54}, and \FKSeq{4.55} are, respectively:
\beqn
d\sigma^{(n+1,0;p)}_{in,i,f}&=&\half
\pdistr{\xii}{c}\left[\pdistr{1-\yi}{\delta_I}+\pdistr{1+\yi}{\delta_I}\right]
\frac{\left(1-\yi^2\right)\xii^2}{(n+1)!}\ampsq^{(n+1,0)}
\nonumber\\*&\times&
\left(\Sfun_{i1}+\Sfun_{i2}\right)F^{(p)}_H J_{n+1}^{n-1}\,
\vol_{(\ep=0)}\,d\xii\,d\yi\,d\Omega_i^{(2)}\,d\phi_{(\ep=0)}\,,
\label{realinfin}
\eeqn
\beqn
~d\hat{\sigma}^{(n+1,0;p)}_{in,+}&=&
\asotpi\left[\frac{(4\pi)^\ep}{\Gamma(1-\ep)}\muoQep\frac{1}{\ep}
-\log\muoQ\right]
\Big(\gamma(a_1)+2C(a_1)\log\xic\Big)d\sigma^{(n,0;p)}
\nonumber\\*&+&
\asotpi\sum_d
\Bigg\{\xi P_{da_1}^<(1-\xi,0)
\left[\pdistr{\xi}{c}\log\frac{S\delta_I}{2\mu^2}+
2\lpdistr{\xi}{c}\right]
\label{dsiginpl}
\\*&&\phantom{\asotpi}
-\xi P_{da_1}^{\prime<}(1-\xi,0)\pdistr{\xi}{c}
-K_{da_1}(1-\xi)\Bigg\}\,
d\sigma^{(n,0;p)}\left(d,(1-\xi)k_1\right)\,d\xi\,,
\nonumber
\\
~d\hat{\sigma}^{(n+1,0;p)}_{in,-}&=&d\hat{\sigma}^{(n+1,0;p)}_{in,+}
\left(a_1\longrightarrow a_2,k_1\longrightarrow k_2\right)\,.
\label{dsiginmn}
\eeqn
The quantity in eq.~(\ref{realinfin}) is a genuine $(n+1)$-body term,
and thus \mbox{$3\le p\le n+3$}. On the r.h.s.~of that equation 
factors appear that stem directly from the FKS representation of the 
$(n+1)$-body phase-space in $4-2\epsilon$ dimensions (see \FKSeq{4.4}), 
namely:
\beq
d\phi(i)=\vol\,\xii^{1-2\ep} d\xii\,d\Omega_i^{(3-2\ep)}\,,
\;\;\;\;\;\;\;\;\;\;\;\;
d\phi_{n+1}=d\phi\,d\phi(i)\,,
\label{dphiidef}
\eeq
with
\beqn
d\Omega_i^{(3-2\ep)}&=&\left(1-\yi^2\right)^{-\ep}d\yi\,
d\Omega_i^{(2-2\ep)}\,,
\label{Omegat}
\\
\vol&=&\frac{1}{2(2\pi)^{3-2\ep}}\left(\frac{\sqrt{S}}{2}\right)^{2-2\ep}\,.
\label{voldef}
\eeqn
Thus, in eq.~(\ref{realinfin}) $d\phi_{(\ep=0)}$ is such that the 
measure on the r.h.s.~times the volume factor $\vol_{(\ep=0)}$
is equal to \mbox{$d\phi_{n+1}/\xii$} in four dimensions. 
The initial-state collinear remainders of eqs.~(\ref{dsiginpl}) 
and~(\ref{dsiginmn}) are (quasi-)$n$-body terms, and therefore 
\mbox{$3\le p\le n+2$}.

As was anticipated in sect.~\ref{sec:nbody}, eqs.~(\ref{realsF}),
(\ref{dsiginpl}), and~(\ref{dsiginmn}) have the same IR-singular
structure as their counterparts in ref.~\cite{Frixione:1995ms} 
(bar for the fragmentation factors). By showing that this is the
case also for the contributions of final-state collinear origin,
we shall achieve the sought proof of the cancellation of the IR 
divergences in the presence of fragmentation. 

In order to do this, we start from the analogue of \FKSeq{4.15},
which reads as follows:
\beqn
d\sigma^{(n+1,0;p)}_{out,ij}&=&{\cal D}_i
\frac{\xii^2}{(n+1)!}\ampsq^{(n+1,0)}\Sfunij
\nonumber
\\*&\times&
F^{(p)}_H J_{n+1}^{n-1}\,
\vol\,\left(1-\yi^2\right)^{-\ep}d\xii\,d\yi\,d\Omega_i^{(2-2\ep)}\,d\phi\,,
\label{realoutij}
\eeqn
where 
\beq
{\cal D}_i=\pdistr{\xii}{c}-2\ep\lpdistr{\xii}{c}\,.
\label{Ddef}
\eeq
The first steps are the same as in ref.~\cite{Frixione:1995ms}, and in 
particular one can follow them up to \FKSeq{4.63} in order to achieve the 
splitting into a singular and a non-singular part that appears in \FKSeq{4.64}. 
The finite part, \FKSeq{4.65}, can directly be used also in the fragmentation 
case, through the usual change of notation and the multiplication by the
fragmentation factor. It reads:
\beqn
d\sigma^{(n+1,0;p)}_{out,ij,f}&=&
\pdistr{\xii}{c}\pdistr{1-\yj}{\delta_O}\!\!
\frac{\left(1-\yj\right)\xii^2\xij}{(n+1)!}\ampsq^{(n+1,0)}
\nonumber\\*&\times&
\Sfunij F^{(p)}_H J_{n+1}^{n-1}\,
\vol_{(\ep=0)}^2\,d\xii\,d\yi\,d\Omega_i^{(2)}\,
d\xij\,d\yj\,d\Omega_j^{(2)}\,d\tilde{\phi}_{(\ep=0)}\,.
\label{realoutfin}
\eeqn
This is a genuine $(n+1)$-body term, and thus one has \mbox{$3\le p\le n+3$}. 
By construction, $d\tilde{\phi}_{(\ep=0)}$ is such that the measure 
on the r.h.s.~times the volume factor $\vol_{(\ep=0)}^2$
is equal to \mbox{$d\phi_{n+1}/(\xii\xij)$} in four dimensions. 

As far as the singular part is concerned, it requires a more careful
treatment than the finite one, which constitutes the only non-trivial 
bit of the proof given in this section. To this end, we need to distinguish 
two cases 
in eq.~(\ref{realoutij}):
\begin{itemize}
\item[A.] $p\ne i$ and $p\ne j$;
\item[B.] either $p=i$ or $p=j$.
\end{itemize}
In words, either the fragmenting parton is different from both partons
associated with the $\Sfunij$ functions (case A.), or it is equal to either 
of them (case B.).
Let us first simply count the number of contributions, and 
make sure that cases A.~and B.~exhaust all possibilities.
The correct counting emerges by considering
both the sum over all possible $\Sfun$ functions, and that over the fragmenting
partons. Thus, in the present case:
\beq
S=\sum_{p=3}^{n+3}\sum_{i=3}^{n+3}\sum_{j=3}^{n+3}\left(1-\delta_{ij}\right)=
(n+1)(n+1)n\,,
\label{sumout}
\eeq
where the $\delta_{ij}$ term enforces the condition $i\ne j$ that stems
from the $\Sfun$ functions. By direct computation:
\beq
\left(1-\delta_{ij}\right)
\left(1-\delta_{ip}\right)\left(1-\delta_{jp}\right)=
\left(1-\delta_{ij}\right)
\left(1-\delta_{ip}-\delta_{jp}\right)\,,
\eeq
given that a term $\delta_{ip}\delta_{jp}$ must vanish if $i\ne j$. Thus:
\beq
1-\delta_{ij}=
\left(1-\delta_{ij}\right)
\big[
\left(1-\delta_{ip}\right)\left(1-\delta_{jp}\right)
+\delta_{ip}+\delta_{jp}
\big]\,.
\label{tmp5}
\eeq
The first term inside the square brackets on the r.h.s.~of eq.~(\ref{tmp5})
corresponds to case A., while the sum of the other two terms corresponds
to case B. The number of contributions they are associated with,
defined in analogy to eq.~(\ref{sumout}), is:
\beqn
A&=&\sum_{p=3}^{n+3}\sum_{i=3}^{n+3}\sum_{j=3}^{n+3}
\left(1-\delta_{ij}\right)\left(1-\delta_{ip}\right)
\left(1-\delta_{jp}\right)=
(n+1)n(n-1)\,,
\label{sumoutA}
\\
B&=&\sum_{p=3}^{n+3}\sum_{i=3}^{n+3}\sum_{j=3}^{n+3}
\left(1-\delta_{ij}\right)\left(\delta_{ip}+\delta_{jp}\right)=
2(n+1)n\,.
\label{sumoutB}
\eeqn
The fact that $S=A+B$ follows from eq.~(\ref{tmp5}), and can be checked
directly from the r.h.s.'s of eqs.~(\ref{sumout}), (\ref{sumoutA}), 
and~(\ref{sumoutB}).

We then go back to the treatment of final-state collinear
singularities in standard FKS (sect.~4.4 of ref.~\cite{Frixione:1995ms}).
The idea is that partons $i$ and $j$ (associated with $\Sfunij$)
are combined into what is called, in ref.~\cite{Frixione:1995ms}, 
parton $7$ (here it will be $n+4$) in the intermediate steps of the 
computation. Thus, one arrives at \FKSeq{4.83} where, similarly to
the cases of soft and initial-state collinear singularities, there
is no dependence upon $i$ and $j$ in the reduced cross sections.
Therefore, all $\Sfunij$ contributions are identical after relabeling.
The number of such contributions is:
\beq
\sum_{i=3}^{n+3}\sum_{j=3}^{n+3}\left(1-\delta_{ij}\right)=
(n+1)n\,.
\label{tmp6}
\eeq
When combined with the real-emission symmetry factor, this gives:
\beq
\frac{1}{(n+1)!}\,(n+1)n = \frac{n}{n!}\,.
\label{tmp7}
\eeq
The factor $1/n!$ is then included in the Born-level cross sections;
the factor $n$ at the numerator is cancelled by turning the identical
contributions into a fully-symmetric sum over final-state partons
implicit in \FKSeq{4.86}. That sum is what appears as sum over $j$ 
in \FKSeq{4.88}.

Now consider case A. Owing to its definition (the Kronecker delta's
in eq.~(\ref{sumoutA})), which implies
that no dependence upon $i$ and $j$ enters the fragmentation part,
the manipulations of sect.~4.4 of ref.~\cite{Frixione:1995ms} do
apply to the present case as well, up to \FKSeq{4.83}. At this point,
however, the fragmentation factor interferes with the counting of
contributions and the relabeling. Eq.~(\ref{tmp6}) must
be replaced by eq.~(\ref{sumoutA}) and therefore, rather than
eq.~(\ref{tmp7}), one has:
\beq
\frac{1}{(n+1)!}\,(n+1)n(n-1) = \frac{n(n-1)}{n!}\,.
\label{tmp8}
\eeq
After including the factor $1/n!$ into the Born cross sections,
one is therefore left with $n(n-1)$ contributions, that must account
for the sum over the fragmenting partons and the sum over $j$ on
the r.h.s.~of \FKSeq{4.88}. The fact that this is really so stems
from observing that the definition of case A.~implies that, after
relabeling, the index of the fragmenting parton can take any value 
different from that of the parton associated with the colour factor
${\cal Z}(a)$ (see appendix~A of ref.~\cite{Frixione:1995ms}), which
is $j$ in \FKSeq{4.88}. Furthermore, we must take into account that, 
after relabeling, the index $p$ of the fragmenting parton is such that
\mbox{$3\le p\le n+2$}. These two facts amount to saying that the
sum over fragmentation contributions and a fully symmetrised final
state must be:
\beq
\sum_{p=3}^{n+2}\sum_{j=3}^{n+2}\left(1-\delta_{jp}\right)=
n(n-1)\,,
\eeq
which is exactly the numerator of eq.~(\ref{tmp8}). This proves that
in case A.~the FKS result of \FKSeq{4.88} still holds (with the usual 
changes of notation), with an extra factor of fragmentation origin:
\beq
\left(1-\delta_{jp}\right)F^{(p)}_H
\label{idF1}
\eeq
that must be inserted under the two sums over $j$ on the r.h.s.~of that
equation, for any given fragmenting parton $p$. Eq.~(\ref{idF1}) explicitly
enforces the conditions that all partons except $j$ may fragment.

Let us now turn to case B. To this end, consider the singular part
that emerges from eq.~(\ref{realoutij}) after the decomposition analogous
to that of \FKSeq{4.64}; the quantity of interest will be (see eq.~(\ref{tmp5})
and the comment that follows it):
\beq
{\cal Y}_{ij}=\sum_{a_ia_j}\left(
d\sigma^{(n+1,0;i)}_{out,+,ij}+d\sigma^{(n+1,0;j)}_{out,+,ij}
\right).
\label{Yijdef}
\eeq
This can be further written as (see \FKSeq{4.66}--\FKSeq{4.69}):
\beq
{\cal Y}_{ij}\equiv {\cal Y}_i+{\cal Y}_j=-\frac{(2\delta_O)^{-\ep}}{\ep}\,
\delta(1-\yj)d\yj\, {\cal D}_i\left({\cal A}_i+{\cal A}_j\right)\, d\mu\,,
\label{Yij}
\eeq
where ${\cal D}_i$ is given in eq.~(\ref{Ddef}) and
\beqn
{\cal A}_i+{\cal A}_j&=&
\sum_{a_ia_j}\frac{(1-\yj)\xii^2}{(n+1)!}\ampsq^{(n+1,0)}\Sfunij
\left(F^{(i)}_H+F^{(j)}_H\right) J_{n+1}^{n-1}\,,
\label{Adef}
\\
d\mu&=&\vol^2\,\xij^{1-2\ep}\left(1-\yi^2\right)^{-\ep}d\xii\,d\xij\,
d\yi\,d\Omega_i^{(2-2\ep)}\,d\Omega_j^{(2-2\ep)}\,d\tilde{\phi}\,.
\label{dmudef}
\eeqn
As the notation suggests, in eq.~(\ref{Adef}) the terms ${\cal A}_i$
and ${\cal A}_j$ are proportional to $F^{(i)}_H$ and $F^{(j)}_H$,
respectively. We now perform a change of variables inspired by
\FKSeq{4.70}, with
\beqn
(\xia,\xib)\;\;&\longrightarrow&\;\;(\xinpf,z)\,,
\\
\xia=z\xinpf\,,\;\;\;\;
\xib=(1-z)\xinpf\;\;&\Longrightarrow&\;\;
d\xia d\xib = \xinpf d\xinpf dz\,.
\eeqn
The indices $\alpha$ and $\beta$ are such that 
\mbox{$\{\alpha,\beta\}=\{i,j\}$}, with the two possible choices
denoted as follows:
\beqn
&&c_i\,:\;\;\;\;\;\;(\alpha,\beta)=(i,j)\,,
\label{cidef}
\\
&&c_j\,:\;\;\;\;\;\;(\alpha,\beta)=(j,i)\,,
\label{cjdef}
\eeqn
so that:
\beqn
\left.d\mu\right|_{c_i}&\!\!=\!\!&\vol^2\,(1-z)^{1-2\ep}\,\xinpf^{2-2\ep}
\left(1-\ya^2\right)^{-\ep}d\xinpf\,dz\,d\ya\,d\Omega_\alpha^{(2-2\ep)}\,
d\Omega_\beta^{(2-2\ep)}\,d\tilde{\phi}\,,\phantom{aaa}
\label{dmui}
\\
\left.d\mu\right|_{c_j}&\!\!=\!\!&\vol^2\,z^{1-2\ep}\,\xinpf^{2-2\ep}
\left(1-\yb^2\right)^{-\ep}d\xinpf\,dz\,d\yb\,d\Omega_\alpha^{(2-2\ep)}\,
d\Omega_\beta^{(2-2\ep)}\,d\tilde{\phi}\,;\phantom{aa}
\label{dmuj}
\eeqn
eq.~(\ref{dmuj}) coincides with \FKSeq{4.80}.
Furthermore, by using \MadFKSeq{4.17}, \FKSeq{4.58}, \FKSeq{4.71},
and \FKSeq{4.72} (and by neglecting the $\Delta$ term in the latter, 
owing to its vanishing upon azimuthal integration), we obtain:
\beqn
&&\left.\delta(1-\yj){\cal A}_i\right|_{c_i}=\delta(1-\yb)
\sum_{a_\alpha a_\beta} \frac{4\pi\as\mu^{2\ep}}{(n+1)!}
\left(\frac{\sqrt{S}}{2}\right)^{-2}\frac{z}{1-z}
\label{limAi}
\\*&&\phantom{aaaaaa}\times
P_{a_\alpha S(a_\alpha,a_\beta)}^<(z,\ep)\,
\ampsq^{(n,0)}(S(a_\alpha,a_\beta))\, h(z) F^{(\alpha)}_H(z)\, 
J_n^{n-1}\,,
\nonumber
\\
&&\left.\delta(1-\yj){\cal A}_j\right|_{c_j}=\delta(1-\ya)
\sum_{a_\alpha a_\beta} \frac{4\pi\as\mu^{2\ep}}{(n+1)!}
\left(\frac{\sqrt{S}}{2}\right)^{-2}\frac{1-z}{z}
\label{limAj}
\\*&&\phantom{aaaaaa}\times
P_{a_\alpha S(a_\alpha,a_\beta)}^<(z,\ep)\,
\ampsq^{(n,0)}(S(a_\alpha,a_\beta))\, h(1-z) F^{(\alpha)}_H(z)\, 
J_n^{n-1}\,.\phantom{aa}
\nonumber
\eeqn
By construction, the reduced matrix element $\ampsq^{(n,0)}$ features the 
parton that splits into the pair $(i,j)$, whose flavour is equal to 
$S(a_\alpha,a_\beta)$, and which is explicitly indicated in eqs.~(\ref{limAi})
and~(\ref{limAj}) -- the dependence on the other partons, being 
irrelevant in what follows, is omitted.
Note that some minor abuse of notation concerns the fragmentation factor,
since according to the general notation introduced in sect.~\ref{sec:not}
the argument of $F^{(p)}_H$ is a momentum and not a fraction of a momentum,
and thus should read (see \FKSeq{4.71}) $F^{(\alpha)}_H(z k_{n+4})$. However,
given that in what follows only the dependence upon $z$ matters, that on
$k_{n+4}$ is dropped from the notation. Finally, observe that the
Altarelli-Parisi kernels in eqs.~(\ref{limAi}) and~(\ref{limAj}) are
identical; this is because the collinear limit of the $(n+1)$-body matrix 
elements can be written (in a fully equivalent manner) by factoring
either of the two following kernels:
\beq
P_{a_iS(a_i,a_j)}^<\left(\frac{E_i}{E_i+E_j}\right)\,,\;\;\;\;\;\;
P_{a_jS(a_i,a_j)}^<\left(\frac{E_j}{E_i+E_j}\right)\,.
\eeq
We have used the former for the change of variable $c_i$ (applied to
the term ${\cal A}_i$), and the latter for the change of variable $c_j$
(applied to the term ${\cal A}_j$). Conversely, the arguments of the
function $h$ are different, since that function results from the collinear
limit of the $\Sfunij$ function, which has the unique form given in
\MadFKSeq{4.17}.

In the collinear limit enforced by $\delta(1-\yj)$, the expressions
of the measures given in eqs.~(\ref{dmui}) and~(\ref{dmuj}) simplify
as well. Firstly, observe that, thanks to \FKSeq{4.71}, in such a limit 
the polar and azimuthal angles of parton $\alpha$ in eq.~(\ref{dmui}) 
(parton $\beta$ in eq.~(\ref{dmuj})) coincide with those of the 
splitting parton (labeled by $n+4$ here, and by $7$ in 
ref.~\cite{Frixione:1995ms}). Thus:
\beqn
\left.\delta(1-\yj)d\mu\right|_{c_i}&\!\!=\!\!&\delta(1-\yb)\vol^2\,
(1-z)^{1-2\ep}dz\,\xinpf\,\xinpf^{1-2\ep}d\xinpf
\nonumber
\\*&&\phantom{aaa}\times
\left(1-\ynpf^2\right)^{-\ep}\,d\ynpf\,d\Omega_{n+4}^{(2-2\ep)}\,
d\Omega_\beta^{(2-2\ep)}\,d\tilde{\phi}
\nonumber
\\*&\!\!=\!\!&
\delta(1-\yb)\vol\,(1-z)^{1-2\ep}dz\,d\Omega_\beta^{(2-2\ep)}\,
\xinpf\, d\phi(n+4)\,d\tilde{\phi}
\nonumber
\\*&\!\!=\!\!&
\delta(1-\yb)\vol\,(1-z)^{1-2\ep}dz\,d\Omega_\beta^{(2-2\ep)}\,
\xinpf\,d\phi_n\,,
\label{dmuit}
\eeqn
where we have used eqs.~(\ref{dphiidef}) and~(\ref{Omegat}). The quantity:
\beq
d\phi_n\equiv
d\phi_n\left(k_1,k_2\to \setS{k_l}{3}{n+4}{ij}\right)
\label{dphicout}
\eeq
is the reduced $n$-body phase space, which originates from the 
$(n+1)$-body one by ``combining'' (in the collinear limit) momenta 
$k_i$ and $k_j$ into the momentum $k_{n+4}$ of the mother parton;
note that eq.~(\ref{dphicout}) coincides with \FKSeq{4.82}.
We can now observe that in the collinear limit the dependence on
the azimuthal variables of parton $\beta$ are trivial (see
eq.~(\ref{limAi})); therefore, we can integrate over them and
use \FKSeq{4.42}, whence eq.~(\ref{dmuit}) becomes:
\beq
\left.\delta(1-\yj)d\mu\right|_{c_i}=
\delta(1-\yb)\,\frac{2\pi^{1-\ep}}{\Gamma(1-\ep)}\,
\vol\,(1-z)^{1-2\ep}dz\,
\xinpf\,d\phi_n\,.
\label{dmuifin}
\eeq
By performing the same manipulations (and by taking into account that the
roles of $\alpha$ and $\beta$ are swapped) we obtain from 
eq.~(\ref{dmuj}):
\beq
\left.\delta(1-\yj)d\mu\right|_{c_j}=
\delta(1-\ya)\,\frac{2\pi^{1-\ep}}{\Gamma(1-\ep)}\,
\vol\,z^{1-2\ep}dz\,
\xinpf\,d\phi_n\,.
\label{dmujfin}
\eeq
Let us now first consider the contribution ${\cal Y}_j$ to eq.~(\ref{Yij});
we can use the results of eqs.~(\ref{limAj}) and~(\ref{dmujfin}), use
the Dirac delta to get rid of the $\ya$ integration, and obtain:
\beqn
{\cal Y}_j&=&-\frac{(2\delta_O)^{-\ep}}{\ep}\,
\frac{4\pi\as\mu^{2\ep}}{(n+1)!}\frac{2\pi^{1-\ep}}{\Gamma(1-\ep)}\,\vol\,
\left(\frac{\sqrt{S}}{2}\right)^{-2}\left(\xinpf {\cal D}_\beta\right)
(1-z)z^{-2\ep}
\nonumber
\\*&\times&
\sum_{a_\alpha d}P_{a_\alpha d}^<(z,\ep)\,
\ampsq^{(n,0)}(d)\, h(1-z) F^{(\alpha)}_H(z)\, 
J_n^{n-1}\,dz\,d\phi_n\,,
\label{Yj}
\eeqn
where we have used the fact that:
\beq
\sum_{a_\beta}f(S(a_\alpha,a_\beta))=\sum_d f(d)\;\;\;\;\;\;\;\;
\forall\,a_\alpha\,,
\eeq
for any function $f()$. Owing to the fact that the change of variables
$c_j$ of eq.~(\ref{cjdef}) is identical to that of \FKSeq{4.70}, the
quantity \mbox{$(\xinpf {\cal D}_\beta)$} that appears in eq.~(\ref{Yj})
is re-written by using \FKSeq{4.76}--\FKSeq{4.79}:
\beqn
\xinpf {\cal D}_\beta&=&\Dz(z)-2\ep\Do(z)
\nonumber
\\*&=&
\pdistr{1-z}{+}+\log\frac{\xinpf}{\xic}\,\delta(1-z)
\nonumber
\\*&-&
2\ep\Bigg[\lppdistr{1-z}{+}+\log\xinpf\pdistr{1-z}{+}
\nonumber
\\*&&\phantom{-2\ep\Bigg[}
+\half\left(\log^2\xinpf-\log^2\xic\right)\delta(1-z)\Bigg].
\label{Dbeta}
\eeqn
Now we consider the contribution ${\cal Y}_i$ to eq.~(\ref{Yij}); 
by proceeding analogously to what was done above, using eqs.~(\ref{limAi}) 
and~(\ref{dmuifin}), we obtain:
\beqn
{\cal Y}_i&=&-\frac{(2\delta_O)^{-\ep}}{\ep}\,
\frac{4\pi\as\mu^{2\ep}}{(n+1)!}\frac{2\pi^{1-\ep}}{\Gamma(1-\ep)}\,\vol\,
\left(\frac{\sqrt{S}}{2}\right)^{-2}\left(\xinpf {\cal D}_\alpha\right)
(1-z)^{-2\ep}z
\nonumber
\\*&\times&
\sum_{a_\alpha d}P_{a_\alpha d}^<(z,\ep)\,
\ampsq^{(n,0)}(d)\, h(z) F^{(\alpha)}_H(z)\, 
J_n^{n-1}\,dz\,d\phi_n\,.
\label{Yi}
\eeqn
It is important to note that the change of variables $c_i$ of 
eq.~(\ref{cidef}) is {\em not} the same as that of FKS. Hence,
the identity of eq.~(\ref{Dbeta}) must not be used to handle the
term $(\xinpf {\cal D}_\alpha)$ that appears in eq.~(\ref{Yi}).
It is clear than an analogous expression exists that can be applied
to the present case; however, this is not necessary. The reason is
the following: thanks to the presence of
\beq
h(z) F^{(\alpha)}_H(z)
\eeq
eq.~(\ref{Yi}) is finite both at $z=0$ (because of $F^{(\alpha)}_H(z)$)
and at $z=1$ (because of $h(z)$). Therefore, the subtraction terms in
${\cal D}_\alpha$ are identically equal to zero, which implies that 
${\cal D}_\alpha$ is a regular function (and not a distribution),
whence:
\beqn
h(z) F^{(\alpha)}_H(z)\left(\xinpf {\cal D}_\alpha\right)&=&
h(z) F^{(\alpha)}_H(z)\frac{\xinpf}{\xia}\left(1-2\ep\log\xia\right)
\nonumber
\\*&=&
h(z) F^{(\alpha)}_H(z)\frac{\xinpf\xia^{-2\ep}}{\xia}+{\cal O}(\ep^2)
\nonumber
\\*&=&
h(z) F^{(\alpha)}_H(z) z^{-1-2\ep}\xinpf^{-2\ep}+{\cal O}(\ep^2)\,.
\eeqn
By substituting this expression into eq.~(\ref{Yi}) we obtain:
\beqn
{\cal Y}_i&=&-\frac{(2\delta_O)^{-\ep}}{\ep}\,
\frac{4\pi\as\mu^{2\ep}}{(n+1)!}\frac{2\pi^{1-\ep}}{\Gamma(1-\ep)}\,\vol\,
\left(\frac{\sqrt{S}}{2}\right)^{-2}
(1-z)^{-2\ep}z^{-2\ep}\xinpf^{-2\ep}
\nonumber
\\*&\times&
\sum_{a_\alpha d}P_{a_\alpha d}^<(z,\ep)\,
\ampsq^{(n,0)}(d)\, h(z) F^{(\alpha)}_H(z)\, 
J_n^{n-1}\,dz\,d\phi_n\,.
\label{Yi2}
\eeqn
Eq.~(\ref{Yi2}) is fairly similar to eq.~(\ref{Yj}). One can in fact
show that such a similarity is even closer than it appears at a first
look. One starts by observing that:
\beqn
&&(1-z)^{-2\ep}\xinpf^{-2\ep}=1-2\ep\left[\log(1-z)+\log\xinpf\right]
+{\cal O}(\ep^2)
\\*&&\phantom{aaaa}=
(1-z)\left\{\frac{1}{1-z}-2\ep\left[\frac{\log(1-z)}{1-z}+
\log\xinpf\frac{1}{1-z}\right]\right\}
+{\cal O}(\ep^2).
\nonumber
\eeqn
Then, one uses again the fact that $h(z)$ damps singularities at $z=1$. 
Thus:
\beqn
&&(1-z)^{-2\ep}\xinpf^{-2\ep}\,h(z)=h(z)(1-z)\Bigg\{
F\delta(1-z)+\pdistr{1-z}{+}
\label{tmp3}
\\*&&\phantom{aaaaaaaa}
-2\ep\left[\lppdistr{1-z}{+}+
\log\xinpf\pdistr{1-z}{+}\right]\Bigg\}
+{\cal O}(\ep^2)\,,
\nonumber
\eeqn
for any $F$ (which is in general a $(4-2\ep)$-dimensional coefficient). 
The quantity in curly brackets on the r.h.s.~of eq.~(\ref{tmp3})
has the same form as the r.h.s.~of eq.~(\ref{Dbeta}), and can be
made identical to it by suitably choosing the coefficient $F$
(namely, by setting it equal to the coefficient of $\delta(1-z)$ 
in eq.~(\ref{Dbeta})). This implies:
\beq
(1-z)^{-2\ep}\xinpf^{-2\ep}\,h(z)=
(1-z)\Big(\Dz(z)-2\ep\Do(z)\Big)h(z)
+{\cal O}(\ep^2)\,.
\eeq
Thus, eq.~(\ref{Yi2}) becomes:
\beqn
{\cal Y}_i&=&-\frac{(2\delta_O)^{-\ep}}{\ep}\,
\frac{4\pi\as\mu^{2\ep}}{(n+1)!}\frac{2\pi^{1-\ep}}{\Gamma(1-\ep)}\,\vol\,
\left(\frac{\sqrt{S}}{2}\right)^{-2}
(1-z)z^{-2\ep}
\nonumber
\\*&\times&
\Big(\Dz(z)-2\ep\Do(z)\Big)
\sum_{a_\alpha d}P_{a_\alpha d}^<(z,\ep)\,\ampsq^{(n,0)}(d)\, 
\nonumber
\\*&\times&
h(z) F^{(\alpha)}_H(z)\, 
J_n^{n-1}\,dz\,d\phi_n\,.
\label{Yi3}
\eeqn
In this way, it is apparent that ${\cal Y}_i$ in eq.~(\ref{Yi3}) has
exactly the same form as ${\cal Y}_j$ in eq.~(\ref{Yj}) (because of
eq.~(\ref{Dbeta})), {\em except} for the fact that the former features 
a factor $h(z)$, while the latter features a factor $h(1-z)$. This
implies that the sum \mbox{${\cal Y}_i+{\cal Y}_j$}, which is the quantity 
of interest (see eq.~(\ref{Yij})), will feature:
\beq
h(z)+h(1-z)\equiv 1\,,
\eeq
thanks to \MadFKSeq{4.23}. After some algebra, and by using 
eq.~(\ref{voldef}), one arrives at:
\beqn
{\cal Y}_{ij}&=&
-\asotpi\left(\frac{1}{\epb}-\log\frac{S\delta_O}{2\mu^2}\right)
\frac{(1-z)z^{-2\ep}}{(n+1)!}
\nonumber
\\*&\times&
\Big(\Dz(z)-2\ep\Do(z)\Big)
\sum_{a_\alpha d}P_{a_\alpha d}^<(z,\ep)\,\ampsq^{(n,0)}(d)\, 
\nonumber
\\*&\times&
F^{(\alpha)}_H(z)\, 
J_n^{n-1}\,dz\,d\phi_n\,.
\label{Yij2}
\eeqn
Note, for future reference, that:
\beq
\frac{1}{\epb}-\log\frac{S\delta_O}{2\mu^2}=
\frac{(4\pi)^\ep}{\Gamma(1-\ep)}\,\muoQep\,
\left(\frac{1}{\ep}-\log\frac{S\delta_O}{2Q^2}\right),
\label{epbdef}
\eeq
with the pre-factor on the r.h.s.~being the one that is conventionally
factored out in FKS formulae (see e.g.~\FKSeq{3.1} and \FKSeq{4.88}).

Eq.~(\ref{Yij2}) is the analogue of \FKSeq{4.83}, and it is apparent
that these two equations are rather similar. In order to proceed, one
needs to perform the same operations as in \FKSeq{4.86}: namely, to
sum over all $(i,j)$ pairs, and to relabel the parton indices.
We have already gone through this process earlier in this section,
when we have dealt with case A. In the case B.~we are discussing here,
the number of contributions is given by the r.h.s.~of eq.~(\ref{sumoutB}),
{\em divided by two}. This is because in eq.~(\ref{sumoutB}) the
overall factor of $2$ comes from the fact that the results stemming from
the $\delta_{ip}$ and $\delta_{jp}$ terms are identical. However, the 
quantity ${\cal Y}_{ij}$ already includes both of them (see 
eq.~(\ref{Yijdef})), whence the necessity of dividing by two the
r.h.s.~of eq.~(\ref{sumoutB}) for a correct counting. The bottom line
is that the number of contributions is the same as in FKS (eq.~(\ref{tmp6})),
and we can therefore proceed as done there as far as the symmetry factors
are concerned. Note that, at variance with case A., in case B.~the sum
over fragmenting partons is no longer present.

There is a further subtlety associated with relabeling that we must mention.
The fragmenting parton $\alpha$ and its mother parton $n+4$ have degenerate
momenta in the collinear limit, \mbox{$k_\alpha=zk_{n+4}$}. In 
eq.~(\ref{Yij2}) there is a sum over the flavour ($d$) of the mother 
parton, which of course does not affect the kinematics: $k_{n+4}$ is
the same for all terms in such a sum. By construction (see 
eq.~(\ref{dphicout})) this momentum is generated in the reduced
phase space $d\phi_n$. Finally, keep in mind that the rescaled momentum
$\xinpf$ enters the expressions of $\Dz(z)$ and $\Do(z)$ (see 
eq.~(\ref{Dbeta})). Upon relabeling, while the flavours of the 
fragmenting parton and of the mother will need to be kept distinct,
one can assign the same label to these two partons, thus resulting
in a more compact notation (no confusion being possible)\footnote{This 
is exactly the same situation as the one relevant to initial-state
branchings: the partons entering the real-emission and Born processes
are different and have in general different flavours, but have degenerate
kinematics and are both labelled by indices $1$ or $2$.}. At the end of the 
day, we can therefore write:
\beqn
\sum_{ij}{\cal Y}_{ij}&=&
-\asotpi\left(\frac{1}{\epb}-\log\frac{S\delta_O}{2\mu^2}\right)
\frac{(1-z)z^{-2\ep}}{n!}
\nonumber
\\*&\times&
\sum_k \Big(\Dz(z)-2\ep\Do(z)\Big)
\sum_{a_k d}P_{a_k d}^<(z,\ep)\,\ampsq^{(n,0)}(d)\, 
\nonumber
\\*&\times&
F^{(k)}_H(z)\, 
J_n^{n-1}\,dz\,d\phi_n\,,
\label{Yij3}
\eeqn
where one understands that the expressions of $\Dz(z)$ and $\Do(z)$ 
are given in eq.~(\ref{Dbeta}) with the formal replacement
\beq
\xinpf\;\longrightarrow\;\xik=\frac{2E_k}{\sqrt{S}}
\label{xinp4}
\eeq
there. From this point onwards, $k_k$ is the momentum of the mother
parton (because is the one that appears in the phase space), with the 
momentum of the fragmenting parton being equal to $zk_k$ by construction.
 
It is manifest that eq.~(\ref{Yij3}) does {\em not} have the same
singular structure as \FKSeq{4.88}. This is what one expects,
because we have so far neglected the contribution of the final-state
collinear counterterms. These are given in the second term on the
r.h.s.~of eq.~(\ref{cnt1}), and we simply have to write them in
the same form as eq.~(\ref{Yij3}), which is straightforward to do.
Firstly, we observe that the counterterm of eq.~(\ref{cnt1}) is relevant
to a single parton that fragments; in keeping with eq.~(\ref{HmFxsec}),
we have therefore to sum over all possible fragmenting partons, 
as well as over their flavours.
Secondly, eq.~(\ref{cnt1}) originates from eq.~(\ref{xscnt}),
where the fragment{\em ed} parton momentum is fixed, and the 
fragment{\em ing} parton momentum is obtained from the former by
a $1/\zeta$ rescaling. As discussed above, eq.~(\ref{Yij3}) is such 
that one fixes the parent momentum; thus, we shall use eq.~(\ref{cnt1})
by formally replacing \mbox{$k\to \zeta k$} (we shall also rename
$\zeta$ as $z$, for consistency with eq.~(\ref{Yij3})). In this way,
the final-state counterterms to be added to eq.~(\ref{Yij3}) read
as follows:
\beqn
d\sigma_{out}^{(cnt)}&=&
\asotpi\sum_k\sum_{a_k d}
\left(\frac{1}{\epb}P_{a_k d}(z,0)-K_{a_k d}^{\rm FF}(z)\right)\,
\nonumber
\\*&\times&
\frac{1}{n!}\ampsq^{(n,0)}(d)\, 
F^{(k)}_H(z)\, J_n^{n-1}\,dz\,d\phi_n\,.
\label{Ycnt}
\eeqn
Note that, precisely as in the case of initial-state collinear
singularities, the AP kernel in eq.~(\ref{Ycnt}) is not the same as
that in eq.~(\ref{Yij3}): the former is in four dimensions but includes
the contribution at $z=1$, while the latter is in $4-2\ep$ dimensions
but lives in the $z<1$ space. Thus, we shall now manipulate eq.~(\ref{Yij3}) 
in order to show that it features a singularity which has the same form
as that in eq.~(\ref{Ycnt}). By expanding eq.~(\ref{Yij3}) in $\ep$, and 
using \FKSeq{4.52} one obtains:
\beqn
\sum_{ij}{\cal Y}_{ij}&=&-\asotpi
\sum_k \sum_{a_k d}
(1-z)
\Bigg[\frac{1}{\epb}\Dz(z)P_{a_k d}^<(z,0)+\Dz(z)P_{a_k d}^{\prime<}(z,0)
\nonumber
\\*&&\phantom{aa}
-\left(\Dz(z)\left(\log\frac{S\delta_O}{2\mu^2}+2\log z\right)+2\Do(z)
\right)P_{a_k d}^<(z,0)\Bigg]
\nonumber
\\*&\times&
\frac{1}{n!}\ampsq^{(n,0)}(d)\,F^{(k)}_H(z)\,J_n^{n-1}\,dz\,d\phi_n\,.
\label{Yij4}
\eeqn
By exploiting the identities given in \FKSeq{4.48} and \FKSeq{4.49},
and using the definition of $\Dz(z)$ (see eq.~(\ref{Dbeta})), we have:
\beqn
(1-z)\Dz(z)P_{a_k d}^<(z,0)&=&
P_{a_k d}(z,0)
\nonumber
\\*&-&
\Bigg(\gamma(a_k)-2C(a_k)\log\frac{\xik}{\xic}\Bigg)
\delta_{a_k d}\,\delta(1-z)\,.\phantom{aa}
\eeqn
By multiplying both members of this equation by the factor \mbox{$1/\epb$} 
that appears in eq.~(\ref{Yij4}), and by using the definition of
$\epb$ (see eq.~(\ref{epbdef})), one obtains:
\beqn
&&\frac{1}{\epb}\,(1-z)\Dz(z)P_{a_k d}^<(z,0)=
\frac{1}{\epb}\,P_{a_k d}(z,0)
\nonumber
\\*&&\phantom{aaa}
-\frac{1}{\ep}\,\frac{(4\pi)^\ep}{\Gamma(1-\ep)}\,\muoQep\,
\Bigg(\gamma(a_k)-2C(a_k)\log\frac{\xik}{\xic}\Bigg)
\delta_{a_k d}\,\delta(1-z)
\nonumber
\\*&&\phantom{aaa}
+\log\muoQ\,
\Bigg(\gamma(a_k)-2C(a_k)\log\frac{\xik}{\xic}\Bigg)
\delta_{a_k d}\,\delta(1-z)\,.\phantom{aa}
\eeqn
By replacing the r.h.s.~of this equation into eq.~(\ref{Yij4}), one
sees that the pole term proportional to the AP kernel has exactly the
same form, except for an overall sign, as that in eq.~(\ref{Ycnt}),
and therefore drops in the sum. Thus:
\beqn
&&\sum_{ij}{\cal Y}_{ij}+d\sigma_{out}^{(cnt)}=\asotpi
\left(\frac{1}{\ep}\,\frac{(4\pi)^\ep}{\Gamma(1-\ep)}\,\muoQep
-\log\muoQ\right)
\nonumber
\\*&&\phantom{aaaa}
\times\sum_k\sum_{a_k}
\Bigg(\gamma(a_k)-2C(a_k)\log\frac{\xik}{\xic}\Bigg)
\frac{1}{n!}\ampsq^{(n,0)}(d)\,F^{(k)}_H(1)\,J_n^{n-1}\,d\phi_n
\nonumber
\\*&&\phantom{aa}
+\asotpi\sum_k\sum_{a_k d} (1-z)
\Bigg[-\Dz(z)P_{a_k d}^{\prime<}(z,0)-\frac{K_{a_k d}^{\rm FF}(z)}{1-z}
\nonumber
\\*&&\phantom{aaaa}
+\left(\Dz(z)\left(\log\frac{S\delta_O}{2\mu^2}+2\log z\right)+2\Do(z)
\right)P_{a_k d}^<(z,0)\Bigg]
\nonumber
\\*&&\phantom{aaaaaaaaaa}\times
\frac{1}{n!}\ampsq^{(n,0)}(d)\,F^{(k)}_H(z)\,J_n^{n-1}\,dz\,d\phi_n\,.
\label{Yij5}
\eeqn
Let us consider eq.~(\ref{Yij5}). The sum over
$a_k$ is what remains of the sums over $a_i$ and $a_j$ in eq.~(\ref{Yijdef}).
According to the definitions of parton- and hadron-level cross sections
given in eqs.~(\ref{mFxsec}) and~(\ref{HmFxsec}) respectively, it belongs
to the latter; hence, in the parton-level quantity defined by eq.~(\ref{Yij5})
one fixes $a_k$ and does not sum over it. It is then clear that, apart
from the fragmentation factor $F^{(k)}_H(1)$, the divergent parts
of eq.~(\ref{Yij5}) and of \FKSeq{4.88} are identical. As far as the
fragmentation factor is concerned, we can employ the identity:
\beq
F^{(k)}_H(1)=\delta_{kp}F^{(p)}_H(1)\,.
\label{idF2}
\eeq
When the r.h.s.~of this equation is substituted into eq.~(\ref{Yij5}), the 
divergent part of the latter is complementary to the divergent part of the 
contribution of case A.~(compare eqs.~(\ref{idF1}) and~(\ref{idF2})).
Thus, regardless of the
value of $p$ (the label of the fragmenting parton) the union of case 
A.~and case B.~is identical to the divergent part of \FKSeq{4.88}, 
up to the fragmentation factor. This is therefore the same structure
as that of all of the other NLO contributions; we have therefore proven
that the cancellation of the IR singularities is fully achieved, and that
it proceeds similarly to what happens in the standard FKS case.

In summary, the contribution of final-state origin can be written as follows:
\beq
\sum_{ij}d\sigma^{(n+1,0;p)}_{out,ij}=
\sum_{ij}d\sigma^{(n+1,0;p)}_{out,ij,f}+
d\sigma^{(n+1,0;p)}_{out,+,\delta}+
d\hat\sigma^{(n+1,0;p)}_{out,+,z}\,.
\label{outfin}
\eeq
The first term on the r.h.s.~of eq.~(\ref{outfin}) is the analogue of 
\FKSeq{4.65} and is given by eq.~(\ref{realoutfin}).
The other two terms emerge from the manipulations performed above, and
can obtained from case A.~and eq.~(\ref{Yij5}). Remember that they are 
$n$-body terms, and thus \mbox{$3\le p\le n+2$}. The quantity 
$d\sigma^{(n+1,0;p)}_{out,+,\delta}$ is the analogue of \FKSeq{4.88},
while $d\hat\sigma^{(n+1,0;p)}_{out,+,z}$ has no final-state analogue
in the standard FKS treatment, being the finite remainder of the 
subtraction of the collinear counterterms. Case A.~contributes only to 
$d\sigma^{(n+1,0;p)}_{out,+,\delta}$, in the form dictated by
\FKSeq{4.88} (up to fragmentation factors). The contributions of
case B.~to $d\sigma^{(n+1,0;p)}_{out,+,\delta}$ are those proportional
to $F^{(k)}_H(1)$ in eq.~(\ref{Yij5}); some are already explicit,
but others have to be obtained from the $\delta(1-z)$ terms contained
in $\Dz(z)$ and $\Do(z)$ -- in order to do so, use has to be made
of \FKSeq{4.48} and \FKSeq{4.49}. Finally, the terms not proportional 
to $\delta(1-z)$ in eq.~(\ref{Yij5}) are collected into
$d\hat\sigma^{(n+1,0;p)}_{out,+,z}$.

After some algebra, and by using eq.~(\ref{xinp4}), the results read 
as follows (\mbox{$3\le p\le n+2$}):
\beqn
&&d\sigma^{(n+1,0;p)}_{out,+,\delta}=
\asotpi\frac{(4\pi)^\ep}{\Gamma(1-\ep)}\muoQep\frac{1}{\ep}
\nonumber
\\*&&\phantom{d\sigma^{(n+1,0;p)}_{out,+,\delta}=aaa}\times
\sum_{k=3}^{n+2}\left[\gamma(a_k)-2C(a_k)\log\frac{2E_k}{\xic\sqrt{S}}\right]
d\sigma^{(n,0;p)}
\nonumber
\\*&&\phantom{aa}
+\asotpi\sum_{k=3}^{n+2} 2C(a_k)
\left[\log\frac{S\delta_O}{2Q^2}\log\frac{2E_k}{\xic\sqrt{S}}
+\log^2\frac{2E_k}{\sqrt{S}}-\log^2\xic
\right]
d\sigma^{(n,0;p)}
\nonumber
\\*&&\phantom{aa}
-\asotpi\sum_{k=3}^{n+2}\Bigg\{
\gamma(a_k)\left[\left(1-\delta_{kp}\right)
\left(\log\frac{S\delta_O}{2Q^2}+2\log\frac{2E_k}{\sqrt{S}}\right)
+\delta_{kp}\log\muoQ\right]
\nonumber
\\*&&\phantom{aa-\asotpi\sum_{k=3}^{n+2}\Bigg\{}
-\gamma^\prime(a_k)\left(1-\delta_{kp}\right)\Bigg\}
d\sigma^{(n,0;p)}\,,
\label{dsigoutd}
\eeqn
and:
\beqn
&&d\hat\sigma^{(n+1,0;p)}_{out,+,z}=\asotpi\sum_d (1-z)
\Bigg\{-\pdistr{1-z}{+}P_{a_p d}^{\prime<}(z,0)-
\frac{K_{a_p d}^{\rm FF}(z)}{1-z}
\nonumber
\\*&&\phantom{\;}
+\left[\pdistr{1-z}{+}\left(\log\frac{S\delta_O}{2\mu^2}+
2\log\frac{2zE_p}{\sqrt{S}}\right)+
2\lppdistr{1-z}{+}\right]P_{a_p d}^<(z,0)\Bigg\}
\nonumber
\\*&&\phantom{aaaaaaaaaa}\times
\frac{1}{n!}\ampsq^{(n,0)}(d)\,F^{(p)}_H(z)\,J_n^{n-1}\,dz\,d\phi_n\,.
\label{dsigoutz}
\eeqn
As was already said, the divergent part of eq.~(\ref{dsigoutd}) is identical,
up to the fragmentation factor, to that of \FKSeq{4.88}. As far as the
finite part is concerned, the $\log Q^2$ term is again identical to that 
of \FKSeq{4.88}; this must be so, since the Ellis-Sexton-scale dependence 
of the present contribution must be compensated by that of the soft and 
virtual ones, which in turn depend on the presence of fragmentation 
factors only in a trivial manner. The $C(a_k)$ bit is also identical to 
that of ref.~\cite{Frixione:1995ms}, while the $\gamma(a_k)$ and 
$\gamma^\prime(a_k)$ ones are not (owing to the conditions enforced by
$\delta_{kp}$). A simple explanation for this fact is that the $\gamma$'s
are ``integrated'' quantities: in the absence of fragmentation a full
integration over the $i$ and $j$ degrees of freedom is possible, but 
this is not the case when fragmentation is present, and thus there is an
amount of talk-to between eqs.~(\ref{dsigoutd}) and~(\ref{dsigoutz}). For 
what concerns the latter equation, note its similarity with the finite parts 
of \FKSeq{4.54} or \FKSeq{4.55}. There is an extra logarithmic term in 
eq.~(\ref{dsigoutz}) which features the energy ($zE_p$) of the fragmenting 
parton: this is the related to the ``missing'' $\gamma$ and $\gamma^\prime$
contributions in eq.~(\ref{dsigoutd}) to which we have alluded above.

We shall soon return to comment on this extra logarithmic term. Before
doing that, it is convenient to re-cast eq.~(\ref{dsigoutz}) in a slightly
different form, so that its similarities with \FKSeq{4.54} and 
\FKSeq{4.55} can be made more explicit. We observe that, while in the 
initial-state case we have used distributions whose subtractions are
controlled by the parameter $\xic$, the final-state cross section is so 
far expressed in terms of standard plus distributions.
We can amend the situation by means of the identities
\beqn
\pdistr{1-z}{+}&=&\pdistr{1-z}{c}+\log(1-z_c)\,\delta(1-z)\,,
\label{omzc}
\\
\lppdistr{1-z}{+}&=&\lppdistr{1-z}{c}+\half\log^2(1-z_c)\,\delta(1-z)\,,
\label{lomzc}
\eeqn
for any $0\le z_c<1$, and having defined:
\beq
\int dz\,f(z)\left(\frac{g(z)}{1-z}\right)_c=
\int_0^1 dz\,\frac{g(z)}{1-z}\,\big(f(z)-f(1)\stepf(z-z_c)\big)\,.
\eeq
The replacement of eqs.~(\ref{omzc}) and~(\ref{lomzc}) into 
eq.~(\ref{dsigoutz}) generates terms of the same form as those
that appear in eq.~(\ref{dsigoutd}) (i.e.~that factorize the Born
without any $z$ convolution), and are proportional to $C(a_p)$ owing 
to \FKSeq{4.49}; thus, they can conveniently be included in the definition
of $d\sigma^{(n+1,0;p)}_{out,+,\delta}$. In order to write them explicitly, 
we choose:
\beq
z_c=1-\xic\,.
\eeq
Some trivial algebra shows that $d\sigma^{(n+1,0;p)}_{out,+,\delta}$ now 
reads as follows:
\beqn
&&d\sigma^{(n+1,0;p)}_{out,+,\delta}=
\asotpi\frac{(4\pi)^\ep}{\Gamma(1-\ep)}\muoQep\frac{1}{\ep}
\label{dsigoutd2}
\\*&&\phantom{d\sigma^{(n+1,0;p)}_{out,+,\delta}=aaa}\times
\sum_{k=3}^{n+2}\left[\gamma(a_k)-2C(a_k)\log\frac{2E_k}{\xic\sqrt{S}}\right]
d\sigma^{(n,0;p)}
\nonumber
\\*&&\phantom{aa}
+\asotpi\sum_{\stackrel{k\ne p}{k=3}}^{n+2}\Bigg\{ 2C(a_k)
\left(\log\frac{S\delta_O}{2Q^2}\log\frac{2E_k}{\xic\sqrt{S}}
+\log^2\frac{2E_k}{\sqrt{S}}-\log^2\xic
\right)
\nonumber
\\*&&\phantom{aa\asotpi\sum_{\stackrel{k\ne p}{k=3}}^{n+2}\Bigg\{aa}
+\gamma^\prime(a_k)-\gamma(a_k)
\left(\log\frac{S\delta_O}{2Q^2}+2\log\frac{2E_k}{\sqrt{S}}\right)
\Bigg\}
d\sigma^{(n,0;p)}
\nonumber
\\*&&\phantom{aa}
+\asotpi\Bigg\{
2C(a_p)\log\frac{2E_p}{\sqrt{S}}\log\frac{\sqrt{S}E_p\delta_O\xic^2}{Q^2}
\nonumber
\\*&&\phantom{aa\asotpi\Bigg\{aa}
-\log\muoQ\Big(\gamma(a_p)+2C(a_p)\log\xic\Big)
\Bigg\}d\sigma^{(n,0;p)}\,,
\nonumber
\eeqn
while:
\beq
d\hat\sigma^{(n+1,0;p)}_{out,+,z}={\rm eq.}~(\protect\ref{dsigoutz})
\Big[()_+\;\longrightarrow ()_c\Big]\,.
\label{dsigoutz2}
\eeq
The structure of eq.~(\ref{dsigoutd2}) is interesting. Its finite part
for $k\ne p$ is the same as before, i.e.~identical to its counterpart
in \FKSeq{4.88} or \MadFKSeq{4.6}. Conversely, the finite contribution 
associated with the fragmenting parton features a \mbox{$\log\mu^2/Q^2$}
term which is identical to an analogous piece of initial-state origin
(see \FKSeq{4.54}, \FKSeq{4.55}, and the first line of \MadFKSeq{4.6}).
It is the other term that {\em seemingly} does not have an initial-state 
analogue. However, as for the logarithmic term in eq.~(\ref{dsigoutz}) 
or~eq.~(\ref{dsigoutz2}) which features the energy $zE_p$, this is an 
accident due to kinematics, and in particular to the fact that FKS cross 
sections are written in the rest frame of the incoming partons, where
the energies of the latter are \mbox{$E_1=E_2=\sqrt{S}/2$}. We have in
fact explicitly verified that, by repeating the same computations that
has led to \FKSeq{4.54} and \FKSeq{4.55} without assigning a specific
value to $E_i$, $i=1,2$, we arrive at the same results as in
ref.~\cite{Frixione:1995ms}, bar for two extra terms. The first of these 
reads as follows:
\beq
2\log\frac{2E_i}{\sqrt{S}}\,\xi P_{da_i}^<(1-\xi,0)\pdistr{\xi}{c}\,,
\label{tmp100}
\eeq
and must be inserted in the curly brackets on the r.h.s. \FKSeq{4.54}
and \FKSeq{4.55}, thus rendering these two equations fully analogous
to their final-state counterpart, eq.~(\ref{dsigoutz2}). The second term
is such that the finite part of the Born-like term in \FKSeq{4.54} 
and \FKSeq{4.55} reads:
\beq
2C(a_i)\log\frac{2E_i}{\sqrt{S}}\log\frac{\sqrt{S}E_i\delta_I\xic^2}{Q^2}
-\log\muoQ\Big(\gamma(a_i)+2C(a_i)\log\xic\Big)\,,
\label{tmp10}
\eeq
which is identical (up to the obvious replacements $i\to p$ and
$\delta_I\to\delta_O$) with the one that appears in the last two
lines of eq.~(\ref{dsigoutd2}). As they must, eqs.~(\ref{tmp100})
and~(\ref{tmp10}) vanish by setting \mbox{$E_i=\sqrt{S}/2$}, thus
recovering the results of ref.~\cite{Frixione:1995ms}.

We have therefore found that the FKS short-distance cross section
of initial- and final-state collinear origin have a larger degree
of similarity in the presence of fragmentation than in the inclusive
case. This need not be surprising, since it is only in the former case
that both initial- and final-state partons are kinematically constrained
by ``external'' objects (the PDFs and FFs, respectively). Having said
that, observe that by construction that scale $\mu$ that appears in 
eqs.~(\ref{dsigoutd}) and~(\ref{dsigoutz}) (or eqs.~(\ref{dsigoutd2}) 
and~(\ref{dsigoutz2})) is the scale that enters the FFs, which in principle 
can be different from the factorisation scale used in the PDFs.

\subsection{Summary\label{sec:sumFF}}
In this section, we have revisited the FKS subtraction by applying it to the 
cases in which one final-state parton is fragmented. By following as closely 
as possible the procedure of the original paper~\cite{Frixione:1995ms},
we have proven the cancellation of the IR singularities that emerge in the
intermediate steps of the computations, and obtained in the process 
the IR-finite short-distance cross sections. The forms of the latter 
are rather similar to those relevant to the inclusive (i.e.~not fragmented)
case, with the most significant differences due to the contribution of
final-state collinear origin.

The origin of the final results, and the formulae where they appear,
are the following: Born (eq.~(\ref{BornFF})), virtual (eq.~(\ref{virtFF})), 
soft (eq.~(\ref{realsF})), subtracted initial state (eq.~(\ref{realinfin})), 
initial-state collinear remainders (eqs.~(\ref{dsiginpl}) and~(\ref{dsiginmn})),
subtracted final state (eq.~(\ref{realoutfin})), and final-state collinear
remainders (eqs.~(\ref{dsigoutd}) and~(\ref{dsigoutz}), or
eqs.~(\ref{dsigoutd2}) and~(\ref{dsigoutz2})).

As a check of the correctness of the IR-finite parts of the above 
formulae, we have applied them to the computation of the initial conditions
of the perturbative $b$-quark fragmentation function, finding full
agreement with ref.~\cite{Mele:1990cw} (note the erratum of 
that paper).

\section{The complex-mass scheme\label{sec:cmscheme}}

\subsection{Introduction\label{CMSintro}}
A long-standing difficulty of perturbative computations of scattering
amplitudes in Quantum Field Theories is that of the handling of unstable 
short-lived particles. The situation is complicated by the fact that 
contributions from such particles can spoil gauge invariance and unitarity, 
and their treatment possibly necessitates relaxing strict 
fixed-order accuracy. In particular, when carrying out NLO EW computations 
in the SM, care is needed when considering processes that involve the 
intermediate massive vector bosons $Z$ and $W^\pm$, the Higgs boson, 
and the top quark. 

A Feynman diagram that features the propagator of an unstable particle
$P$ with mass $M$ and virtuality $p^2$ diverges when $p^2\to M^2$. In
those cases where $p^2$ is associated with a physical observable (e.g.~the
$\epem$ invariant mass in a $Z$ decay) such a divergence can be avoided by
means of final-state cuts, but this is not desirable in general, since
it prevents one from studying the kinematically-dominant pole region.
Thus, a universal solution entails a regularisation of the propagator.
The ones which are best motivated from the physical viewpoint all stem
from the Dyson summation of the geometric series that results from the
insertion of two-point 1PI graphs; this leads to the propagator\footnote{We
simplify the notation by considering only the case of a scalar particle.
It should be clear that, for the arguments of this section, this implies
no loss of generality.}$^,$\footnote{\label{ft:IRreg}Throughout this section
we understand the $+i0$ prescription of the Feynman propagators.}:
\beq
G_D(p^2)= -i \Big[p^2-M_0^2+\Sigma(p^2)\Big]^{-1},
\label{GD}
\eeq
with $M_0$ the bare mass of $P$, and $\Sigma(p^2)$ equal (possibly up to a 
factor of $i$) to the 1PI contribution. Because of the presence of an imaginary 
part in the denominator, eq.~(\ref{GD}) achieves the sought regularisation. 
Unfortunately, one cannot naively Dyson-sum all $P$ propagators in
an arbitrarily complicated Feynman diagram. Among other things, by
effectively setting the pole mass of the propagator of $P$ to be 
different from the corresponding Lagrangian mass parameter, one 
might violate gauge invariance~\cite{Stuart:1991xk,Stuart:1991cc,
Sirlin:1991rt,Sirlin:1991fd}. However, eq.~(\ref{GD}) does suggest
that a better-defined mass (and width) should indeed be associated
with the position of the complex pole $\bar{p}^2$:
\beq
\bar{p}^2-M_0^2+\Sigma(\bar{p}^2)=0\;\;\;\;\;\;
\Longrightarrow\;\;\;\;\;\;
\bar{p}^2=\bar{M}^2-i\bar{\Gamma}\bar{M}\,.
\label{CMSpole}
\eeq
In fact, note that the appeal of eq.~(\ref{CMSpole}) is that it reads the
same when expressed in terms of either bare/unrenormalised quantities or
renormalised ones: the position of the pole must not change under 
renormalisation, lest the analiticity properties of the S matrix be
changed too (see e.g.~chapter 10 of ref.~\cite{Sterman:1994ce}). 

Equation~(\ref{GD}) also suggests an alternative way (still
potentially gauge-violating) in which the $P$ propagator can be 
regularised, namely:
\beq
G_R(p^2)=-i\Big[p^2-M^2 + i\Gamma M\Big]^{-1},
\label{GF}
\eeq
with $M$ the on-shell mass, $\Gamma$ the total decay width of $P$, 
and where one exploits the fact:
\beq
\Im\Big(\Sigma(p^2=M^2)\Big)=\Gamma M\,,
\label{SigtoGam}
\eeq
which follows directly from the optical theorem.
At the level of squared amplitudes, eq.~(\ref{GF}) will lead to
a Breit-Wigner (BW) form:
\beq
BW(p^2)=\frac{1}{(p^2-M^2)^2+\Gamma^2M^2}\,.
\label{BreitWigner}
\eeq
In the timelike region $p^2\equiv s>0$, i.e.~when $P$ becomes resonant
at $s\simeq M^2$, the term $\Gamma^2M^2$ prevents $BW(s)$ from diverging.
Furthermore, a BW function admits the following expansion in terms of 
distributions:
\beq
BW(s)=\frac{\pi}{\Gamma M}\delta(s-M^2)+
{\cal P}\left(\frac{1}{(s-M^2)^2}\right)-
\frac{\pi\Gamma M}{2}\delta(s-M^2)\frac{\partial^2}{\partial s^2}+
\ord\left(\left(\Gamma / M\right)^2\right),
\label{BreitWignerExpanded}
\eeq
where by ${\cal P}$ we denote the principal-value operator. Through
eq.~(\ref{BreitWignerExpanded}) one can study the impact of off-shell 
effects (where $s\ne M^2$), and address gauge violations. For example, by 
keeping the first term on the r.h.s.~of eq.~(\ref{BreitWignerExpanded}) 
one works in the well-known (and gauge invariant) narrow-width 
approximation.

Equation~(\ref{GF}) is also employed in the context of the so-called
pole approximation~\cite{Stuart:1991xk,Stuart:1991cc,Aeppli:1993cb,
Aeppli:1993rs}, where a suitable subset of amplitudes (typically
associated with NLO corrections), stripped of the propagator of
eq.~(\ref{GF}), are expanded around $p^2\simeq M^2$; the leading term
of such an expansion, which can be shown to lead to a gauge-invariant
result, is then kept.
A systematic generalisation of the pole approximation can be achieved
within the unstable-particle effective theories~\cite{Chapovsky:2001zt,
Beneke:2003xh,Beneke:2004km}, which at one-loop accuracy are equivalent
to the former. Here, one exploits the fact that at high energies there
is a natural hierarchy established by the relationship $E^2\gg M\Gamma$,
which allows one to separate the production and decay mechanisms, and to
achieve a formal expansion in $\Gamma/M$ in the pole region $p^2\simeq M^2$.

Neither the pole approximation nor the unstable-particle EFTs are apt to
study the off-shell region, where non-resonant contributions might become
important, and the transition between the off- and on-shell regions.
Furthermore, automated calculations for arbitrary processes present 
non-trivial problems in these approaches (beginning with the fact that, 
in the latter one, the construction of the underlying theory model is 
a very involved operation). Conversely, these issues are absent if one 
works in the complex mass (CM henceforth) scheme~\cite{Denner:1999gp,
Denner:2005fg}. Thus, the CM scheme is the strategy of choice in \aNLOs\ for 
dealing with unstable particles, and we shall limit our discussion to it 
in the present paper.

The core idea of the CM scheme, which stems from the observations
related to eq.~(\ref{CMSpole}), is to modify the renormalisation conditions 
of the theory, yielding complex-valued renormalised parameters that include 
the masses of the unstable particles, and also potentially a subset of the
coupling constants. By construction, this procedure leaves unaltered all
algebraic relations realizing gauge invariance. At the same time, it
naturally regularises unstable-particle propagators, which assume
the same functional form as in eq.~(\ref{GF}).

At the LO, such a modification is rather innocuous, since the analytic
continuation of LO amplitudes, which involve only rational expressions of
kinematic invariants, is unambiguous. At the NLO, however, the extension of 
the on-shell renormalisation condition~\cite{Denner:1991kt} presents many
subtleties in Quantum Field Theory, such as the proof of perturbative
unitarity investigated in ref.~\cite{Bauer:2012gn,Denner:2014zga}, and its
applicability beyond NLO. In what follows, we shall limit ourselves to
discussing the more pragmatic concerns of assessing the correctness of the 
CM implementation in \aNLOs, as well as listing and providing the necessary 
ingredients for guaranteeing the formal NLO EW accuracy of our results.

\subsection{Complex mass scheme formulation\label{sec:cms_formula}}
Let us first point out that when referring generically to the CM scheme 
we actually understand the common properties of a class of schemes, whose 
members differ in the choice made for the independent input parameters.
This implies that, when only a specific member of this class is relevant,
we may characterise it more precisely as, for example, CM $G_\mu$ scheme,
or CM $\aem(m_Z)$ scheme, and so forth.

As was anticipated in sect.~\ref{CMSintro}, at the LO the use of the CM
scheme simply amounts to redefining the mass of all unstable particle 
fields, according to eq.~(\ref{CMSpole}):
\beq
m^2 = \bar{M}^2-i\bar{\Gamma}\bar{M}\,.
\label{complexMass}
\eeq
All derived parameters must then be expressed in terms of the complex masses
of eq.~(\ref{complexMass}) and of the independent inputs, thereby possibly
acquiring an imaginary part. In the SM, this is for example (but not 
exclusively) the case of $G_\mu$ (in the $\aem(m_Z)$ scheme), of $\aem$ 
(in the $G_\mu$ scheme), of the cosine of the Weinberg angle, and of the 
Yukawa couplings.

At the NLO, one needs to properly define the renormalisation conditions.
Given the similarities between the CM scheme and the standard on-shell
(OS henceforth) one, we start by recalling those that in the latter
are relevant to the self-energies. We start by introducing the mass
and wave-function counterterms:
\beqn
M^2&=&M_0^2-\delta M^2\,,
\nonumber\\*
Z&=&1-\delta Z\,,
\label{RenormalizedSelfEnergies}
\\*
\Sigma_\ren(p^2) &=&\Sigma_\unren(p^2) - \delta M^2 + 
\big(p^2-M^2\big)\delta Z\,,
\nonumber
\eeqn
where we have denoted by $\Sigma_\ren$ and $\Sigma_\unren$ the renormalised
and unrenormalised self-energy, respectively, of the unstable 
particle\footnote{Where possible, we shall adhere to the notation convention
established in eqs.~(\ref{complexMass}) and~(\ref{RenormalizedSelfEnergies}),
within which lowercase (uppercase) symbols are associated with Lagrangian
parameters in the CM (OS) scheme. Note that, in general, $\bar{M}\ne M$.}.
Since in the OS scheme the renormalised mass must remain real, the 
renormalisation conditions need only to take into account the real part 
of the self-energy\footnote{See footnote~\ref{ft:IRreg}, which is especially 
relevant to the evaluation of self energies on the mass shell.}:
\beqn
\Re\left[\Sigma_\ren(p^2)\right]\left |_{p^2=M^2} \right. &=& 0\,,
\label{OSrencon1}
\\
\lim_{p^2\rightarrow M^2} \frac{1}{p^2-M^2}\,
\Re\!\left[\Sigma_\ren(p^2)\right] &=& 1\,,
\label{OSrencon2}
\eeqn
which yield the following expressions for the counterterms:
\beqn
\Re\left[\Sigma_\ren(p^2=M^2)\right] = 0
\;\;\;\;\;\;&\Longrightarrow&\;\;\;\;\;\;
\delta M^2=\Re\left[ \Sigma_\unren(p^2 = M^2) \right]\,,
\label{OScntM}
\\*
\Re\left[\Sigma_\ren^\prime(p^2=M^2)\right]=0
\;\;\;\;\;\;&\Longrightarrow&\;\;\;\;\;\;
\delta Z = - \Re \left[\Sigma_\unren^\prime(p^2 = M^2) \right]\,,
\label{OScntZ}
\eeqn
where we have employed the usual shorthand notation:
\beq
\Sigma_{\ren/\unren}^\prime(q^2)=
\left.\frac{\partial\Sigma_{\ren/\unren}(p^2)}{\partial p^2}
\right|_{\,p^2=q^2}
\eeq
for any $q^2$.

We point out that the real part that appears in 
eqs.~(\ref{OSrencon1})--(\ref{OScntZ}) is trivial for stable particles 
(since for them the self energies are real quantities), but is necessary 
when dealing with unstable particles, and thus encompasses all situations.
Indeed, for unstable particles the one-loop $\Sigma$ develops an imaginary 
absorptive part that can be related, through the optical theorem, to the total 
LO decay width (see eq.~(\ref{SigtoGam})), and this happens independently 
of whether the free propagator is assigned a zero or a non-zero width. 
It is then the zero-width case that implies that the use of the real part 
is not academic, because the OS scheme can be employed to carry out 
NLO calculations\footnote{The explicit expressions for the NLO OS 
counterterms relevant to the EW sector of the SM can be found e.g.~in 
ref.~\cite{Denner:1991kt}.} for processes that either feature only 
stable particles, or where unstable particles are present but finite-width 
effects can be neglected -- the typical situation being that of a 
non-resonant $t$-channel exchange, for which the renormalisation 
conditions of eqs.~(\ref{OScntM}) and~(\ref{OScntZ}) are directly 
relevant (conversely, non-vanishing widths pose the problems already 
discussed in sect.~\ref{CMSintro}, thus rendering the OS scheme
an option not viable {\em in general} in this case). We shall further 
discuss the consistency of renormalisation procedures when setting
particle widths equal to zero in sect.~\ref{sec:zeroTopWidthNonZeroWwidth}.

We now turn to the CM scheme, which allows one to address the finite-width 
scenario. Fundamentally, in such a scheme one imposes on-shell-type 
renormalisation conditions that involve both the real and the imaginary part 
of self-energies. The ensuing UV counterterms feature an imaginary component
that order by order helps avoid any double counting between the non-zero 
width and the bubble insertions in internal propagators.
We write the analogues of eqs.~(\ref{RenormalizedSelfEnergies}) 
as follows, by taking eq.~(\ref{complexMass}) into account:
\beqn
\bar{M}^2-i\bar{\Gamma}\bar{M}\,\equiv\,
m^2&=&M_0^2-\delta m^2\,,
\nonumber\\*
z&=&1-\delta z\,,
\label{CMrenSig}
\\*
\Sigma_\ren(p^2) &=&\Sigma_\unren(p^2) - \delta m^2 + 
\big(p^2-m^2\big)\delta z\,.
\nonumber
\eeqn
One then generalises eqs.~(\ref{OScntM}) and~(\ref{OScntZ}) as follows:
\beqn
\Sigma_\ren(p^2=\bar{M}^2-i\bar{\Gamma}\bar{M}) = 0
\;\;\;\;\;\;&\Longrightarrow&\;\;\;\;\;\;
\delta m^2=\Sigma_\unren(p^2 = \bar{M}^2-i\bar{\Gamma}\bar{M})\,,
\label{CMOScntM}
\\*
\Sigma_\ren^\prime(p^2=\bar{M}^2-i\bar{\Gamma}\bar{M})=0
\;\;\;\;\;\;&\Longrightarrow&\;\;\;\;\;\;
\delta z = - \Sigma_\unren^\prime(p^2 = \bar{M}^2-i\bar{\Gamma}\bar{M}) \,.
\label{CMOScntZ}
\eeqn
These definitions, together with those relevant to coupling renormalisation
(which we do not report here, owing to their being functionally identical 
to those relevant to the OS scheme), ensure that by working in 
the CM scheme one can proceed analogously to what is done in other 
renormalisation schemes. For example, we observe that the renormalised
Lagrangian expressed in terms of renormalised parameters is equal to
the bare Lagrangian expressed in terms of bare parameters. Furthermore,
while all the derived parameters might acquire an imaginary part as was
already the case at the LO, they do so without spoiling the relationships
among them which are constrained by gauge invariance. Finally, we note
that~\cite{Stuart:1991xk,Sirlin:1991fd}:
\beq
M^2=\bar{M}^2+\bar{\Gamma}^2+\ord(\aem^3)=
\bar{M}^2+\ord(\aem^2)\,.
\label{MvsMbar}
\eeq
By writing the rightmost equality in eq.~(\ref{MvsMbar}) we have assumed 
\mbox{$\bar{\Gamma}=\ord(\aem)$}, which may appear in contradiction with 
the fact that $\bar{\Gamma}$ is regarded as an independent input parameter.
What we understand here is that such an input is associated with a given
value of $\aem$ (in general, the value of the independent coupling relevant 
to the chosen scheme). If we were to measure $\bar{\Gamma}$ for progressively 
smaller values of $\aem$, we should obtain a series tending linearly to zero. 
Likewise, this is the meaning we associate with saying that the difference 
between $M^2$ and $\bar{M}^2$ is of $\ord(\aem^2)$. Given the actual very 
slow running of $\aem$, these remarks do not play a major role in physics
simulations, but must be taken into account when considering some of the
mathematical properties of the CM scheme, which we have to exploit in its
\aNLOs\ implementation, and which we now turn to discussing.

In what follows, we shall need to consider some $\aem\to 0$ limits. It is 
therefore convenient to assume to work in the CM $\aem(m_Z)$ scheme, where 
$\aem$ is real. There is no loss of generality in this, since in other CM 
schemes one would instead consider the limit in the real-valued input 
coupling from which $\aem$ is derived. By using eqs.~(\ref{CMrenSig})
and~(\ref{CMOScntM}), one obtains:
\beq
\Im [m^2] = -\bar{\Gamma}\bar{M} =
- \Im [{\delta m^2}] = 
- \Im [{\Sigma_\unren(p^2 = \bar{M}^2-i\bar{\Gamma}\bar{M})}].
\label{CM_width_definition}
\eeq
We can now exploit eq.~(\ref{MvsMbar}), the optical theorem 
(eq.~(\ref{SigtoGam})) applied to the one-loop self-energy, and the fact 
that at this order the imaginary part of such self-energy is finite.
By expanding the rightmost term of eq.~(\ref{CM_width_definition})
around $p^2=M^2$, and by keeping only the dominant term, we obtain:
\beq
\bar{\Gamma}\bar{M}=
\Gamma^{(0)} M  + \ord(\aem^2)
\;\;\;\;\;\;\;\;\Longrightarrow\;\;\;\;\;\;\;\;
\bar{\Gamma}=
\Gamma^{(0)} + \ord(\aem^2)\,,
\label{CMwdefE}
\eeq
with $\Gamma^{(0)}$ the LO total decay width. Eq.~(\ref{CMwdefE}) complements
our previous comment about $\bar{\Gamma}$. Namely, in the context of numerical
tests not only this quantity must vanish linearly with $\aem$, but also
it must do so in such a way to guarantee that:
\beq
\lim_{\alpha\to 0}\;\frac{\bar{\Gamma}}{\Gamma^{(0)}} = 1\,.
\label{CMGalim}
\eeq
We shall show the importance of eq.~(\ref{CMGalim}) in 
appendix~\ref{sec:CMtest}, where we shall present a technique for 
some systematic and automated tests of CM-scheme implementations.

Equation~(\ref{MvsMbar}) and the ``perturbative'' vanishing of
$\bar{\Gamma}$ are also directly relevant to the proper definition
of the mass and wave-function counterterms, eqs.~(\ref{CMOScntM}) 
and~(\ref{CMOScntZ}) -- here, we shall consider only the former in order 
to give a definite example. The self-energy $\Sigma$ typically features 
branch cuts in the $p^2$ complex plane; hence, its analytical continuation 
must be performed with care, and we shall discuss this issue in 
sect.~\ref{AnalyticContinuationUVCT}. Here, we limit 
ourselves to pointing out that, because of the strict connection
between the OS and CM schemes, a natural cross-check on the final result
is to verify that the two expressions of the bare mass, as derived in the 
OS scheme (eq.~(\ref{OScntM})) and in the CM scheme (eq.~(\ref{CMOScntM})), 
differ by terms beyond the current perturbative accuracy (thus, in our 
case, beyond NLO). Specifically:
\beqn
&&\left( m^2+\delta m^2 \right) - \left( M^2 + \delta M^2\right) =
\nonumber \\*&&\phantom{aaa}
\Big( \bM^2-i\bGa\bM + \Sigma_\unren(p^2 = \bM^2-i\bGa\bM)\Big) - 
\Big(M^2+\Re\left[\Sigma_\unren(p^2 = M^2)\right] \Big) = 
\nonumber \\*&&\phantom{aaa}
\Big(\bM^2 - M^2\Big) + \Big(\Sigma_\unren(p^2 = \bM^2-i\bGa\bM)
-i\bGa\bM - \Re\left[\Sigma_\unren(p^2 = M^2)\right]\Big) 
\nonumber \\*
&&\phantom{aaaaaa}
\stackrel{{\rm NLO}}{=}
\;\;\ord(\aem^2)\,.
\label{RiemannSheetDefCheck}
\eeqn
The non-zero higher-order terms in the rightmost side of this equation
originate from the fact that, when truncating the perturbative series,
contributions beyond the current accuracy are still present, owing to
the non-zero widths of unstable particles in the CM scheme.
It is crucial to understand that, for the rightmost equality in
eq.~(\ref{RiemannSheetDefCheck}) to be correct, the properties of 
eq.~(\ref{MvsMbar}), eq.~(\ref{CMwdefE}), and $\Gamma^{(0)}=\ord(\aem)$ 
must hold. Such properties are enforced when one checks 
eq.~(\ref{RiemannSheetDefCheck}) numerically (possibly in an indirect 
manner, through the mass counterterms that enter UV-finite expressions).
Equation~(\ref{RiemannSheetDefCheck}) can only hold for one particular 
choice of Riemann sheet in the evaluation of the logarithms present in 
the self-energy functions. Contrary to what is customary done, we do 
not circumvent this issue by Taylor-expanding $\Sigma$, but instead 
evaluate the complete self-energies in the appropriate Riemann sheet. 

We conclude this section by mentioning the fact that, as far as the SM
is concerned, all renormalisation counterterms have been derived and 
implemented by hand in two \UFO\ models (distributed with the \aNLOs\
release), relevant to the CM $\aem(m_Z)$ and CM $G_\mu$ schemes (see 
sect.~\ref{sec:alphaComplexPhase} for further details), respectively. 
Thus, as is the case for all NLO-grade \UFO\ models, these include in
particular all UV and $R_2$ counterterms necessary for \MadLoop\ to compute 
an arbitrary one-loop amplitude\footnote{Instructions for obtaining a 
standalone code for such one-loop evaluations with \MadLoop\ can be found at: 
\url{http://cp3.irmp.ucl.ac.be/projects/madgraph/wiki/MadLoopStandaloneLibrary}.}.

\subsection{Definition of mass and wavefunction UV counterterms in the CM
  scheme\label{AnalyticContinuationUVCT}}
As was just discussed in sect.~\ref{sec:cms_formula}, UV-renormalisation
conditions in the CM scheme imply, in particular, the necessity of evaluating
massive-particle self energies at \mbox{$p^2=\bar{M}^2-i\bar{\Gamma}\bar{M}$}
(see e.g.~eqs.~(\ref{CMOScntM}) and~(\ref{CMOScntZ})). Thus, owing to the
presence on the extra (w.r.t.~that of the OS scheme) imaginary part
\mbox{$-\bar{\Gamma}\bar{M}$} in this mass-shell condition, analytical
continuation might lead one to compute logarithms (or any other multi-valued
function) in Riemann sheets different from the first. The goal of this
section is to discuss the strategy put in place in \aNLOs\ in order
to tackle this problem.

\begin{figure}[t]
\begin{center}
  \includegraphics[width=0.45\textwidth]{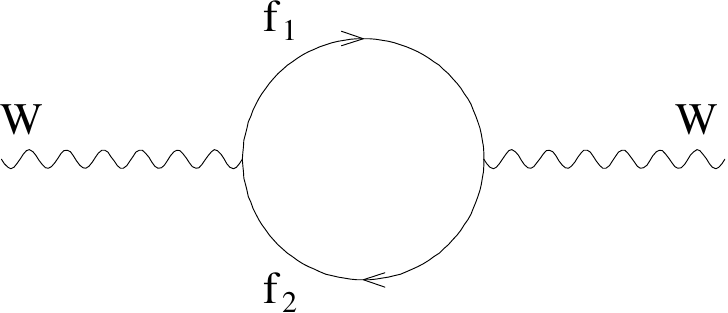}
\hskip 0.8truecm
  \includegraphics[width=0.45\textwidth]{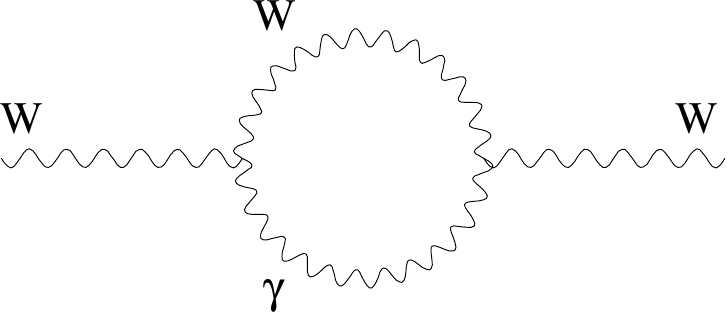}
\end{center}
\caption{\label{fig:bubbles} 
Fermion (left panel) and photon-$W$ (right panel) bubble graphs relevant
to the $W$ self-energy one-loop correction, that contribute to
$\Sigma_{\unren,T}^{f_1f_2}$ (see eq.~(\ref{SigffW})) and to
$\Sigma_{\unren,T}^{\gamma W}$ (see eq.~(\ref{SiggaWW})), respectively.
}
\end{figure}
One example relevant to the computation of a logarithm can be readily given 
by considering the one-loop fermion contribution to the $W$ self energy in 
the SM, depicted in the left panel of fig.~\ref{fig:bubbles}, whose 
unrenormalised transverse part reads as follows on the CM-scheme 
mass shell\footnote{We follow the \ml\ conventions, according to which
$\epsilon$-dependent finite pre-factors are not expanded, and are left
understood -- hence, no $\gamma_{\rm E}$ or $\log 4\pi$ terms appear
on the r.h.s..}:
\beqn
\Sigma_{\unren,T}^{f_1f_2}(\bMW^2-i\bGaW\bMW)&=&
\frac{\aem\NC^{f}\,(\bMW^2-i\bGaW\bMW)}{12\pi s_W^2}
\left[\frac{1}{\epsilon}+\frac{5}{3}\right.
\nonumber\\*&&\phantom{aaaaaaaaa}
-\left.\log_{-1}\!\left(-\frac{\bMW^2-i\bGaW\bMW}{\mu^2}\right)\right]\,.
\label{SigffW}
\eeqn
Here, $s_W$ is the complex-valued sine of the Weinberg angle, $\NC^{f}$
is the number of colours relevant to the $SU_c(3)$ representation to which
fermions $f_1$ and $f_2$ belong, and by $\mu$ we denote the mass scale
that needs to be introduced in the context of dimensional regularisation.
We have also employed the logarithm in the second negative Riemann
sheet, defined according to the general formula:
\beq
\log_k z=\log_0 z + 2\pi k i\,,\;\;\;\;\;\;\;\;\;\;
k\in{\mathbb Z}\,,
\label{eq:logdef}
\eeq
where by $\log_0{z}$ we have denoted the principal-value 
or first-Riemann-sheet definition of the logarithm with a branch
cut on the negative real axis:
\beq
\log_0{z}=\log\abs{z}+i\,{\rm arg}_0\,z\,,
\;\;\;\;\;\;\;\;\;\;
-\pi<{\rm arg}_0\,z\le\pi\,,
\eeq
whence:
\beq
\log_k{z}=\log\abs{z}+i\,{\rm arg}_k\,z\,,
\;\;\;\;\;\;\;\;\;\;
(2k-1)\pi<{\rm arg}_k\,z\le (2k+1)\pi\,.
\eeq
Since in practice we shall end up using only either first or second
Riemann sheet logarithms, we shall adopt the shorthand notation:
\beq
\log{z}\equiv \log_0{z}\,,
\;\;\;\;\;\;\;\;\;\;
\log_{\pm}z\equiv \log_{\pm 1}z\,.
\label{logshort}
\eeq
The use of $\log_{-}()$ in eq.~(\ref{SigffW}) instead of $\log()$ guarantees
that the result for $\Sigma_{\unren,T}^{f_1f_2}$ is obtained consistently
with the Feynman-propagator prescription it stems from. Ultimately,
together with similarly consistent treatments of the other contributions
to the self-energy (which we shall discuss below), this leads to 
eq.~(\ref{RiemannSheetDefCheck}) being fulfilled.

In general, the situation is complicated by the possible presence of 
multiple scales and/or unstable particles that circulate in the loops.
The solution proposed in ref.~\cite{Denner:2005fg} is to Taylor-expand
in $p^2$ around $\bM^2$ any self-energy $\Sigma(p^2)$ relevant to
the computation of UV counterterms. Given that the latter are evaluated
on the complex-mass pole of eq.~(\ref{complexMass}), this would be 
essentially identical to an expansion in $\bGa$ around $\bGa=0$,
were it not for the possible dependence on $\bGa$ due to sources different
from the mass shell of the particle whose self-energy is being computed
(we shall give below an example of this). We point out that, in order to
be NLO-accurate (assuming that \mbox{$\ord(\bGa/\bM)=\ord(\aem)$}), at least 
the first two terms of the Taylor expansion must be retained, in order to
get rid of a contribution $\sim 1/(i\bGa\bM)$ due to an intermediate
propagator being on-shell in the resonant region.

The Taylor-expansion technique has some tricky aspects, in particular due 
to the possible sensitivity to soft kinematics of loop propagators. In order 
to illustrate such aspects, we consider another contribution to the transverse 
part of the $W$ self energy, namely that due to a $(\gamma,W)$ loop (see 
the right panel of fig.~\ref{fig:bubbles}). For our current purposes,
it is sufficient to deal with the $B_0$ part of such a contribution:
\beq
\Sigma_{\unren,T}^{\gamma W}(\bMW^2-i\bGaW\bMW)\supset
\left.B_0\left(p^2,0,\bMW^2-i\bGaW\bMW\right)
\right|_{p^2\to\bMW^2-i\bGaW\bMW}\,.
\label{SiggaWW}
\eeq
We remind the reader that the second and third arguments of the
$B_0$ function correspond to the masses of the two particles that 
circulate in the loop. By using the explicit expression for
$B_0$ one obtains: 
\beqn
&&\frac{1}{i\pi^2}\!\left.B_0\left(p^2,0,\bMW^2-i\bGaW\bMW\right)
\right|_{p^2\to\bMW^2-i\bGaW\bMW}
\nonumber\\*&&\phantom{aaaa}
=\frac{1}{\epsilon}+2+
\log{\frac{\mu^2}{\bMW^2-i\bGaW\bMW}}
\nonumber\\*&&\phantom{aaaaaaaa}
+\left.\frac{\bMW^2-i\bGaW\bMW-p^2}{p^2}
\log{\frac{\bMW^2-i\bGaW\bMW-p^2-i0}{\bMW^2-i\bGaW\bMW}}
\right|_{p^2\to \bMW^2-i\bGaW\bMW}
\nonumber\\*&&\phantom{aaaa}
=\frac{1}{\epsilon}+2+
\log{\frac{\mu^2}{\bMW^2-i\bGaW\bMW}}\,.
\label{SiggaWWfull}
\eeqn
Conversely, the first-order Taylor expansion\footnote{It is understood
that the $B_0$ and $B_0^\prime$ functions on the r.h.s.~of eq.~(\ref{B0exp})
are evaluated in the first Riemann sheet, i.e.~the same relevant to setting
$\bGaW=0$ in the third arguments of those functions. This is in fact what 
renders the Taylor-expansion procedure meaningful from a physics viewpoint.
\label{ft:Tayexp}} (where 
\mbox{$B_0^\prime(p^2)=\partial B_0(p^2)/\partial(p^2/\bMW^2)$}):
\beqn
&&B_0\left(p^2,0,\bMW^2-i\bGaW\bMW\right)=
B_0\left(\bMW^2,0,\bMW^2-i\bGaW\bMW\right)
\label{B0exp}
\\&&\phantom{aaaaaaaaaaaaaaaa}
+\left(\frac{p^2-\bMW^2}{\bMW^2}\right)\!
B_0^\prime\left(\bMW^2,0,\bMW^2-i\bGaW\bMW\right)+
\ord\left(\left(\frac{p^2-\bMW^2}{\bMW^2}\right)^2\right)\,,
\nonumber
\eeqn
leads to (with $p^2\to\bMW^2-i\bGaW\bMW$):
\beqn
&&\Sigma_{\unren,T}^{\gamma W,(1)}(\bMW^2-i\bGaW\bMW)\supset
B_0\left(\bMW^2,0,\bMW^2-i\bGaW\bMW\right)
\label{SiggaWWexp1}
\\&&\phantom{aaaaaaaaaaaaaaaa}
-i\frac{\bGaW}{\bMW}\,B_0^\prime\left(\bMW^2,0,\bMW^2-i\bGaW\bMW\right)+
\ord\left(\left(\frac{\bGaW}{\bMW}\right)^2\right)\,,
\nonumber
\eeqn
Equation~(\ref{SiggaWWexp1}) explicitly shows one of the features
of the Taylor expansion mentioned above, namely that it is not
exactly equivalent to expanding in $\bGaW$, owing to the remaining
dependence upon this quantity in the third argument of $B_0$.
Equations~(\ref{SiggaWWfull}) and~(\ref{SiggaWWexp1}) lead to:
\beqn
&&\Sigma_{\unren,T}^{\gamma W}(\bMW^2-i\bGaW\bMW)-
\Sigma_{\unren,T}^{\gamma W,(1)}(\bMW^2-i\bGaW\bMW)\supset
\nonumber\\*&&\phantom{aaaaaa}
-\frac{\pi^2\bGaW\left[\bMW-i\bGaW\log
{\left(-\frac{i\bGaW}{\bMW-i\bGaW}\right)}\right]}{\bMW^2}+
\ord\left(\left(\frac{\bGaW}{\bMW}\right)^2\right)
\nonumber\\*&&\phantom{aaaaaa}
=\frac{\pi^2\bGaW}{\bMW}+\ord
\left(\left(\frac{\bGaW}{\bMW}\right)^2\right)\,.
\label{SiggaWWdiff1}
\eeqn
The difference in eq.~(\ref{SiggaWWdiff1}) is thus of NLO, which
implies that the Taylor expansion of eq.~(\ref{B0exp}) is not sufficient
to obtain a correct result at this perturbative order. The problem stems
from the derivative of the logarithm that appears in the third line of
eq.~(\ref{SiggaWWfull}), that induces a contribution proportional to
$1/(\bGaW\bMW)$ when $p^2=\bMW^2$, and ultimately from the fact that, 
in the $\bGaW\to 0$ region relevant to the Taylor expansion, one becomes
sensitive to the branch point of the logarithm. This is also the reason 
why the situation does not change if considering higher-order terms in the
Taylor expansion. Indeed, one can show that an $n^{th}$ order 
expansion leads to:
\beq
\Sigma_{\unren,T}^{\gamma W}(\bMW^2-i\bGaW\bMW)-
\Sigma_{\unren,T}^{\gamma W,(n)}(\bMW^2-i\bGaW\bMW)\supset
-\frac{\pi^2\bGaW}{n\bMW}+
\ord\left(\left(\frac{\bGaW}{\bMW}\right)^2\right).
\label{SiggaWWdiffn}
\eeq
The problem exposed above has indeed been pointed out in 
ref.~\cite{Denner:2005fg}, and a pragmatic solution proposed there
is that of adding the ``missing term'' (i.e.~the r.h.s.~of 
eq.~(\ref{SiggaWWdiffn})) back to the expanded counterterms.
It is clear that a straightforward Taylor expansion works for all
graphs where one does not cross branch cuts of logarithms and other
multi-valued functions when $\bGa\to 0$.
On the other hand, such an expansion requires that $\bGa$ can be viewed
as a small parameter (which is not necessarily equivalent to saying that
$\ord(\bGa)=\ord(\aem)$). This is certainly true in the SM, but not
necessarily so in arbitrary new-physics theories.

The obvious alternative to Taylor-expanding is to keep the full 
self-energy expressions, and to figure out the appropriate analytical
continuation of the OS results which becomes necessary when imposing
CM-scheme renormalisation conditions. This is equivalent to being
able to choose the appropriate Riemann sheets where the multi-valued 
functions that appear in the OS results are to be computed. Such an 
approach has the advantage of being immediately applicable to any 
BSM theory, regardless of its mass spectrum and width settings.

In order to pursue this strategy, we start by reminding the reader
that for the computation of mass and wavefunction UV counterterms
we are concerned only with $1$-point and $2$-point scalar integrals,
and the former do not pose any problems. As far as the latter are
concerned, the basic results we need to consider are the 
following~\cite{tHooft:1978jhc} (see also ref.~\cite{Ellis:2007qk}):
\beqn
\frac{1}{i\pi^2}\,B_0(p^2,0,0)&=&
\frac{1}{\epsilon}+2-\log{\frac{-p^2-i0}{\mu^2}}\,,
\label{B0zz}
\\
\frac{1}{i\pi^2}\,B_0(p^2,0,\mu_2^2)&=&
\frac{1}{\epsilon}+2+\log{\frac{\mu^2}{\mu_2^2}}}
+\frac{\mu_2^2-p^2}{p^2}\log{\frac{\mu_2^2-p^2-i0}{\mu_2^2}
\,,
\label{B0zm}
\\
\frac{1}{i\pi^2}\,B_0(p^2,\mu_1^2,\mu_2^2)&=&
\frac{1}{\epsilon}+2-
\log{\frac{p^2-i0}{\mu^2}}+\sum_{i=\pm}
\Big[\gamma_i \log{\frac{\gamma_i-1}{\gamma_i}}-\log{(\gamma_i-1)}\Big]\,,
\label{B0mm}
\eeqn
with
\beqn
\gamma_{\pm}&=&\half\left(\gamma_0\pm\sqrt{\gamma_0^2-4\gamma_1}\right)\,,
\label{gammapm}
\\
\gamma_0&=&1+\frac{\mu_1^2}{p^2}-\frac{\mu_2^2}{p^2}\,,
\label{gamma0}
\\
\gamma_1&=&\frac{\mu_1^2}{p^2}-\frac{i0}{p^2}\,,
\label{gamma1}
\eeqn
and where $p^2$ and $\mu_i$ ($i=1,2$) are the virtuality of the incoming
particle and the masses of the particles that circulate in the loop,
respectively. Equations~(\ref{B0zz})--(\ref{B0mm}) are derived by assuming 
$p^2$, $\mu_1$, and $\mu_2$ to be real-valued parameters; both the square
roots and the logarithms are meant to be evaluated in the first Riemann sheet.
These results can be thus directly applied to an OS-renormalisation procedure 
by setting $p^2=M^2$, $\mu_1=M_1$, and $\mu_2=M_2$, where $M$, $M_1$, and 
$M_2$ are the relevant OS masses.

The next step is to show that eqs.~(\ref{B0zm}) and~(\ref{B0mm}) still
apply to the case of complexified loop masses\footnote{This fact has
already been used in the case of the Taylor expansion, where it is
crucial (see footnote~\ref{ft:Tayexp}). A quick, if not fully rigorous,
way to argue that this is the case is to observe that, by giving a
{\em negative} imaginary part to the masses of the particles that circulate
in an one-loop bubble diagram, the signs of the imaginary parts of the 
$+i0$-regulated Feynman propagators do not change.} (in our notation, this
implies setting \mbox{$\mu_i^2=m_i^2\equiv \bM_i^2-i\bGa_i\bM_i$}).
This is immediately obvious in the case of eq.~(\ref{B0zm}), and requires
only a slightly more elaborate proof for eq.~(\ref{B0mm}), which we refrain
from showing here. We understand the assumptions $\bM>0$, $\bM_i>0$, 
and $\bGa_i>0$.

Therefore, in order to arrive at the results that one needs to use
in a CM-scheme computation, the only thing that remains to be done
is to show how to analytically continue the expressions given above for 
real and positive $p^2$ to complex $p^2$ values -- since we need to evaluate
the self energies at \mbox{$p^2=m^2\equiv\bM^2-i\bGa\bM$}. This then
entails figuring out whether at $p^2=\bM^2-i\bGa\bM$ the square roots
and logarithms can still be evaluated in the first Riemann sheet, or
other Riemann sheets are also necessary.

We proceed as follows. Let
\beq
f(T(\bga))
\label{elfun}
\eeq
denote any of the elementary functions that appear in 
eqs.~(\ref{B0zz})--(\ref{B0mm}), with explicit or implicit dependencies 
on the following quantities:
\beq
p^2=\bM^2-i\bga\bM\,;\;\;\;\;\;\;\;\;
\mu_i^2=\bM_i^2-i\bGa_i\bM_i\;\;\;{\rm or}\;\;\;
\mu_i^2=0\,;\;\;\;\;\;\;\;\;
\mu^2\,.
\eeq
Therefore, the only non-trivial cases are those where $f()$ is either
a logarithm or a square root. Furthermore, the latter case can be derived
from the former, since for any complex number $z$ and a given real number
$a$ one {\em defines}:
\beq
z^a=\exp\left(a\,{\rm Log}\,z\right)\,,
\label{powerdef}
\eeq
where ${\rm Log}\,z$ is the logarithm function whose codomain is the
full Riemann surface. Then:
\beq
{\rm if}~~{\rm Log}\,z\in{\cal R}_k\;\;\;\;\;\;
\Longrightarrow\;\;\;\;\;\;
{\rm Log}\,z=\log_k z\,,
\eeq
with $\log_k z$ given in eq.~(\ref{eq:logdef}), and ${\cal R}_k$ the 
$(\abs{k}+1)^{th}$ Riemann sheet (positive or negative) defined as the 
complex plane with a branch cut on the negative real axis. The fundamental 
property of the logarithm function needed here is the direct result of its 
analytic continuation along a curve $C$ and of the monodromy theorem, namely:
\beq
{\rm Log}\,z={\rm Log}\abs{z}+i\Big[\Im\left({\rm Log}\,z_0\right)+
\Delta_C {\rm Arg}\,z\Big]\,.
\label{Logprop}
\eeq
Here, $C$ is an oriented curve that starts from the arbitrary complex
number $z_0$ and arrives at the arbitrary complex number $z$. By
\mbox{$\Delta_C {\rm Arg}\,z$} we have denoted the variation of the
multi-valued ${\rm Arg}\,z$ function along $C$. By construction,
\mbox{${\rm Arg}\,z={\rm arg}_k\,z$} if \mbox{${\rm Arg}\,z\in {\cal R}_k$}.
One also has the properties:
\beq
\Delta_C {\rm Arg}\,z=\Delta_{C^\prime} {\rm Arg}\,z\,,
\;\;\;\;\;\;\;\;\;\;
\Delta_C {\rm Arg}\,z=-\Delta_{C^{-1}} {\rm Arg}\,z\,.
\label{DeltaArgprop}
\eeq
Here, $C^\prime$ is a curve with the same endpoints as $C$ that
can be obtained by a continuous deformation of $C$ without passing
through the origin $z=0$; $C^{-1}$ is the same curve as $C$, with
the opposite orientation.

Given eq.~(\ref{elfun}), we call the (infinite) set of complex
numbers:
\beq
{\cal T}=\Big\{\big(\Re T(\bga),\Im T(\bga)\big)\in {\mathbb C}\,\Big|\,
0\le\bga\le\bGa\Big\}\,,
\label{trjdef}
\eeq
the {\em trajectory} of $T$; in other words, we regard the argument of 
the function $f()$ in eq.~(\ref{elfun}) as a curve in the complex plane,
parametrised by $\bga$. The idea is then the following: to each elementary
function $f()$ that appears in eqs.~(\ref{B0zz})--(\ref{B0mm}), we apply
eq.~(\ref{Logprop}) by identifying $C$ with ${\cal T}$, $z_0$ with the
endpoint of ${\cal T}$ at $\bga=0$, and $z$ with the endpoint of ${\cal T}$ 
at $\bga=\bGa$. Therefore:
\begin{itemize}
\item
The endpoint of ${\cal T}$ at $\bga=0$ corresponds to the known 
first-Riemann sheet forms of eqs.~(\ref{B0zz})--(\ref{B0mm}),
whence \mbox{$\Im\left({\rm Log}\,z_0\right)={\rm arg}_0\,T(0)$}
in eq.~(\ref{Logprop}). The physical meaning of such an endpoint 
is an OS-scheme-like one if $\bGa_1=\bGa_2=0$. 
\item
The endpoint of ${\cal T}$ at $\bga=\bGa$ is relevant to the CM scheme.
Thus from eq.~(\ref{Logprop}) with the setting of the previous item 
we obtain ${\rm Log}\,T(\bGa)$, namely the analytical continuation
sought. If $f()\equiv\sqrt{()}$, eq.~(\ref{powerdef}) 
is finally used.
\end{itemize}
We further observe that, given eqs.~(\ref{B0zz})--(\ref{gamma1}), 
${\cal T}$ has a zero winding number\footnote{The procedure 
proposed here is valid also for trajectories with non-zero winding 
numbers; those simply entail the use of logarithms with values in Riemann
sheets different from the first or second ones.} around the origin. This 
implies that the final, analytically-continued form of $B_0$ will feature 
either first- or second- (both positive and negative) Riemann sheet logarithms.

We remark that multi-valued functions can be nested; an explicit
example is given in eq.~(\ref{B0mm}), where the arguments of the
logarithms feature a square root. In such a case, we proceed in an
iterative manner. Namely, we first deal with the inner function (the
square root), and determine whether the first or the second Riemann
sheet is to be used (for each $\bga\in [0,\bGa]$); then, we apply the 
procedure to the logarithms. We point out that, in this way, the trajectories 
relevant to the logarithms include the information on the Riemann sheet 
employed to evaluate the square roots which, among other things, guarantees 
that such trajectories are continuous.

\begin{figure}[t]
\begin{center}
  \includegraphics[width=0.6\textwidth,angle=270]{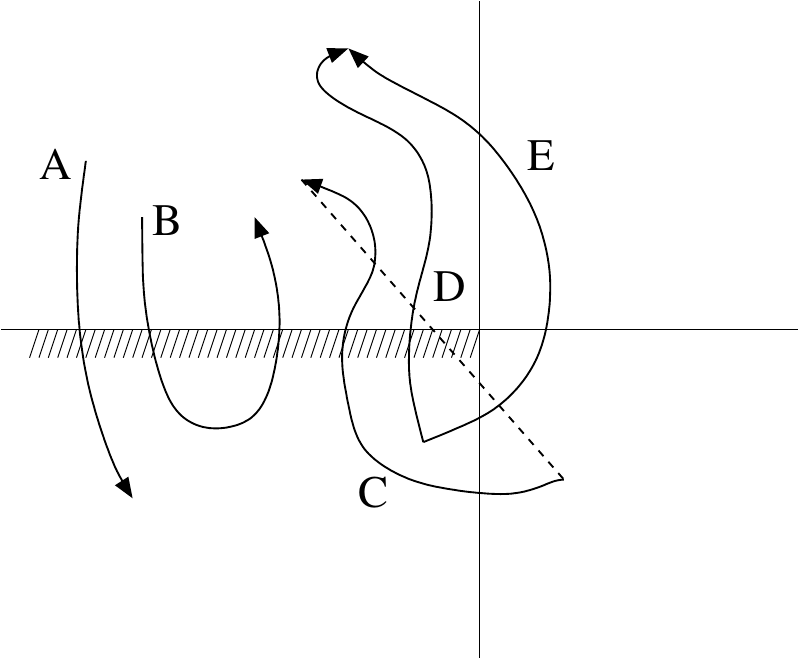}
\end{center}
\caption{\label{fig:trj} 
Examples of trajectories, for a function with a branch cut on the negative
real axis. The arrow indicates the direction one moves in from $\bga=0$
to $\bga=\bGa$. See the text for details.
}
\end{figure}
The procedure advocated above\footnote{To the best of our understanding,
this is analogous to the integral-level one of ref.~\cite{Passarino:2010qk},
at least for a class of trajectories (loosely identified as ``monotonic''
in the paper).} implies that, starting from $\bga=0$, we follow 
the trajectory by increasing $\bga$ till the endpoint
$\bga=\bGa$ is reached, counting the number of times in which we
cross the branch cut of $f()$; for each of them, the direction
in which the cut is crossed needs also to be considered. If we
denote by $n_{+-}$ ($n_{-+}$) the number of such crossings from
the positive to the negative (negative to positive) imaginary part,
then:
\beq
{\rm Log}\,T(\bGa)=\log_k T(\bGa)\,,
\;\;\;\;\;\;\;\;
k=n_{+-}-n_{-+}\,.
\label{kvsnpm}
\eeq
Equation~(\ref{kvsnpm}) is then used in eq.~(\ref{powerdef}) when
$f()\equiv\sqrt{()}$.

A few idealized examples of trajectories are given in fig.~\ref{fig:trj}, 
where the branch cut on the negative real axis is represented by a shaded
region, and to be definite we choose $f()=\log()$. According to 
eq.~(\ref{kvsnpm}), in the case of trajectory A we shall need to turn 
$\log()$ into $\log_+()$. For both trajectory C and D $\log()$ 
will have to be turned into $\log_-()$, while in the cases of 
trajectories B and E one will end up using the principal-value logarithm, 
although for different reasons (in the case of B, the branch cut is 
crossed twice, in opposite directions; no crossings occur for E).
Note the different behaviour of D and E, in spite of the fact that
these trajectories have the same endpoints.

Figure~\ref{fig:trj} can also be employed to sketch a couple of shortcuts
to following the complete trajectory (which is a numerically involved
procedure). We point out that, although neither of these can be used 
in a fully general case for reasons that will be explained below, 
they are nevertheless pedagogically useful; furthermore, they give 
the correct results in a very relevant physics situation (the SM). 
In the {\em endpoint method}, only
the endpoints of ${\cal T}$ are considered. If $\Re T(0)<0$ and
$\Re T(\bGa)<0$, then either $\log_+()$ (when $\Im T(0)>0$ and
$\Im T(\bGa)<0$) or $\log_-()$ (when $\Im T(0)<0$ and $\Im T(\bGa)>0$)
must be employed. Conversely, if either $\Re T(0)>0$ and $\Re T(\bGa)>0$,
or $\Im T(0)$ and $\Im T(\bGa)$ have the same sign, then $\log()$ is used 
instead. These criteria give the correct results for trajectories
A, B, and D. However, they also imply that there are cases, where both 
the real and the imaginary parts have opposite signs at the two endpoints, 
which cannot be addressed with this method -- an example is that of
trajectory C. Finally, in other cases (such as that of trajectory E)
the method just gives an incorrect result.
As a refinement of the endpoint method the {\em straight-trajectory method}
can be used. This entails replacing the trajectory with a segment
that connects the two endpoints, and then proceeding as before (which
is then simply equivalent to finding the intersection of such a 
segment with the real axis). Thanks to the leftmost identity in 
eq.~(\ref{DeltaArgprop}), this method is guaranteed to give the correct
result, provided that one is able to understand whether the continuous
deformation from the trajectory to the segment does not cross
the origin. This is the case for all of the trajectories in 
fig.~\ref{fig:trj}, except for E -- in the latter case, the 
straight-trajectory method would lead to an incorrect result.
It is important to note that both the endpoint and the straight-trajectory
methods cannot self-diagnose a failure; some further information on the 
complete trajectory is necessary. Thus, they cannot be reliably used in
the context of an arbitrary model with arbitrary parameter assignments.

The trajectory approach leads straightforwardly from eqs.~(\ref{B0zz})
and~(\ref{B0zm}) to the following results: 
\beqn
&&\phantom{aaaaaaaaa}\;
\frac{1}{i\pi^2}\,B_0(\bM^2-i\bGa\bM,0,0)=
\frac{1}{\epsilon}+2-\log_{-}{\frac{-\bM^2+i\bGa\bM}{\mu^2}},
\\
&&\frac{1}{i\pi^2}\,B_0(\bM^2-i\bGa\bM,0, \bM_2^2-i\bGa_2 \bM_2)=
\frac{1}{\epsilon}+2+\log{\frac{\mu^2}{\bM_2^2-i\bGa_2\bM_2}}
\\*&&\phantom{aaa}
+\frac{\bM_2^2-i\bGa_2\bM_2-\bM^2+i\bGa\bM}{\bM^2-i\bGa\bM}
\nonumber
\\*&&\phantom{aaaaaaaaaaa}\times
\left\{\begin{array}{ll}
\log_{-}\frac{\bM_2^2-i\bGa_2\bM_2-\bM^2+i\bGa\bM}{\bM_2^2-i\bGa_2 \bM_2}
&\;\;\;\;\Longleftrightarrow\;\;\;\;
\bM>\bM_2
\;\;\;{\rm and}\;\;\;\bGa\bM_2>\bGa_2\bM,
\\
\log\frac{\bM_2^2-i\bGa_2\bM_2-\bM^2+i\bGa\bM}{\bM_2^2-i\bGa_2 \bM_2}
&\;\;\;\;\phantom{\Longleftrightarrow}\;\;\;\;\;
{\rm otherwise}.
\\
\end{array}
\right.
\nonumber
\eeqn
Conversely, eq.~(\ref{B0mm}) renders an analytical formulation impractical
(although possible), and it is best to resort to numerical methods. In
what follows, we present sample results obtained by setting masses
and widths as specified in table~\ref{tab:trj} -- note that 
$\bGa=\bM$\footnote{The trajectory associated with any value
$\bGa<\bM$ is simply a subset of the trajectory relevant to $\bGa=\bM$.
This choice of a very large width value is thus simply a practical way
to address all situations of interest within a single study.}.
\begin{table}
\begin{center}
\begin{tabular}{ccccc}
\toprule
 & $\bM^2/\bM_1^2$ & $\bM_2^2/\bM_1^2$ & $\bGa_1/\bM_1$ & $\bGa_2/\bM_2$ \\
\midrule
A & 0.5  & 1 & 0.1 & 0.1 \\
B & 1.88 & 1 & 0.1 & 0.1 \\
C & 2.8  & 1 & 0.1 & 0.1 \\
D & 4.2  & 1 & 0.1 & 0.1 \\
E & 5.9  & 2 & 0.8 & 0.1 \\
\bottomrule
\end{tabular}
\caption{\label{tab:trj}
Mass and width settings relevant to illustrative studies of the 
trajectory method. For all of these cases we also set $\bGa=\bM$.
The trajectories we consider are dimensionless, hence there is no 
need to specify a reference mass scale. 
}
\end{center}
\end{table}
In fig.~\ref{fig:sqrtlog}, we show the trajectories of:
\beq
\gamma_0^2-4\gamma_1\,,
\;\;\;\;\;\;\;\;\;\;\;\;
\frac{\gamma_+-1}{\gamma_+}\,,
\label{sampletrj}
\eeq
which correspond to the argument of the square root that appears
in eq.~(\ref{gammapm}) and to the argument of one of the logarithms
in eq.~(\ref{B0mm}), respectively. The quantities $\gamma_0$, $\gamma_1$,
and $\gamma_+$ are defined in eqs.~(\ref{gamma0}), (\ref{gamma1}), 
and~(\ref{gammapm}), respectively.
\begin{figure}[t]
\begin{center}
  \includegraphics[width=0.45\textwidth]{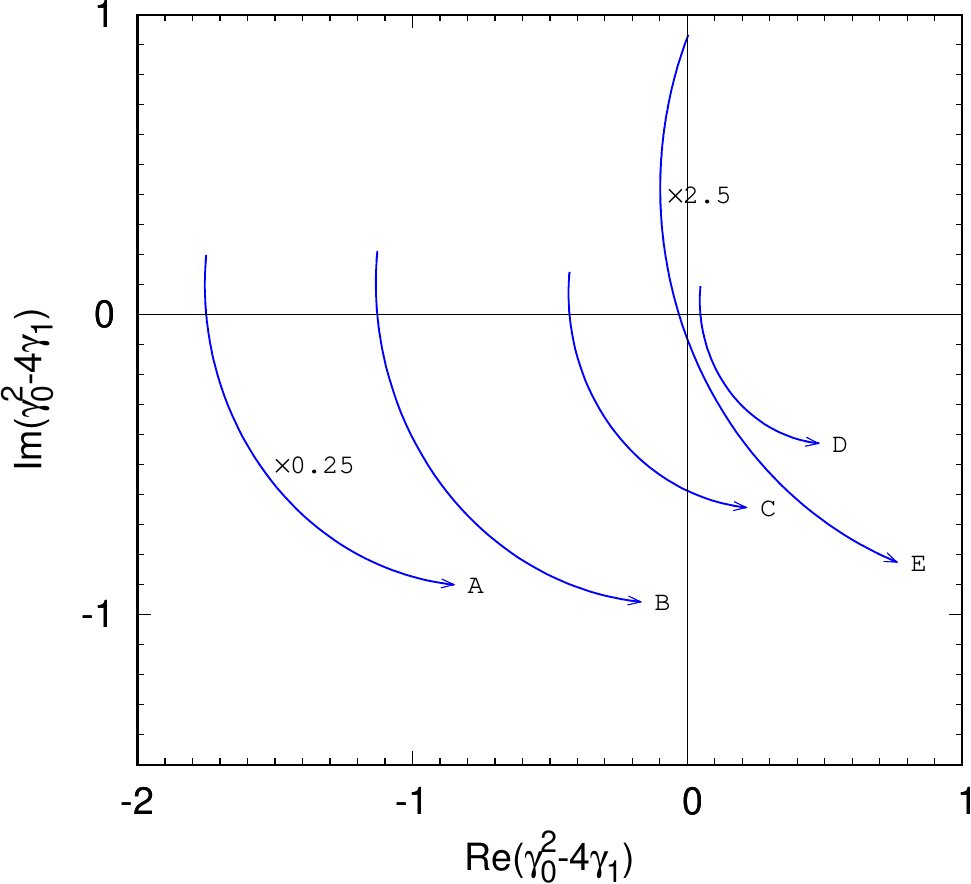}
\hskip 0.8truecm
  \includegraphics[width=0.45\textwidth]{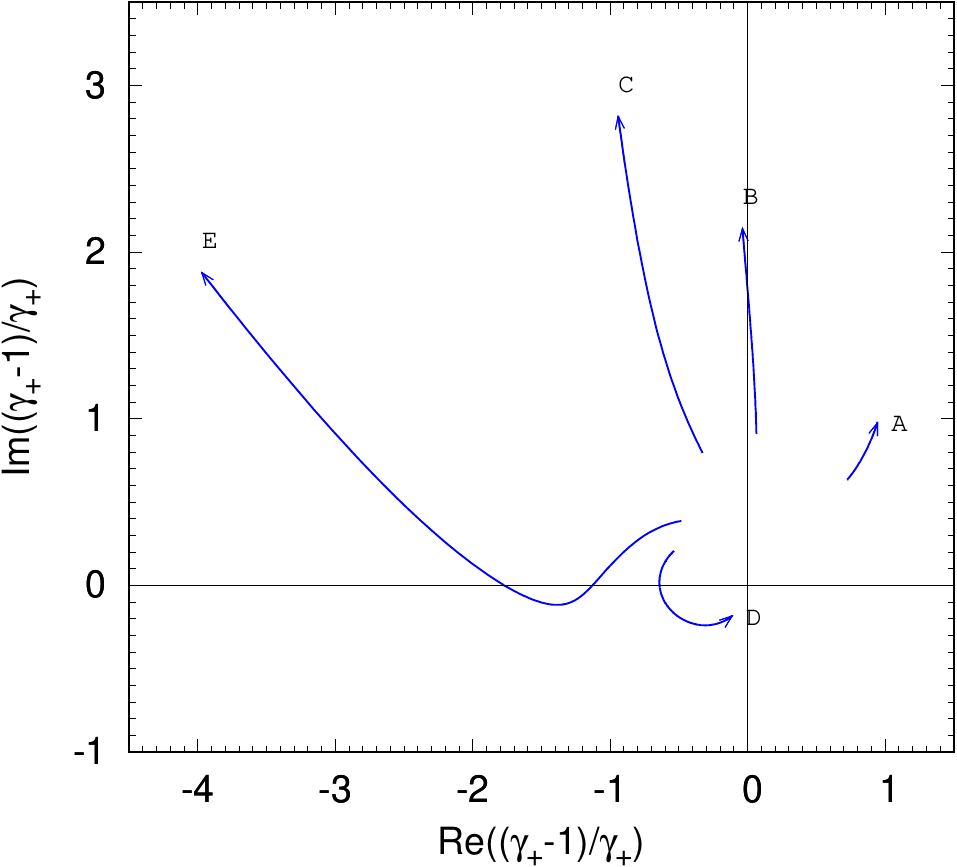}
\end{center}
\caption{\label{fig:sqrtlog} 
Trajectories of eq.~(\ref{sampletrj}) for the five configurations of 
table~\ref{tab:trj}. In the left panel, trajectories relevant to 
configurations A and E are multiplied by the numerical factors reported 
in the labels in order for them to fit into the layout, and to improve 
visibility; note that $\Re(T(\bga=0))>0$ for configuration E.
}
\end{figure}
As far as the left panel of fig.~\ref{fig:sqrtlog} is concerned, we see that 
for all configurations bar D the square root ends up being computed in the 
second Riemann sheet. The endpoint method gives the correct results for 
trajectories A, B, and D, fails for E, and cannot be applied in the case of 
C as explained before. The straight-trajectory method gives the correct 
results in all cases except for E, where it fails. Conversely, the 
trajectories of the argument of the logarithm in eq.~(\ref{sampletrj}), 
depicted in the right panel, show that in all cases $\log()$ must be used, 
except for D that requires the use of $\log_+()$. For all trajectories, both 
the endpoint and the straight-trajectory methods give the correct results.

\begin{figure}[t]
\begin{center}
  \includegraphics[width=0.45\textwidth]{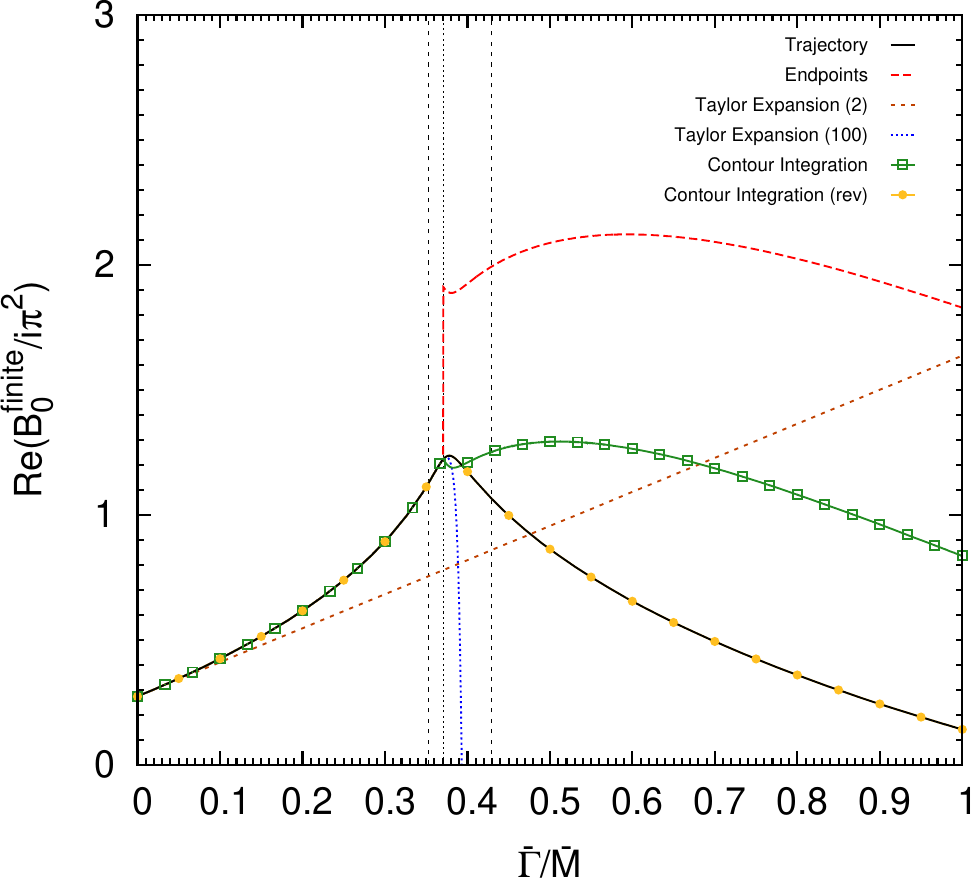}
\hskip 0.8truecm
  \includegraphics[width=0.45\textwidth]{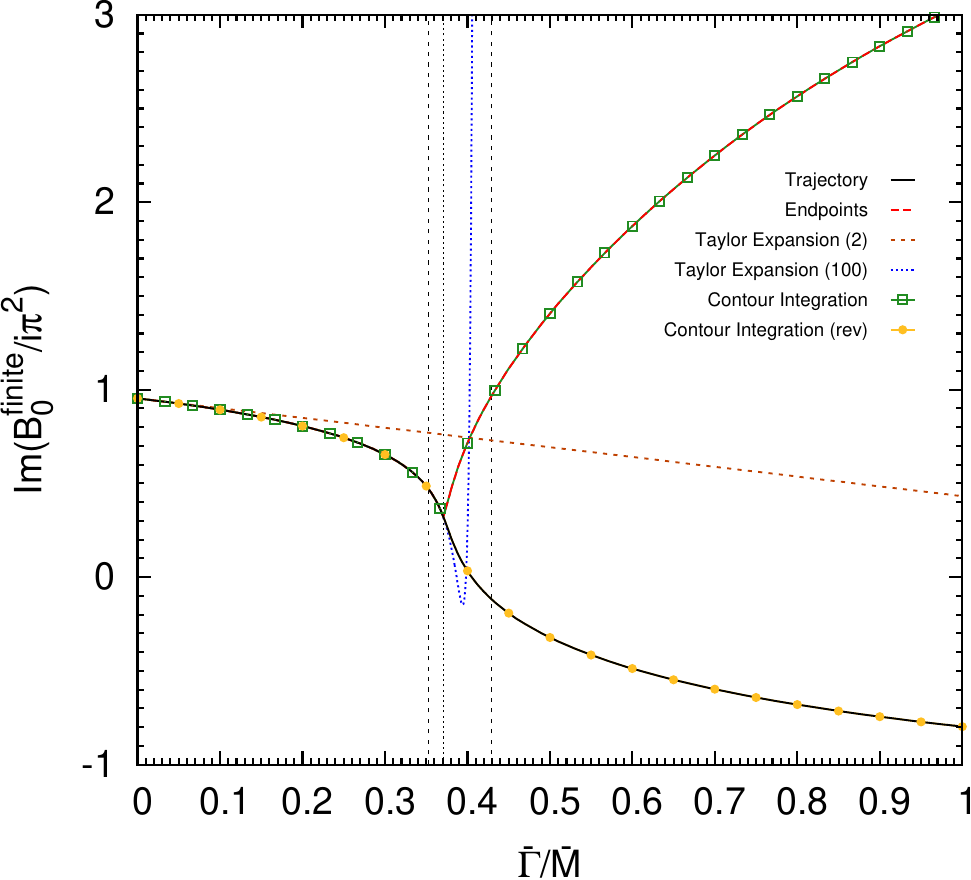}
\end{center}
\caption{\label{fig:B0res} 
Results for the real (left panel) and imaginary (right panel) part of the
UV-finite contribution to the r.h.s.~eq.~(\ref{B0mm}) (i.e.~the non-divergent
component of $B_0/(i\pi^2)$), as a function of $\bGa$. The other
widths and the masses are chosen according to configuration E in
table~\ref{tab:trj}. See the text for details.
}
\end{figure}
We now consider the case of configuration E in table~\ref{tab:trj} (except 
for $\bGa$, which is allowed to vary in the range \mbox{$0<\bGa<\bM$}),
and present the corresponding results for the finite part of $B_0$  in 
fig.~\ref{fig:B0res} (the left panel displays the real part of $B_0/(i\pi^2)$,
the right panel the imaginary one) as a function of $\bGa/\bM$; we 
have also set $\mu^2=\bM_1^2$. Several
curves appear in fig.~\ref{fig:B0res}, each of them obtained by computing
$B_0$ with different approaches and approximations. The black solid line
is the prediction of the trajectory method. The endpoint-method result
is shown as a red long-dashed curve. By Taylor expanding as explained
earlier in this section we obtain the brown short-dashed and blue dotted
curves, that correspond to keeping the first two\footnote{Point-wise
matrix-element comparisons between the trajectory-method results and 
those of the two-term Taylor expansion have been reported in sect.~7.3.2 
of ref.~\cite{Bendavid:2018nar}.} and the first one hundred terms 
in the expansion, respectively. Furthermore, the green line overlaid 
with open boxes is the result of the contour integration proposed in 
ref.~\cite{Passarino:2010qk}, while the yellow curve overlaid with full 
circles is analogous to the former, but with an alternative definition
of the contour (which we dub ``revised''). Although strictly necessary only 
for $\bGa>\bGa_{\!\!\sqrt{\;}}$, with $\bGa_{\!\!\sqrt{\;}}\simeq 0.37\bM$,
we have employed the revised contour also for $\bGa<\bGa_{\!\!\sqrt{\;}}$
where, as is apparent from fig.~\ref{fig:B0res}, its results are identical
to those of the contour of ref.~\cite{Passarino:2010qk}. The interested
reader can find more details on the revised contour in appendix~\ref{sec:trj}.
Finally, the vertical lines indicate the $\bGa$ values at which the trajectories
associated with configuration E cross the negative real axis. The leftmost 
and rightmost ones (at $\bGa\simeq 0.35\bM$ and $\bGa\simeq 0.43\bM$, 
respectively) correspond to the trajectory of \mbox{$(\gamma_+-1)/\gamma_+$},
while the central one (at $\bGa=\bGa_{\!\!\sqrt{\;}}$) corresponds to 
the trajectory of \mbox{$\gamma_0^2-4\gamma_1$}. 

We see from fig.~\ref{fig:B0res} that, for very small $\bGa$ values, all 
methods give identical results. By increasing $\bGa$, the Taylor expansion 
limited to two terms departs quickly from the other predictions, which are 
essentially on top of each other up to $\bGa\lesssim\bGa_{\!\!\sqrt{\;}}$. At
$\bGa=\bGa_{\!\!\sqrt{\;}}$ the endpoint method is unable to figure out 
correctly the appropriate Riemann sheet for the calculation of the square root
(see the left panel of fig.~\ref{fig:sqrtlog}), and thus differs significantly
from the full-trajectory result for all values $\bGa\ge\bGa_{\!\!\sqrt{\;}}$.
This is also the region where the Taylor expansion breaks down, irrespective
of how many terms are kept, and where the contour integration of
ref.~\cite{Passarino:2010qk} leads to a cusp-like behaviour in $\bGa$
for both the real and imaginary parts of $B_0$. Conversely, by using the
revised version of the contour, the resulting predictions are
in perfect agreement with those of the full-trajectory approach
described earlier.

We conclude this section with a few remarks. We point out that in
the SM trajectories are necessarily ``short'' (since $\bGa\ll\bM$),
and start from points located close to the real axis (since
$\bGa_i\ll\bM_i$). This is the reason why, apart from exceptional 
cases (such as that of eq.~(\ref{B0zm})), the Taylor expansion approach 
can be shown to be a reasonable one; likewise, the endpoint method is 
expected to yield the correct analytical continuation (also thanks to the 
fact that the SM mass spectrum 
is not close to being degenerate)\footnote{It is then clear that neither
the Taylor expansion nor the endpoint method should be blindly used in
a model with the same Lagrangian as the SM, but with parameters different
(i.e.~not within errors) w.r.t.~those measured in actual experiments.}. 
Thus, for the CM-scheme SM \UFO\ models shipped with \aNLOs,
we have limited ourselves to implementing the latter approach. We stress,
however, that it is crucial to maintain the full flexibility of
the trajectory method in view of simulations within models with
arbitrary assignments of masses and width, in keeping with the 
philosophy that underpins \aNLOs. In fact, it is relatively 
easy to automate the study of eq.~(\ref{trjdef}), e.g.~starting with 
a coarse discretisation of the parameter range \mbox{$\bga\in[0,\bGa]$},
which can be refined if necessary (for example when nearing the branch
cut). We present a couple of possible approaches in appendix~\ref{sec:trj},
one of which we have used to obtain the results presented in this
section.

\subsection{About the phase of $\alpha$ in the complex-mass $G_\mu$
  scheme\label{sec:alphaComplexPhase}}
As was already discussed, derived couplings can potentially become
complex when they depend on couplings and/or masses that acquire an imaginary
part as a result of the renormalisation conditions in the CM scheme. In 
particular, this is the case of the EW coupling $\aem$ in the CM $G_\mu$ 
scheme, which reads:
\beq
\aem = \frac{\sqrt{2} G^{(G_\mu)}_\mu m_W^2 (m_Z^2 - m_W^2)}{\pi m_Z^2}\,.
\label{aemGmusch}
\eeq
Here, we have denoted by $G^{(G_\mu)}_\mu$ the renormalized Fermi constant, 
whose superscript $^{G_\mu}$ reminds one that we are working in the CM $G_\mu$ 
scheme, where $G^{(G_\mu)}_\mu$ is {\em real-valued}. The complex values of 
the $Z$ and $W$ boson masses, defined according to eq.~(\ref{complexMass}),
imply that $\aem$ has a non-zero phase. 

In the context of NLO computations this fact is problematic, since it
may lead to uncancelled IR singularities in the context of subtraction
procedures. In order to illustrate this point with a simple example,
we consider selected virtual and real-emission contributions to the
$\NLOth$ corrections (i.e., of $\ord(\as^2\aem^2)$) to the process 
\mbox{$q\bq\to q^\prime\bq^\prime g$} --- such contributions are
shown in fig.~\ref{fig:qqqqgdiags}. At the level of matrix elements,
\begin{figure}[h!]
\centering
\subfloat[]{\includegraphics[trim=10 110 10 130,clip,scale=0.24]
    {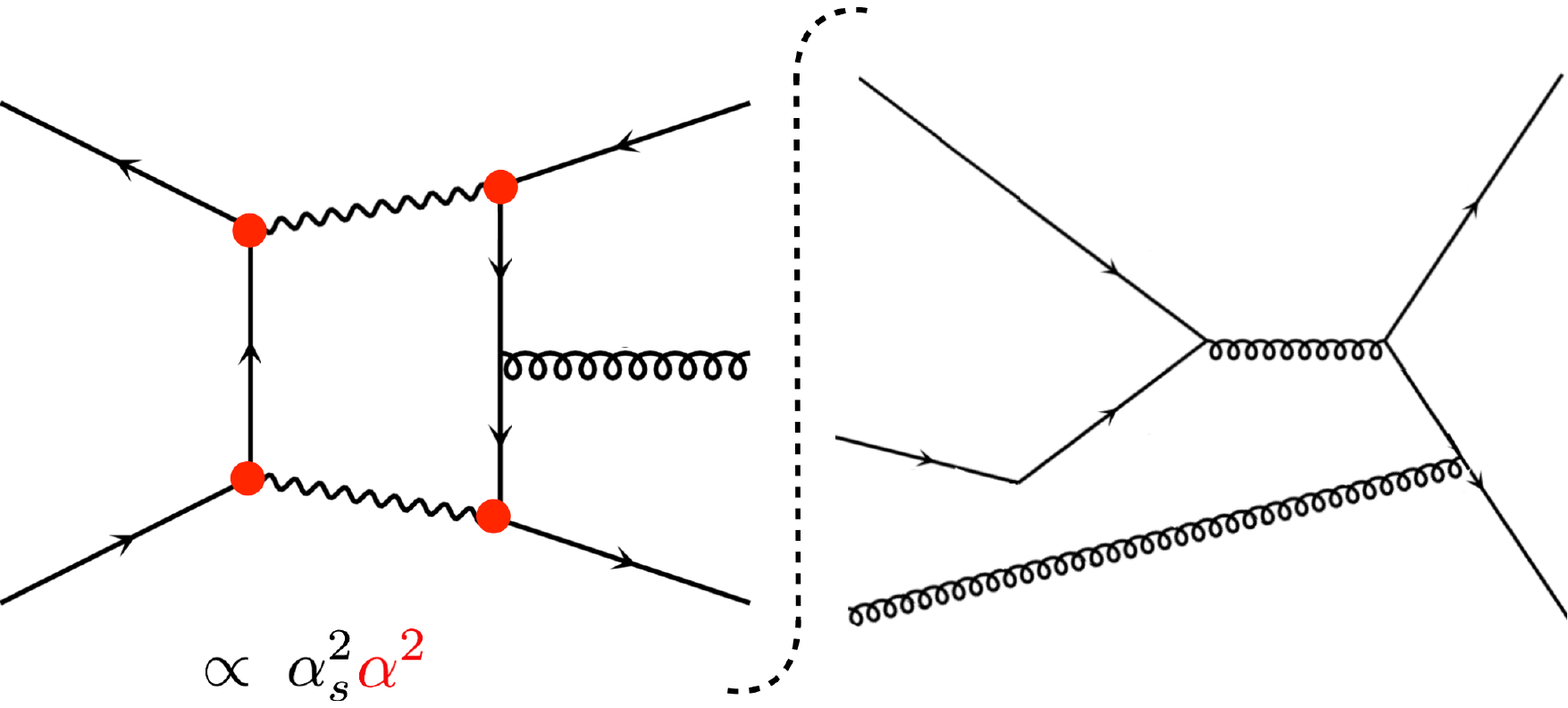}}\hskip 1.0truecm
\subfloat[]{\includegraphics[trim=10 110 10 130,clip,scale=0.24]
    {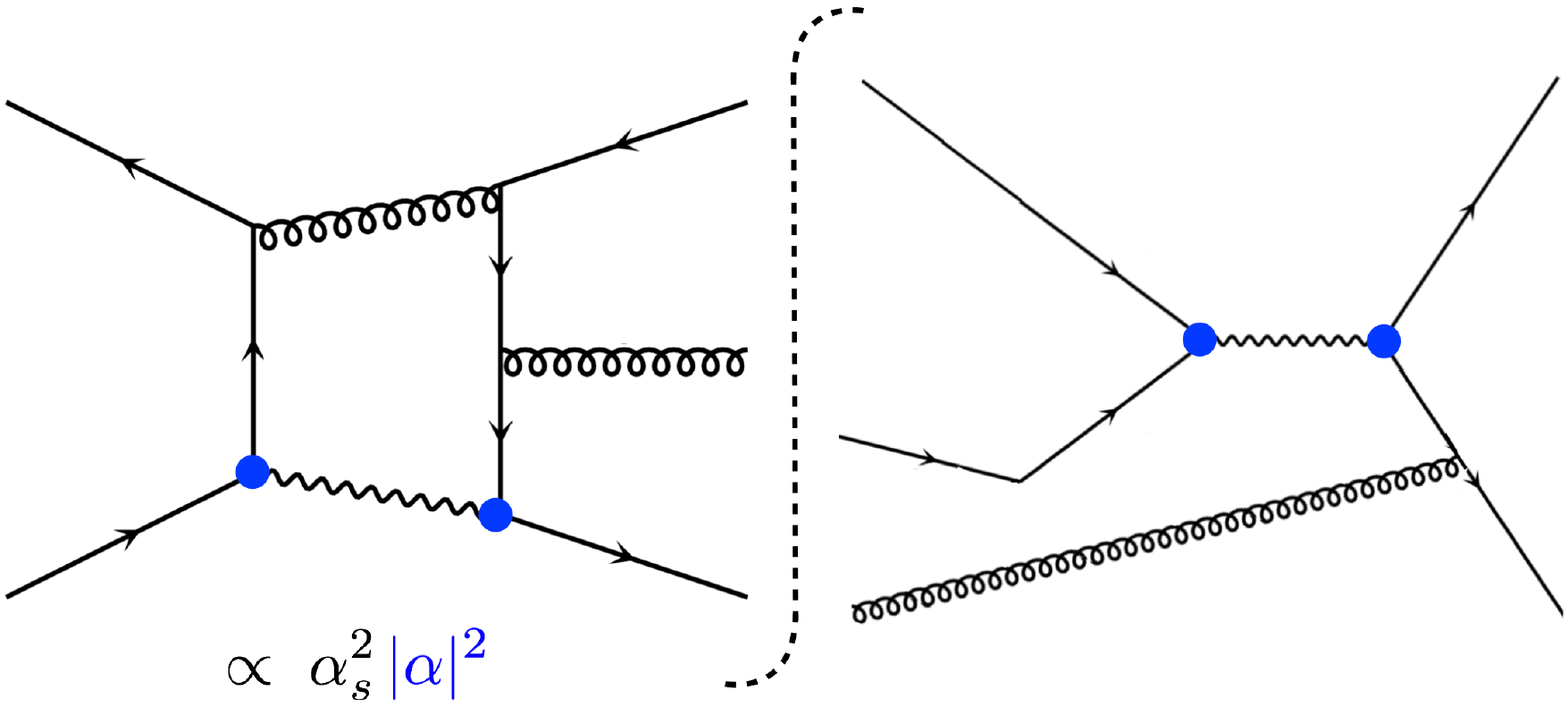}}\\
\subfloat[]{\includegraphics[trim=10 110 10 130,clip,scale=0.24]{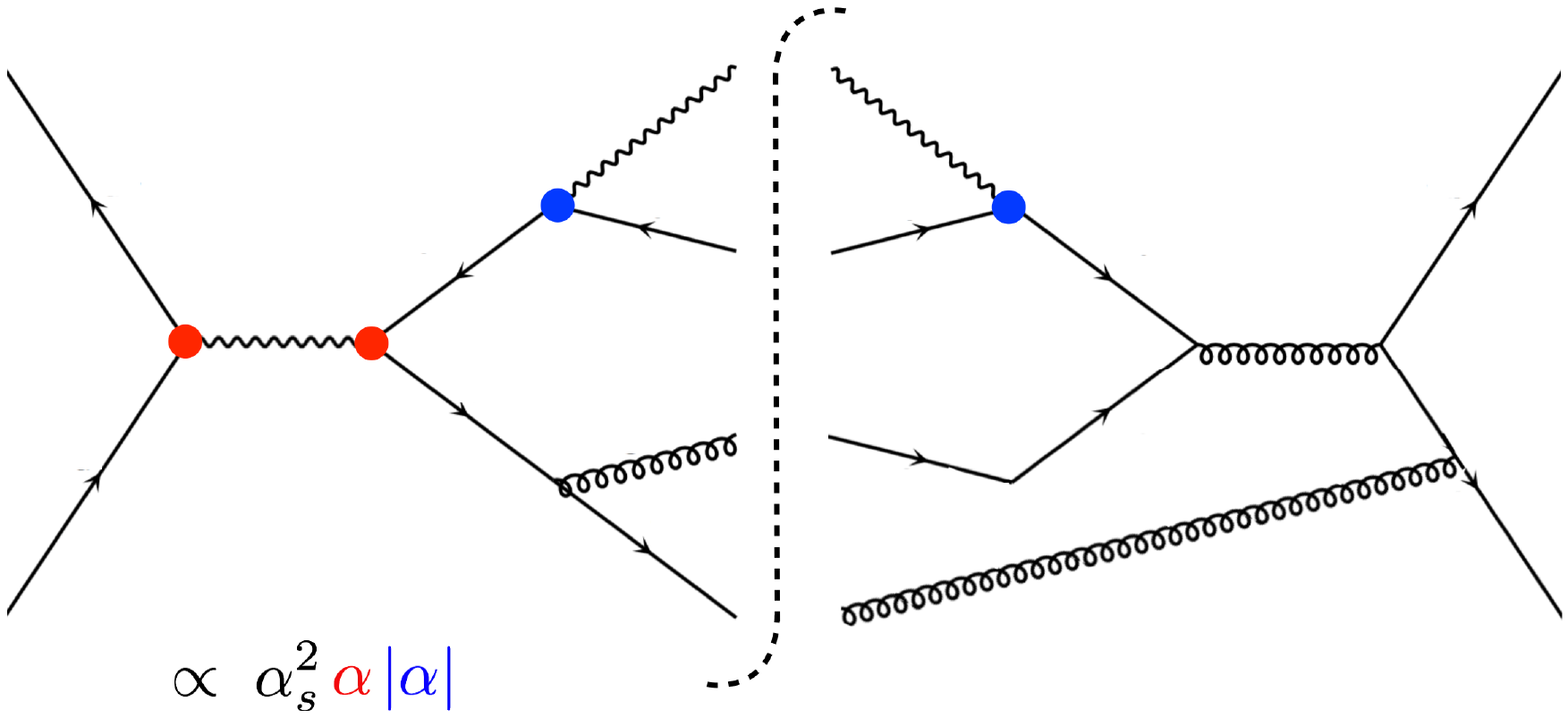}}
\caption{\label{fig:qqqqgdiags} 
Representative virtual (panels (a) and (b)) and real (panel (c)) contributions
to the $q\bq\to q^\prime\bq^\prime g$ partonic cross section. The figure
highlights the coupling-constant combinations relevant to each
contribution. See the text for details.}
\end{figure}
one will need to consider (see sect.~\ref{sec:ME}) the real parts of the 
diagrams of fig.~\ref{fig:qqqqgdiags}. We sketchily write these as follows:
\beqn
\ampsq_{(a)}&\propto&
\Re\left[\gs e^4 L_{(a)}\left(\gs^3 B_{(a)}\right)^\star\right]=
\as^2\Re\left[\aem^2 L_{(a)}B_{(a)}^\star\right]\,,
\label{c1}
\\
\ampsq_{(b)}&\propto&
\Re\left[\gs^3 e^2 L_{(b)}\left(\gs e^2 B_{(b)}\right)^\star\right]=
\as^2\abs{\aem}^2\Re\left[L_{(b)}B_{(b)}^\star\right]\,,
\label{c2}
\\
\ampsq_{(c)}&\propto&
\Re\left[\gs e^3 R_{(cL)}\left(\gs^3 e R_{(cR)}\right)^\star\right]=
\as^2\abs{\aem}\Re\left[\aem\, R_{(cL)}R_{(cR)}^\star\right]\,,
\label{c3}
\eeqn
for the (a), (b), and (c) panels respectively. We have denoted by 
$L_{()}$, $B_{()}$, $R_{()}$ the one-loop, Born-level, and real-emission-level
corresponding amplitudes. One expects that the IR singularities, of both 
QCD and QED origin, present in the real matrix elements of eq.~(\ref{c3})
will be ({\em partly}) cancelled by their counterparts in the virtual
matrix elements of eqs.~(\ref{c1}) and~(\ref{c2}). In subtraction 
procedures this cancellation is achieved by subtracting from the real matrix
elements some suitable local counterterms, which are then added back in
an integrated form to the virtual matrix elements. 
Such a form features a Born-level matrix element,
times a kernel that depends on kinematic quantities and on colour or charge 
factors, times the coupling relevant to the singular branching, i.e.~$\aem$ 
in the QED. Thus, for the process we are considering we can write this
integrated kernel symbolically as follows:
\beq
\int {\cal I}=\as^2\,\aem\,K\,
\Re\left[e^{2-n}B_1\left(e^n B_2\right)^\star\right]+\ldots\,,
\label{c4}
\eeq
with $0\le n\le 2$ a number that depends on the specific Born-level
amplitudes $B_1$ and $B_2$. It is therefore apparent that there is a 
mismatch due to $\alpha$ between eq.~(\ref{c4}), and eqs.~(\ref{c1}) 
and~(\ref{c2}). Even by bearing in mind that eq.~(\ref{c4}) has been
worked out by assuming that $\aem$ is real, the different ways in which
$\aem$ enters eqs.~(\ref{c1}) and~(\ref{c2}) imply that the generalisation
of eq.~(\ref{c4}) for complex-valued $\aem$ is involved, if at all possible.

We note that this problem is not specific to the computation of some
subleading term such as the $\NLOth$ of the present example. Suppose that 
one studies the same process, but in the pure-QED case, by replacing gluons
with photons. Equations~(\ref{c1})--(\ref{c4}) become then:
\beqn
\ampsq_{(a)}^{\QED}&\propto&
\abs{\aem}^3\Re\left[\aem L_{(a)}B_{(a)}^\star\right]\,,
\label{c1b}
\\
\ampsq_{(b)}^{\QED}&\propto&
\abs{\aem}^3\Re\left[\aem L_{(b)}B_{(b)}^\star\right]\,,
\label{c2b}
\\
\ampsq_{(c)}^{\QED}&\propto&
\abs{\aem}^4\Re\left[R_{(cL)}R_{(cR)}^\star\right]\,,
\label{c3b}
\\
\int {\cal I}^{\QED}&=&\aem\abs{\aem}^3 K\,
\Re\Big[B_1 B_2^\star\Big]\,.
\label{c4b}
\eeqn
While it appears that eq.~(\ref{c4b}) can be modified so as to have
the same singular behaviour as both eq.~(\ref{c1b}) and~(\ref{c2b}),
by doing so it would not any longer be equal to the integral of its 
local counterpart, which is employed in the subtraction of eq.~(\ref{c3b}).

In order to address the issue we have just discussed, what is commonly done 
in the CM $G_\mu$ scheme is to use, in place of of eq.~(\ref{aemGmusch}),
a real-valued $\alpha^{\rm Re}$:
\beq
\aem^{\rm Re}=\frac{\sqrt{2}}{\pi} G^{(G_\mu)}_\mu 
\left|\frac{m_W^2 (m_Z^2 - m_W^2)}{m_Z^2}\right|\,.
\label{aemGmusch2}
\eeq
We point out that the choice of an absolute value in eq.~(\ref{aemGmusch2})
is arbitrary, and e.g.~employing instead the real value of the r.h.s.~of
eq.~(\ref{aemGmusch}) is possible, as it would serve the same purpose. 
However, any such solution is not really appealing, in that it might spoil 
a gauge relation by higher-order terms. In particular, derived parameters 
other than $\aem$ can assume different values, depending on whether they
are expressed directly in terms of all of the input parameters of 
the CM $G_\mu$ scheme, or by using those inputs with the exception
of $G^{(G_\mu)}_\mu$, replaced\footnote{In other words, $G^{(G_\mu)}_\mu$ 
is employed here solely to define $\aem^{\rm Re}$, which is then 
utilised to evaluate all derived EW parameters.} by $\aem^{\rm Re}$. 
In view of the problem posed by IR singularities it is the latter 
approach (and not the former one, which is effectively equivalent to 
making use of an inconsistent set of input parameters) that we must 
adopt in order to prevent gauge relations from being broken.

Another way to remedy this situation, equivalent at the NLO to the
one we have just discussed, is that of turning $G^{(G_\mu)}_\mu$ into a 
complex parameter $\oG^{(G_\mu)}_\mu$, which differ from the former 
solely by a phase. Such a phase is chosen  so that eq.~(\ref{aemGmusch}), 
with \mbox{$G^{(G_\mu)}_\mu\to \oG^{(G_\mu)}_\mu$}, yields 
$\aem^{\rm Re}$ as given in eq.~(\ref{aemGmusch}). Explicitly:
\beqn
\oG^{(G_\mu)}_\mu &=& 
\aem^{\rm Re} \frac{\pi m_Z^2}{\sqrt{2} m_W^2 (m_Z^2 - m_W^2)} = 
G^{(G_\mu)}_\mu \left | \frac{m_W^2 (m_Z^2 - m_W^2)}{m_Z^2} \right | 
\frac{m_Z^2}{m_W^2 (m_Z^2 - m_W^2)}  
\nonumber \\*&=& 
G^{(G_\mu)}_\mu\;e\;^{ i {\rm Arg}
\left[\frac{m_Z^2}{m_W^2 (m_Z^2 - m_W^2)}\right] }
\label{aemGmuschExp}
\eeqn
This effectively amounts to working in another scheme of the CM class,
which we dub $\oG_\mu$ scheme and where one uses $\oG^{(G_\mu)}_\mu$
in place of $G^{(G_\mu)}_\mu$. The independent input parameter is still 
$G^{(G_\mu)}_\mu\equiv\abs{\oG^{(G_\mu)}_\mu}$, the Fermi constant.
In this way, $\oG^{(G_\mu)}_\mu$ becomes analogous to the complex
vector boson masses $m_Z$ and $m_W$, which acquire non-zero phases
given the real-valued inputs \mbox{$(\bar{M}_Z,\bar{\Gamma}_Z)$}
and \mbox{$(\bar{M}_W,\bar{\Gamma}_W)$} (with non-null widths).
By construction, and at variance with the $G_\mu$ scheme, in the
$\oG_\mu$ scheme one obtains exactly the same results independently
of whether one uses 
\mbox{$\{G^{(G_\mu)}_\mu,\bar{M}_Z,\bar{\Gamma}_Z,\ldots\}$} or
\mbox{$\{\aem^{\rm Re},\bar{M}_Z,\bar{\Gamma}_Z,\ldots\}$}
as input parameters.

In conclusion, the real-valued $\aem$ of eq.~(\ref{aemGmusch2})
emerges automatically in the $\oG_\mu$ scheme, where one reshuffles
higher-order terms in order to trade the phase of the coupling
$\aem$ for a suitable phase of $\oG^{(G_\mu)}_\mu$. As a result, 
one retains both order-by-order IR pole cancellations and gauge-invariant 
expressions of derived SM couplings. The numerical predictions presented
in sect.~\ref{sec:Res} have been obtained in the $\oG_\mu$ scheme,
i.e.~are identical to those of the $G_\mu$ scheme with $\aem=\aem^{\rm Re}$
used everywhere.

\subsection{About enforcing zero widths for final-state 
unstable particles\label{sec:zeroTopWidthNonZeroWwidth}}
In order to guarantee the unitarity of the $S$ matrix, 
one must consider stable particles in the final state (see 
e.g.~ref.~\cite{Veltman:1963th}). Therefore, the widths of potentially 
unstable final-state particles must be set equal to zero; in the CM scheme, 
this implies that the corresponding fields are then renormalised with the 
same conditions as in the OS scheme.
On the other hand, it has been already argued in sect.~\ref{CMSintro}
that, for processes that feature unstable-particle {\em resonances}, 
it is desirable to resum the relevant 1PI insertions, thereby
obtaining natural regulators in the propagators, which prevent the 
numerical integration from diverging in certain regions of the phase space.
Because of this, the situation arises where one must assign non-zero widths 
to the unstable particles that appear as resonances in a given process,
and simultaneously set equal to zero the widths of final-state unstable
particles. We shall denote the sets composed of such particles by
$P_{\rm res}$ and $P_{\rm fs}$, respectively. Whenever these two
sets $P_{\rm res}$ and $P_{\rm fs}$ overlap, one cannot perform 
the computation consistently, and the decays of the final-state unstable 
particles must be included\footnote{This statement relies on the
implicit assumption that the pole-mass regions of the resonances 
are integrated over.}. This is for example the case of the top 
quark in the process\footnote{It is crucial that EW corrections are driven
by the same interactions through which the top decays. As is well known,
pure-QCD NLO corrections to $tW$ associated production can be 
computed also with stable tops, although admittedly in a 
non-trivial manner.} $p p \rightarrow t W^- \bb$, or of the $Z$ boson in 
the process $p p \rightarrow Z e^+ e^-$.

In this section we discuss the case in which the sets $P_{\rm res}$ 
and $P_{\rm fs}$ do not overlap, and consider the question of whether
a consistent NLO computation within the CM scheme is possible. In order 
to be definite, it may be useful to refer to an explicit example, which
we shall take to be the $\NLOth$ corrections to the process $pp\to t\bt j$ 
(this corresponds to $\ord(\as^2\aem^2)$) due to real-emission contributions 
such as \mbox{$u\bar{d}\to t \bar{t} (W^{+*} \rightarrow)u\bar{d}$} or
\mbox{$u\bar{u}\to t \bar{t} (Z^*\rightarrow)d\bar{d}$} . The latter 
require non-zero $W$- and $Z$-boson widths, while the width of the top 
quark must be set equal to zero -- thus, \mbox{$P_{\rm fs}=\{t,\bt\}$} and 
\mbox{$P_{\rm res}=\{W^+,Z\}$}. Given that the top quark or antiquark 
does not appear as a resonance in this process, one could anticipate 
that this width assignment is not problematic. 

More generally, our conclusions are the following. In the context of
a CM-scheme calculation, it is always possible to impose OS-type 
renormalisation conditions on potentially-unstable particles, provided
that the latter {\em only} appear in the final state and not as
intermediate resonances that may go on shell.

The previous seemingly trivial statement crucially depends, among other 
things, on a correct interpretation of the operators $\tilde{\Re}$ and 
$\dagger$, as introduced and used in ref.~\cite{Denner:1991kt}. We shall 
comment on these aspects in what follows. For what concerns $\tilde{\Re}$, 
in particular, we point out that in a straightforward application of the
CM scheme this operator is simply irrelevant, while in the OS scheme,
because of the absence of any complex-valued couplings, the operator
$\Re$ is sufficient -- it is only in a ``mixed'' CM-OS setup that
$\tilde{\Re}$ becomes necessary.

Let us consider the particular contribution\footnote{At variance with 
what is done elsewhere and in order to avoid the proliferation of subscripts,
in this section we denote renormalised quantities by means of a hat symbol.}
$\hat{\Gamma}^{t}_{(W^+,b)}(p^2)$ to the top quark renormalised irreducible
two-point function $\hat{\Gamma}^{t}(p^2)$ due to a $(W^+,b)$ bubble graph.
We choose it to exemplify a general situation with analytical expressions 
which are as simple as possible -- the conclusions apply universally.
Such a contribution reads as follows:
\beqn
\hat{\Gamma}^{t}(p)\supset 
\hat{\Gamma}^{t}_{(W^+,b)}(p)&=& 
i(\slashed{p}-M_t)
\label{topRenormalizedPropagator}
\\*&+&
i\frac{1}{s_W^2} \left[
\slashed{p}w_-\hat{\Sigma}^{t,L}_{(W^+,b)}(p^2)
+\slashed{p}w_+\hat{\Sigma}^{t,R}_{(W^+,b)}(p^2)
+M_t \hat{\Sigma}^{t,S}_{(W^+,b)}(p^2)
\right],
\nonumber
\eeqn
where $s_W$ is the sine of the Weinberg angle, $M_t$ the (OS) top-quark mass 
(see sect.~\ref{sec:cms_formula}), and \mbox{$w_\pm=(1\pm\gamma_5)/2$}. 
The application of {\em OS-renormalisation conditions} for a massive 
fermion, sketched in sect.~\ref{sec:cms_formula} and reported explicitly 
in ref.~\cite{Denner:1991kt}, leads to the following expression for 
the top quark wavefunction renormalisation counterterm:
\beqn
\delta{Z_t^{L/R}} \supset \delta_{(W^+,b)}{Z_t^{L/R}} &=&
-\tilde{\Re}\left[\Sigma^{t,L/R}_{(W^+,b)}(M_t^2)\right]
\label{topWFCounterTerm}
\\*&-& 
M_t^2 \left. \frac{\partial}{\partial p^2} \tilde{\Re} 
\left[\Sigma^{t,L}_{(W^+,b)}(p^2)+\Sigma^{t,R}_{(W^+,b)}(p^2) 
+ 2 \Sigma^{t,S}_{(W^+,b)}(p^2) \right]\right|_{p^2=M_t^2}.
\nonumber
\eeqn
According to ref.~\cite{Denner:1991kt}, the notation $\tilde{{\Re}}$ 
understands that the real part must be taken of only the loop functions,
and not of the factorised couplings\footnote{Strictly speaking, this 
is actually a generalisation of what is {\em explicitly} done in 
ref.~\cite{Denner:1991kt}, where only CKM matrix elements are mentioned,
but it is obvious that the basic idea of ref.~\cite{Denner:1991kt} is 
respected. This still leaves one with some freedom in the definition
of $\tilde{{\Re}}$, which we exploit as detailed later -- see 
footnote~\ref{ft:Rtil}.}. 
The expressions for the unrenormalised self-energy functions 
$\Sigma^{t,S/L/R}_{(W^+,b)}$ (where we assume a diagonal CKM matrix) 
read as follows:
\beqn
\Sigma^{t,S}_{(W^+,b)}(p^2)&=&0\,,
\label{topSE1}
\\
\Sigma^{t,L}_{(W^+,b)}(p^2)&=& 
\frac{\aem}{2\pi} \frac{1}{s_W^2}\left[\frac{1}{4}
\frac{1}{\epsilon_{\rm UV}} + \kappa^{L}_{\Re,\rm fin}(p^2)
+i\kappa^{L}_{\Im,\rm fin}(p^2)\right]\,,
\label{topSE2}
\\
\Sigma^{t,R}_{(W^+,b)}(p^2)&=&
\frac{\aem}{2\pi}\frac{M_t^2}{m_W^2 s_W^2} 
\left[\frac{1}{8}\frac{1}{\epsilon_{\rm UV}}
+\kappa^{R}_{\Re,\rm fin}(p^2)+
i\kappa^{R}_{\Im,\rm fin}(p^2)\right]\,,
\label{topSE3}
\eeqn
where the coefficients $\kappa(p^2)$ are finite and real. As it has been
already noted in sect.~\ref{sec:cms_formula}, the optical theorem relates the
onshell coefficients \mbox{$\kappa^{L/R}_{\Im,\rm fin}(p^2=M_t^2)$} to the
perturbative LO width of the top quark, independently of any input value
$\Gamma_t$\footnote{\label{ft:Rtil}
Note that the finite widths of the particles running
in the loop can also contribute an imaginary part to the self-energies of
unstable particles. Such contributions are always beyond NLO accuracy and can
be finite, as in $\delta_{(W^+,b)}{Z_t}$, but also UV divergent as in
$\delta_{(W^+,\gamma)}{Z_{W^+}}$. We have chosen to define $\tilde{\Re}$ so
that it removes the {\em complete imaginary finite} part (irrespective of
its origin) of the loop functions defined with couplings factored out, but not
its singular part; thus retaining the exact IR poles cancellation at fixed
order.}.  This absorptive part of the loop function is precisely what is
intended to be removed by the $\tilde{\Re}$ operator in
eq.~(\ref{topWFCounterTerm}).

By substituting eqs.~(\ref{topSE1})--(\ref{topSE3}) into
eq.~(\ref{topWFCounterTerm}), we find the following expressions
for the top quark wavefunction counterterms\footnote{Notation-wise,
we have: $\delta Z_t^{L/R}({\rm our~work})\equiv\delta
Z_{33}^{L/R}({\rm ref.}$~{\protect\cite{Denner:1991kt}}$)$.
See also eq.~(\ref{Zdag}) and the discussion related to it.}:
\beqn
\delta_{(W^+,b)}{Z_t^{L}}&=&
-\frac{\aem}{2\pi}\frac{1}{s_W^2} \left[
\frac{1}{4\epsilon_{\rm UV}} +\kappa^{L}_{\Re,\rm fin}+
M_t^2\frac{\partial}{\partial p^2}
\left( \kappa^{L}_{\Re,\rm fin} + \frac{M_t^2}{m_W^2}\,
\kappa^{R}_{\Re,\rm fin} \right)\right]_{p^2=M_t^2}, 
\\
\delta_{(W^+,b)}{Z_t^{R}}&=& 
-\frac{\aem}{2\pi}\frac{M_t^2}{m_W^2s_W^2} \left[ 
\frac{1}{8\epsilon_{\rm UV}} +\kappa^{R}_{\Re,\rm fin}
+M_t^2\frac{\partial}{\partial p^2} 
\left(\frac{m_W^2}{M_t^2}\,\kappa^{L}_{\Re,\rm fin} + 
\kappa^{R}_{\Re,\rm fin} \right)\right]_{p^2=M_t^2}.\phantom{aaaaa}
\eeqn
In the context of our case-study computation of the $\rm{NLO}_3$ 
correction to \mbox{$pp\to t\bar{t}j$}, the OS renormalisation conditions
apply to the top quark only and the factors $\aem s_W^{-2}$ and 
$\aem\,m_W^{-2} s_W^{-2}$ take the complex values assigned by the CM 
renormalisation conditions. It is therefore crucial that the real part
$\tilde{{\Re}}$ does not apply to the factorised couplings, so as 
to guarantee the cancellation of UV divergent terms. However, the finite
imaginary absorptive part \mbox{$i \kappa^{L/R}_{\Im,\rm fin}(p^2=M_t^2)$} 
must not be included in the top mass OS counterterm $\delta_{(W^+,b)}{M_t}$, 
since the latter is the fixed-order counterpart of the width regulator that
appears in $G_R^{(t)}(p^2)$ and that is set to zero there just in force of
the OS conditions. In fact, this selective behaviour of the operator 
$\tilde{\Re}$ is mandatory in order to maintain gauge relations (see 
eqs.~(3.30) and~(3.31) in ref.~\cite{Denner:1991kt}), guaranteeing the 
fermion-flavour universality of the coupling constant $\aem$. We finally
point out that the same logic demands that the operator $\tilde{\Re}$ 
act as the identity on the CKM matrix elements.

The procedure exemplified above by the top quark is unchanged when applied 
to other unstable particles. In particular, the operator $\tilde{\Re}$ is 
exactly as that defined before, and it thus acts as the identity on any 
complex-valued couplings that self-energy diagrams may factorise, while
removing the imaginary finite part (see footnote~\ref{ft:Rtil}) of the
loop functions. It is interesting to consider the case in which 
this mixed OS and CM scheme is used for the vector bosons $W^\pm$ 
and $Z$ in the SM, for example in the process $p p \rightarrow Z e^+ \nu_e$.
Here, the $Z$ boson is renormalised in the OS scheme (whence the
$\tilde{\Re}$  operator removes the imaginary part of its self-energy
loop function), and the $W$ boson in the CM one. The Weinberg angle 
(which may be seen as just a representative of the class of all derived 
couplings) is then expressed directly in terms of the complex mass 
$m_W^2$  and of the {\em real-valued} mass $M_Z^2$ (see eq.~(\ref{cWexp})), 
and similary for its counterterm (see eq.~(\ref{cwCounterterm})). 

We point out that the observations made above for the top quark in 
principle apply also to the wavefunction renormalisation constants 
$\delta Z_q^{L/R}$ for a massless fermions $q$. However, their 
contributions to the one-loop matrix-element end up being added 
incoherently:
\beqn
2\Re\!\left[\delta Z_q \abs{\amp^{(0)}}^2\right]&=&
2\Re\!\left[\delta Z_{q}^L \abs{\amp^{(0)}_{L}}^2\right]+
2\Re\!\left[\delta Z_{q}^R \abs{\amp^{(0)}_R}^2\right]
\nonumber\\*&=&
2\abs{\amp^{(0)}_{L}}^2 \Re\!\left[\delta Z_{q}^L\right]+
2\abs{\amp^{(0)}_{R}}^2 \Re\!\left[\delta Z_{q}^R\right]\,,
\eeqn
with $\amp^{(0)}_{L/R}$ is the Born amplitude for the production of a
left- or right-handed massless quark $q$. Thus, the imaginary part of 
$\delta Z_q^{L/R}$ is irrelevant in practice for one-loop computations.  
This is not the case for massive fermions, because they can undergo a chirality
flip, whence the imaginary part of $\delta Z_t^{L/R}$ does contribute 
to the one-loop matrix elements of processes featuring final-state top quarks.
It is therefore important to perform any complex-conjugate operation
in a correct manner.

In view of what has just been discussed,
we conclude this section by remarking that we find the $\dagger$ operator
as introduced in ref.~\cite{Denner:1991kt} to be potentially misleading. 
To give one example, eq.~(3.21) therein reads:
\beq
\delta Z_{ii}^{L/R\,\dagger} = \delta Z_{ii}^{L/R}\,,
\eeq
and may appear inconsistent since in general
\mbox{$\delta Z_{ii}^{L/R\,\star} \ne \delta Z_{ii}^{L/R}$}. For this reason, 
we stress here that we understand any quantity $\delta Z_{ij}^{L/R\,\dagger}$ 
to be an independent renormalisation constant associated with the 
antiquarks $\bar{i}$ and $\bar{j}$, and not as a derived
parameter obtained by (possibly) transposed complex conjugate of 
$\delta Z_{ij}^{L/R}$.

To underscore this fact, we prefer to use the notation:
\beq
\delta Z^{L/R}_{\bar{i}\bar{j}}({\rm our~work}) \equiv 
\delta Z_{ij}^{L/R\,\dagger}
(\mbox{\textrm ref.~\protect\cite{Denner:1991kt}})\,.
\label{Zdag}
\eeq
Thus, \mbox{$\delta Z_{\bar{t}}({\rm our~work})\equiv
\delta Z_{33}^\dagger({\rm ref}.$~\protect\cite{Denner:1991kt}$)$},
whence $\delta Z_{\bar{t}}^{L/R}\ne\delta Z_t^{L/R\,\star}$. This makes 
it more explicit that the quantities $\delta Z^{L/R}_{\bar{i}\bar{j}}$ are
independently derived from the renormalisation conditions of the
{\em anti}-quark fields, identical to the quark ones but with the following
substitution of Dirac spinors: $u(p)\rightarrow \bar{v}(p)$ and
$\bar{u}(p)\rightarrow v(p)$.

\section{Results\label{sec:Res}}
In this section we present illustrative results relevant to several 
hadroproduction processes. We start by discussing in sect.~\ref{sec:setup}
the setup for the \aNLOs\ runs -- the model used, the input parameters,
and the definition of the basic observables. In sect.~\ref{sec:resNLOEW}
we give our predictions for total rates and some selected differential
distributions, by considering processes that display a significant 
diversity in their final states\footnote{As was mentioned at the end
of sect.~\ref{sec:gen}, features related to FFs are not publicly released
yet, and therefore we do not consider processes with tagged photons. Photons
are treated on equal footing with QCD partons (see appendix~\ref{sec:tech}),
and either enter a jet or help define a dressed charged lepton.}. For such 
processes, we compute
\beq
\Sigma_{\LOo}\,,\;\;\;\;\;\;\;\;\Sigma_{\NLOt}\,,
\label{proc1}
\eeq
that is, the leading LO and second-leading NLO (i.e.~NLO EW) contributions. 
We do not include the HBR cross sections (see sect.~\ref{sec:gen})
in the latter. Finally, in sect.~\ref{sec:restt} we study the production
of a $t\bt$ pair, possibly in association with either a heavy boson ($Z$, 
$W^+$, and $H$) or a light jet, again at both the fully-inclusive and the 
differential level. In the case of $t\bt(+B,j)$ production, we shall report 
the complete LO and NLO results, namely:
\beqn
&&\Sigma_{\LOi}\,,\;\;\;\;\;\;\;\;\;\;i=1,2,3,4\,,
\label{proc2}
\\*&&
\Sigma_{\NLOi}\,,\;\;\;\;\;\;\;\;i=1,2,3,4,5\,,
\label{proc3}
\eeqn
with $\Sigma_{\LOf}$ and $\Sigma_{\NLOfv}$ being relevant only to $t\bt j$
production. As for the case of eq.~(\ref{proc1}), HBR contributions have 
not been included.

We stress that the goal of this section is to document the achievement
of the full automation in \aNLOs\ of mixed-coupling calculations, and the 
typical usage of the code. Therefore, we did not choose our settings
on a process-by-process basis, as one would do if the primary concern
were a phenomenology study, but rather preferred to impose the same
conditions on all processes, so as to facilitate direct comparisons among 
them. Note that while the phenomenological impact of HBR cross 
sections should always be assessed (see e.g.~refs~\cite{Ciafaloni:2006qu,
Baur:2006sn,Bell:2010gi,Stirling:2012ak,Chiesa:2013yma,Frixione:2014qaa,
Frixione:2015zaa,Czakon:2017wor}, and refs.~\cite{Christiansen:2014kba,
Krauss:2014yaa,Bauer:2016kkv} for shower-type approaches)
in the context of a realistic analysis, this paper emphasises the 
technicalities of genuine NLO automated computations. Therefore HBR results,
that are obtained by means of IR-finite tree-level calculations which
are straightforward to carry out with LO-type runs in \aNLOs, are
unimportant here.

Likewise, as far as systematics are concerned one of the key aspects 
for the present paper is that of a good control over the numerical
accuracy of the final results. This is non-trivial, owing to the expected 
strong numerical hierarchy among the different contributions to the LO and 
NLO cross sections in a mixed-coupling expansion. We shall therefore limit
ourselves to reporting the errors associated with the Monte Carlo
integrations over the phase spaces. Thus we shall neglect the 
uncertainties due to renormalisation-scale, factorisation-scale,
and PDFs variations. However, we point out that such uncertainties
can be obtained with \aNLOs\ at no extra computational costs, thanks
to the reweighting procedure introduced in ref.~\cite{Frederix:2011ss}
which has been extended to the mixed-coupling scenario -- see
appendix~\ref{sec:RGE}.

\subsection{Setup of the calculation\label{sec:setup}}
The \UFO~\cite{Degrande:2011ua} model we use, {\tt loop\_qcd\_qed\_sm\_Gmu},
is included in the standard \aNLOs\ package. It contains the UV and $R_2$ 
counterterms relevant to NLO QCD and EW corrections, the latter in the
$\oG_\mu$ scheme (see sect.~\ref{sec:alphaComplexPhase}). 
The model features five massless quark flavours, sets the CKM matrix equal 
to the identity, and is compatible with the usage of both the OS and the 
CM schemes for all massive particles ($W^\pm$, $Z$, $H$, and 
top quark). Prior to process generation, we therefore execute
the following commands:

\vskip 0.25truecm
\noindent
~~\prompt\ {\tt ~set~complex\_mass\_scheme~true}

\noindent
~~\prompt\ {\tt ~import~model~loop\_qcd\_qed\_sm\_Gmu}

\noindent
~~\prompt\ {\tt ~define p = g d d\~{} u u\~{} s s\~{} c c\~{} b b\~{} a}

\noindent
~~\prompt\ {\tt ~define j = g d d\~{} u u\~{} s s\~{} c c\~{} b b\~{} a}

\vskip 0.25truecm
\noindent
The latter two instructions tell the code that the photon (denoted by the
symbol {\tt a}) must be considered part of the proton and of the jets, in 
keeping with the democratic approach -- more details are given in 
appendix~\ref{sec:tech}.

We consider $pp$ collisions at a center of mass energy of 13 TeV 
(LHC Run II). The values of masses and widths of the two heavy 
vector bosons are:
\beqn
&&
M_Z^\BW=91.1876~\gev\,,\;\;\;\;\;\;\;\;\;\;\;
\Gamma_Z^\BW=2.4952~\gev\,,
\label{ZmGBW}
\\*
&&
M_W^\BW=80.385~\gev\,,\;\;\;\;\;\;\;\;\;\;\;\;\;
\Gamma_W^\BW=2.0850~\gev\,.
\label{WmGBW}
\eeqn
defined, according to the PDG~\cite{Patrignani:2016xqp}, by using a
Breit-Wigner (OS-like) lineshape approach. The values in eqs.~(\ref{ZmGBW})
and~(\ref{WmGBW}) can be converted into those used as inputs to the CM 
scheme (generically denoted by $\bar{M}$ and $\bar{\Gamma}$, see 
eq.~(\ref{complexMass})) as follows~\cite{Stuart:1991xk,Sirlin:1991fd,
Kniehl:1998fn,Kniehl:2002wn,Grassi:2001bz,Passarino:2010qk,
Patrignani:2016xqp}:
\beq
\bM=M^\BW\Big/\sqrt{1+\big(\Gamma^\BW/M^\BW\big)^2}\,,\;\;\;\;\;\;\;\;
\bGa=\Gamma^\BW\Big/\sqrt{1+\big(\Gamma^\BW/M^\BW\big)^2}.
\label{BWtoCM}
\eeq
In the cases of the top quark and the Higgs boson, the input values
to the CM scheme are:
\beqn
&&
\bM_t=173.34~\gev\,,\;\;\;\;\;\;\;\;\bGa_t=1.36918~\gev\,,
\\
&&
\bM_H=125~\gev\,.
\eeqn
In the case of the Higgs, given the smallness of its SM width 
($\Gamma_H^{\rm SM}=4.07~\mev$) the choice $\Gamma_H=0$ is 
always an excellent approximation from a physics viewpoint, that allows
one to compute Higgs production in various channels without bothering with
its decays\footnote{This fact is often abused in BSM models with
large-width scalars, where a ``Higgs cross section'' is not necessarily
a meaningful concept.}. Therefore, we have always assumed $\Gamma_H=0$,
except for the $pp\to e^+\nu_e jj$ and $pp\to e^+e^-jj$ processes, where 
$\Gamma_H=\Gamma_H^{\rm SM}$ has been employed\footnote{These 
processes receive contributions from a subset of one-loop diagrams that 
feature an $s$-channel Higgs boson propagator. At the orders that we are
considering, eq.~(\ref{proc1}), such diagrams are interfered with
LO diagrams that do not have Higgs propagators, thus leading to
integrable singularities even when $\Gamma_H=0$. Therefore, we have
set $\Gamma_H\ne 0$ in these cases solely to improve the behaviour
of the numerical integration.}.

We remind the reader that the self-consistency of the calculation (see also
sect.~\ref{sec:zeroTopWidthNonZeroWwidth}) demands that, in processes where 
any massive particle ($W$, $Z$, top quark, and Higgs) is treated as stable, 
i.e.~is left undecayed, its width has to be set equal to zero, and it must
not appear as an intermediate resonance that may go on shell. In the case 
of the vector bosons, we set $\Gamma_Z^\BW=0$ and $\Gamma_W^\BW=0$ 
before employing eq.~(\ref{BWtoCM}) to obtain the CM-scheme inputs. 

The value of the Fermi constant as extracted from the muon decays, 
which is an input in the theory model adopted here, is set equal to:
\beq
G_\mu=1.16639\times 10^{-5}~\gev^{-2}\,.
\label{Gmu}
\eeq
The EW coupling $\alpha$ is derived from eq.~(\ref{Gmu}) and from the  $W$ 
and $Z$ CM-scheme masses as explained in sect.~\ref{sec:alphaComplexPhase} -- 
see in particular eq.~(\ref{aemGmusch2}) and the discussion given there. 
The PDFs are the central ones of the {\tt LUXqed\_plus\_PDF4LHC15\_nnlo\_100}
set~\cite{Butterworth:2015oua,Manohar:2016nzj}, extracted from
LHAPDF6~\cite{Buckley:2014ana} with number 82000; these are 
associated with
\beq
\as(\mz)=0.118\,,
\eeq
with a three-loop running (which is performed by LHAPDF6). 
The renormalisation and factorisation scales have been set as follows:
\beq
\muR=\muF=\frac{\Ht}{2}=\half\sum_i\sqrt{\pti^2+m_i^2}\,,
\eeq
where the sum is extended to all final-state particles at the parton
level (i.e.~prior to possibly combining them into jets or dressed leptons).

In order to define final-state objects in an IR-safe way, we apply the
following set of minimal selection cuts.
\begin{itemize}
\item All photons that are within $\Delta R_{f^{\pm}\gamma} \le 0.1$ of a 
  charged fermion (either a quark or a lepton), 
  are recombined with that fermion (by definition, $\Delta
  R_{ij}=\sqrt{(\Delta\phi_{ij})^2+(\Delta\eta_{ij})^2}$). If there is
  more than one charged fermion candidate, the photon is recombined
  with the one that yields the smallest $\Delta R_{f^{\pm}\gamma}$. The
  four-momentum assigned to each dressed fermion is the sum of the
  momenta of the original fermion and the photon. Hence, after
  recombination, these dressed fermions possibly feature a non-zero mass.
\item Charged dressed leptons are required to have $\pt(l^{\pm})>10~\gev$ 
and $|\eta(l^{\pm})|<2.5$. Pairs of opposite-sign same-flavour leptons must 
also satisfy minimum $\Delta R$-distance and invariant-mass cuts: 
$\Delta R_{l^{\pm}l^{\mp}}>0.4$ and $m_{l^{\pm}l^{\mp}}>30~\gev$.
\item Jets are reconstructed from all massless QCD partons
  (i.e.~gluons and quarks, the latter might be a recombined
  quark-photon pair) and from photons that were not recombined with
  charged fermions. They are defined with the anti-$\kt$
  algorithm~\cite{Cacciari:2008gp} (as implemented in
  \FJ~\cite{Cacciari:2011ma}) with $R=0.4$, and subject to the conditions
  $\pt(j)>30~\gev$, $|\eta(j)|<4.5$.
\item All jets so obtained are kept in the analysis (in other words,
photon-jets~\cite{Frederix:2016ost} are not tagged as such). Leptons
and jets are not required to be separated in $\Delta R$.
\end{itemize}

\noindent
The results that follow are obtained by integrating
simultaneously all of the contributions we have considered, in an
iterative manner till a target accuracy is reached or surpassed. 
By introducing the shorthand keywords \LO\ and \NLO\ as follows:
\beq
\LO\equiv\Sigma_{\LOo}\,,\;\;\;\;\;\;\;\;
\NLO\equiv\Sigma_{\LOo}+\Sigma_{\NLOt}
\label{NLOdefNLOEW}
\eeq
for the processes of sect.~\ref{sec:resNLOEW} (see eq.~(\ref{proc1})), and:
\beq
\LO\equiv\sum_{i=1}^{n_{\LO}}\Sigma_{\LOi}\,,\;\;\;\;\;\;\;\;
\NLO\equiv\sum_{i=1}^{n_{\LO}}\Sigma_{\LOi}+\sum_{i=1}^{n_{\NLO}}\Sigma_{\NLOi}
\label{NLOdeftt}
\eeq
for the processes of sect.~\ref{sec:restt} (see eqs.~(\ref{proc2}) 
and~(\ref{proc3}), with $n_{\LO}=3$ and $n_{\NLO}=4$ for $t\bt(+B)$ 
production ($n_{\LO}=4$ and $n_{\NLO}=5$ for $t\bt j$ production), 
the target accuracy is defined to be a relative MC error 
equal to $0.02\%$ on the total NLO cross section,
within cuts (NLO being that of eqs.~(\ref{NLOdefNLOEW})
and~(\ref{NLOdeftt})). We remind the reader that \aNLOs\ determines
automatically the number of phase-space points sampled in each MC
iteration to achieve a given accuracy -- more details on this item
can be found in sect.~2.4.3 of ref.~\cite{Alwall:2014hca}. The
overall runtime to compute all of the results presented in this section
is a couple of weeks on $\mathcal{O}(200)$~CPUs. 

As was mentioned in sect.~\ref{sec:gen}, \ml\ can choose dynamically
which integral-reduction module to employ; this is done in an order that
is pre-defined by the user. In the current \aNLOs\ version, the default
order is the following. One starts with double-precision
arithmetic, and \nin\ is used first. If the internal numerical stability 
tests are not passed (see sect.~2.4.2 of ref.~\cite{Alwall:2014hca}), 
\coll\ is used instead. If that also fails to provide a stable result, 
\ct\ is finally adopted. Yet another unstable result entails the use
of quadruple-precision computations, which are available in both \nin\ 
and \ct\ (called again by \ml\ in this order, if necessary). The \nin\ 
and \ct\ integral-reduction modules obtain the scalar integrals from
\OneLoop~\cite{vanHameren:2010cp}. For the processes 
considered in this paper, we have found that, with the accuracy
as specified above, an overall (i.e.~relevant to all of the processes
combined) negligible amount of $\mathcal{O}(100)$ phase-space points 
have required quadruple-precision calculations, all of which were then 
deemed to be numerically stable.

\subsection{NLO EW corrections\label{sec:resNLOEW}}
In this section we present the leading LO and second-leading NLO
(i.e.~NLO EW) results for a variety of processes, whose complete
list can be found in the first column of table~\ref{tab:1}. The
second column of that table reports instead the \aNLOs\ commands used 
to generate those processes. These adhere to the general syntax reported 
in sect.~\ref{sec:gen}; note in particular the keywords\footnote{The keyword 
{\tt [QED]} is conventional, and it implies that both electromagnetic and weak 
effects (i.e.~the complete $\ord(\aem)$ corrections) are taken into account, 
since both are included in the {\tt loop\_qcd\_qed\_sm\_Gmu} model. 
Restrictions to the QED-only or weak-only cases can be achieved by 
adopting a simpler theory model (for those processes for which these 
restrictions are meaningful).\label{ft:QED}} that determine which 
coupling-constant combinations are considered in the calculations, 
according to eqs.~(\ref{LOsynt}) and~(\ref{NLOsynt}).

We start by looking at fully-inclusive rates, obtained with the conditions
and acceptance cuts given in sect.~\ref{sec:setup}. The third and fourth
columns in table~\ref{tab:1} report the LO and NLO results, defined 
according to eq.~(\ref{NLOdefNLOEW}). The fifth column displays 
instead the fractional correction (given in percent) due to
NLO EW effects, i.e.:
\beq
\delta_{\rm EW}=\frac{\Sigma_{\NLOt}}{\Sigma_{\LOo}}=
\frac{\NLO}{\LO}-1\,.
\label{delEW}
\eeq
As was anticipated in sect.~\ref{sec:setup}, all of the uncertainties
reported in the three rightmost columns in table~\ref{tab:1} are MC
integration errors; as one can see, in absolute value they are 
almost always well below the per-mille level\footnote{The largest 
fractional error (still a mere $1.1\cdot 10^{-3}$ on the NLO cross section) 
affects $HHW^+$ production. We have checked that this is dominated by the 
opening at the NLO of a new $t$-channel configuration where an initial-state 
photon couples directly to the $W^+$. This channel is not mapped ideally 
by our phase-space parametrisation.}.

Table~\ref{tab:1} confirms the well-known fact that NLO EW effects
to fairly inclusive observables are mostly negative, and rather small
in absolute value (a few percent). Several (but not all) of the 
triple-boson production processes constitute an exception to the
latter rule, the largest correction being that associated with the 
$HHW^+$ final state, where \mbox{$\delta_{\rm EW}=-12.8\%$}. For
such processes, the four largest $\delta_{\rm EW}$'s are all negative
and of $\ord(-10\%)$, while the largest positive correction is
\mbox{$\delta_{\rm EW}=6.2\%$} in $W^+W^-W^+$ production.

It should not come as a complete surprise that the corrections
are typically larger for processes with large transferred momenta
and several final-state bosons,
such as in the case of triple heavy boson production. Indeed, two of the
most significant contributions to EW corrections, the EW Sudakov logarithms 
and the initial-state photon-induced terms, are relatively enhanced in this
kinematics regime and for these final states. In the differential 
distributions plotted and discussed below, we shall touch upon these 
effects in more detail.

\begin{landscape}
\begin{table}
\begin{center}
\begin{small}
\begin{tabular}{llr@{$\,\,\pm\,\,$}lr@{$\,\,\pm\,\,$}lr@{$\,\,\pm\,\,$}l}
\toprule
Process & Syntax & \multicolumn{4}{c}{Cross section (in pb)} &  \multicolumn{2}{c}{Correction (in \%)}\\
 && \multicolumn{2}{c}{LO} &  \multicolumn{2}{c}{NLO} & \multicolumn{2}{c}{} \\
\midrule
$pp \to e^+ \nu_e$ & \verb|p p > e+ ve QCD=0 QED=2 [QED]| & $5.2498 $&$ 0.0005\,\cdot 10^{3}$ & $5.2113 $&$ 0.0006\,\cdot 10^{3}$ & $-0.73 $&$ 0.01$\\
$pp \to e^+ \nu_e j$ & \verb|p p > e+ ve j QCD=1 QED=2 [QED]| & $9.1468 $&$ 0.0012\,\cdot 10^{2}$ & $9.0449 $&$ 0.0014\,\cdot 10^{2}$ & $-1.11 $&$ 0.02$\\
$pp \to e^+ \nu_e jj$ & \verb|p p > e+ ve j j QCD=2 QED=2 [QED]| & $3.1562 $&$ 0.0003\,\cdot 10^{2}$ & $3.0985 $&$ 0.0005\,\cdot 10^{2}$ & $-1.83 $&$ 0.02$\\
$pp \to e^+ e^-$ & \verb|p p > e+ e- QCD=0 QED=2 [QED]| & $7.5367 $&$ 0.0008\,\cdot 10^{2}$ & $7.4997 $&$ 0.0010\,\cdot 10^{2}$ & $-0.49 $&$ 0.02$\\
$pp \to e^+ e^-j$ & \verb|p p > e+ e- j QCD=1 QED=2 [QED]| & $1.5059 $&$ 0.0001\,\cdot 10^{2}$ & $1.4909 $&$ 0.0002\,\cdot 10^{2}$ & $-1.00 $&$ 0.02$\\
$pp \to e^+ e^-jj$ & \verb|p p > e+ e- j j QCD=2 QED=2 [QED]| & $5.1424 $&$ 0.0004\,\cdot 10^{1}$ & $5.0410 $&$ 0.0007\,\cdot 10^{1}$ & $-1.97 $&$ 0.02$\\
$pp \to e^+ e^- \mu^+ \mu^-$ & \verb|p p > e+ e- mu+ mu- QCD=0 QED=4 [QED]| & $1.2750 $&$ 0.0000\,\cdot 10^{-2}$ & $1.2083 $&$ 0.0001\,\cdot 10^{-2}$ & $-5.23 $&$ 0.01$\\
$pp \to e^+ \nu_e \mu^- \bar{\nu}_{\mu}$ & \verb|p p > e+ ve mu- vm~ QCD=0 QED=4 [QED]| & $5.1144 $&$ 0.0007\,\cdot 10^{-1}$ & $5.3019 $&$ 0.0009\,\cdot 10^{-1}$ & $+3.67 $&$ 0.02$\\
$pp \to H e^+ \nu_e$ & \verb|p p > h e+ ve QCD=0 QED=3 [QED]| & $6.7643 $&$ 0.0001\,\cdot 10^{-2}$ & $6.4914 $&$ 0.0012\,\cdot 10^{-2}$ & $-4.03 $&$ 0.02$\\
$pp \to H e^+ e^-$ & \verb|p p > h e+ e- QCD=0 QED=3 [QED]| & $1.4554 $&$ 0.0001\,\cdot 10^{-2}$ & $1.3700 $&$ 0.0002\,\cdot 10^{-2}$ & $-5.87 $&$ 0.02$\\
$pp \to H j j$ & \verb|p p > h j j QCD=0 QED=3 [QED]| & $2.8268 $&$ 0.0002\,\cdot 10^{0}$ & $2.7075 $&$ 0.0003\,\cdot 10^{0}$ & $-4.22 $&$ 0.01$\\
$pp \to W^+W^-W^+$ & \verb|p p > w+ w- w+ QCD=0 QED=3 [QED]| & $8.2874 $&$ 0.0004\,\cdot 10^{-2}$ & $8.8017 $&$ 0.0012\,\cdot 10^{-2}$ & $+6.21 $&$ 0.02$\\
$pp \to Z Z W^+$ & \verb|p p > z z w+ QCD=0 QED=3 [QED]| & $1.9874 $&$ 0.0001\,\cdot 10^{-2}$ & $2.0189 $&$ 0.0003\,\cdot 10^{-2}$ & $+1.58 $&$ 0.02$\\
$pp \to ZZZ$ & \verb|p p > z z z QCD=0 QED=3 [QED]| & $1.0761 $&$ 0.0001\,\cdot 10^{-2}$ & $0.9741 $&$ 0.0001\,\cdot 10^{-2}$ & $-9.47 $&$ 0.02$\\
$pp \to HZZ$ & \verb|p p > h z z QCD=0 QED=3 [QED]| & $2.1005 $&$ 0.0003\,\cdot 10^{-3}$ & $1.9155 $&$ 0.0003\,\cdot 10^{-3}$ & $-8.81 $&$ 0.02$\\
$pp \to HZW^+$ & \verb|p p > h z w+ QCD=0 QED=3 [QED]| & $2.4408 $&$ 0.0000\,\cdot 10^{-3}$ & $2.4809 $&$ 0.0005\,\cdot 10^{-3}$ & $+1.64 $&$ 0.02$\\
$pp \to HHW^+$ & \verb|p p > h h w+ QCD=0 QED=3 [QED]| & $2.7827 $&$ 0.0001\,\cdot 10^{-4}$ & $2.4259 $&$ 0.0027\,\cdot 10^{-4}$ & $-12.82 $&$ 0.10$\\
$pp \to HHZ$ & \verb|p p > h h z QCD=0 QED=3 [QED]| & $2.6914 $&$ 0.0003\,\cdot 10^{-4}$ & $2.3926 $&$ 0.0003\,\cdot 10^{-4}$ & $-11.10 $&$ 0.02$\\
$pp \to t \bar{t} W^+$ & \verb|p p > t t~ w+ QCD=2 QED=1 [QED]| & $2.4119 $&$ 0.0003\,\cdot 10^{-1}$ & $2.3025 $&$ 0.0003\,\cdot 10^{-1}$ & $-4.54 $&$ 0.02$\\
$pp \to t \bar{t} Z$ & \verb|p p > t t~ z QCD=2 QED=1 [QED]| & $5.0456 $&$ 0.0006\,\cdot 10^{-1}$ & $5.0033 $&$ 0.0007\,\cdot 10^{-1}$ & $-0.84 $&$ 0.02$\\
$pp \to t \bar{t} H$ & \verb|p p > t t~ h QCD=2 QED=1 [QED]| & $3.4480 $&$ 0.0004\,\cdot 10^{-1}$ & $3.5102 $&$ 0.0005\,\cdot 10^{-1}$ & $+1.81 $&$ 0.02$\\
$pp \to t \bar{t} j$ & \verb|p p > t t j QCD=3 QED=0 [QED]| & $3.0277 $&$ 0.0003\,\cdot 10^{2}$ & $2.9683 $&$ 0.0004\,\cdot 10^{2}$ & $-1.96 $&$ 0.02$\\
$pp \to jjj$ & \verb|p p > j j j QCD=3 QED=0 [QED]| & $7.9639 $&$ 0.0010\,\cdot 10^{6}$ & $7.9472 $&$ 0.0011\,\cdot 10^{6}$ & $-0.21 $&$ 0.02$\\
$pp \to t j$ & \verb|p p > t j QCD=0 QED=2 [QED]| & $1.0613 $&$ 0.0001\,\cdot 10^{2}$ & $1.0539 $&$ 0.0001\,\cdot 10^{2}$ & $-0.70 $&$ 0.02$\\
\bottomrule
\end{tabular}
\caption{\label{tab:1}\protect
Processes considered in sect.~\ref{sec:resNLOEW}. The second column reports
the \aNLOs\ syntax used to generate them. The third and fourth columns display
the fully-inclusive results for the quantities defined in 
eq.~(\ref{NLOdefNLOEW}). The fifth column shows the results for the fractional
correction defined in eq.~(\ref{delEW}). All uncertainties are due to
MC integration errors.}
\end{small}
\end{center}
\end{table}
\end{landscape}

We now turn to presenting a few sample differential distributions. 
We have collected in a single figure the results relevant to processes 
characterised by similar final states. All of the figures have
the same layout, constituted by a main panel and an inset.
In the main panel, LO and NLO results (defined according to
eq.~(\ref{NLOdefNLOEW})) are shown as dashed and solid histograms,
respectively. The histograms are normalised so that the content of
each bin is the total cross section in that bin. The inset displays
the predictions for \mbox{$1+\delta_{\rm EW}$}, which is therefore
equal to the ratio of the NLO over LO results presented in the main panel.

\vskip 0.2truecm
\noindent
{\bf $\blacklozenge$ Vector boson (plus jets) production}

\noindent
In left panel of fig.~\ref{fig:Vjets} the scalar sum of transverse
momenta, $\Ht$, is shown for $W^{+^*}$ production, possibly in association
with jets. The processes we have considered are:
\beq
pp\longrightarrow e^+ \nu_e\,,
\;\;\;\;\;\;\;\; 
pp\longrightarrow e^+ \nu_e j\,,
\;\;\;\;\;\;\;\;
pp\longrightarrow e^+ \nu_e jj\,.
\label{Wjprocs}
\eeq
In order to be definite, we have restricted ourselves to presenting
results only for the cases of positively charged leptons. Obviously,
the generation of $W^{-^*}+$jets processes is fully analogous to that
carried out here. NLO EW corrections to $W+$jets production have already 
appeared in the literature in various approximations~\cite{Dittmaier:2001ay,
Kuhn:2007cv,Hollik:2007sq,Denner:2009gj,Denner:2012ts,Kallweit:2014xda, 
Chiesa:2015mya,Kallweit:2015dum}; we shall not compare those predictions 
with ours in this paper\footnote{Comparisons of results obtained with
then-private \aNLOs\ versions with those computed by means of other tools
have been previously reported in ref.~\cite{Badger:2016bpw} (for $t\bt H$ 
production) and ref.~\cite{Bendavid:2018nar} (for the processes of
eq.~(\ref{4lprocs})).}.

The contributions to $\Ht$ by the final-state particles that appear
in eq.~(\ref{Wjprocs}) are due to the positron (possibly dressed with 
a photon), to the missing transverse momentum (set equal, by using MC truth, 
to the transverse momentum of the neutrino), and to any reconstructed jets.
\begin{figure}[t]
\begin{center}
  \includegraphics[width=0.495\textwidth]{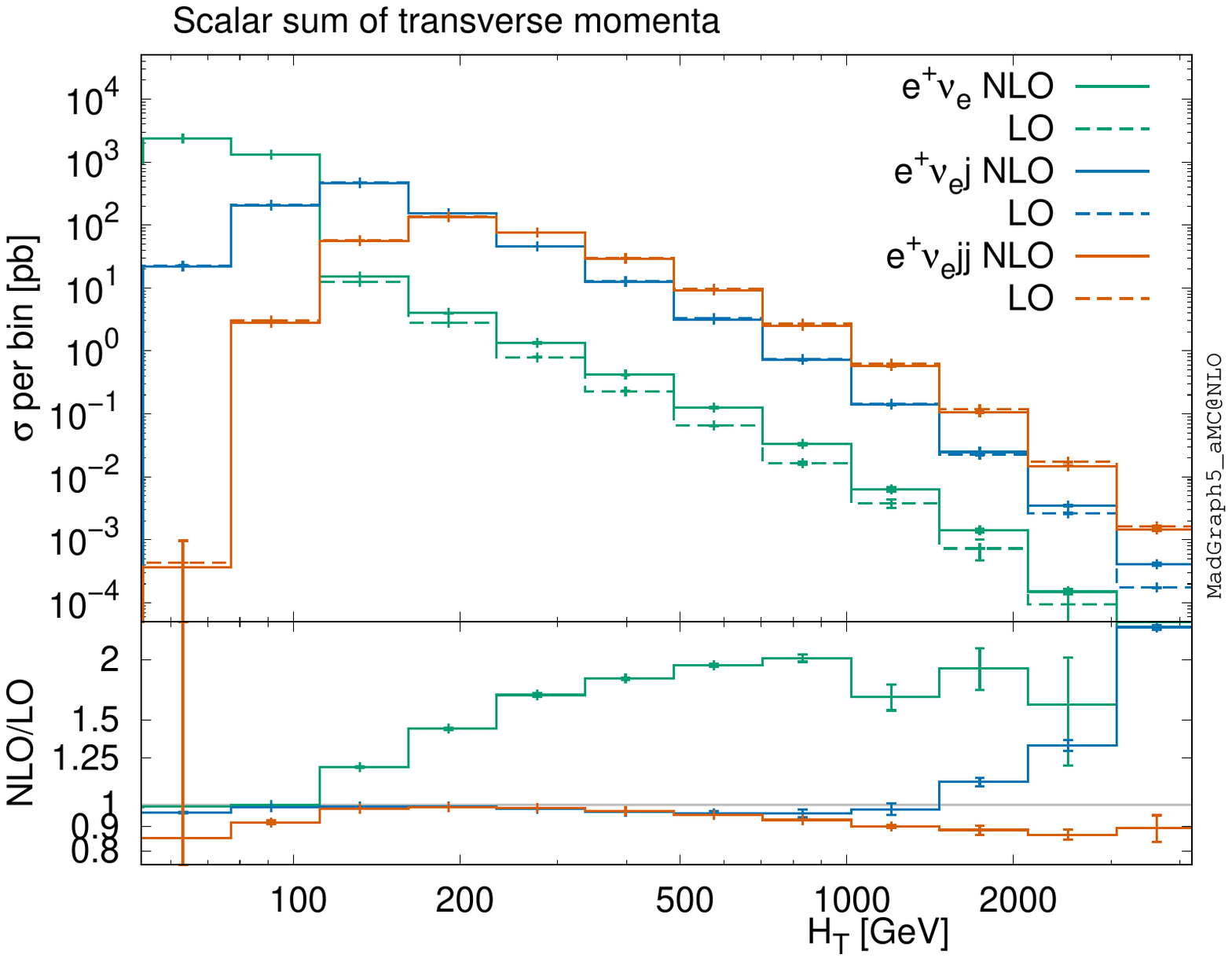}
  \includegraphics[width=0.495\textwidth]{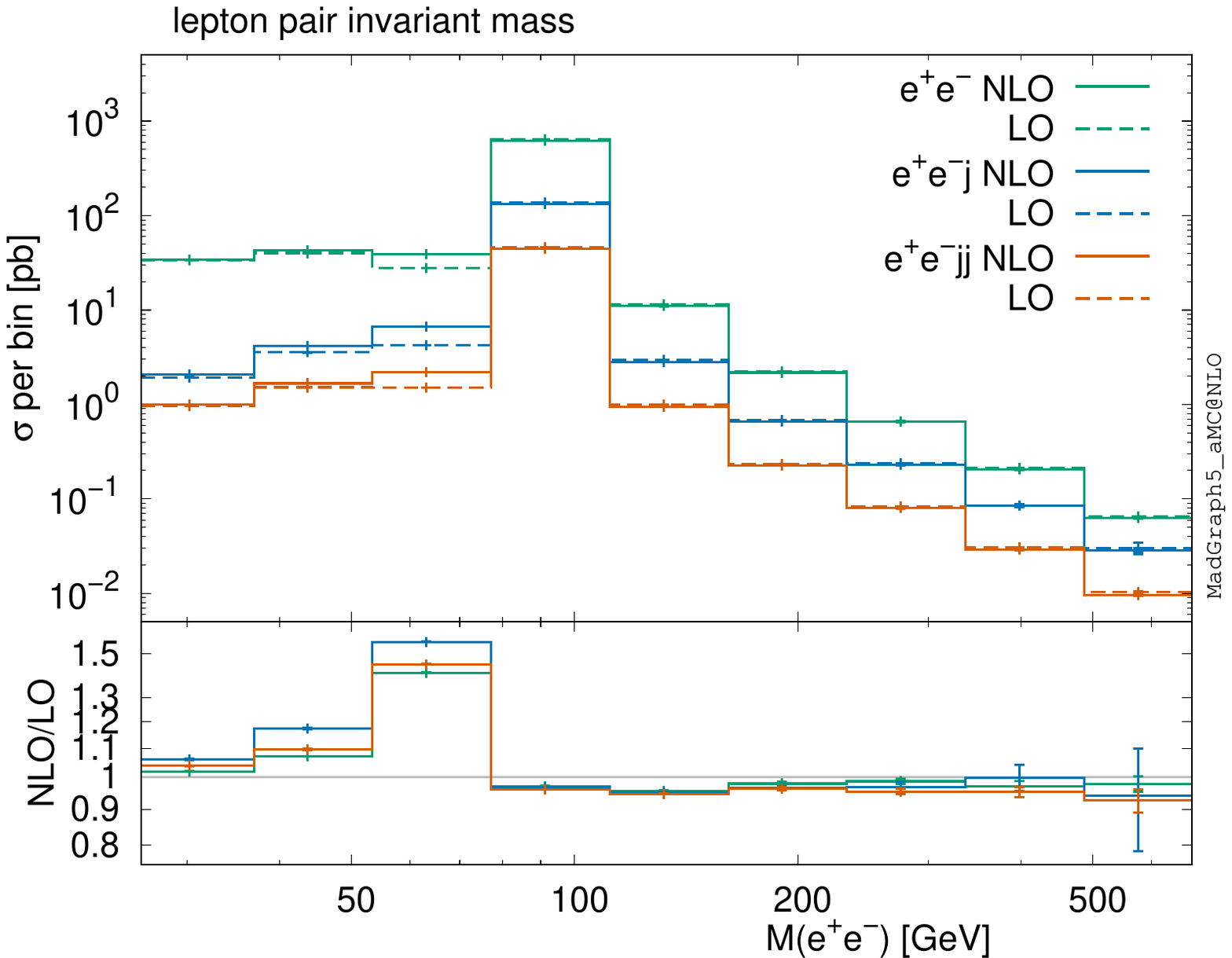}
\end{center}
\caption{\label{fig:Vjets}$\Ht$ in $W^{+^*}\!\!+\,$jets production (left 
panel), and dressed-lepton pair invariant mass in $Z^*+\,$jets production 
(right panel).}
\end{figure}
For the simplest process in eq.~(\ref{Wjprocs}), $pp\to e^+\nu_e$, at the LO 
and in the narrow width approximation it would be kinematically impossible
to have $\Ht>M_W$. The inclusion of finite-width effects for the $W$
boson opens up this region of phase space, which is thus populated exclusively
by configurations where the $W$ boson is off-shell. Therefore, the cross
section for $\Ht>M_W$ remains very suppressed at the LO, as is evident
from the green dashed curve in the main frame. When NLO corrections are 
included, real emissions contribute to this region also when the $W$ is 
on-shell, leading to very large and positive corrections -- the $K$ factor,
shown in the lower inset, is equal to about $2$ for $\Ht\gtrsim 500~\gev$.
It is again because of the outsize impact of the real corrections that
the $K$ factor for the process $pp \to e^+\nu_ej$ grows rapidly and
assumes large values in the region $\Ht\gtrsim 1000~\gev$. There, one
is dominated by configurations where there are two hard, back-to-back jets, 
and the $W$ boson has a small $\pt$ relative to either jet.
These kinds of enhancement due to kinematic features that become available
only at a certain perturbative order beyond the leading one are of course
not peculiar to EW corrections, and indeed they have often been studied in 
QCD as well (see e.g.~refs.~\cite{Frixione:1992pj,Butterworth:2008iy, 
Bauer:2009km,Denner:2009gj,Rubin:2010xp}). Eventually, by increasing
the jet multiplicity, the complexity of the final state is sufficiently
large already at the Born level, and such effects are pushed towards
highly-constrained corners of the phase space. For example, in the
case we are considering here, the EW corrections to $pp\to e^+ \nu_e jj$
behave as one naively expects them to do, becoming negative and growing
with $\Ht$ (as is typical in a regime dominated by EW Sudakovs).

On the other hand, it is in the comparison between processes that
differ by final-state jet multiplicities where the QCD and EW cases
deviate from each other. This is because in pure QCD the real corrections
to $pp\to e^+ \nu_e$ are identical to the Born of $pp\to e^+ \nu_e j$.
Hence, in kinematic regions dominated by real emissions (e.g.~at
large $\Ht$ in the present case) one expects the NLO result of the former 
process to be rather close to the LO result of the latter one. The same 
argument holds when comparing $pp\to e^+ \nu_e j$ with $pp\to e^+ \nu_e jj$. 
But when one computes EW corrections instead of QCD ones, these considerations 
are no longer valid: the Born-level jet(s) of $pp\to e^+ \nu_e j(j)$ is (are)
still predominantly emerging from a QCD branching, while the one(s) that 
contributes at the NLO to $pp\to e^+ \nu_e (j)$ production has (have) 
a rate governed by EW effects, and therefore much smaller.

NLO subleading contributions to $pp\to e^+\nu_ej$ feature 
doubly-resonant diagrams relevant to $pp\to W^{+^*}Z^*$ and 
$pp\to W^{+^*}W^{-^*}$ production, with the $Z^*$ and $W^{-^*}$ ``decaying''
hadronically. Among such subleading terms, the one with the largest $\as$
power ($\NLOt$) is due to the interference of these di-boson diagrams with QCD 
non-resonant ones. This, and the related fact that several of these interference
contributions are identically equal to zero owing to their colour structures,
implies that their presence in our calculations poses neither numerical
nor phenomenological problems. When they are squared, di-boson doubly-resonant 
diagrams enter only the $\NLOth$ term (that is of $\ord(\aem/\as)$ 
relative to the $\NLOt$ one), which we do not consider 
here\footnote{In an analogous manner, $pp\to W^{+^*}Z^*j$ and  
$pp\to W^{+^*}W^{-^*}j$ diagrams enter the LO subleading contributions 
to $pp\to e^+\nu_ejj$ production, starting at $\LOt$ where they interfere
with QCD non-resonant diagrams, and at the squared level in the $\LOth$ term.
Because of eq.~(\ref{NLOdefNLOEW}), these terms do not concern us.}.

We now turn to discussing $Z^*(+{\rm jets})$ production; as in the case of 
$W^{+*}(+{\rm jets})$ production, no comparisons will be given in this
paper with previous NLO EW results~\cite{Baur:2001ze,Dittmaier:2009cr,
Kuhn:2005az,Denner:2011vu,Hollik:2015pja,Denner:2014ina,Kallweit:2015dum}. 
The plot on the right panel of fig.~\ref{fig:Vjets} presents the invariant 
mass of the charged-lepton pair in the processes:
\beq
pp\longrightarrow \epem\,,
\;\;\;\;\;\;\;\; 
pp\longrightarrow \epem j\,,
\;\;\;\;\;\;\;\;
pp\longrightarrow \epem jj\,.
\label{Zjprocs}
\eeq
The invariant mass is computed for dressed leptons (according to the
definition given in sect.~\ref{sec:setup}). All of the three processes 
of eq.~(\ref{Zjprocs}) clearly display the expected $Z$-boson mass peak. 
The flatness of the histograms to the left of the peak is to a certain 
extent an artifact of our plotting choice (a logarithmic $x$ axis) -- 
we stress that the contributions due to an $s$-channel photon exchange
(as well as those without any $s$-channel ``decay'' to the $\epem$ pair),
which rapidly grow with decreasing invariant masses, are included in our 
results.

From the inset of the right panel of fig.~\ref{fig:Vjets}, we see that
the NLO EW corrections are fractionally the largest just below the $Z$-boson
peak. Indeed non-soft wide-angle (i.e.~out of the dressed-lepton cone)
real-emission photon radiation from the 
final state leptons decreases the reconstructed invariant mass w.r.t.~that
determined by the bare leptons, thus essentially shifting contributions from 
the $Z$-boson peak to the bins to its left\footnote{Any effects due to the 
finite $Z$-boson width are subdominant.}. On the other hand, the narrowness 
of the $\Delta R$ cone that defines the dressed leptons renders it much harder
to shift the bare-lepton mass to the right of the peak, which would occur by 
dressing either leptons with a photon emitted by any non-leptonic hard 
particle.

Furthermore, in the region \mbox{$M(\epem)<90~\gev$} the EW corrections 
are larger (and significantly so in the two bins to the left of the $Z$ peak)
for $pp\to e^+e^-j$ production than for $pp\to e^+e^-jj$ and 
$pp\to e^+e^-$ production. There is a hierarchy among these NLO effects in the 
latter two processes as well, which is visible given our MC uncertainties, 
but marginal in absolute value. We point out that the relative behaviours of
the corrections affecting different processes are in part controlled by the 
chosen acceptance cuts. More specifically, when one considers $pp\to e^+e^-j$ 
at the LO, the requirement that the jet have $\pt(j)>30~\gev$ implies 
$\pt(\epem)>30~\gev$. This renders it very likely that the leptons cuts 
$\pt(l^\pm)>10~\gev$ are trivially satisfied. In turn, this implies that
when either lepton emits a real photon (a configuration that, as we have 
argued before, is responsible for giving the dominant contribution to NLO 
corrections to the left of the $Z$ peak), lepton cuts will still be easily 
passed. On the other hand, in the case of $pp\to e^+e^-$, at the LO 
$\pt(\epem)=0$.
Therefore, lepton cuts imposed on real-emission configurations where the
photon does not belong to a dressed lepton will be more stringent than
those imposed on the associated LO configurations, and in this way NLO
corrections will be relatively more suppressed than for $pp\to e^+e^-j$
production. Finally, the case of $pp\to e^+e^-jj$ in an intermediate
one between the other two processes. Note that it is still comparatively 
easy to have a LO $\pt(\epem)\simeq 0$ configuration with two hard back-to-back
jets, in which case the mechanism advocated for $pp\to e^+e^-$ production
plays a role here as well.

\vskip 0.2truecm
\noindent
{\bf $\blacklozenge$ Four-lepton production, and VBF Higgs and 
associated production}

\noindent
In the left panel of fig.~\ref{fig:4lH} we present the transverse momentum 
of the hardest same-flavour lepton pair for the processes:
\beq
pp\longrightarrow e^+e^-\mu^+\mu^-\,,
\;\;\;\;\;\;\;\; 
pp\longrightarrow e^+\nu_e\mu^-\bar{\nu}_{\mu}\,.
\label{4lprocs}
\eeq
We have chosen the two processes in eq.~(\ref{4lprocs}) in order to
be definite, as representatives of the class of reactions with four
final-state leptons; both have been studied before~\cite{Gieseke:2014gka,
Billoni:2013aba,Biedermann:2016yvs,Biedermann:2016lvg,Biedermann:2016guo,
Kallweit:2017khh}. In fact, without any additional complications, 
\aNLOs\ is able to deal with any process that belongs
to this class, regardless of the particular flavour and charge combinations.

In detail, the definitions of the $\pt(ll)$ (relevant to $pp\to\epem\mpmm$)
and $\pt(l\nu)$ (relevant to $pp\to e^+\nu_e\mu^-\bar{\nu}_{\mu}$)
observables are the following. For the former, one uses dressed leptons;
the $e^+e^-$ and $\mu^+\mu^-$ pairs transverse momenta are then computed, 
and the largest of the two is set equal to $\pt(ll)$. In the latter case, 
charged leptons are again dressed first; then, the transverse momenta of 
the $e^+\nu_e$ and $\mu^-\bar{\nu}_{\mu}$ pairs are computed (by using the 
MC truth information to find the neutrinos), and the largest of the two is 
set equal to $\pt(l\nu)$.
\begin{figure}[t]
\begin{center}
  \includegraphics[width=0.495\textwidth]{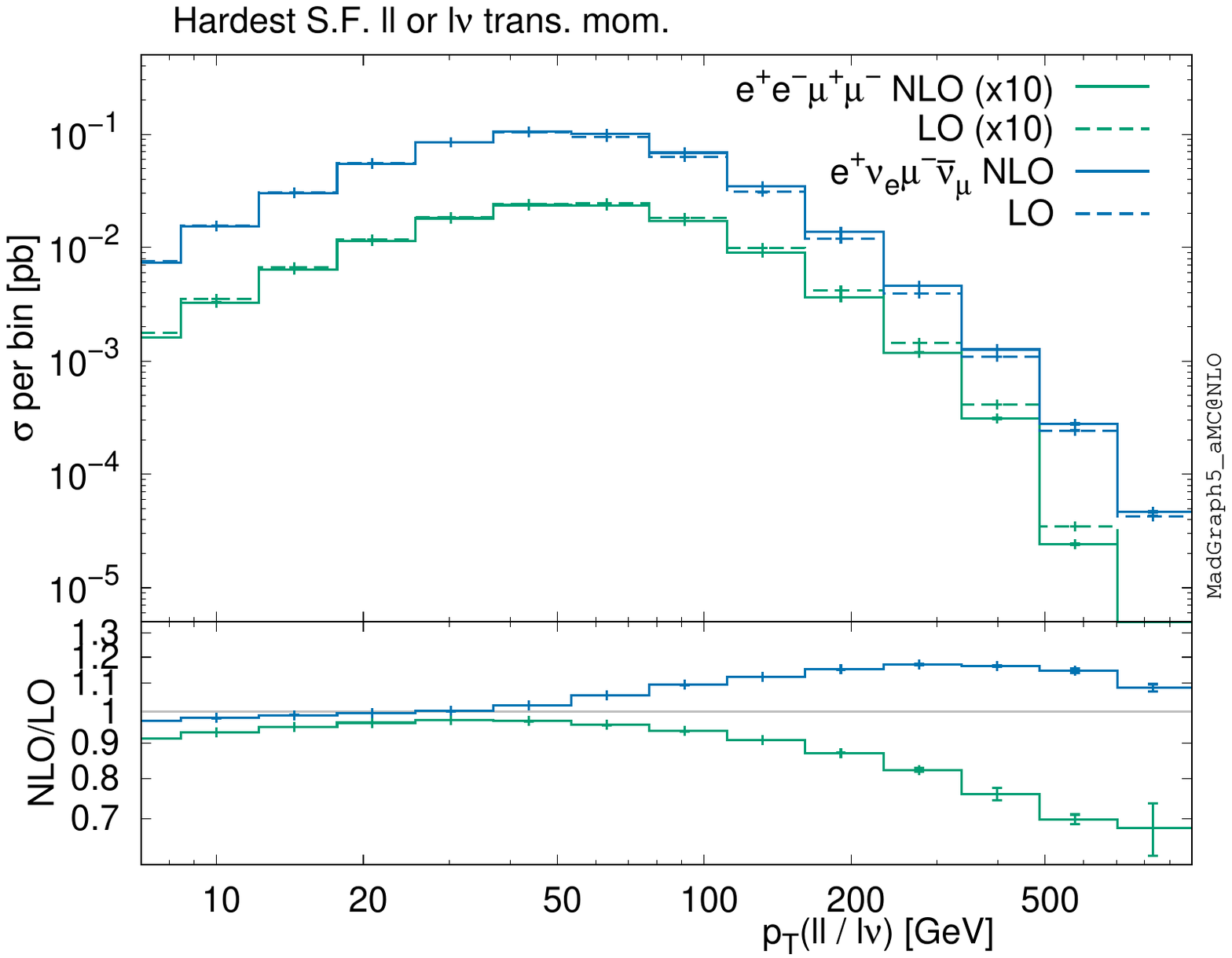}
  \includegraphics[width=0.495\textwidth]{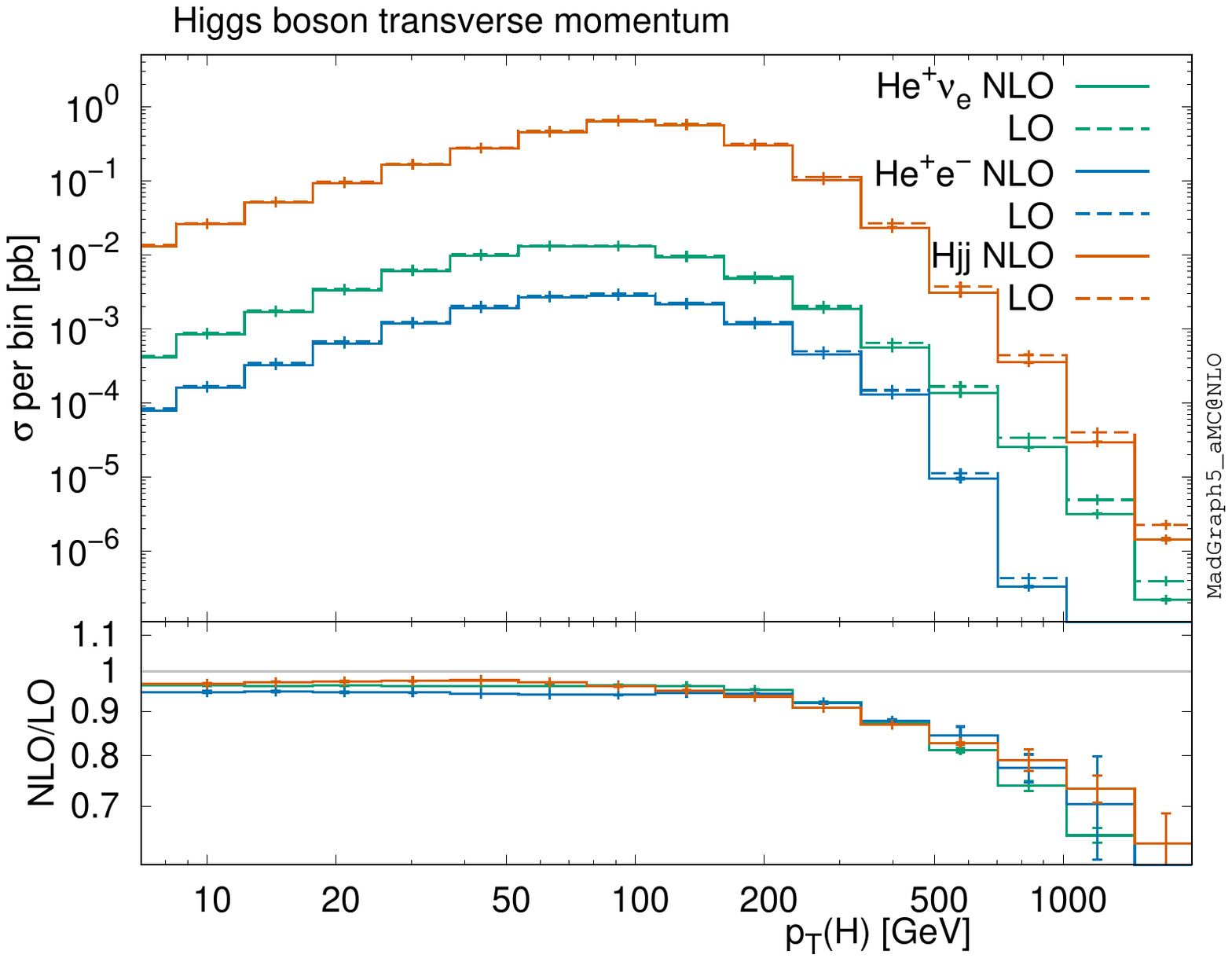}
\end{center}
\caption{\label{fig:4lH}
Transverse momentum of the hardest same-flavour dressed-lepton pair in
the processes of eq.~(\ref{4lprocs}) (left panel), and Higgs transverse 
momentum in the processes of eq.~(\ref{Hprocs}) (right panel).
}
\end{figure}
The NLO EW corrections behave rather differently for the two
processes. While for the four charged lepton process they display the
typical Sudakov behaviour at high $\pt$, for the other process the 
corrections  are positive and growing for $\pt\gtrsim 40~\gev$, starting
to decrease only towards $\pt\simeq 400~\gev$. We point out that the
two processes have significant differences in their underlying mechanisms.
Firstly, although both $2l2\nu$ and $4l$ production are dominated by 
di-boson resonant contributions (namely, di-$W$ and di-$Z$, respectively),
it is only the former case that features diagrams with $t$-channel spin-one
exchanges (thus enhanced at large momentum transfers). These appear in
$\gamma\gamma$-initiated processes, owing to the direct $\gamma W^+W^-$ 
coupling. Secondly, partonic processes such as 
$\gamma q\to W^{+*}W^{-*}q^\prime$ that give rise to $2l2\nu$ final 
states may be enhanced at large lepton-pair $\pt$'s owing to quasi-collinear 
$q^*\to W^*q^\prime$ splittings (see e.g.~ref.~\cite{Frixione:1992pj}). While
a similar mechanism also occurs in $4l$ production, in that case its
effects are balanced by a stronger suppression than in the case of $2l2\nu$ 
production\footnote{The overall impact of quasi-collinear enhancements
on observable cross sections ultimately depends on the interplay
between their kinematics characteristics, the partonic matrix elements,
and PDF effects -- see e.g.~refs.~\cite{Frixione:1993yp,Baglio:2013toa} 
for discussions on this point.}.
Finally, at the NLO $2l2\nu$ production features a real-emission 
contribution due to an underlying $tW$ doubly-resonant mechanism,
which might induce very large corrections driven by the top-quark
on-shell region. In practice, this does not happen -- the resonant
channels are always associated with an initial-state $\gamma b$ 
(or $\gamma\bb$) pair, and are 
thus significantly suppressed by parton luminosities, so that the
``convergence'' of the perturbative series is not spoiled. We note 
that the same remark applies to the computation of the NLO {\em QCD} 
corrections to $W^+W^-$ production where the effects, although 
larger~\cite{Gehrmann:2014fva} than those relevant to the NLO EW case 
considered here, are still under control\footnote{The presence of 
resonant contributions does not constitute a problem {\em per se}; other 
quantities entering the cross sections, particularly the couplings and the 
parton luminosities, play a crucial role as well. Probably the best known among
the problematic cases is that of the NLO QCD corrections to $tW$ production, 
which receives an enormous $gg\to t\bt$-channel enhancement. We point out 
that it is this underlying $tW-t\bt$ mechanism that renders the {\em NNLO} 
QCD calculation of the $W^+W^-$ cross section prone to large corrections.
Although solutions can be devised (see footnote~\ref{ft:subres} for
some examples) that eliminate or reduce the impact of resonant 
contributions, they are always arbitrary to a certain extent,
and it is difficult to quantify the uncertainties associated with them.
Hence, they are best avoided if not deemed to be strictly necessary.}.

The right panel of fig.~\ref{fig:4lH} displays the transverse
momentum of the Higgs boson that emerges from the processes:
\beq
pp\longrightarrow Hjj\,,
\;\;\;\;\;\;\;\; 
pp\longrightarrow He^+e^-\,,
\;\;\;\;\;\;\;\; 
pp\longrightarrow He^+\nu_e\,.
\label{Hprocs}
\eeq
The EW corrections to these processes have appeared in the literature in 
various approximations~\cite{Ciccolini:2003jy,Denner:2011id,
Granata:2017iod,Ciccolini:2007jr,Ciccolini:2007ec}. Obviously, the study of 
$He^-\bar{\nu}_e$ production, which we have chosen to ignore, is identical
to that of the rightmost process in eq.~(\ref{Hprocs}).

The cross sections for $pp\to He^+e^-$ and $pp\to He^+\nu_e$ receive
(dominant) contributions from the underlying resonant production of Higgs 
in association with a $Z^*$ and a $W^{+*}$ boson, respectively. These 
mechanisms do also contribute to the leftmost process in eq.~(\ref{Hprocs}),
in which case the vector bosons ``decay'' hadronically rather than 
leptonically. By weighting with the different hadronic and leptonic 
vector-boson branching ratios, with a back-of-the-envelope calculation
one can compare the total rates for $Hjj$ production with the sum of
those relevant to $He^+e^-$ plus $He^+\nu_e$ production, and find
that the former is about four times larger than the latter sum.
This is a first naive indication that $pp\to Hjj$ production is
dominated by a VBF mechanism, rather than by the associated production
ones; clearly, a much better evidence for this comes from studying
the topology of the outgoing jets.

The NLO EW corrections behave rather similarly for the three processes in 
eq.~(\ref{Hprocs}), and are about $\ord(-5\%)$ at small transverse momentum, 
\mbox{$\pt(H)\lesssim 200~\gev$}. They increase in absolute value
with $\pt(H)$ and remain negative, as is typical for these types of
corrections in regions of phase space dominated by EW Sudakov
logarithms.

\vskip 0.2truecm
\noindent
{\bf $\blacklozenge$ Triple-boson production}

\noindent
In the left panel of fig.~\ref{fig:3v} we present the transverse
momentum of the hardest vector boson in triple vector boson 
production. This class of processes has been considered previously
in refs.~\cite{Nhung:2013tfu,Yong-Bai:2015xna,Hong:2016aek,
Yong-Bai:2016sal,Dittmaier:2017bnh}; we focus on:
\beq
pp\longrightarrow W^+W^-W^+\,,
\;\;\;\;\;\;\;\; 
pp\longrightarrow ZZW^+\,,
\;\;\;\;\;\;\;\; 
pp\longrightarrow ZZZ\,.
\label{3vprocs}
\eeq
The computations relevant to final states which are obtained by those
in eq.~(\ref{3vprocs}) by charge conjugation, namely $W^-W^+W^-$
and $ZZW^-$, do not pose any additional complications, and are thus
ignored here. Conversely, $pp\to ZW^-W^+$ is more problematic, since it 
receives contributions from real-emission diagrams with an $s$-channel top 
quark (i.e.~from an underlying $t^*W^-Z$ or $\bar{t}^*W^+Z$ production 
mechanism). Thus, while technically this process is doable in our setup
by setting the top width equal to its physical value in order to 
prevent the matrix elements from diverging on the top resonance
(see sect.~\ref{sec:zeroTopWidthNonZeroWwidth}), potentially it still poses 
the problems common to all processes which, at the NLO, ``interfere'' with
a top-induced ``background'' (such as instabilities in the numerical 
integration caused by extremely large $K$ factors). We have already discussed
an example ($W^{+*}W^{-*}$ production, eq.~(\ref{4lprocs})) where such
an interference in practice does not lead to any issues at the 
perturbative orders we are interested in. However, the case 
of $ZW^-W^+$ production is much more involved, and therefore we prefer to
postpone its study to when \aNLOs\ will feature an automated treatment
of the subtraction or removal of resonant contributions, with
procedures analogous to those already considered in the literature
in different contexts\footnote{\label{ft:subres}The procedures that are
being implemented in \aNLOs\ are fully local in the phase-space of final-state 
particles, such as those of refs.~\cite{Frixione:2008yi,White:2009yt,
Weydert:2009vr,Re:2010bp,Binoth:2011xi,GoncalvesNetto:2012yt,Gavin:2013kga,
Gavin:2014yga,Demartin:2016axk}. Global~\cite{Tait:1999cf,Zhu:2001hw,
Cascioli:2013wga,Gehrmann:2014fva} or semi-local~\cite{Beenakker:1996ch,
Berger:2003sm,Dao:2010nu,Hollik:2012rc} approaches are not suited to automated 
observable-independent short-distance computations.}.
Another, simpler, solution is that of performing the computation in a 
scheme with four flavours. This will not be done here, but it is feasible 
with the present version of \aNLOs\ (we note that a 4FS restriction of the
OS model is available, while its CM counterpart has still to be 
constructed)\footnote{Another possibility in the context of a five-flavour 
computation is that of adding a dedicated integration channel for each 
of the new resonant contributions that open at the NLO level.}.
\begin{figure}[t]
\begin{center}
  \includegraphics[width=0.495\textwidth]{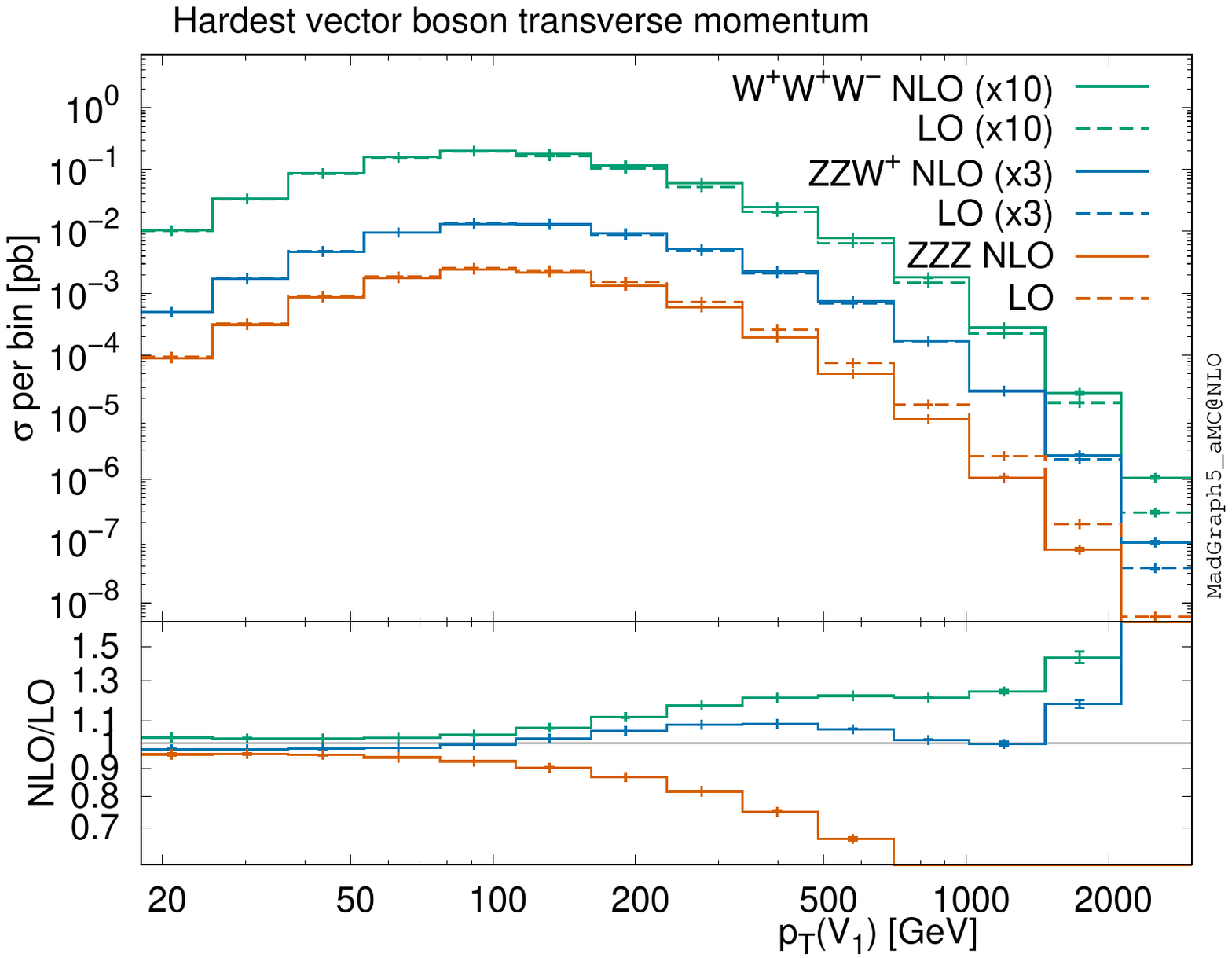}
  \includegraphics[width=0.495\textwidth]{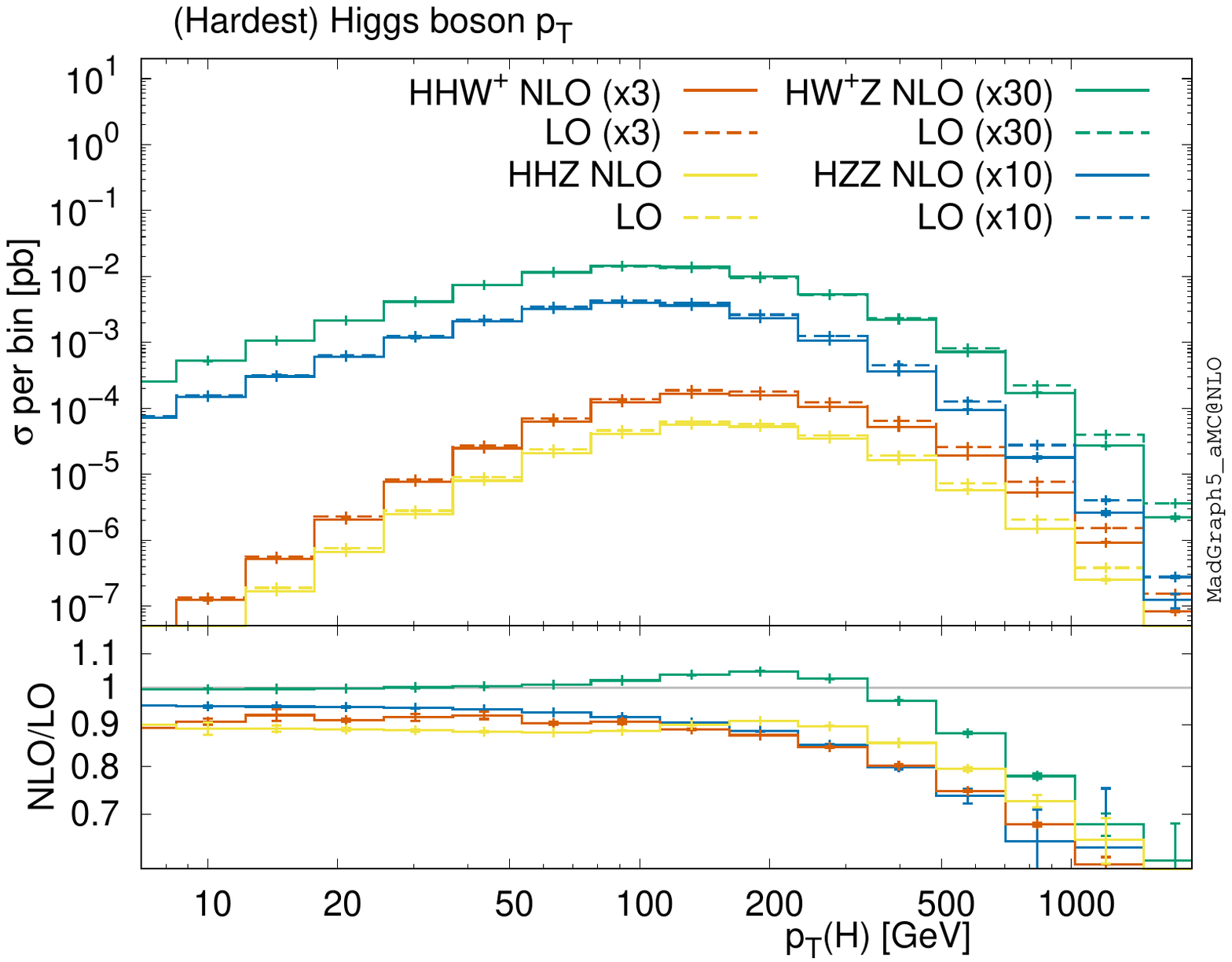}
\end{center}
\caption{\label{fig:3v} Transverse momentum of the hardest vector boson 
in the processes of eq.~(\ref{3vprocs}) (left panel), and transverse 
momentum of the Higgs boson in the processes of eq.~(\ref{3bprocs}) 
(right panel). Some of the histograms in the main frames are rescaled
as indicated in order to enhance their visibility.}
\end{figure}

From the inset in left panel of fig.~\ref{fig:3v}, we see that $ZZZ$ 
production exhibits the typical behaviour of NLO EW corrections, which 
are small at small transverse momentum, and grow in absolute 
value with $\pt$. The other two processes in eq.~(\ref{3vprocs}) 
display a more intricate behaviour, owing to a combination of effects: the 
virtual Sudakov corrections, which decrease the rates; and the positive 
enhancement of the cross section, due to the presence of photon in
the initial state. 
The latter are related to the direct coupling of $W$'s with photons,
and entail the opening at the NLO of new $t$-channel contributions
which are absent in $pp\to ZZZ$ production.

In the right panel of fig.~\ref{fig:3v} we consider the
transverse momentum of the hardest Higgs boson in the following
heavy-boson production processes:
\beq
 pp\longrightarrow HZZ\,,
\;\;\;\;\;\;\;\; 
pp\longrightarrow HW^+Z\,,
\;\;\;\;\;\;\;\; 
pp\longrightarrow HHW^+\,,
\;\;\;\;\;\;\;\; 
pp\longrightarrow HHZ\,.
\label{3bprocs}
\eeq
The NLO EW corrections for these processes are computed here for the first 
time. For reasons fully analogous to those that apply to triple vector-boson 
production (reported in the discussion below eq.~(\ref{3vprocs})), 
the process $pp\to HW^+W^-$, for which top-quark resonant contributions 
appear at the NLO, is not studied in what follows.
Triple-$H$ production is loop-induced, and therefore is also ignored.
As in all of the other cases treated so far, processes obtained by means of
charge conjugation from those of eq.~(\ref{3bprocs}) can be generated
without problems by \aNLOs, but have not been considered here.

For inclusive rates (see table~\ref{tab:1}) the NLO EW corrections are 
$-9\%$ for $pp \to HZZ$, $-11\%$ for $pp\to HHZ$, and $-13\%$ for 
$pp \to HHW^+$, while for $pp \to HW^+Z$ they are a positive $1.6\%$. 
At the differential level, all of the four processes display the typical 
behaviour of EW corrections (i.e.~negative and growing in absolute value
with $\pt$) at large transverse momenta; however, the $\pt$ values for which 
these effects become dominant do depend on the specific process. In particular, 
as is the case for the inclusive rates, it is $HW^+Z$ production that
stands apart, since up to relatively large transverse momenta
($\pt\simeq 200~\gev$) the negative contributions due to the EW
Sudakovs (which are present in the other three processes as well)
are compensated by positive contributions. Among these, the dominant
one is driven by a quasi-collinear enhancement stemming from 
$\gamma q\to HW^+q^*(\to Zq)$, a mechanism fully analogous to that 
already advocated for the second process in eq.~(\ref{4lprocs}), and 
that cannot be present in the other three processes in eq.~(\ref{3bprocs}).
Finally, we notice that (smaller) differences between the triple-boson
processes of eq.~(\ref{3bprocs}) can be induced by virtual corrections,
owing to the different ways in which the bosons enter the one-loop
diagrams (chiefly, by being directly attached to the heavy-quark loop, 
or by resulting from the branching of a parent particle that is directly 
attached to the loop).

\vskip 0.2truecm
\noindent
{\bf $\blacklozenge$ Associated top-quark, and jet production}

\noindent
In the left panel of fig.~\ref{fig:rest} we consider the transverse
momentum of the $t\bt$ pairs in the following processes:
\beq
pp\longrightarrow t\bt W^+\,,
\;\;\;\;\;\;\;\; 
pp\longrightarrow t\bt Z\,,
\;\;\;\;\;\;\;\; 
pp\longrightarrow t\bt H\,.
\label{ttVprocs}
\eeq
These have been studied before in the literature~\cite{Frixione:2014qaa, 
Yu:2014cka,Frixione:2015zaa,Badger:2016bpw,Denner:2016wet}, also
with a then-private version of \aNLOs. The $\pt(t\bt)$ distributions behave 
as is typical for EW corrections dominated by EW Sudakov logarithms. Full
agreement with our previous results~\cite{Frixione:2014qaa,Frixione:2015zaa}
is found, which constitutes a further cross-check of the full automation
of the mixed-coupling expansion achieved in the current version of \aNLOs. 
The processes of eq.~(\ref{ttVprocs}) are also considered in
sect.~\ref{sec:restt}, where we include all of the LO and NLO terms, 
as anticipated in eqs.~(\ref{proc2}) and~(\ref{proc3}).
\begin{figure}[t]
\begin{center}
  \includegraphics[width=0.495\textwidth]{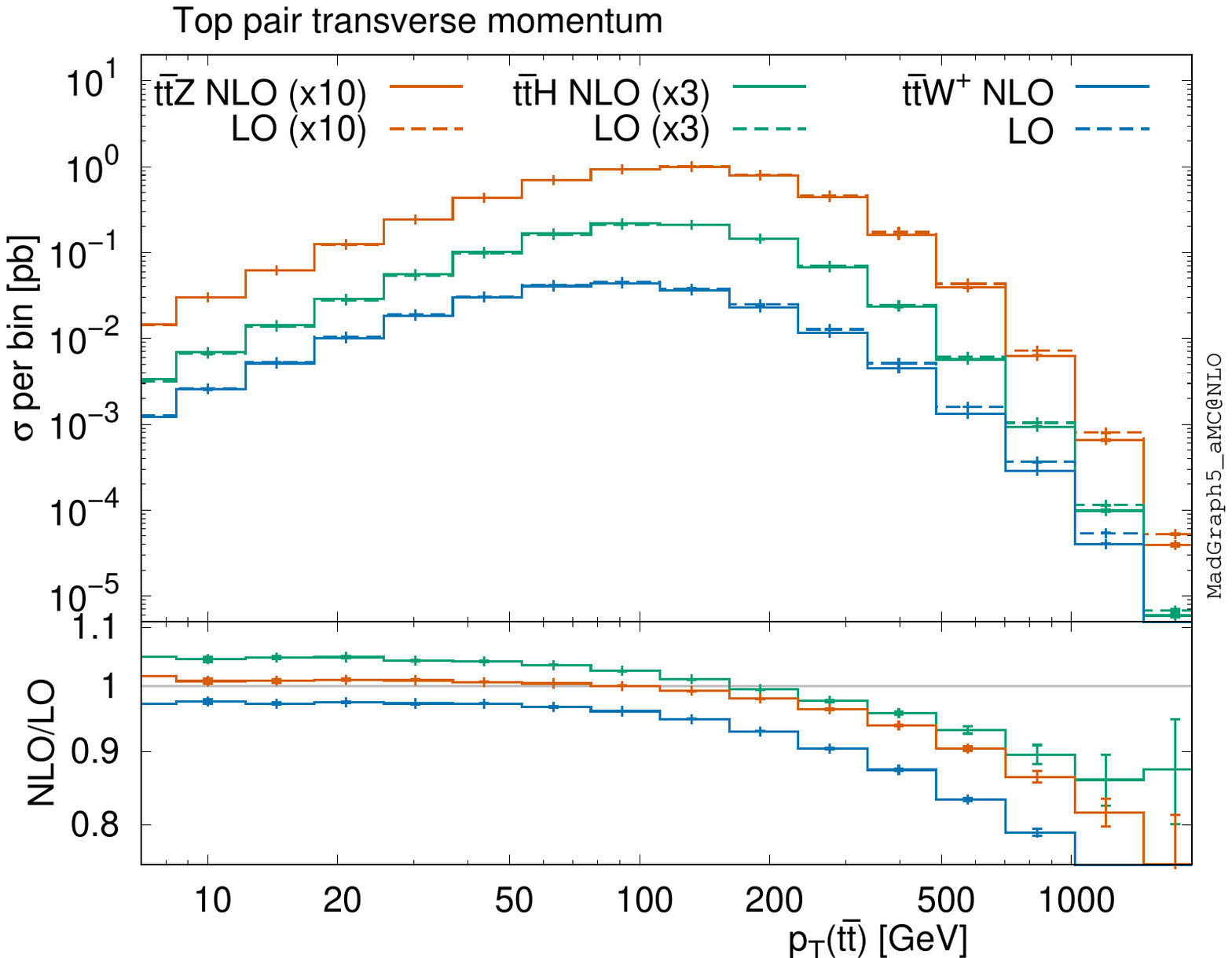}
  \includegraphics[width=0.495\textwidth]{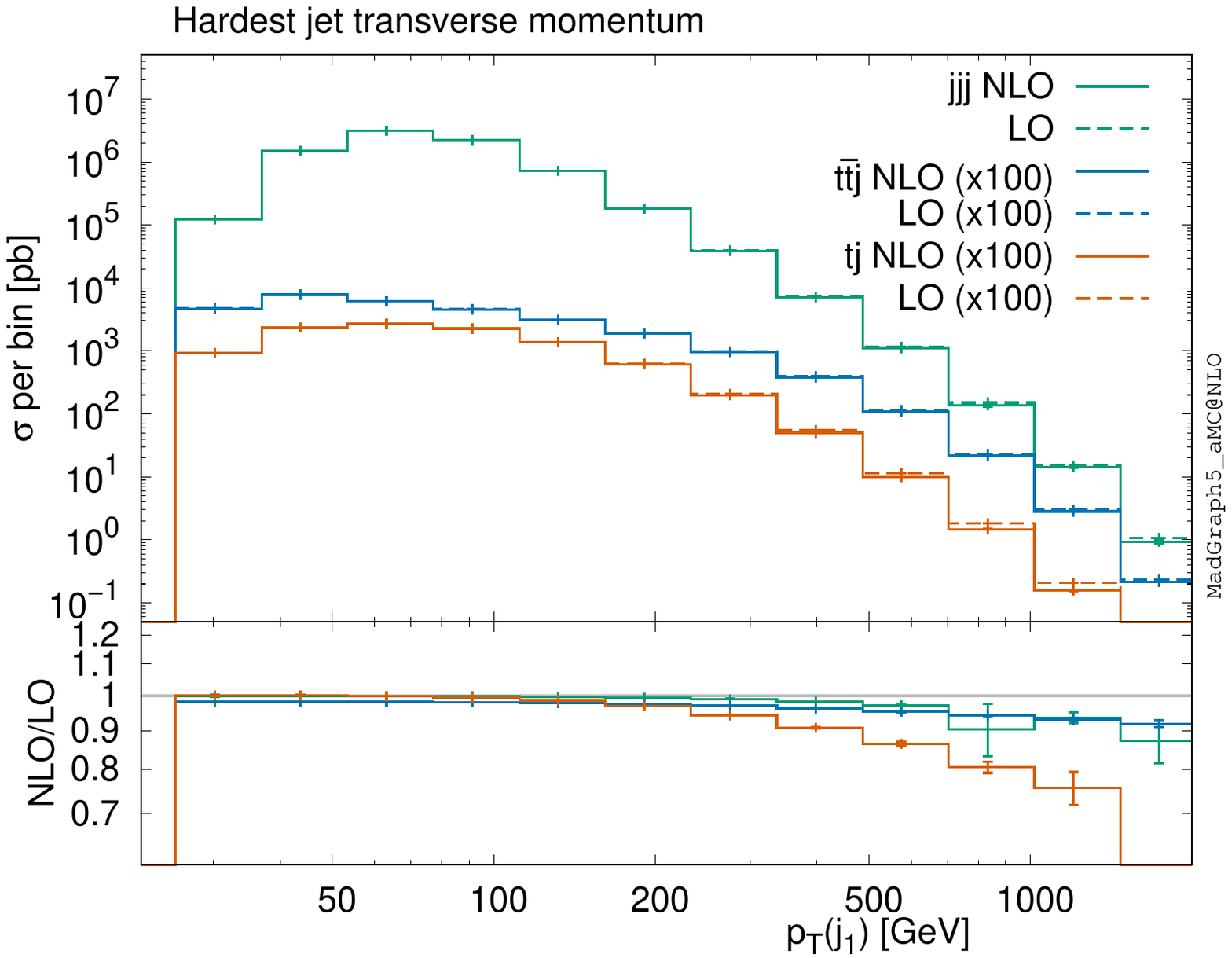}
\end{center}
\caption{\label{fig:rest} Top-pair transverse momentum in the processes
of eq.~(\ref{ttVprocs}) (left panel), and transverse momentum of the
hardest jet in the processes of eq.~(\ref{jprocs}) (right panel). 
Some of the histograms in the main frames are rescaled as indicated 
in order to enhance their visibility.
}
\end{figure}

On the right panel of fig.~\ref{fig:rest} we show the transverse
momentum of the hardest jet in triple jet, single-top, and $t\bt j$ 
production:
\beq
pp\longrightarrow jjj\,,
\;\;\;\;\;\;\;\; 
pp\longrightarrow tj\,,
\;\;\;\;\;\;\;\; 
pp\longrightarrow t\bt j\,.
\label{jprocs}
\eeq
As the notation suggests, in the single-top process we do not include 
single {\em anti}-top production. 

NLO EW corrections to triple-jet production are computed here for the 
first time. As fig.~\ref{fig:rest} shows, we find them to be small
for this observable, but not entirely negligible at the upper end of
the considered transverse momentum range ($\sim 1$~TeV), where they
are of $\ord(-10\%)$. Up to small differences, they thus exhibit the
same pattern as the EW corrections to the inclusive transverse momentum 
in dijet production~\cite{Dittmaier:2012kx,Frederix:2016ost}. We have
verified that similar effects are present in the second- and third-hardest 
jet $\pt$'s. Conversely, the impact of NLO EW corrections is seen to
be essentially negligible on any of the two- and three-jet invariant masses 
(up to 4~TeV) that can be constructed from the three-hardest jet momenta.

The single-top process of eq.~(\ref{jprocs}) includes both $t$- and
$s$-channel mechanisms. Its NLO EW corrections have been computed before 
in the context of supersymmetric extensions of the SM~\cite{Beccaria:2006ir, 
Mirabella:2008gj,Bardin:2010mz} (a soft approximation has been employed in
ref.~\cite{Beccaria:2006ir} to deal with real-emission contributions). 
We find that EW corrections follow the typical pattern of cross-section
suppression at large transverse momenta, due to Sudakov effects. As far
as $t\bt j$ production is concerned (which will also be discussed in
sect.~\ref{sec:restt}), the impact of the EW corrections is significantly
smaller in this $\pt$ range than for single-top.

\subsection{Complete NLO corrections to $pp\longrightarrow t\bt(+V,H)$
and $pp\longrightarrow t\bt j$ production\label{sec:restt}}
In this section we focus on top pair production, possibly in
association with either a heavy boson:
\beq
pp\longrightarrow t\bt\,,
\;\;\;\;\;\;\;\; 
pp\longrightarrow t\bt Z\,,
\;\;\;\;\;\;\;\; 
pp\longrightarrow t\bt W^+\,,
\;\;\;\;\;\;\;\; 
pp\longrightarrow t\bt H\,,
\label{ttVprocs2}
\eeq
or an extra light jet:
\beq
pp\longrightarrow t\bt j\,.
\label{ttjproc}
\eeq
Since we consider all of the LO and NLO contributions, eqs.~(\ref{proc2}) 
and~(\ref{proc3}), we have generated these processes in \aNLOs\ by using the
following commands:

\vskip 0.25truecm
\noindent
~~\prompt\ {\tt ~generate p p > t t\~{} QED=2 QCD=2 [QCD QED]}

\noindent
~~\prompt\ {\tt ~generate p p > t t\~{} z QED=3 QCD=2 [QCD QED]}

\noindent
~~\prompt\ {\tt ~generate p p > t t\~{} w+ QED=3 QCD=2 [QCD QED]}

\noindent
~~\prompt\ {\tt ~generate p p > t t\~{} h QED=3 QCD=2 [QCD QED]}

\noindent
~~\prompt\ {\tt ~generate p p > t t\~{} j QED=3 QCD=3 [QCD QED]}

\vskip 0.25truecm
\noindent
The syntax of these commands has already been discussed in sect.~\ref{sec:gen}.
We point out that in the case of $t\bt j$ production at these perturbative
orders massless leptons must also be included in the definition of both 
the {\tt p} and {\tt j} multiparticles, in keeping with what is explained 
in appendix~\ref{sec:tech}. This can be done by executing the following 
commands:
\vskip 0.25truecm
\noindent
~~\prompt\ {\tt ~define p = p e+ e- mu+ mu- ta+ ta-}

\noindent
~~\prompt\ {\tt ~define j = p}

\vskip 0.25truecm
\noindent
immediately after the {\tt p} and {\tt j} definitions given at the
beginning of sect.~\ref{sec:setup}, and before the process-generation
command. The computation of $t\bt W^-$ production would not pose 
any additional problem w.r.t.~that of $pp\to t\bt W^+$; it is not
carried out here. The results for all the LO and NLO terms have 
already been computed with a private version of \aNLOs\ for the 
$pp\to t\bt$ and $pp\to t\bt W^+$ processes, and presented in 
refs.~\cite{Czakon:2017wor,Frederix:2017wme}, respectively
(in the latter paper, predictions for $pp\to t\bt t\bt$ are
reported as well). Recently,
the NLO corrections to $t\bt j$ production, bar for photon-induced 
processes, have been computed in ref.~\cite{Gutschow:2018tuk}.
The complete NLO corrections for $pp\to t\bt Z$ and $pp\to t\bt H$
are given here for the first time.

We start by considering total rates, which we report in table~\ref{tab:blobs}.
The first row displays the $\LOo$ contributions to the cross sections,
given in pb. Rows 2--9 present instead all of the other contributions,
as fractions over the $\LOo$ one, namely:
\beqn
&&\frac{\Sigma_{\LOi}}{\Sigma_{\LOo}}\,,\;\;\;\;\;\;\;\;\;\;
i=2,3,4\,,
\\
&&\frac{\Sigma_{\NLOi}}{\Sigma_{\LOo}}\,,\;\;\;\;\;\;\;\;
i=1,\ldots 5\,;
\eeqn
note that $\Sigma_{\LOf}$ and $\Sigma_{\NLOfv}$ are identically equal 
to zero for all processes bar that of eq.~(\ref{ttjproc}).
As for all the results shown so far, the uncertainties are solely
associated with MC integration errors. We point out that the predictions
of table~\ref{tab:blobs} have been generated independently from those
reported in sect.~\ref{sec:resNLOEW} (see in particular table~\ref{tab:1} 
and fig.~\ref{fig:rest}), and are therefore slightly different 
from the latter (while being statistically compatible with them) --
see the discussion immediately before eq.~(\ref{NLOdefNLOEW}). 
As expected, for fully inclusive rates
all contributions apart from the \LOone\ and \NLOone\ ones are small, with
the exception of the \NLOthree\ term (and, to a smaller extent, of the 
\NLOtwo\ one as well) in $t\bt W^+$ production -- this
constitutes a $+12\%$ correction of the \LOone\ cross section, and can 
be understood as due to the opening of a $tW$ scattering process, as was 
already suggested in ref.~\cite{Dror:2015nkp,Frederix:2017wme}. More in 
details, $\Sigma_{\NLOth}$ and $\Sigma_{\NLOt}$ are equal to about $+7.6$\%
and $-2.9\%$, respectively, of the total NLO cross section for such
a process.

We now turn to presenting predictions for selected differential
distributions in figs.~\ref{fig:tt_blobs}--\ref{fig:ttj_blobs}.
These figures have all the same layout. In the main frame there
are eight histograms (ten in fig.~\ref{fig:ttj_blobs}).
The solid back one is the sum of all the three (four) $\LOi$ 
and four (five) $\NLOi$ contributions (as in sect.~\ref{sec:resNLOEW}, 
these are given as cross sections per bin), which has been denoted by 
``NLO'' in  eq.~(\ref{NLOdeftt}). The three (four) dashed/dot-dashed ones 
are the $\Sigma_{\LOi}$ terms (green for $\LOo$, blue for $\LOt$, red for 
$\LOth$, and yellow for $\LOf$ in fig.~\ref{fig:ttj_blobs}), while
the four (five) solid/dotted ones show the $\Sigma_{\NLOi}$ terms (green 
for $\NLOo$, blue for $\NLOt$, red for $\NLOth$, yellow for $\NLOf$, and 
light blue for $\NLOfv$ in fig.~\ref{fig:ttj_blobs}). A dot-dashed or 
dotted pattern is used when the corresponding result is negative -- what 
is displayed on the figure is then the absolute value of the cross section.
In the lower insets of the figures, the ratios are shown of the individual 
$\LOi$ and $\NLOi$ contributions over the total NLO result, with the same 
patterns as those used in the main frames.
\begin{landscape}
\begin{table}
\begin{center}
\begin{small}
\begin{tabular}{lr@{$\,\,\pm\,\,$}lr@{$\,\,\pm\,\,$}lr@{$\,\,\pm\,\,$}lr@{$\,\,\pm\,\,$}lr@{$\,\,\pm\,\,$}l}
\toprule
 & \multicolumn{2}{c}{$pp \to t \bar{t}$} & \multicolumn{2}{c}{$pp \to t \bar{t} Z$} & \multicolumn{2}{c}{$pp \to t \bar{t} W^+$} & \multicolumn{2}{c}{$pp \to t \bar{t} H$} & \multicolumn{2}{c}{$pp \to t \bar{t} j$}\\
\midrule
\LOone & \multicolumn{2}{c}{$4.3803 \pm 0.0005\,\cdot 10^{2}$~pb} & \multicolumn{2}{c}{$5.0463 \pm 0.0003\,\cdot 10^{-1}$~pb} & \multicolumn{2}{c}{$2.4116 \pm 0.0001\,\cdot 10^{-1}$~pb} & \multicolumn{2}{c}{$3.4483 \pm 0.0003\,\cdot 10^{-1}$~pb} & \multicolumn{2}{c}{$3.0278 \pm 0.0003\,\cdot 10^{2}$~pb}\\
\LOtwo & $+0.405 $&$ 0.001$~\% & $-0.691 $&$ 0.001$~\% & $+0.000 $&$ 0.000$~\% & $+0.406 $&$ 0.001$~\% & $+0.525 $&$ 0.001$~\%\\
\LOthree & $+0.630 $&$ 0.001$~\% & $+2.259 $&$ 0.001$~\% & $+0.962 $&$ 0.000$~\% & $+0.702 $&$ 0.001$~\% & $+1.208 $&$ 0.001$~\%\\
\LOfour & \multicolumn{2}{c}{ } & \multicolumn{2}{c}{ } & \multicolumn{2}{c}{ } & \multicolumn{2}{c}{ } & $+0.006 $&$ 0.000$~\%\\
\NLOone & $+46.164 $&$ 0.022$~\% & $+44.809 $&$ 0.028$~\% & $+49.504 $&$ 0.015$~\% & $+28.847 $&$ 0.020$~\% & $+26.571 $&$ 0.063$~\%\\
\NLOtwo & $-1.075 $&$ 0.003$~\% & $-0.846 $&$ 0.004$~\% & $-4.541 $&$ 0.003$~\% & $+1.794 $&$ 0.005$~\% & $-1.971 $&$ 0.022$~\%\\
\NLOthree & $+0.552 $&$ 0.002$~\% & $+0.845 $&$ 0.003$~\% & $+12.242 $&$ 0.014$~\% & $+0.483 $&$ 0.008$~\% & $+0.292 $&$ 0.007$~\%\\
\NLOfour & $+0.005 $&$ 0.000$~\% & $-0.082 $&$ 0.000$~\% & $+0.017 $&$ 0.003$~\% & $+0.044 $&$ 0.000$~\% & $+0.009 $&$ 0.000$~\%\\
\NLOfive & \multicolumn{2}{c}{ } & \multicolumn{2}{c}{ } & \multicolumn{2}{c}{ } & \multicolumn{2}{c}{ } & $+0.005 $&$ 0.000$~\%\\
\bottomrule
\end{tabular}
\caption{\label{tab:blobs} Cross sections for the five $t\bt+X$ processes 
  of eqs.~(\ref{ttVprocs2}) and~(\ref{ttjproc}), resulting from the 
  setup described in section~\ref{sec:setup}. The uncertainties quoted
  are of statistical nature only, originating from the Monte Carlo integration
  over the phase space. The subleading LO and NLO contributions are given as
  percentage fractions of \LOone.
}
\end{small}
\end{center}
\end{table}
\end{landscape}

\begin{figure}[t]
\begin{center}
  \includegraphics[width=0.6\textwidth]{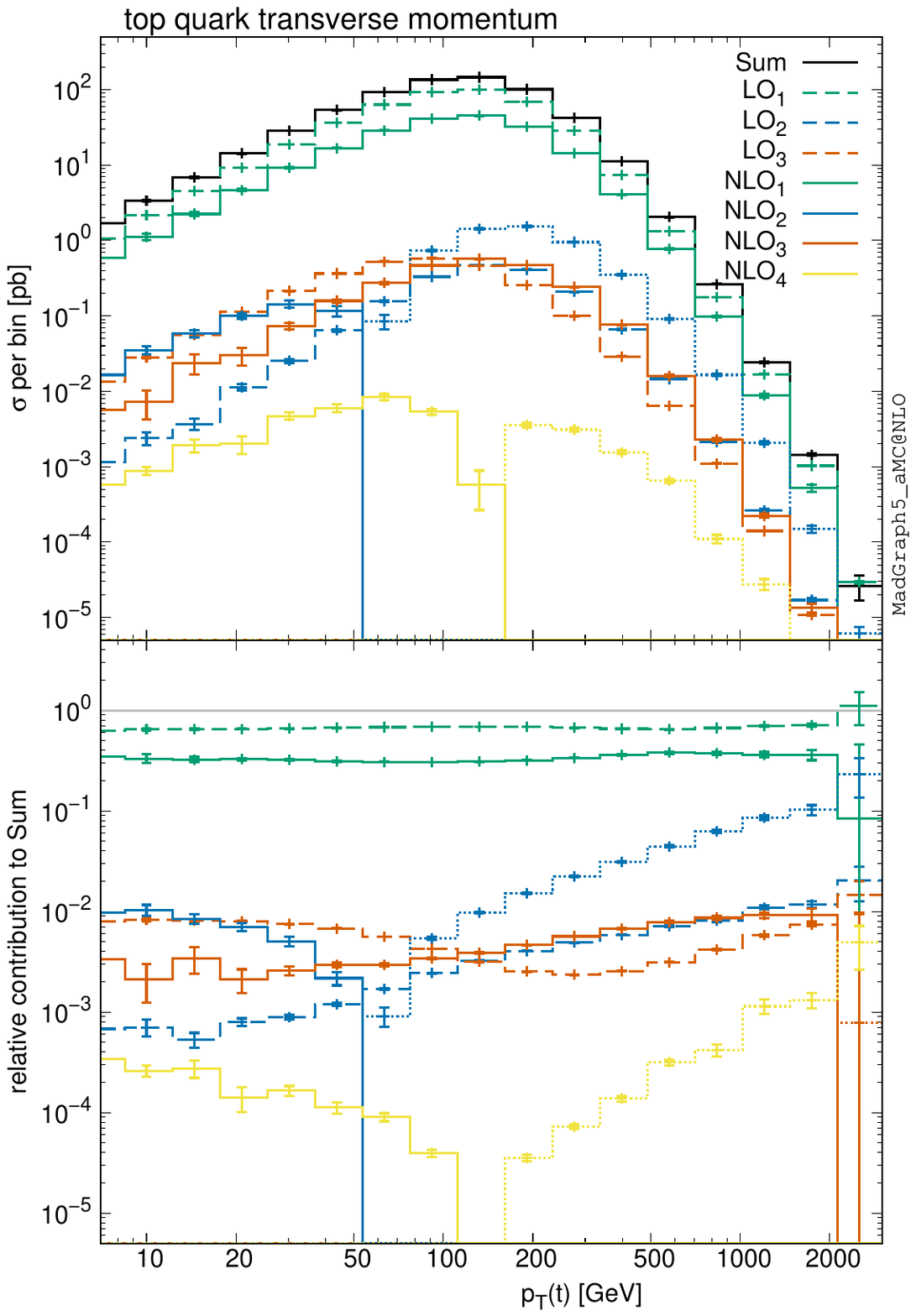}
\end{center}
\caption{\label{fig:tt_blobs}
Transverse momentum of the top quark in $pp\to t\bt$ production. 
}
\end{figure}
The transverse momentum of the top quark in $pp\to t\bt$ production
is presented in fig.~\ref{fig:tt_blobs}. The $\LOo$ contribution
is dominant in the whole $\pt(t)\le 3$~TeV range considered,
accounting for at least 60\% of the total NLO cross section. The second
largest contribution is the $\NLOo$ one, that constitutes a correction
of the $\LOo$ term equal to about 40\% of the latter, with a rather mild 
dependence on $\pt(t)$ -- needless to say, this is well know from standard 
NLO QCD results. All of the other contributions are small and equal to at 
most 1\% of the $\LOo$ one. The exception is the $\NLOt$ term, which
monotonically decreases with $\pt(t)$, becoming negative at around
$\pt(t)\simeq 50~\gev$, and growing up to $-10$\% when $\pt(t)\to 1$~TeV.
This example confirms that, when an accuracy at the percent level is
required, corrections subleading w.r.t.~the dominant QCD ones must
be computed. In the particular case of $\pt(t)$, and up to 3~TeV, the
only contribution that can be safely neglected is the $\NLOf$ one,
which remains everywhere below the per-mille level w.r.t.~$\LOo$.
However, it is important to point out that this conclusion is both
process- and observable-specific.

\begin{figure}[t]
\begin{center}
  \includegraphics[width=0.6\textwidth]{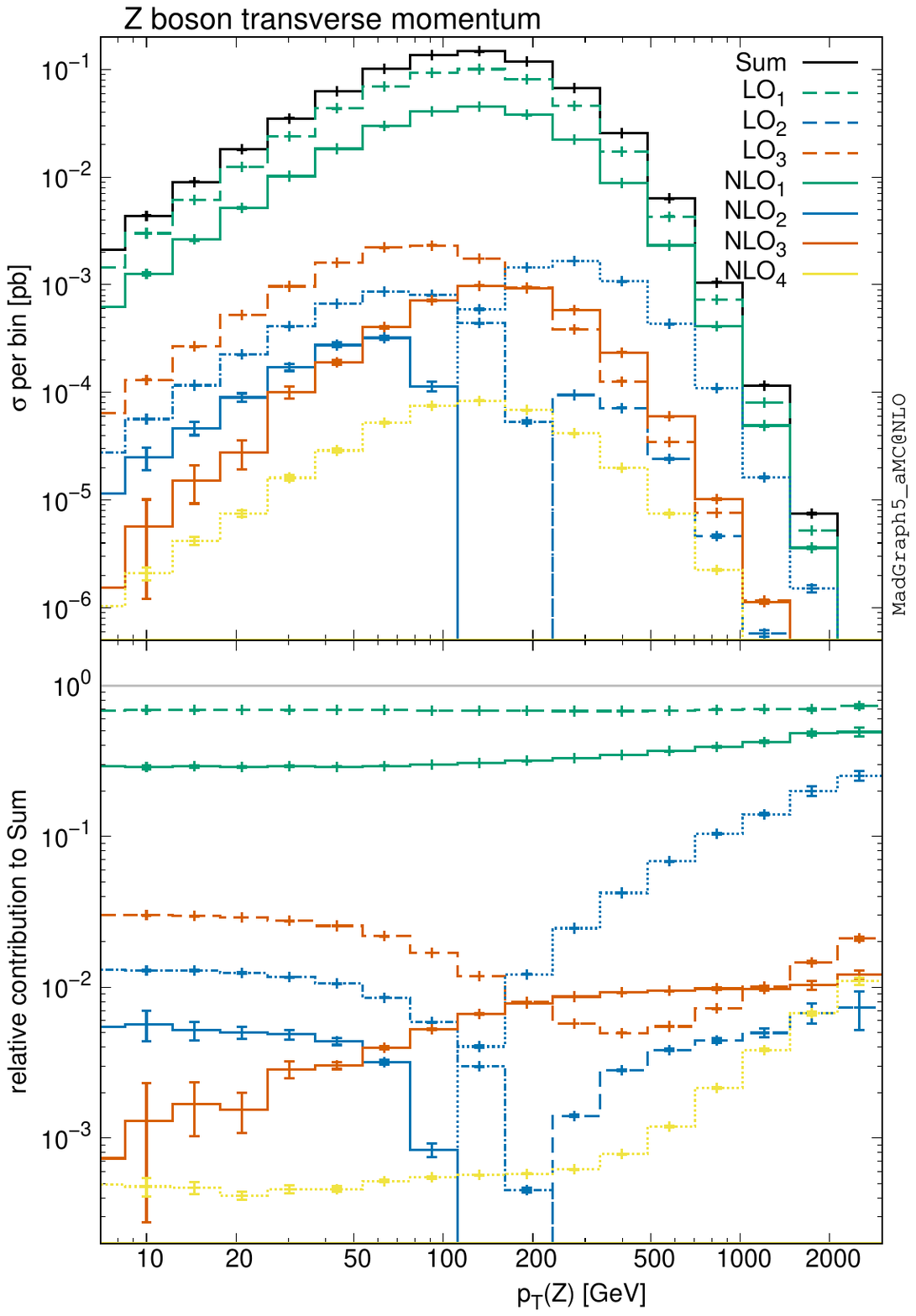}
\end{center}
\caption{\label{fig:ttz_blobs}
Transverse momentum of the $Z$ boson in $pp\to t\bt Z$ production. 
}
\end{figure}
In fig.~\ref{fig:ttz_blobs} we show the transverse momentum of the $Z$
boson in $pp\to t\bt Z$ production. The $\LOo$ contribution is again
dominant, at the level of 70\% of the total NLO cross section. For
$\pt(Z)\lesssim 0.5$~TeV the NLO QCD corrections ($\NLOo$) essentially
account for the remaining 30\% of the rate. However, for larger transverse
momenta the $\NLOt$ contribution decreases very rapidly towards negative
values, which can be as large as $-25$\% of the total at $\pt(Z)\simeq 3$~TeV.
There is thus a significant cancellation between $\NLOo$ and $\NLOt$, since
the former also grows (towards larger positive values) with increasing
$\pt$'s, but slower than the latter. In general, the pattern of the impact
of the subleading terms is an interesting one, in that it systematically
violates the hierarchy one would naively expect on the basis of a simple
coupling-constant counting. For example, at small $\pt$'s the largest
contribution among the subleading ones is that due to $\LOth$, that 
amounts to about 2.5\% of the total NLO rate, followed by $\LOt$ (equal
to about $-1$\% of the total). Moving towards larger $\pt$'s the NLO
subleading terms become increasingly important. Apart from the case of
$\NLOt$, which we have already discussed, it is worth noting at 
$\pt(Z)\gtrsim 2$~TeV we have $\Sigma_{\LOth}>\abs{\Sigma_{\NLOf}}\simeq
\Sigma_{\NLOth}>\Sigma_{\LOt}$, with all these contributions being
relatively close to each other and thus featuring non-negligible
cancellations (since $\Sigma_{\NLOf}<0$).

\begin{figure}[t]
\begin{center}
  \includegraphics[width=0.6\textwidth]{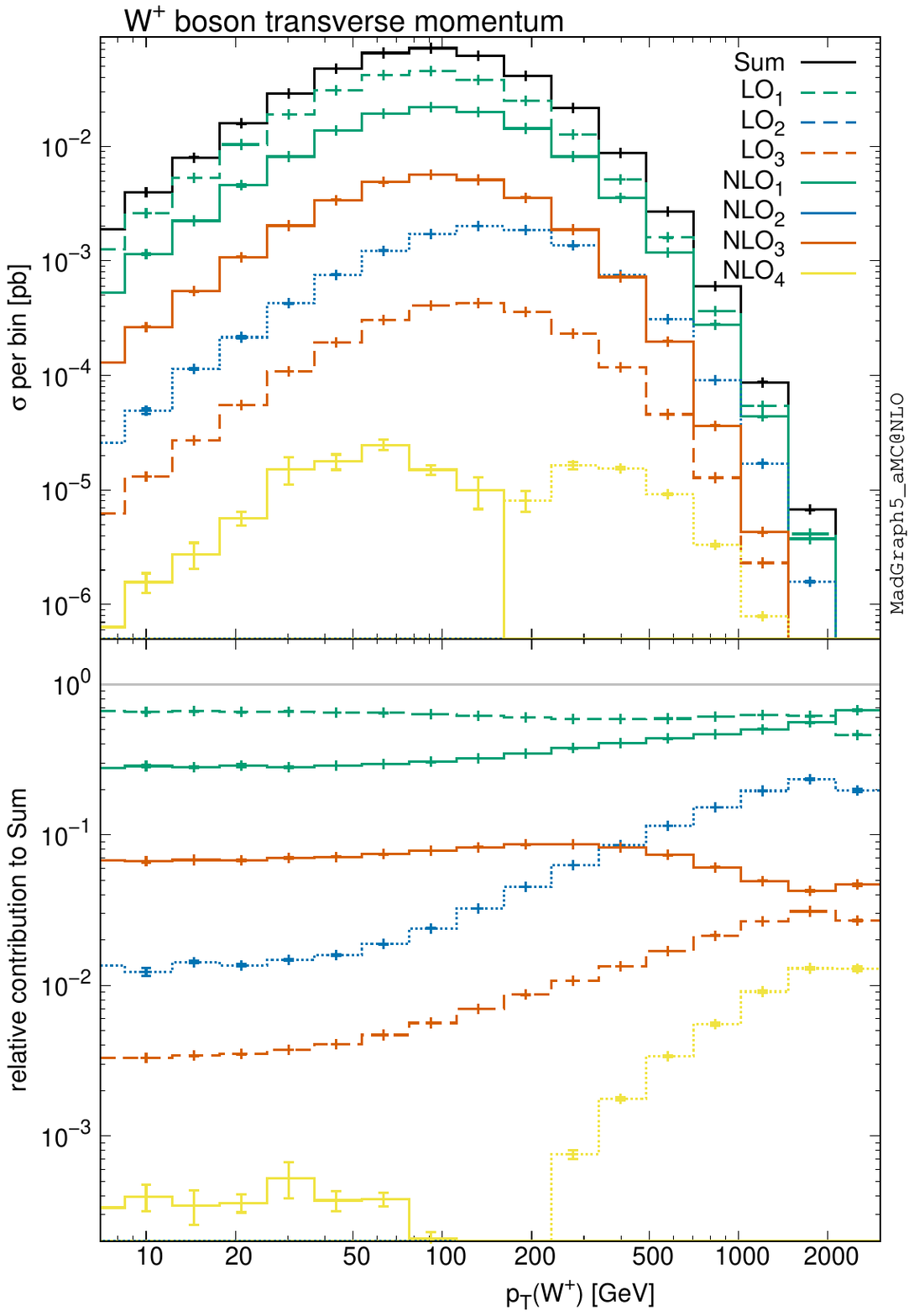}
\end{center}
\caption{\label{fig:ttw_blobs}
Transverse momentum of the $W^+$ boson in $pp\to t\bt W^+$ production. 
}
\end{figure}
The transverse momentum of the hard $W^+$ boson in $pp\to t\bt W^+$ production
is presented in fig.~\ref{fig:ttw_blobs}. As was the case for $Z$ transverse
momentum of fig.~\ref{fig:ttz_blobs}, QCD-induced mechanisms are responsible
for the dominant contributions to the cross section, at both the LO ($\LOo$) 
and the NLO ($\NLOo$). However, there are also notable differences
w.r.t.~the case of $Z$ associated production. More
specifically, we observe what follows. Firstly, the $\Sigma_{\LOt}$ term
is identically equal to zero because of colour. Secondly, for $\pt$'s in 
the TeV region, the $\Sigma_{\NLOo}$ contribution is comparable to or 
larger than the $\Sigma_{\LOo}$ one. Thirdly, for $\pt(W)\lesssim 400~\gev$
the largest subleading term is $\NLOth$, and in particular 
$\Sigma_{\NLOth}>\Sigma_{\NLOt}$. This is the manifestation, at the
differential level, of what has been already observed in the case of
fully inclusive rates in table~\ref{tab:blobs}. More details on
this process can be found in ref.~\cite{Frederix:2017wme}.

\begin{figure}[t]
\begin{center}
  \includegraphics[width=0.6\textwidth]{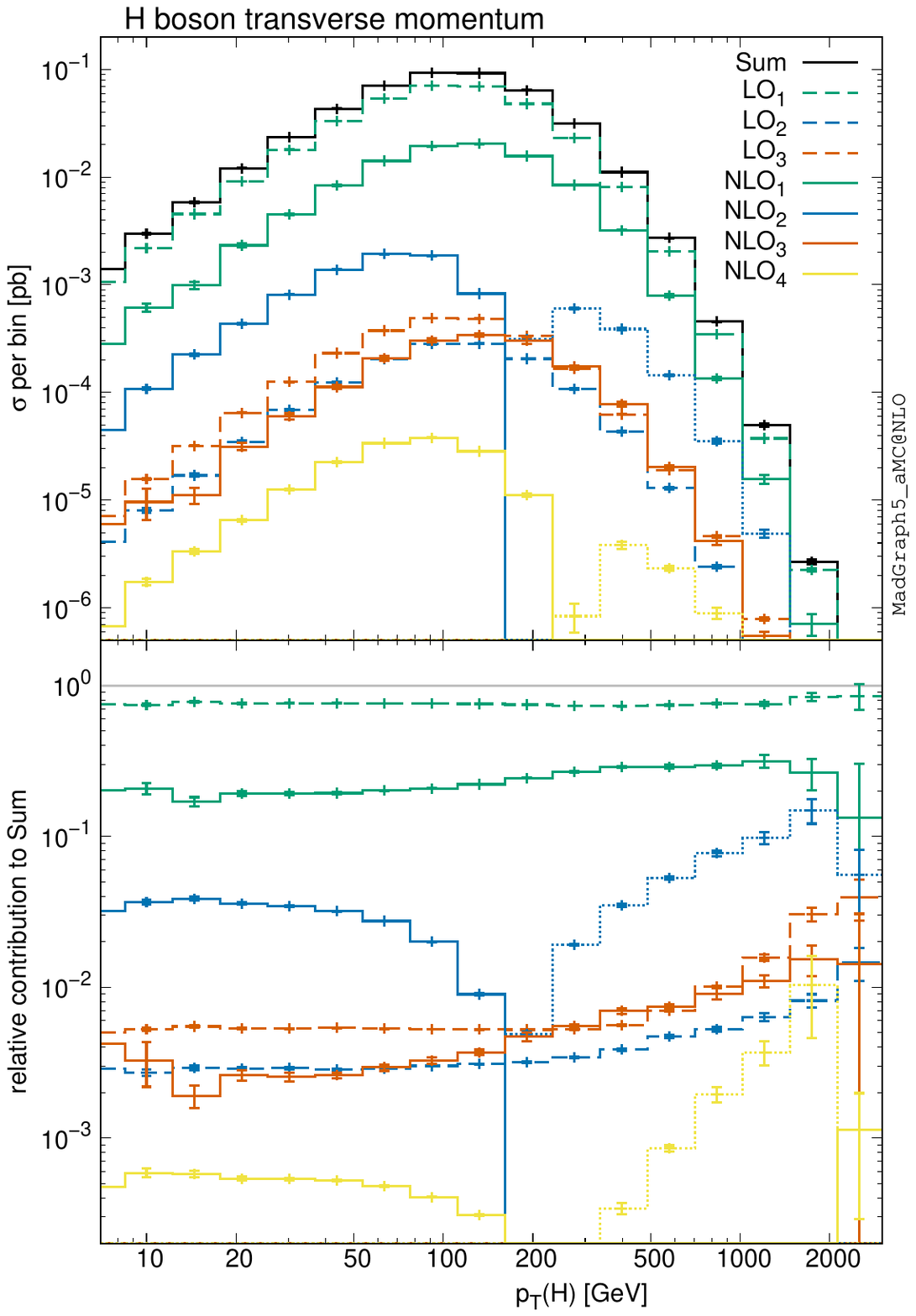}
\end{center}
\caption{\label{fig:tth_blobs}
Transverse momentum of the Higgs boson in $pp\to t\bt H$ production. 
}
\end{figure}
We consider $t\bt H$ production in fig.~\ref{fig:tth_blobs},
where we display the transverse momentum of the Higgs boson. Apart
from the very large $\pt(H)$'s, in this case subleading contributions do 
tend to be numerically subleading. All of them, apart from $\Sigma_{\NLOt}$, 
are well below $1\%$ of the total NLO rate for $\pt(H)\lesssim 1$~TeV.
As was already observed in fig.~\ref{fig:rest}, the NLO EW corrections 
($\NLOt$) are positive ($3-4\%$) at small $\pt$'s, but become negative
at around $\pt(H)\simeq 150~\gev$, and approach the $-10$\% level in 
the TeV range.

\begin{figure}[t]
\begin{center}
  \includegraphics[width=0.6\textwidth]{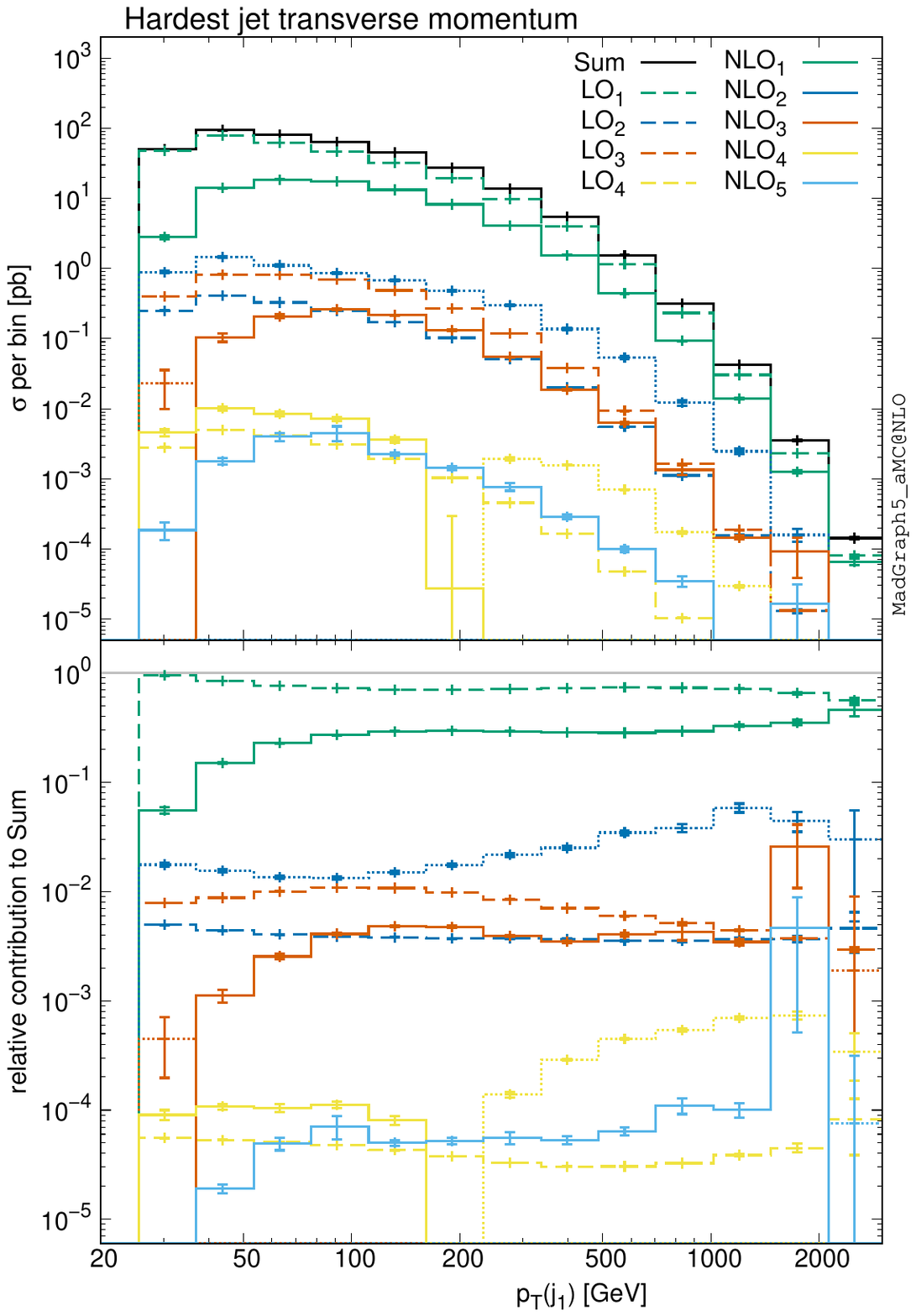}
\end{center}
\caption{\label{fig:ttj_blobs}
Transverse momentum of the hardest jet in $pp\to t\bt j$ production. 
}
\end{figure}
Finally, the transverse momentum of the hardest jet in $pp\to t\bt j$ 
production is presented in fig.~\ref{fig:ttj_blobs}. Consistently
with the results of table~\ref{tab:blobs}, the dominant contributions
are (in this order) $\LOo$, $\NLOo$, and $\NLOt$. As is evident from
the plot, in particular from the lower inset, the relative impact of
the former NLO contribution increases with $\pt$ -- being equal to 
about 5\% of the total cross section at the threshold, growing
significantly immediately afterwards, and reaching a value of
about 30\% for $\pt(j_1)\gtrsim 150$~GeV. As far as $\Sigma_{\NLOt}$
is concerned, it is equal to about $-2$\% of the total cross section at 
the threshold. It decreases slightly up to $\pt(j_1)\sim 100$~GeV, and
then increases (in absolute value) significantly, to reach values of
$\ord(-5\%)$ at $\pt(j_1)\sim 1$~TeV. As was already observed in several
of the cases discussed so far in this section, the hierarchy among the
various contributions does not really follow the one based on naive
coupling-constant counting (apart from the two dominant contributions),
with large violations associated with increasingly subleading terms.

\section{Conclusions\label{sec:conc}}
In this paper, we have studied a number of topics relevant to the perturbative 
computations that are accurate to NLO in the simultaneous expansion in two 
coupling constants, which we refer to as mixed-coupling scenario, with 
the final goal of applying our findings to the automation of such 
computations in the framework of \aNLO. In order to be definite, and
given its importance for the current and future physics collider
programmes, we have explicitly discussed the QCD+EW case, which offers
the additional advantage of exposing the problems posed by an infrared
sector which is by far and large maximally involved. However, the validity
of our treatment is not restricted to those theories, and importantly this
remark applies to the upgraded public version of the \aNLO\ code, which we 
release in conjunction with this paper. In particular, mixed-coupling 
capabilities do not impair one of the guiding principles that underpin 
our software, namely that the characteristics specific to a given theory 
are not hardwired in the code, but loaded dynamically into it as part of a 
model that is fully under the user's control. One example not related to 
EW corrections has been presented in ref.~\cite{Degrande:2016hyf}, in the 
context of charged-Higgs studies.

In order to exemplify the procedure adopted to subtract the infrared
singularities in a mixed-coupling expansion, we have explicitly extended
the relevant formulae of the FKS method to the QCD+QED case. We have
also shown how to re-formulate the FKS subtraction in presence of
final-state tagged particles, represented by means of fragmentation
functions. In doing so, we have proven the cancellation of the 
dimensionally-regulated singularities that emerge in the intermediate 
steps of the computations in an observable- and process-independent
way, and expressed the final results in terms of IR-finite short-distance
partonic cross sections. These features are thus fully analogous to their
counterparts in the inclusive non-tagged case. We have also discussed
several aspects of interest, both formal and relevant to their practical
implementation, of the complex-mass scheme approach to theories with 
unstable particles, so that they can be readily applied also to cases
other than the Standard Model, characterised by arbitrary mass spectra
and possibly large decay widths.

We have computed the total cross sections and selected differential
distributions for a large number of processes at the 13-TeV LHC. We
have always included NLO EW effects, and in the case of $t\bt$ production
(typically in association with other objects) we have actually given
the complete LO and NLO predictions in QCD+EW. Several of these results 
are presented here for the first time. From a phenomenological viewpoint,
one of the most interesting consequences of our study is the fact that
the numerical impact of the various subleading terms is more often than 
not impossible to predict with simple counting arguments based on the
hierarchy of the couplings. 
Needless to say, the extent to which these subleading effects (and
the NLO EW corrections, for that matter) will be important in
practice will depend on the ability of the experiments to collect
data samples with sufficiently large statistics in the relevant
corners of the phase space

Having the LHC phenomenology in mind, the primary application of
this paper in the near future will be that of the computation of
QCD+EW corrections in a systematic manner for all hadroproduction
processes of interest. It is thus important to
bear in mind the current limitations of \aNLO, which will be
removed in later versions; we point out that such limitations
are due to a lack of either implementation work or phenomenological
information, since at the formal level all of the necessary theoretical
ingredients have been given either here or in previous publications.
Firstly, only fixed-order predictions can be obtained, since NLO+PS
capabilities are not publicly released, in spite of the fact that the
MC counterterms appropriate for the MC@NLO modified subtraction of
QCD-like QED showers are available. This stems from both the necessity
of thorough tests against current collider data of the models employed 
in MCs for QED showers, and some interesting issues posed by the presence 
of tagged light particles. Secondly, we have not implemented the FKS 
subtraction with fragmentation functions. We plan to do so starting from
theoretically-motivated functions associated with photons and leptons,
in the hope that this will help their extraction from actual data.
We also point out that the NLO predictions automatically computed
by the code are in the additive scheme; this is mandatory, since the
calculations are not restricted to NLO QCD and NLO EW effects, but might
include further subleading terms. If a multiplicative-scheme result
is desired, one must combine manually the $\Sigma_{\NLOo}$ and
$\Sigma_{\NLOt}$ cross sections given in output by \aNLO. Finally, 
applications relevant to $\epem$ collisions will require the implementation
of ISR and beamstrahlung effects.

\section*{Acknowledgements}
We are indebted to our \aNLOs\ co-authors (J.~Alwall, F.~Maltoni, 
O.~Mattelaer, T.~Stelzer, and P.~Torrielli) for a very fruitful collaboration.
We are grateful to A.~Djouadi, J.~Kalinowski, M.~M\"uhlleitner, and M.~Spira
for their permission to embed \HDecay\ in the \aNLOs\ package.
SF thanks the CERN TH Division for the hospitality during the course of
this work, and P.~Nason for discussions on the results of 
ref.~\cite{Mele:1990cw}. 
RF, VH and DP thank the Munich Institute for Astro- and Particle Physics
(MIAPP) of the DFG cluster of excellence ``Origin and Structure of the
Universe'' for its support.
SF, VH, and HSS are grateful for the support from the organisers of the
Les Houches 2017 ``Physics at TeV Colliders'' workshop, where
interesting discussions related to this work took place.
HSS thanks A.~Denner for discussions about the complex-mass scheme.
MZ thanks the LPTHE Paris, where he was employed during the early
stages of this paper.
RF and DP are supported by the Alexander von Humboldt Foundation in the
framework of the Sofja Kovalevskaja Award Project ``Event Simulation for the
Large Hadron Collider at High Precision''.
The work of VH has been supported by the ERC grant 694712 ``pertQCD''.
HSS is supported by the ILP Labex (ANR-11-IDEX-0004-02, ANR-10-LABX-63).
The work of MZ has been supported by the Netherlands National Organisation
for Scientific Research (NWO) and in part by the European Union's Horizon
2020 research and innovation programme under the Marie Sklodovska-Curie
grant agreement No 660171.

\appendix

\section{Altarelli-Parisi kernels and FKS charge factors
in QED\label{sec:QEDkern}}
We start by reporting the $4-2\epsilon$ dimensional forms of the
unpolarised unregularised ($z<1$) one-loop Altarelli-Parisi (AP henceforth) 
QED kernels:
\beqn
P_{ff}^{\QED<}(z,\ep)&=&e(f)^2\left(\frac{1+z^2}{1-z}-\ep\,(1-z)\right)\,,
\label{APqq}
\\*
P_{\gamma f}^{\QED<}(z,\ep)&=&e(f)^2\left(\frac{1+(1-z)^2}{z}-\ep\,z\right)\,,
\label{APgq}
\\*
P_{f\gamma}^{\QED<}(z,\ep)&=&\nC(f)\,e(f)^2
\left(z^2+(1-z)^2-2\ep\,z(1-z)\right)\,.
\label{APqg}
\eeqn
Equations~(\ref{APqq})--(\ref{APqg}) have to be supplemented with:
\beq
P_{\gamma\gamma}^{\QED<}(z,\ep)=0\,,
\label{APgg}
\eeq
owing to the absence of a three-photon vertex in QED. Equations~(\ref{APqq}) 
and~(\ref{APgq}) display the expected consistency condition:
\beq
P_{\gamma f}^{\QED<}(z,\ep)=P_{ff}^{\QED<}(1-z,\ep)\,.
\eeq
As anticipated in sect.~\ref{sec:ME}, the Casimirs that emerge from
eqs.~(\ref{APqq}) and~(\ref{APgg}) are the same as those in 
eq.~(\ref{CcasQED}).

The AP splitting functions in QED must still obey eqs.~(4.48) and~(4.49)
of ref.~\cite{Frixione:1995ms}, namely:
\beqn
P_{ab}^{\QED}(z,0)&=&\frac{(1-z)P_{ab}^{\QED<}(z,0)}{(1-z)_+}
+\gamma_{\QED}(a)\delta_{ab}\delta(1-z)\,,\phantom{aaa}
\label{APid1}
\\
2C_{\QED}(a)\delta_{ab}\delta(1-z)&=&\delta(1-z)(1-z)P_{ab}^{\QED<}(z,0)\,.
\label{APid2}
\eeqn
By using eq.~(\ref{APqq}) in eq.~(\ref{APid2}), one obtains again
eq.~(\ref{CcasQED}). Conversely, eq.~(\ref{APid1}) can be exploited
to derive the $\gamma_{\QED}$ factors. In order to do so, one can use
the momentum conservation condition and the QED evolution equations
for PDFs. Namely:
\beq
\sum_a\int dx\, x\, f_a(x)=1\,,
\label{momcons}
\eeq
with the sum extended over all flavours. Eq.~(\ref{momcons}) leads to:
\beqn
&&0=\frac{\partial}{\partial\log\mu^2}\,1=\sum_a\int dx\, x\, \sum_b 
P_{ab}\otimes f_b(x)
\nonumber
\\&&\phantom{aaaaaa}
=\sum_{ab}\int dxdydz\, x\,\delta(x-yz)P_{ab}(y)f_b(z)
\nonumber
\\&&\phantom{aaaaaa}
=\sum_{b}\int dz z\left(\sum_{a} \int dy y
P_{ab}(y)\right) f_b(z)\,.
\eeqn
Since this must hold for any set of PDFs, it implies:
\beq
\sum_{a} \int dy y P_{ab}(y)=0\;\;\;\;\;\;\;\;\forall\,b\,.
\label{momcons2}
\eeq
Eq.~(\ref{momcons2}) can be solved explicitly by expanding the AP
kernels perturbatively. At the NLO, the following formula suffices:
\beq
P_{ab}(z)=\frac{\as}{2\pi}P_{ab}^{\QCD}(z)+\frac{\aem}{2\pi}P_{ab}^{\QED}(z)+
{\cal O}(\as^2,\as\aem,\aem^2)\,.
\label{APexp}
\eeq
By considering the terms proportional to $\aem$ in eq.~(\ref{momcons2}),
one arrives at:
\beqn
&&\int dy\,y \Big(P_{\gamma q_i}^{\QED}(y)+P_{q_iq_i}^{\QED}(y)\Big)=0\,,
\\
&&\int dy\,y \left(P_{\gamma l_i}^{\QED}(y)+P_{l_il_i}^{\QED}(y)\right)=0\,,
\\
&&\int dy\,y \left(\sum_i P_{q_i\gamma}^{\QED}(y)+
\sum_i P_{l_i\gamma}^{\QED}(y)+P_{\gamma\gamma}^{\QED}(y)\right)=0\,,
\eeqn
where the range of the index $i$ must include both fermions and antifermions.
By using the forms of the QED AP kernels given before, explicit computations
lead to the following results:
\beqn
\gamma_{\QED}(q)&=&\frac{3}{2}\,e(q)^2\,,
\label{gmmQEDq}
\\
\gamma_{\QED}(l)&=&\frac{3}{2}\,e(l)^2\,,
\label{gmmQEDl}
\\
\gamma_{\QED}(\gamma)&=&-\frac{2}{3}\left(
\NC\sum_{i=1}^{\NF}e(q_i)^2
+\sum_{i=1}^{\Nl}e(l_i)^2\right)\,.
\label{gmmQEDg}
\eeqn
Owing to the absence of QED contributions to gluon branchings at this order,
we also set
\beq
\gamma_{\QED}(g)=0\,;
\label{gmmQCDg}
\eeq
note that $C_{\QED}(g)=0$ from eq.~(\ref{CcasQED}). The QCD $\gamma$'s 
factors are given in \MadFKSeq{4.7} and \MadFKSeq{4.8} for 
strongly-interacting particles (bear in mind that they are denoted here 
by $\gamma_{\QCD}(\ident)$). We also need to define such factors for 
photons and leptons, and obviously:
\beq
C_{\QCD}(l)=C_{\QCD}(\gamma)=\gamma_{\QCD}(l)=\gamma_{\QCD}(\gamma)=0\,.
\eeq
These and eq.~(\ref{gmmQCDg}) serve the sole purpose of expressing
short-distance cross sections in a compact way.

In the FKS formulae for Born-like final-state remainders, there appear
factors denoted by (with the present notation) $\gamma^\prime_{\QCD}$.
These are given in \MadFKSeq{4.9} and \MadFKSeq{4.10} for gluons and
quarks respectively. As was done above, we also need to set:
\beq
\gamma^\prime_{\QCD}(l)=\gamma^\prime_{\QCD}(\gamma)=0\,.
\eeq
It is clear that an FKS subtraction of QED singularities will lead to
similar quantities. One can identify them with the ${\cal O}(\ep)$ 
non-logarithmic term of ${\cal Z}(\ident)$ with a minus sign in front 
-- see eqs.~(A.10) and~(A.11) of ref.~\cite{Frixione:1995ms}. 
Explicit computations of ${\cal Z}(q)$, ${\cal Z}(l)$, and
${\cal Z}(\gamma)$ with AP QED kernels lead to:
\beqn
\gamma_{\QED}^\prime(q)&=&\left(\frac{13}{2}-\frac{2\pi^2}{3}\right)e(q)^2\,,
\label{gmmpQEDq}
\\
\gamma_{\QED}^\prime(l)&=&\left(\frac{13}{2}-\frac{2\pi^2}{3}\right)e(l)^2\,,
\label{gmmpQEDl}
\\
\gamma_{\QED}^\prime(\gamma)&=&\frac{23}{6}\,\gamma_{\QED}(\gamma)\,,
\label{gmmpQEDg}
\eeqn
and trivially $\gamma_{\QED}^\prime(g)=0$.

\noindent
The results of this appendix imply, in particular:
\beqn
P_{\gamma\gamma}^{\QED}(z)&=&\gamma_{\QED}(\gamma)\,\delta(1-z)\,,
\\
\gamma_{\QED}(\gamma)&=&2\pi\beta_0^{\QED}\,,
\\
\gamma_{\QCD}(g)&=&2\pi\beta_0^{\QCD}\,,
\eeqn
with the normalization (note the sign of the QED coefficient):
\beqn
\frac{\partial\as(\mu^2)}{\partial\log\mu^2}&=&-\beta_0^{\QCD}\as^2(\mu^2)+
{\cal O}(\as^3)\,,
\label{asrun}
\\
\frac{\partial\aem(\mu^2)}{\partial\log\mu^2}&=&-\beta_0^{\QED}\aem^2(\mu^2)+
{\cal O}(\aem^3)\,.
\label{aemrun}
\eeqn

\section{RGE invariance\label{sec:RGE}}
In order re-instate in the cross sections the separate dependence 
on the different hard scales, let $\muR$ be the QCD renormalisation 
scale, $\mua$ the QED renormalisation scale, and $\muF$ the factorisation 
scale. Then, the rule given at the beginning of appendix C of 
ref.~\cite{Frederix:2009yq} is modified as follows:

\vskip 0.3truecm
\noindent
\begin{center}
\begin{minipage}{0.85\textwidth}
{\em 
If $\muF\ne\muR$ or $\muF\ne\mua$, all the formulae for the short-distance
cross sections given in this paper must be computed with $\mu=\muF$, 
except for the argument of $\as$, which must be set equal to $\muR$, 
and for the argument of $\aem$, which must be set equal to $\mua$.}
\end{minipage}
\end{center}

\vskip 0.3truecm
\noindent
Because of this, two extra terms must be added on the r.h.s.~of 
eq.~(\ref{factTH3}) at the level of cross sections (that is, convoluted
with PDFs and summed over $p$ and $q$):
\beq
C_{(p,q)}\log\frac{\muF^2}{\muR^2}+D_{(p,q)}\log\frac{\muF^2}{\mua^2}\,.
\label{extra}
\eeq
By neglecting terms beyond NLO, one arrives at the analogue of
\MadFKSeq{C.4}:
\beq
\sum_{pq}\left(\frac{\partial}{\partial\log\muR^2}
d\sigma_{(p,q)}^{(B,n)}-C_{(p,q)}\right)=0\,.
\label{murin}
\eeq
By using eq.~(\ref{asrun}), one sees that a sufficient condition
for the solution of eq.~(\ref{murin}) is:
\beq
C_{(p,q)}=-2\pi(p-1)\beta_0^{\QCD}\,\frac{\as(\muR^2)}{2\pi}\,
d\sigma_{(p-1,q)}^{(B,n)}\,,
\label{Csol}
\eeq
for any $p\ge 1$ and $q$. Analogously, the independence of $\mua$
leads to:
\beq
D_{(p,q)}=-2\pi(q-1)\beta_0^{\QED}\,\frac{\aem(\mua^2)}{2\pi}\,
d\sigma_{(p,q-1)}^{(B,n)}\,,
\label{Dsol}
\eeq
for any $p$ and $q\ge 1$. Eqs.~(\ref{Csol}) and~(\ref{Dsol}) show that
both $C_{(p,q)}$ and $D_{(p,q)}$ are proportional to $\as^p\aem^q$,
which justifies the notation adopted.

In ref.~\cite{Frederix:2009yq} the RGE for $\muF$ was used to 
insert the dependence on multiple scales into the virtual corrections
computed with a single scale; the same operation can be performed here.
By keeping only terms of LO and NLO, from eqs.~(\ref{factTH3})
and~(\ref{extra}) we obtain:
\beqn
&&\frac{\partial d\sigma_{\sss H_1H_2}}{\partial\log\muF^2}=
\left(\frac{\partial f^{(H_1)}}{\partial\log\muF^2}\star f^{(H_2)}
+f^{(H_1)}\star \frac{\partial f^{(H_2)}}{\partial\log\muF^2}\right)
\star\sum_{pq} d\sigma_{(p,q)}^{(B,n)}
\nonumber\\&&\phantom{aaa}
+f^{(H_1)}\star f^{(H_2)}\star\sum_{pq}
\frac{\partial}{\partial\log\muF^2}
\left(
d\bar{\sigma}_{(p,q)}^{(n+1)}+d\sigma_{(p,q)}^{(C,n)}+
d\sigma_{(p,q)}^{(V,n)}\right)
\nonumber\\&&\phantom{aaa}
+f^{(H_1)}\star f^{(H_2)}\star\sum_{pq}
\left(C_{(p,q)}+D_{(p,q)}\right)=0\,.
\label{factTH3d}
\eeqn
The terms that feature the derivative of the PDFs are dealt with
the AP equations and the following identity, where we explicitly
indicate initial-state flavours:
\beq
\frac{\partial}{\partial\log\muF^2}f_a\star d\sigma_{ab}=
P_{ad}\otimes f_d\star d\sigma_{ab}=
f_a\star P_{da}\star d\sigma_{db}\,.
\label{PDFev}
\eeq
By using the perturbative expansion of the AP kernels given
in eq.~(\ref{APexp}), each of the first two terms on the r.h.s.~of
eq.~(\ref{factTH3d}) generates two contributions, corresponding
to the QCD and QED evolution of the relevant incoming leg.
From eqs.~(\ref{Kdef}) and~(\ref{degnpo}), one sees that the 
derivative w.r.t.~$\muF$ of the degenerate $(n+1)$-body terms
is determined by (${\rm T}=\QCD,\QED$):
\beqn
\frac{\partial}{\partial\log\muF^2}\,{\cal K}^{\Tt}&=&-
\xi P^{\Tt<}(1-\xi)\xidistr{c}
\nonumber\\&=&
-\xi P^{\Tt<}(1-\xi)\left[\xidistr{+}-\delta(\xi)\log\xicut\right].
\label{tmp}
\eeqn
By using the identities of eqs.~(\ref{APid1}) and~(\ref{APid2})
in eq.~(\ref{tmp}), one has:
\beqn
\frac{\partial}{\partial\log\muF^2}{\cal K}_{ab}^{\Tt}&=&
-P_{ab}^{\Tt}(1-\xi)+
\Big(\gamma_{\Tt}(a)+2C_{\Tt}(a)\log\xicut\Big)\delta_{ab}\delta(\xi)\,.
\eeqn
The first term on the r.h.s.~of this equation cancels the contributions
due to PDFs evolution (see eq.~(\ref{PDFev})). The other term cancels
the contribution due to $d\sigma_{(p,q)}^{(C,n)}$ in eq.~(\ref{factTH3d}),
which results from the first line on the r.h.s.~of eq.~(\ref{Qdef}).
This implies that a sufficient condition for eq.~(\ref{factTH3d}) to
be satisfied is:
\beq
\frac{\partial}{\partial\log\muF^2}d\sigma_{(p,q)}^{(V,n)}
+C_{(p,q)}+D_{(p,q)}=0
\label{soldsigv}
\eeq
for any $(p,q)$, which generalises \MadFKSeq{C.8}. Eq.~(\ref{soldsigv})
can now be employed in the same way as \MadFKSeq{C.8} in appendix~C of 
ref.~\cite{Frederix:2009yq}, starting from the analogue of \MadFKSeq{C.9}:
\beq
\vampsqnl_{(p,q){\sss FIN}}(\muR^2,\mua^2,\muF^2,Q^2)=
\as^p(\muR^2)\aem^q(\mua^2)\hvampsqnl_{(p,q){\sss FIN}}(\muF^2,Q^2)\,,
\eeq
while \MadFKSeq{C.10} is still valid. Therefore, if given a one-scale
finite virtual contribution $v_{(p,q)}(M^2)$ such that:
\beq
\vampsqnl_{(p,q){\sss FIN}}(M^2,M^2,M^2,M^2)=v_{(p,q)}(M^2)\,,
\eeq
we can obtain the analogue of \MadFKSeq{C.13}:
\beqn
&&\vampsqnl_{(p,q){\sss FIN}}(\muR^2,\mua^2,\muF^2,Q^2)=
\as^p(\muR^2)\aem^q(\mua^2)
\frac{v_{(p,q)}(Q^2)}{\as^p(Q^2)\aem^q(Q^2)}
\\&&\phantom{aaaaaa}
+2\pi\left((p-1)\beta_0^{\QCD}\ampsqnt_{(p-1,q)}
+(q-1)\beta_0^{\QED}\ampsqnt_{(p,q-1)}\right)\log\frac{\muF^2}{Q^2}\,.
\nonumber
\eeqn
For all practical purposes, the scale dependence of $\aem$ can be
neglected. In the formulae above, this is achieved by formally
setting $\beta_0^{\QED}=0$, and by replacing $\aem(\mua)$ with a
constant value appropriate to the EW renormalisation scheme that
is being employed.

\section{Symmetry factors in FKS\label{sec:symm}}
The symmetry factor associated with the $(m+2)$-body matrix elements in
FKS formulae is always equal to $m!$, regardless of the
flavours of the $m$ final-state partons (see eq.~(\ref{mFxsec})).
To see why this is so, one starts from the identity\footnote{The
arguments of the matrix elements in eq.~(\ref{Mperm}) would best be
written as ordered sets, since we understand that, for any $l$, parton $l$ 
with flavour $a_l$ has momentum $\bar{k}_l$ on the l.h.s.~and 
momentum $\bar{k}_{\sigma(l)}$ on the r.h.s. However, since we are
about to prove that in all cases the final state can be regarded as
fully symmetric, we thought it unnecessary to introduce an ordered-set
notation just for the sake of the present discussion.}:
\beq
\sum_{\set{a_l}{1}{m+2}}
\ampsq^{(m,L)}\left(\set{a_l}{1}{m+2};\set{\bar{k}_l}{1}{m+2}\right)=
\sum_{\set{a_l}{1}{m+2}}
\ampsq^{(m,L)}\left(\set{a_l}{1}{m+2};
\set{\bar{k}_{\sigma(l)}}{1}{m+2}\right)\,,
\label{Mperm}
\eeq
for any given kinematic configuration $\{\bar{k}_l\}$, \mbox{$1\le l\le m+2$}.
In eq.~(\ref{Mperm}), $\sigma$ denotes any permutation of $m$ objects,
defined so that it acts (possibly) non-trivially only on indices
\mbox{$3\le l\le m+2$}:
\beq
\sigma(1)=1\,,\;\;\;\;
\sigma(2)=2\,,\;\;\;\;
\{3,\ldots m+2\}=\{\sigma(3),\ldots\sigma(m+2)\}\,.
\label{sig12}
\eeq 
It is straightforward to convince oneself that eq.~(\ref{Mperm}) holds 
true. To this end, consider a given set of $m+2$ flavour values,
\mbox{$\{f_1,\ldots f_{m+2}\}$}, and a given permutation $\sigma$.
Owing to the fact that the sums over flavours in eq.~(\ref{Mperm}) 
extend over all possible values, there exists on the l.h.s.~of that
equation a contribution where parton $i$ has flavour and momentum equal to:
\beq
\left(f_{\sigma^{-1}(i)},\bar{k}_i\right)\,.
\label{tmp11}
\eeq
By introducing an index $j=\sigma^{-1}(i)$, eq.~(\ref{tmp11}) can be 
re-written as follows:
\beq
\left(f_j,\bar{k}_{\sigma(j)}\right)\,,
\eeq
which therefore is manifestly identical to a contribution to the
r.h.s.~of eq.~(\ref{Mperm}). The same argument allows one to
start from a term on the r.h.s.~of eq.~(\ref{Mperm}), and associate
it with a term on the l.h.s.~of that equation, thus concluding the proof

Equation~(\ref{Mperm}) is essentially the definition of a fully-symmetric
final state, which therefore entails a symmetry factor equal to $m!$. 
However, this does not yet imply that:
\beq
\frac{1}{m!}\sum_{\set{a_l}{1}{m+2}}
\ampsq^{(m,L)}\left(\set{a_l}{1}{m+2};\set{\bar{k}_l}{1}{m+2}\right)
\label{sum1}
\eeq
is equal to the quantity that enters the definition\footnote{We assume
for the moment that all other factors in the definition of the cross
sections are invariant under permutations of final-state partons, as
in the standard FKS formulation. Other cases, including that of fragmentation,
will be treated shortly.} of the short-distance cross sections, namely:
\beq
\sum^\star_{\set{a_l}{1}{m+2}}\frac{1}{S(\set{a_l}{3}{m+2})}\,
\ampsq^{(m,L)}\left(\set{a_l}{1}{m+2};\set{\bar{k}_l}{1}{m+2}\right).
\label{sum2}
\eeq
Here, $\sum^\star$ restricts the sum over all non-redundant flavour
combinations (i.e.~all of the partonic subprocesses). The symmetry factor
is:
\beq
S(\set{a_l}{3}{m+2})=
\prod_{i=1}^k n_i!\;,
\label{nisym}
\eeq
where we have assumed that the $m$-body final state is partitioned into 
$k$ sets, each  of which composed of $n_i$ identical particles 
($i=1,\ldots k$), so that:
\beq
\sum_{i=1}^k n_i = m\,.
\eeq
In order to show that eq.~(\ref{sum1}) is identical to eq.~(\ref{sum2}),
we first re-write the former as follows:
\beqn
&&\frac{1}{m!}\sum_{\set{a_l}{1}{m+2}}
\ampsq^{(m,L)}\left(\set{a_l}{1}{m+2};\set{\bar{k}_l}{1}{m+2}\right)=
\label{sum10}
\\*&&\phantom{aaaaaaaaa}
\frac{1}{m!}\sum^\star_{\set{a_l}{1}{m+2}} f(\set{a_l}{3}{m+2})\,
\ampsq^{(m,L)}\left(\set{a_l}{1}{m+2};\set{\bar{k}_l}{1}{m+2}\right)\,,
\nonumber
\eeqn
where $f(\set{a_l}{3}{m+2})$ is the number of identical contributions
to the sum of eq.~(\ref{sum1}). Such a number is equal to the number of 
permutations of $m$ objects that do {\em not} leave invariant its $k$ subsets 
composed of $n_i$ identical objects (because all ordered sets left invariant 
by a permutation that acts non-trivially only on identical 
elements are counted once in flavour sums such as those that appear
in eq.~(\ref{Mperm})). To compute this, start from the total number
of permutations of $m$ objects, which is equal to $m!$. Choose the set 
generated by one of them, and consider its subset labelled by $k=1$; each 
of the $n_1!$ permutations that operates within this subset is equal to the 
identity. This argument is valid for {\em each} of the original permutations 
chosen, which implies that the number of permutations that do not act as 
the identity on the $k=1$ subset is $m!/n_1!$. By repeating this argument
for each of the $k$ subsets, one arrives at the number sought:
\beq
f(\set{a_l}{3}{m+2})=\frac{m!}{\prod_{i=1}^k n_i!}\,.
\label{indip}
\eeq
By inserting this result into eq.~(\ref{sum10}) and by taking
eq.~(\ref{nisym}) into account, one proves that eqs.~(\ref{sum1}) 
and~(\ref{sum2}) are indeed identical.

Let us now consider the expression:
\beq
X=\sum_{\set{a_l}{3}{m+2}}
\ampsq^{(m,L)}\left(\set{a_l}{1}{m+2}\right)
g(a_p)\, d\phi_m\,,
\label{sumgap}
\eeq
for any function $g(a)$ (which can possibly also depend on the four-momentum
of parton $a$) and index \mbox{$3\le p\le m+2$}. Examples of 
eq.~(\ref{sumgap}) are the hadron-level fragmentation cross
section, eq.~(\ref{HmFxsec}), or cross sections one arrives at in
the intermediate steps of the FKS procedure, such as \FKSeq{4.20}.
By applying to eq.~(\ref{sumgap}) the same arguments made before 
in this appendix one arrives at:
\beq
X=\sum_{\set{a_l}{3}{m+2}}
\ampsq^{(m,L)}\left(\set{a_l}{1}{m+2}\right)
g(a_{p^\prime})\, d\phi_m\,,
\label{tmp4}
\eeq
for any \mbox{$3\le p^\prime\le m+2$}, $p^\prime\ne p$. Here, there is a 
subtlety which is worth stressing: at variance with eq.~(\ref{Mperm}),
that is fully local in the momentum space, the fact that eqs.~(\ref{sumgap})
and~(\ref{tmp4}) are identical understands a phase-space integration
(whence the presence of the phase-space $d\phi_m$ in these equations),
owing to the possible momentum dependence of the function $g()$. This
is obviously not restrictive, since all IR-safe observables are obtained
by an integration over the phase-space of the relevant short-distance
quantities. The arbitrariness of $p^\prime$ in eq.~(\ref{tmp4}) thus 
leads to:
\beq
\sum_{\set{a_l}{3}{m+2}}
\ampsq^{(m,L)}\left(\set{a_l}{1}{m+2}\right)
g(a_p)\, d\phi_m =
\frac{1}{m}\sum_{p=3}^{m+2}\sum_{\set{a_l}{3}{m+2}}
\ampsq^{(m,L)}\left(\set{a_l}{1}{m+2}\right)
g(a_p)\, d\phi_m\,.
\label{sumgap2}
\eeq
The factor $1/m$ on the r.h.s.~of eq.~(\ref{sumgap2}) implies that
the symmetry factor associated with quantities such as $X$ defined in
eq.~(\ref{sumgap}) {\em summed over} $p$ is indeed equal to $m!$, 
since $m!=m(m-1)!$, and $(m-1)!$ is the symmetry factor of
$X$ with $p$ fixed. We point out that the identity in eq.~(\ref{sumgap2})
is used not only when dealing with fragmentation cross sections, but
also in the manipulation of the cross sections that emerge in the
intermediate steps of the computation in the standard FKS case -- an
explicit example will be given later in this appendix.

In the context of fragmentation cross sections, eq.~(\ref{sumgap2}) 
can also be understood in a less formal way by using simple physics
considerations. Consider an $m$-gluon final state. A gluon is fragmented,
and one is thus left with $(m-1)$ gluons, with an associated symmetry factor
equal to $(m-1)!$. It is clear that it does not matter which particular
gluon fragments; one the other hand, in order to avoid counting the same
contribution more than once, the sum over different fragmenting gluons
need {\em not} be performed: one is sufficient. However, precisely because
all possible fragmentation contributions are identical, one can symbolically
write:
\beq
\frac{1}{(m-1)!}=\frac{1}{m!}\sum_{p=3}^{m+2}\,,
\label{mid}
\eeq
which is what eq.~(\ref{sumgap2}) expresses more precisely\footnote{This
includes the fact that, by summing over flavours, final-state configurations 
that feature different parton species can be treated in the same way as
identical-particle configurations as far as symmetry factors are concerned.}.
Note, finally, that the r.h.s.~of eq.~(\ref{mid}) is also 
consistent with the idea that the sum over fragmentation contributions is 
related to final-state multiplicities. By formally thinking of a gluon as a 
hadron (which can be done by setting the FF equal to a Dirac delta), in an
$m$-body configuration the gluon multiplicity has to be equal to $m$,
and the corresponding symmetry factor equal to $m!$, which is precisely
eq.~(\ref{mid}). 

We conclude this appendix with a remark on the label of the fragmenting
parton, and specifically on its summation range. Owing to eq.~(\ref{HmFxsec}),
one starts with a real-emission contribution (which has an $(n+1)$-body
final state) where \mbox{$3\le p\le n+3$}. After the procedure that leads
to the cancellation of the IR singularities, one is left with several
(quasi-)$n$-body contributions (namely, soft (eq.~(\ref{realsF})), 
initial-state collinear remainders (eqs.~(\ref{dsiginpl}) and~(\ref{dsiginmn})),
and final-state collinear remainders (eqs.~(\ref{dsigoutd}) 
and~(\ref{dsigoutz}), or eqs.~(\ref{dsigoutd2}) and~(\ref{dsigoutz2})))
for which \mbox{$3\le p\le n+2$}. It is important to realise that this
reduced summation range is not imposed in order to be consistent with
the $n$-bodiness of the corresponding contribution, but rather that it
emerges naturally from the computation. In order to show that this is
the case, we consider the soft cross section of eq.~(\ref{realsF}) in
order to be definite.

Firstly, we note that when dealing with the analogue of \FKSeq{4.11},
the fragmentation-specific information is taken into account by using
the identity:
\beq
\sum_{i=3}^{n+3}\sum_{j=1}^{n+3}\delta(\xii)\Sfunij F^{(p)}_H J_{n+1}^{n-1}=
\sum_{i=3}^{n+3}\delta(\xii)\left(1-\delta_{ip}\right)F^{(p)}_H 
J_{n}^{n-1}([i])\,,
\label{tmp1}
\eeq
where we have exploited \MadFKSeq{4.18} and eq.~(\ref{SClimF}). 
The key point of the procedure of sect.~4.2 of ref.~\cite{Frixione:1995ms}
is that the dependence upon the soft parton $i$ is restricted to the
eikonal factors, and disappears from the reduced $n$-body cross sections
(upon relabeling). Crucially, eq.~(\ref{tmp1}) guarantees that this
remains true also in the fragmentation case, thanks to the factor
\mbox{$1-\delta_{ip}$}. Because of this, the flavour $a_p$ of the fragmenting
parton does not play any special role: it appears also in the reduced
matrix elements, where its properties are never used in the manipulations
of sect.~4.2 of ref.~\cite{Frixione:1995ms}. Indeed, the only parton
flavour which needs a specific treatment is that of the soft one, $a_i$,
the sum over which must be carried out explicitly (and turns out to be
trivial owing to a factor $\delta_{a_ig}$ -- see the comment immediately
below \FKSeq{4.18}). 

The bottom line is that also in the present case one arrives at 
\FKSeq{4.20} (with minor notation changes, and the fragmentation
factor $F^{(p)}_H$). At this point, however, some care is required.
In the FKS procedure, one passes from \FKSeq{4.20} to the individual
terms in the sum on the r.h.s.~of \FKSeq{4.25} by means of a relabeling,
i.e.~of a map $R_\sigma$:
\beq
\left(3,\cdots i-1,\cancel{i},i+1,\cdots n+3\right)\;\;
\stackrel{R_\sigma}{\longrightarrow}\;\;
\left(\sigma(3),\cdots\sigma(i-1),\sigma(i),\cdots \sigma(n+2)\right)\,,
\label{relab}
\eeq
with $\sigma$ a permutation of $n$ objects. The relabeling formalises
the independence of \FKSeq{4.20} of $i$, and allows one to see that
the sum on the r.h.s.~of \FKSeq{4.25} amounts to an overall factor
equal to $n+1$ (equal to $4$ in ref.~\cite{Frixione:1995ms}). It is 
the arbitrariness of $\sigma$ that implies that one can put \FKSeq{4.27}
in its fully symmetric form in terms of final-state quantities. One 
starts by writing the Born cross section times\footnote{Note that this 
pre-factor $n$ arises not from the sum over $i$, but from that on one of 
the indices $n$ and $m$ in \FKSeq{4.25}.} e.g.~$nC(a_3)$ (and by choosing 
the colour-linked Born indices e.g.~as $k=3$ and $k=4$), and then 
uses the procedure that leads to eq.~(\ref{sumgap2}) to arrive at the 
sums that appear in \FKSeq{4.27}. 

In the fragmentation case this procedure is unchanged. However, 
the presence of \mbox{$1-\delta_{ip}$} implies that the
overall factor that results from the sum over $i$ in \FKSeq{4.25} is
not equal to $n+1$, but rather to $n$. Therefore, when this is combined
with the symmetry factor of the real-emission matrix elements $(n+1)!$
(i.e.~the $1/4!$ of \FKSeq{4.20}), one obtains $n/(n+1)!$ rather than
$1/n!$ (which in FKS is then interpreted as the Born-level symmetry
factor, and absorbed into the Born cross sections of \FKSeq{4.25}). 
Therefore, in the fragmentation case we can still arrive at
\FKSeq{4.27}, up to two differences: a fragmentation factor
\beq
F^{(p^\prime)}_H\,,\;\;\;\;\;\;\;\;
p^\prime=\sigma\big(\stepf(i-p)p+\stepf(p-i)(p-1)\big)\,,
\label{Fsig}
\eeq
and an overall factor
\beq
\frac{n}{n+1}\,,
\label{ofac}
\eeq
which arises from:
\beq
\frac{n}{(n+1)!}=\frac{1}{n!}\,\frac{n n!}{(n+1)!}=
\frac{1}{n!}\,\frac{n}{n+1}\,,
\eeq
with the $1/n!$ term then included in the Born cross sections as in FKS.
Now observe that by construction (see eq.~(\ref{relab})) the index
of the fragmenting parton in eq.~(\ref{Fsig}) is such that
\mbox{$3\le p^\prime\le n+2$}, while for the original index
\mbox{$3\le p\le n+3$}. Owing to the
symmetry of \FKSeq{4.27}, the property of eq.~(\ref{tmp4}) can be
exploited here as well, to relabel $p^\prime$ as $1$. This 
implies that the sum over $p$ in eq.~(\ref{HmFxsec}) now contains
$n+1$ identical terms. By performing that sum, one gets rid
of the $n+1$ factor in the denominator of eq.~(\ref{ofac}). The remaining
factor $n$ there is then cancelled by performing the procedure that
leads to eq.~(\ref{sumgap2}) (by what is there the factor $1/m$),
procedure that allows one to re-instate a fully symmetric form in
terms of the fragmentation factors. This concludes the proof that 
the analogue of \FKSeq{4.27} is given by eq.~(\ref{realsF}).

\section{Process generation and infrared safety\label{sec:tech}}
We remind the reader that the generic expression of a cross
section in a mixed-coupling expansion is given by the master
equation~(\ref{taylor40}). At the LO and NLO, in particular,
this reads as in eqs.~(\ref{SigB}) and~(\ref{SigNLO}), respectively.
The integer numbers $k_0$, $c_s(k_0)$, $c(k_0)$, and $\Delta(k_0)$ are 
process-specific quantities (with \mbox{$k_0=c_s(k_0)+c(k_0)+\Delta(k_0)$}), 
whose {\em Born-level} interpretation is apparent in eq.~(\ref{SigB}), namely:
\begin{itemize}
\item $k_0$ is the overall power of the coupling constants combined
(i.e.~$\as^n\aem^m$ is such that $n+m=k_0$) at the Born level, which is 
the same for all the $\Sigma_{\LOi}$ contributions.
\item $c_s(k_0)$ is the power of $\as$ common to all $\Sigma_{\LOi}$ 
contributions.
\item $c(k_0)$ is the analogue of $c_s(k_0)$, relevant to $\aem$.
\item $\Delta(k_0)+1$ is the number of contributions to the
complete LO cross section $\Sigma^{\rm (LO)}$.
\end{itemize}
As was already anticipated in sect.~\ref{sec:gen}, the typical
\aNLOs\ process-generation command may read as follows:

\vskip 0.25truecm
\noindent
~~\prompt\ {\tt ~generate~p$_1$ p$_2$ > p$_3$ p$_4$ p$_5$ p$_6$ 
QCD=$\nmax$ QED=$\mmax$ [QCD QED]}

\vskip 0.25truecm
\noindent
where {\tt p$_i$} denotes either a particle or a multiparticle label.
The integers $\nmax$ and $\mmax$ help decide which $\Sigma_{\LOi}$ and 
(indirectly) $\Sigma_{\NLOi}$ will be included in the predictions, according
to the constraints given in eqs.~(\ref{LOsynt}) and~(\ref{NLOsynt}) 
for the LO and NLO cross sections, respectively.
$\nmax$ and $\mmax$ can be freely set by the user. However, we point out
that some assignments might not correspond to any physical contribution.
In particular, by comparing eq.~(\ref{LOsynt}) with eq.~(\ref{SigB}),
we obtain the following conditions:
\beqn
&&\nmax\ge c_s(k_0)\,,\;\;\;\;\;\;\;\;
\mmax\ge c(k_0)\,,
\label{cond0}
\\*&&
\max\left(0,c_s(k_0)+\Delta(k_0)-\nmax\right)\le q\le
\min\left(\Delta(k_0),\mmax-c(k_0)\right)\,.
\label{cond1}
\eeqn
By imposing that the $\max$ and $\min$ operators in eq.~(\ref{cond1}) be
non-trivial, and that the $q$ range in the sum on the r.h.s.~of 
eq.~(\ref{SigB}) be non-null, one further obtains:
\beqn
&&\nmax\le c_s(k_0)+\Delta(k_0)\,,\;\;\;\;\;\;\;\;
\mmax\le c(k_0)+\Delta(k_0)\,,
\label{cond2}
\\*&&
\nmax+\mmax\ge
c_s(k_0)+c(k_0)+\Delta(k_0)\equiv k_0\,,
\label{cond3}
\eeqn
which, together with eq.~(\ref{cond0}), give the complete conditions that 
guarantee that there will be at least one Born-level contribution\footnote{We
point out that there is no loss of efficiency if \aNLOs\ is given in input 
values of $\nmax$ and $\mmax$ larger than the upper bounds that appear in
eq.~(\ref{cond2}) -- in that case, the code will simply compute all
the non-null LO and NLO contributions.}.
\begin{figure}[t]
\begin{center}
  \includegraphics[width=0.4\textwidth,angle=270]{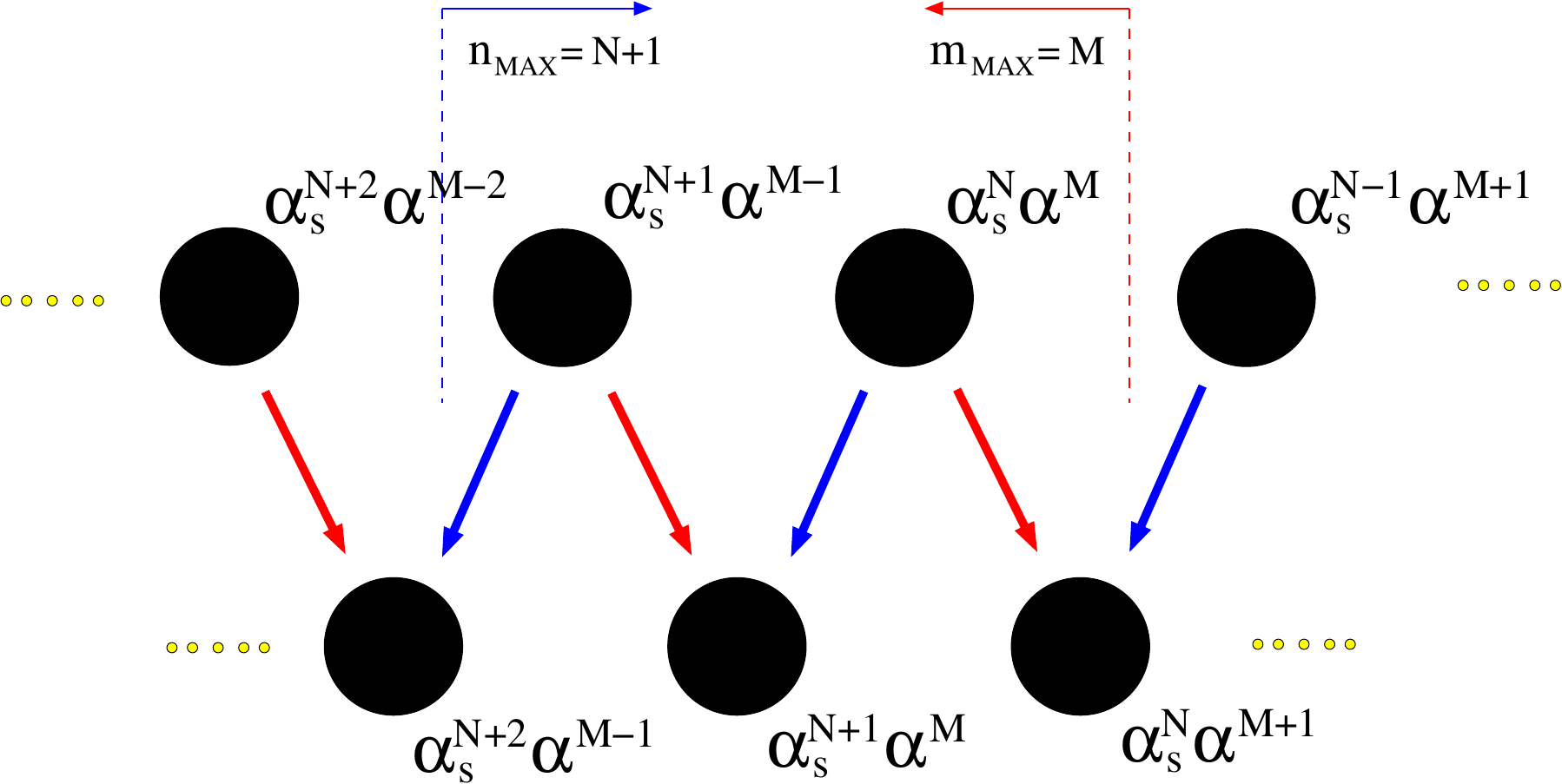}
\end{center}
\caption{\label{fig:blobs} 
Example of QCD (blue, right-to-left arrows) and QED 
(red, left-to-right arrows) corrections to a generic process. 
The small yellow circles indicate the possible presence of further 
cross section contributions. The parameters $\nmax$ and $\mmax$ can 
be freely changed in order to select different Born contributions 
w.r.t.~those depicted here. See the text for details.
}
\end{figure}
We point out that the keywords {\tt QCD=$\nmax$} and {\tt QED=$\mmax$}
may both be omitted. When this is the case, \aNLOs\ generates the process
with the smallest possible number of QED vertices at the Born level
(which is equivalent to using {\tt QED=$\mmax$} with the smallest
$\mmax$ compatible with eqs.~(\ref{cond0}) and~(\ref{cond2})).

In a graphical manner, the generation procedure is shown in the
example of fig.~\ref{fig:blobs}, where each of the black blobs in the upper 
(lower) row represents a $\Sigma_{k,q}$ contribution at the LO (NLO). 
By setting $\nmax=N+1$ as in fig.~\ref{fig:blobs}, one selects 
the set $S_1$ of all the LO contributions which lie to the right of 
the vertical blue dashed line. Conversely, by setting $\mmax=M$, the 
selected set $S_2$ is that of all the LO contributions which lie to the 
left of the vertical red dashed line\footnote{It should be clear that,
by choosing different values for $\nmax$ and/or $\mmax$, the positions
of of the vertical dashed lines change w.r.t.~those shown in 
fig.~\ref{fig:blobs}, and thus one includes a number of blobs in the 
computation which is different from that of the example 
considered here.}. The LO cross section that will enter
the physical predictions is then obtained by summing the contributions 
that belong to the intersection $S_{1\oplus 2}=S_1\cap S_2$ of the two 
sets obtained previously -- hence, these are the terms of 
$\ord(\as^{N+1}\aem^{M-1})$ and $\ord(\as^{N}\aem^{M})$ in the example. 
If the keyword {\tt [QCD]} is used, the set of NLO contributions 
contains all the blobs that can be reached by following the blue 
right-to-left arrow that starts from {\em each} of the LO contributions 
that belong to $S_{1\oplus 2}$ -- hence, these are the terms of 
$\ord(\as^{N+2}\aem^{M-1})$ and $\ord(\as^{N+1}\aem^{M})$ in the example. 
If {\tt [QED]} is used, the procedure is analogous, but one needs to follow 
the red left-to-right arrows -- thus, one obtains the terms of 
$\ord(\as^{N+1}\aem^{M})$ and $\ord(\as^{N}\aem^{M+1})$ in the example. 
Finally, if both {\tt [QCD QED]} are employed, one must follow both types 
of arrows (which is equivalent to the union of the two sets obtained with 
{\tt [QCD]}-only and {\tt [QED]}-only; in the example, this leads to the 
terms of $\ord(\as^{N+2}\aem^{M-1})$, $\ord(\as^{N+1}\aem^{M})$, and 
$\ord(\as^{N}\aem^{M+1})$).

The procedure just described is implemented in \aNLOs\ essentially by 
constraining the powers of the coupling constants. Naively, one might
assume that eqs.~(\ref{LOsynt}) and~(\ref{NLOsynt}) are all that is 
needed, but in fact care must be exercised at the NLO, where one must
relax the strict implementation of such constraints in the case of a 
mixed-coupling expansion. In order to illustrate the problem,
let us consider the process generated by:

\vskip 0.25truecm
\noindent
~~\prompt\ {\tt ~generate p p > t t\~{} QED=0 QCD=2 [QED]}

\vskip 0.25truecm
\noindent
which thus computes the second-leading (``EW'') corrections 
of $\ord(\as^2\aem)$ to the leading Born terms of $\ord(\as^2)$.
Among the real-emission contributions, one finds e.g.~the partonic process:
\beq
\gamma q\;\longrightarrow\;t\bt q\,,
\label{gqttq}
\eeq
with $q$ any massless quark. There are two FKS sectors associated with
eq.~(\ref{gqttq}), relevant to the collinear configurations in which
the outgoing quark is parallel either to the incoming photon or to
the incoming quark. In these configurations, the Born processes
that factorise are:
\beq
\bq q\;\longrightarrow\;t\bt\,,
\;\;\;\;\;\;\;\;
\gamma g\;\longrightarrow\;t\bt\,,
\label{gqttqB}
\eeq
respectively. The corresponding matrix elements are of $\ord(\as^2)$
and $\ord(\as\aem)$. Therefore, in spite of the fact that Born-level
$\ord(\as\aem)$ terms must not contribute to the predictions that 
result from the generation command given above, they must nevertheless
be generated in order for the code to be able to construct the
counterterms that render the cross section IR finite.

The bottom line is the following: at the {\em Born level}, the constraints
of eq.~(\ref{LOsynt}) are applied to the predictions given in output,
as requested by the user. Internally, those of eq.~(\ref{NLOsynt}) are
used instead. More precisely, the Born matrix elements that are generated
for the sole purpose of constructing IR counterterms factorise a
coupling-constant combination $\as^n\aem^m$ that has\footnote{Note that
such matrix elements may not exist. This is always the case when one
computes QCD corrections to the leading Born contribution, and QED
corrections to the most subleading Born contribution.}:
\begin{itemize}
\item If QCD corrections are computed: $n$ one unity larger and $m$ one 
unity smaller than the corresponding exponents associated with the
user-selected Born contribution with the largest $\as$ power;
\item If QED corrections are computed: $n$ one unity smaller and $m$ one 
unity larger than the corresponding exponents associated with the
user-selected Born contribution with the largest $\aem$ power.
\end{itemize}
In the example of fig.~\ref{fig:blobs}, these prescriptions lead to
compute the $\as^{N+2}\aem^{M-2}$ contributions in the case of QCD
corrections (i.e.~the blob left of the leftmost user-selected blob
in the upper row), and those of $\as^{N-1}\aem^{M+1}$ in the case of QED 
corrections (i.e.~the blob right of the rightmost user-selected blob).

We now turn to commenting on the role of the multiparticles in the
context of NLO computations. We remind the reader that in \aNLOs\ a 
multiparticle is a subset of the particles that belong to the loaded 
physics model, which may be defined by the user at runtime. 
For example:

\vskip 0.25truecm
\noindent
~~\prompt\ {\tt ~define p = g d d\~{} u u\~{} s s\~{} }

\vskip 0.25truecm
\noindent
defines the multiparticle {\tt p} to be composed of the gluon and the 
three lightest quarks. Multiparticle names can be freely chosen, but
in practice the two most commonly used ones, namely {\tt p} and {\tt j},
are conventionally associated with the incoming hadrons and the outgoing
``jets'', respectively, and serve to define the elementary (i.e.~those
that are relevant at the level of short-distance cross sections)
components of these objects. Because of this, multiparticles can be 
effectively used at the LO\footnote{Indeed, they have been inherited 
in \aNLOs\ in their current forms from \MadGraph~\cite{Alwall:2011uj}.} 
to control which partonic processes can contribute to the calculation 
of a given observable cross section. For example, with the definition of 
the proton given above, no processes will be taken into account that have at 
least one charm or bottom quark in the initial state. The situation is 
unfortunately more involved at the NLO, as the example of eq.~(\ref{gqttq}) 
shows. If one defines the proton as a multiparticle that does include the 
photon but {\em not} the gluon, one generates (among others) the process 
of eq.~(\ref{gqttq}) and the $q\bq$-initiated Born of eq.~(\ref{gqttqB}),
but excludes the $\gamma g$-initiated Born, which is essential in
order to have an IR-finite cross section. We point out that this 
problem could be avoided by using a different multiparticle definition,
to be used exclusively when constructing IR counterterms. 
We have refrained from implementing this option, because it would lead
to IR-finite, but still unphysical predictions\footnote{Note that this
remarks applies to LO computations as well which, if performed with
multiparticles whose contents are not consistent with the underlying
physics model, are also unphysical, and must thus be used with care.}.

\noindent
The immediate consequence of what we have just said is the following:
\begin{itemize}
\item Whenever EW corrections are considered, the photon must be included
in the definition of the {\tt p} and {\tt j} multiparticles.
\end{itemize}
An explicit example, which applies to the computations performed in
this paper, is given at the beginning of sect.~\ref{sec:setup}.

Besides the mandatory inclusion of photons in multiparticles, there might
be cases when (massless) leptons must be included there as well. 
In general, this is the case when one is interested in computing EW 
corrections to processes that have initial- or final-state photons 
at the Born level in the user-selected contributions. For example,
leptons need not be part of the multiparticles in the calculation
of the EW corrections to the $t\bt$ hadroproduction considered before (the
initial-state photon that appears in eq.~(\ref{gqttqB}) is relevant to
a real-emission, not a Born, contribution). Conversely, they must
be included in the computation of the complete NLO corrections to
dijet hadroproduction, as was done in ref.~\cite{Frederix:2016ost}
where QCD partons, photons, and leptons have been treated democratically.

The discussion given above on IR safety should render it clear
that the inclusion of leptons in the {\tt p} multiparticle serves the
purpose of the correct implementation of collinear subtractions, and 
is thus independent of whether lepton PDFs are non-zero. In fact, although 
examples of sets with non-zero lepton PDFs exist 
(see e.g.~refs.~\cite{Bertone:2015lqa,Bauer:2017isx})
in practice their contributions are in general safely negligible.
Therefore, if we assume $f_{l}^{(H)}(x)\equiv 0$, we may wonder which
Born-level matrix elements with initial-state leptons will be generated 
that will give a null contribution to the cross section, and can therefore
be discarded before generating them in order to increase the efficiency.
This is the case when two initial-state leptons are present. In fact,
if the corresponding matrix element is regarded as a physical Born
contribution, it is equal to zero because of $f_{l}^{(H)}(x)\equiv 0$.
Conversely, if it used to construct an IR counterterm for a real-emission
matrix element, the latter will necessarily have one initial-state lepton,
which will again be equal to zero because of the vanishing of the lepton PDFs.
Processes with a single initial-state lepton ($Xl^\pm \to Y$) will have
to be kept. In fact, although they vanish as Born contributions, they
give the factorised matrix element in real-emission processes of
the type $X\gamma\to l^\mp Y$, which are in general different from zero.

In order to take the above into account, the shell variable\\
{\tt include\_lepton\_initiated\_processes} is made available in
\aNLOs, that helps optimise the generation of lepton-initiated
processes. In particular, when such a variable is set equal to
{\tt False} (its default value), all processes with initial-state
leptons will be discarded, except for those at the Born level that
feature a single lepton, and such that the real-emission processes
associated with them cross that lepton to the final state, and
replace it with an initial-state photon. When the variable above is
set equal to {\tt True}, no process will be discarded.

\begin{table}
\begin{center}
\resizebox{\textwidth}{!}{
\begin{tabular}{llllll}
\toprule
 & \multicolumn{2}{c}{Processes without jets}
 & \multicolumn{2}{c}{Processes with jets}
 & \multirow{2}{*}{Physical objects}\\
 & PDF($q g$) & PDF($q g\gamma$) & PDF($q g$) & PDF($q g\gamma$) & \\
\midrule
\multirow{2}{*}{$i=1$} & \multirow{2}{*}{\texttt{p = q g}} &
\multirow{2}{*}{\texttt{p = q g a}} & \texttt{p = q g} &
\texttt{p = q g a} & $j(qg)$, $\gamma$, $l$, $\nu$,\\
 & & & \texttt{j = q g} & \texttt{j = q g} & massive particles\\
\midrule
\multirow{2}{*}{$i=2$} & \multirow{2}{*}{inconsistent} &
\multirow{2}{*}{\texttt{p = q g a}} & \multirow{2}{*}{inconsistent} &
\texttt{p = q g a} & $j(qg\gamma)$, $l$, $\nu$, \\
 & & & & \texttt{j = q g a} & massive particles\\
\midrule
\multirow{2}{*}{$i\ge 3$} & \multirow{2}{*}{inconsistent} & 
\multirow{2}{*}{\texttt{p = q g a}} & \multirow{2}{*}{inconsistent} &
\texttt{p = q g a l} & $j(qg\gamma l)$, $\nu$,\\
 & & & & \texttt{j = q g a l} & massive particles\\
\bottomrule
\end{tabular}
}
\caption{\label{tab:recs}
Recommendations for the definitions of the {\tt p} and {\tt j}
multiparticles in the computations of the NLO corrections given
in eq.~(\ref{Sconds}). {\tt q} stands for all of the massless quarks and
anti-quarks of the loaded physics model, {\tt g} is the gluon, {\tt a} is 
the photon, and {\tt l} collects all of the massless charged-leptons 
(for example, {\tt l = e+ e- mu+ mu- ta+ ta-}). In the rightmost column,
by $j$, $\gamma$, $l$, and $\nu$ we denote jets, photons, charged leptons,
and neutrinos, defined as explained in the text; $q$ and $g$ denote
light quarks and gluons, respectively.
}
\end{center}
\end{table}
In conclusion, whenever the following NLO corrections are computed:
\beq
\Sigma_{{\rm NLO}_{i-k}}+\ldots +\Sigma_{\NLOi}\,,
\label{Sconds}
\eeq
for any $k$ such that \mbox{$1\le i-k\le i$}, we recommend to define the 
{\tt p} and {\tt j} multiparticles as summarised in table~\ref{tab:recs}.
Note, therefore, that the definitions are dictated by the most subleading 
term among those selected in eq.~(\ref{Sconds}).
We point out that the definitions in table~\ref{tab:recs} stem from
the fact that the current version of \aNLOs\ does not handle tagged
photons and leptons, and thus will be modified when this limitation
will be lifted. By \mbox{PDF($qg$)} we have denoted PDF sets whose 
contents are limited to light quarks and gluons; conversely, 
\mbox{PDF($qg\gamma$)} denotes PDF sets that include light quarks,
gluons, and photons\footnote{In the case where a PDF set features
charged-lepton distributions, such leptons must always be included 
in the definition of {\tt p}. Furthermore, the shell variable
{\tt include\_lepton\_initiated\_processes} must be set equal
to {\tt True} prior to the generation of the process.}. The inclusion
of charged leptons into {\tt p} and {\tt j} summarises the discussion
presented before on IR-safety requirements. Note that it holds regardless
of whether lepton PDFs are zero or non-zero; however, in the former
case the variable {\tt include\_lepton\_initiated\_processes} can be
left to its default value (i.e.~{\tt False}).

In the rightmost column of table~\ref{tab:recs} we list the final-state
objects that can be defined; not surprisingly, their nature depends 
on the most subleading perturbative contribution one considers. 
In particular, jets must be defined democratically in quarks and
gluons ($i=1$), and photons ($i=2$), and charged leptons ($i\ge 3$);
we have symbolically denoted the jets thus obtained by $j(qg)$, 
$j(qg\gamma)$, and $j(qg\gamma l)$, respectively, in order to render
their particle contents explicit. When $i\ge 2$, charged leptons must be 
defined as dressed, i.e.~recombined with nearby photons; photons cannot
be tagged, but only found inside jets (note that it is legitimate to
have a jet composed of a single photon)\footnote{This implies that
photons enter both the jet-finding algorithm and the charged-lepton 
definition. There are at least two IR-safe procedures to accomplish
this. The first one requires to start from the recombination of photons
with charged fermions (both leptons and quarks), and then to reconstruct
jets using gluons, quarks (possibly dressed), and not-recombined
photons. The second procedures starts from the recombination of photons 
with charged leptons, and then defines jets using gluons, quarks, and
not-recombined photons. In the former procedure, there is no need for 
a jet-lepton separation, while in the latter one IR-safety demands
that such a separation be imposed.}. Finally, when $i\ge 3$ in general
all of the light particles must be considered in the jet-finding
procedure; leptons can possibly be defined at the analysis level
as jets that feature a non-zero lepton number.

The recommendations summarised in table~\ref{tab:recs} must be seen
as minimal. While they will always give consistent results if appropriate 
final-state cuts are applied, for specific processes some simplifications 
might be possible for the definitions of both the multiparticles and the 
physical objects.

\section{Technicalities of the complex-mass scheme\label{sec:appcms}}

\subsection{A systematic test of the CM scheme implementation in the 
  off-shell region\label{sec:CMtest}}
Given the relative complexity of the derivation of the UV and $R_2$ 
counterterms for mixed QCD+EW corrections in the SM, as well as the
intricacies of the expansions in the coupling constants that formally
affect the masses, widths, and matrix elements in the CM scheme, 
it is important to provide validation techniques which are complementary 
to the check of IR-poles cancellation (performed by \aNLOs\ for each 
generated process), and probe the parts of the loop amplitudes and of 
the counterterms that are both UV and IR finite.

In this section, we describe one such technique, that we have implemented
in \aNLOs. In essence, it consists in comparing kinematically-local results 
obtained in the CM and OS schemes, checking that they differ by higher-order
terms (i.e.~whose accuracy is higher than that one considers, which is either 
LO or NLO). The rationale is as follows. We have already
illustrated in sect.~\ref{sec:cms_formula} the strict similarities
between the CM and OS schemes. We have also recalled there that OS
computations are possible in the presence of unstable particles,
provided that width effects can be neglected. The latter condition
does not necessarily imply that all widths are set equal to zero,
which is on the other hand the framework we want to work with. Because
of this, we formally introduce a zero-width (ZW henceforth) setup, that 
we define as follows:
\beq
\ampsq^{(L)}_{\text{ZW}}(\{p_k\}) = 
\ampsq^{(L)}_{\text{OS}}(\{p_k\}){\Big |}_{\{\Gamma_r=0\} }
\label{ZWdefinition}
\eeq
for tree-level ($L=0$) and one-loop ($L=1$) matrix elements\footnote{At
variance with what was done in sects.~\ref{sec:Xsec} and~\ref{sec:Frag},
in the notation for the matrix elements we have omitted the index that 
corresponds to the final-state particle multiplicity, since it plays no 
role here.}. The index $r$ numbers the unstable particles, and $\{p_k\}$ 
is a given kinematic configuration. The idea, then, is that for a suitable
off-shell configuration $\{p_k\}$ (i.e.~where the virtualities of all
unstable particles are not close to the corresponding pole masses) the
ZW matrix elements and their CM-scheme counterparts will differ by 
higher-order terms.

The meaning of ``higher orders'' requires a clarification in the context
of a mixed-coupling scenario. Here, it implicitly understands terms
associated with an increasing power in the coupling that governs
unstable-particle decays which, in the SM case, coincides with $\aem$.
In what follows, we thus use $\aem$ in order to be definite, and we assume 
to work in a scheme where $\aem$ is a real number (see 
sect.~\ref{sec:alphaComplexPhase}). 

Therefore, we shall deal with the following expressions for
the ZW and CM-scheme matrix elements:
\beqn
\ampsq^{(L)}_{\text{ZW}}(\{p_k\})&=&
\alpha^b\alpha^L\kappa^{(L)}_{L,\ZW}(\{p_k\})\,, 
\label{MEexps1}
\\
\ampsq^{(L)}_{\text{\CM}}(\{p_k\})&=&
\alpha^b \sum_{i \geq L}\alpha^i\kappa^{(L)}_{i,\CM}(\{p_k\})\,,
\label{MEexps2}
\eeqn
with $b$ the power of $\aem$ associated with the leading Born term.
The dependence on $\as$, or on any other couplings, is left implicit
in the coefficients $\kappa$. The CM-scheme matrix elements on the
l.h.s.~of eq.~(\ref{MEexps2}) are understood to be computed by using
$\bar{M}$ and $\bar{\Gamma}$ (see sect.~\ref{sec:cms_formula}). The 
series that appears on the r.h.s.~of that equation stems from
the perturbative expansion of the $\bar{\Gamma}_r$'s; we point out that, 
for the sake of the present numerical tests, it is crucial that each
$\bar{\Gamma}_r$ obeys eq.~(\ref{CMGalim}). Conversely, given 
eq.~(\ref{MvsMbar}), the fact that $M_r$ (at variance with $\Gamma_r^{(0)}$) 
does not vanish when $\aem\to 0$, and the fact that we are working at either
the LO or the NLO, the quantities $\bar{M}_r$ may be treated as fixed
external parameters in the tests. The off-shell kinematic configuration
used for the tests is such that\footnote{The use of $M_r$ and $\Gamma_r$
in eq.~(\ref{PStest}) in place of $\bar{M}_r$ and $\bar{\Gamma}_r$
would lead to the same results.}: 
\beq
\abs{\left(\sum_{k\in\Omega_r} p_k\right)^2-\bar{M}_r^2}> 
\rho_{\rm min}\bar{\Gamma}_r\bar{M}_r
\;\;\;\;\;\;\;\;\forall\;r\,,
\label{PStest}
\eeq
where $\Omega_r$ denotes the set of the decay products\footnote{Decay 
products may be resonances themselves, as is the case of a $W$ that
emerges from a top-quark decay.} of the $r^{th}$ resonance in the process 
considered, and we set $\rho_{\rm min}=10$ by default. Both the list of 
resonances and a random kinematic configuration that satisfies 
eq.~(\ref{PStest}) are constructed automatically by \aNLOs.

At the tree level, the validation test consists in effectively
verifying the following equality:
\beq
\kappa^{(0)}_{0,\CM} = \kappa^{(0)}_{0,\ZW}\,,
\label{treeCMTest}
\eeq
while at the one-loop level, one also checks that:
\beq
\kappa^{(0)}_{1,\CM}+\kappa^{(1)}_{1,\CM} = \kappa^{(1)}_{1,\ZW}\,.
\label{loopCMTest}
\eeq
Since the actual evaluation of the parameters $\kappa$ in the CM scheme
would involve the analytic Taylor expansion of eq.~(\ref{MEexps2}),
for an automated numerical test is more convenient, and fully equivalent,
to construct the following quantities:
\beqn
\Delta^{(0)}&=&
\lim_{\lambda\rightarrow 0} \left(\left.
\frac{\ampsq^{(0)}_{\text{\CM}}-\ampsq^{(0)}_{\text{ZW}}}
{\lambda\ampsq^{(0)}_{\text{ZW}}}\right|_{\aem=\lambda \aem^{\rm ref}} 
\right)\phantom{aaaaaaaaaaaaaa}
\equiv \lim_{\lambda\rightarrow 0}\delta^{(0)}(\lambda)\,,
\label{D0def}
\\
\Delta^{(1)}&=&
\lim_{\lambda\rightarrow 0} \left (\left.
\frac{\ampsq^{(1)}_{\text{\CM}}+\ampsq^{(0)}_{\text{\CM}}
-\ampsq^{(1)}_{\text{ZW}}-\ampsq^{(0)}_{\text{ZW}}}  
{\lambda^2\ampsq^{(0)}_{\text{ZW}}}\right |_{\aem=\lambda\aem^{\rm ref}} 
\right)
\equiv \lim_{\lambda\rightarrow 0}\delta^{(1)}(\lambda)\,,
\label{D1def}
\eeqn
and verify that that they are both finite real numbers:
\beq
\Delta^{(L)}\in{\mathbb R}\,,
\;\;\;\;\;\;\;\;
\abs{\Delta^{(L)}}<\infty\,,
\;\;\;\;\;\;\;\;
L=0,1\,.
\eeq
In eqs.~(\ref{D0def}) and~(\ref{D1def}), $\lambda$ is a real-valued parameter,
and $\aem^{\rm ref}$ denotes an arbitrary user-specified fixed coupling value,
which serves as a reference.  We stress that the actual value of the coupling
constant used in the computations, $\aem=\lambda \aem^{\rm ref}$, must be
employed {\em everywhere} in the matrix elements, including in the expression
of $\bar{\Gamma}_r(\aem)$ chosen for the unstable particle widths.
The limits on the r.h.s.~of eqs.~(\ref{D0def}) and~(\ref{D1def}) are
computed numerically by \aNLOs\ by choosing progressively smaller values of
$\lambda$ (equally spaced on a logarithmic scale, starting from $\lambda=1$),
and by verifying that the resulting sequences have constant asymptotes 
for $\lambda\to 0$ -- in other words, the sequences must feature
vanishingly small slopes before becoming sensitive to numerical inaccuracies
in matrix element evaluations. 

In the remainder of this section, we shall give an explicit example of
the test described above, by considering the partonic process
\mbox{$u\bar{d}\to c\bar{s}$} (i.e.~one contribution to $W^{+*}$ production)
in the $\aem(m_Z)$ scheme; the quarks are taken to be massless, and
a diagonal CKM matrix is assumed. We start by considering the tree-level 
amplitude $\ampzCM$, which reads\footnote{Overall $i$ factors do not
play a role in what follows, and are thus ignored.}: 
\beqn
\plaat{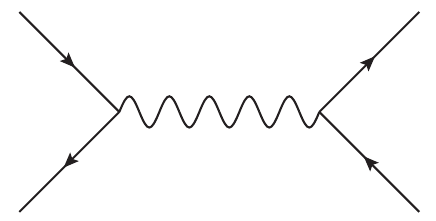}{4}{-25}\equiv\ampzCM=
\vertmuCM(q_W) G^{\mu\nu}_{\CM}(q_W) \vertnuCM (q_W)\,,
\label{TreeLevelDY}
\eeqn
with $\vertmuCM$ the $q \bar{q'}\to W^{*+\mu}$ {\em offshell} current.
The corresponding ZW amplitude, defined in the same way as for matrix
elements (eq.~(\ref{ZWdefinition})), can be obtained from the r.h.s.~of 
eq.~(\ref{TreeLevelDY}) by means of the formal replacement 
\mbox{${\rm CM}\to {\rm ZW}$}. We can now expand each term on
the r.h.s.~of eq.~(\ref{TreeLevelDY}) in series of $\aem$. We
begin with the $W$ propagator $G^{\mu\nu}_{\CM}$:
\beqn
G^{\mu\nu}_{\CM}(q) &=& 
\left(-g^{\mu\nu}+(1-\xi)\frac{q^{\mu}q^{\nu}}{q^2-\xi m_W^2}\right) 
\frac{1}{q^2-m_W^2} 
\nonumber\\
&=& \left(-g^{\mu\nu}+(1-\xi)
\frac{q^{\mu}q^{\nu}}{q^2-\xi \bMW^2+\xi i\bGaW \bMW}\right) 
\frac{1}{q^2-\bMW^2+i\bGaW\bMW} 
\nonumber\\
&=& \left(-g^{\mu\nu}+(1-\xi)
\frac{q^{\mu}q^{\nu}}{q^2-\xi \bMW^2}-
(1-\xi)\frac{\xi i\bGaW\bMW}{q^2-\xi \bMW^2}
\frac{q^{\mu}q^{\nu}}{q^2-\xi \bMW^2}+\ord(\aem^2)\right)
\nonumber\\
&&\times\left[\frac{1}{q^2-\bMW^2}-i\frac{\bGaW\bMW}{(q^2-\bMW^2)^2}+
\ord(\aem^2)\right].
\eeqn
We shall be working in the Feynman gauge ($\xi=1$) for the rest of this
section, in which case the expression above simplifies to:
\beqn
G^{\mu\nu}_{\CM}(q)&=& 
G^{\mu\nu}_{\ZW}(q)+
ig^{\mu\nu}\frac{\bGaW\bMW}{(q^2-\bMW^2)^2}+\ord(\aem^2)
\nonumber\\*
&=&
\label{CMExpandedPropagator}
G^{\mu\nu}_{\ZW}(q)+
ig^{\mu\nu}\frac{\Gamma_W M_W}{(q^2-M_W^2)^2}+\ord(\aem^2)\,,
\eeqn
where
\beq
G^{\mu\nu}_{\ZW}(q)=-\frac{g^{\mu\nu}}{q^2-M_W^2}\,.
\eeq
However, we stress that the conclusions drawn in this section apply 
independently of the chosen gauge.

The offshell current $\vertmuCM$ differs from its ZW counterpart solely
because of the complexified couplings present in the former. We 
choose\footnote{This choice is straightforward in the $\aem(m_Z)$ OS
scheme considered here. We have shown in sect.~\ref{sec:alphaComplexPhase} 
how one should define a real-valued $\alpha$ coupling in the CM $G_\mu$ scheme
as well.} the input values for $\aem$ in the CM and ZW schemes to be real 
and equal to each other. Thus, the {\em only} difference between the 
currents $\vertmuCM$ and $\vertmuZW$ originates from the different values 
of the Weinberg angle, that we denote by $c_W$ and $C_W$, respectively,
in the two computations. We then have:
\beq
\frac{d}{d\aem} \vertmuCM = 
\left (\frac{\partial}{\partial\aem}+
\frac{d c_W}{d \aem}\frac{\partial}{\partial c_W}\right)\vertmuCM\,.
\eeq
The expansion in $\aem$ of the derived parameter $c_W$ reads:
\beq
c_W=\sqrt{\frac{m_W^2}{m_Z^2}}=
\sqrt{\frac{\bMW^2-i\bGaW\bMW}{\bMZ^2-i\bGaZ\bMZ}}=
\underbrace{\frac{M_W}{M_Z}}_{C_W}+
\underbrace{i\,\frac{M_W \Gamma_Z - M_Z \Gamma_W}{2 M_Z^2}}_
{\aem\frac{d c_W }{ d \aem }}
+\,\ord(\aem^2)\,.
\label{cWexp}
\eeq
Equation~(\ref{cWexp}) can be used to derive the expansion of $\vertmuCM$ 
around $\aem=0$ (where $\Gamma_W=\Gamma_Z=0$ and $c_W=C_W$), yielding:
\beqn
\label{CMExpandedME}
\vertmuCM(c_W)&=&\vertmuCM (C_W)+
 \aem\frac{d c_W}{d \aem}\left(\frac{\partial}{\partial c_W}
\vertmuCM(c_W)\right )_{c_W=C_W}+\ord(g\aem^2)
\nonumber\\
&=&\vertmuZW+
\left(i\frac{M_W\Gamma_Z - M_Z\Gamma_W}{2 M_Z^2}\right) 
\frac{\partial}{\partial C_W}\vertmuZW(C_W)+\ord(g\aem^2)
\eeqn
where $g\propto\sqrt{\aem}$ appears in the vertex of the current
$\vertmuCM$, and we have used $\vertmuCM(C_W)=\vertmuZW(C_W)$ that stems
from choosing the same value of $\aem$ in the two schemes. By 
substituting eqs.~(\ref{CMExpandedPropagator}) and~(\ref{CMExpandedME}) 
into eq.~(\ref{TreeLevelDY}) one obtains:
\beqn
\ampzCM-\ampzZW &=&\left\{
\left[1+\left(i\frac{M_W \Gamma_Z - M_Z \Gamma_W}{2 M_Z^2}\right) 
\frac{\partial}{\partial C_W}\right]\vertmuZW 
+\ord(g\aem^2)\right\} 
\nonumber\\&&\times
\left[ G^{\mu\nu}_{\ZW}(q^{\rho})-
ig_{\mu\nu}\frac{\Gamma_WM_W}{(q^2-M_W^2)^2}+\mathcal{O}(\alpha^2)\right] 
\nonumber\\&&\times
\left\{\left[1+\left(i\frac{M_W \Gamma_Z - M_Z \Gamma_W}{2 M_Z^2}\right) 
\frac{\partial}{\partial C_W}\right]\vertnuZW 
+\ord(g\aem^2)\right\} 
\nonumber \\&&-
\vertmuZW G^{\mu\nu}_{\ZW}(q) \vertnuZW 
\nonumber \\
&=&0+\ord(\aem^2)\,.
\label{anCMTtest}
\eeqn
By moving $\ampzZW$ from the l.h.s.~to the r.h.s.~of eq.~(\ref{anCMTtest}),
and then taking the absolute value squared of both sides, one sees that
the difference of the matrix elements that appears in the numerator of 
$\delta^{(0)}(\lambda)$ (see eq.~(\ref{D0def})) is of $\ord(\aem^3)$, 
and thus shows that eq.~(\ref{treeCMTest}) is satisfied. Actually,
since the $\ord(\aem^2)$ term in eq.~(\ref{anCMTtest}) is purely
imaginary, the $\ord(\aem^3)$ contribution to the difference of the
matrix elements in $\delta^{(0)}(\lambda)$ is identically equal to
zero, which implies that in the present example $\Delta^{(0)}=0$. This
is in fact the case for all $2\to 2$ scatterings, which renders the
test proposed here less stringent for these processes. Because of this,
our actual numerical validation has been based on processes with final-state 
multiplicities larger than two, as we shall explicitly show in the following.

The terms of $\ord(\aem^2)$ on the r.h.s.~of eq.~(\ref{anCMTtest}) would 
cause the limit of eq.~(\ref{D1def}) to diverge, were they not canceled by 
analogous terms emerging from the difference of the one-loop matrix elements 
\mbox{$\ampsq^{(1)}_{\rm CM/ZW}=2\Re [\amp^{(1)}_{\rm CM/ZW}
\amp^{\star(0)}_{\rm CM/ZW}]$}. In order to show that this is the
case, we start by noting that most amplitudes are such that the difference 
between the CM and ZW results is at least of $\ord(\aem^3)$. We denote the 
sum of such amplitudes by $\amp^{(1){\rm reg}}_{\rm CM}$; by definition, 
this quantity therefore obeys the following relationship:
\beq
\amp^{(1){\rm reg}}_{\rm CM} = 
\amp^{(1){\rm reg}}_{\rm ZW}+\ord(\aem^3)\,.
\label{RegularLoops}
\eeq
Hence, the contributions of $\amp^{(1){\rm reg}}_{\rm CM}$ and its
ZW counterpart to eq.~(\ref{D1def}) cannot possibly lead to a divergent
limit, and for this reason we call them ``regular''. Note that they
include both unrenormalised one-loop amplitudes and UV counterterms.
As far as the former are concerned, for illustrative purposes we consider 
here explicitly the self-energy insertion:
\beqn
&&\plaat{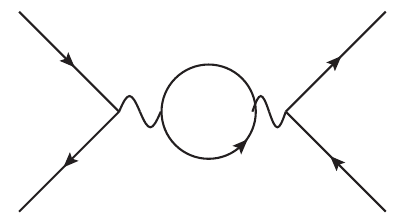}{4}{-25}\equiv
\amp^{(1)\Sigma}_{\rm CM} = 
\vertmuCM G^{\mu\alpha}_{\CM} 
\Sigma_{{\rm CM}}^{\alpha\beta}
G^{\beta\nu}_{\CM}\vertnuCM
\nonumber\\&&\phantom{aaaaa}
=\vertmuZW G^{\mu\alpha}_{\ZW} G^{\beta\nu}_{\ZW} \vertnuZW 
\left[\left(g_{\alpha\beta}-\frac{q_\alpha q_\beta}{q^2}\right)
\Sigma_{T,{\rm CM}}(q^2)+
\frac{q_\alpha q_\beta}{q^2}\,\Sigma_{L,{\rm CM}}(q^2)+\ord(\aem^2)\right]
\nonumber\\&&\phantom{aaaaa}
=\vertmuZW G^{\mu\alpha}_{\ZW} G^{\beta\nu}_{\ZW} \vertnuZW 
\Sigma_{{\rm ZW}}^{\alpha\beta}(q^2)+\ord(\aem^3)\,.
\label{cntself}
\eeqn
Equation~(\ref{cntself}) is based on the fact that the 
self-energy components $\Sigma_{T/L,{\rm CM}}$ of the $W$ boson 
differ from their ZW counterparts only because of the complexified coupling 
$c_W$, which evidently only leads to NLO-subleading terms in $\alpha$. 
Similar consideration hold for {\em all} other unrenormalised one-loop
amplitudes, as well as for the $\aem$ and wave-function UV
counterterms. It follows that the only non-regular contributions to the
full one-loop amplitude are due to the mass and $c_W$ UV counterterms.
Thus, we write:
\beq
\amp^{(1)}_{\rm CM} = \amp^{(1){\rm reg}}_{\rm CM}+
\amp^{(1)\delta{m_W^2}}_{\rm CM}+
\amp^{(1)\delta{c_W}}_{\rm CM}\,.
\label{LoopLevelDY}
\eeq
As far as the contribution due to the mass counterterm is concerned,
its expression can be formally read from the r.h.s.~of eq.~(\ref{cntself})
with the formal replacement 
\mbox{$\Sigma_{{\rm CM}}^{\alpha\beta}\to g^{\alpha\beta}\delta m_W^2$} 
which, in a ZW computation, becomes 
\mbox{$\Sigma_{{\rm ZW}}^{\alpha\beta}\to g^{\alpha\beta}\delta M_W^2$}.
Therefore:
\beqn
\label{sigmaContribs}
&&\amp^{(1)\delta{m_W^2}}_{\rm CM}-\amp^{(1)\delta{M_W^2}}_{\rm ZW}= 
\vertmuZW G^{\mu\alpha}_{\ZW} G^{\alpha\nu}_{\ZW} \vertnuZW 
(\delta{m_W^2} - \delta{M_W^2}) + \ord(\aem^2) 
\\&&\phantom{aaa}
=\vertmuZW\vertnuZW \left(\frac{g_{\mu\nu}}{(q^2-M_W^2)^2} 
(\Sigma(M_W^2-i\Gamma_WM_W) - \Re[\Sigma(M_W^2)])
+\ord(\aem^2)\right) 
\nonumber \\&&\phantom{aaa}
=\vertmuZW\vertnuZW \left(
\frac{i g_{\mu\nu}\Im[\delta{m_W^2}]}{(q^2-M_W^2)^2}
+\ord(\aem^2)\right)
=\vertmuZW\vertnuZW 
\left(\frac{i g_{\mu\nu}M_W \Gamma^{(0)}_W}{(q^2-M_W^2)^2}
+\ord(\aem^2)\right)\,.
\nonumber
\eeqn
Here, we have substituted the mass counterterms with their explicit 
expressions, given in eqs.~(\ref{OScntM}) and~(\ref{CMOScntM}).
In eq.~(\ref{sigmaContribs}) we have used the fact that the real parts 
of the mass counterterms in the ZW and CM setups differ by higher-order
terms, provided that the analytical continuation in the latter scheme
is performed appropriately (see eq.~(\ref{RiemannSheetDefCheck}) 
and sect.~\ref{AnalyticContinuationUVCT}). As for the imaginary part of 
$\Sigma(M_W^2-i\Gamma_WM_W)$, it has been expressed in terms of the
LO $W$-boson total decay width $\Gamma_W^{(0)}$ by using 
eq.~(\ref{CM_width_definition}). This underscores the importance of 
employing an expression for $\bar{\Gamma}_W$ which obeys
eq.~(\ref{CMGalim}) when evaluating eq.~(\ref{D1def}) numerically.

We now turn to considering the second non-regular contribution
on the r.h.s.~of eq.~(\ref{LoopLevelDY}), namely that stemming from 
the UV counterterm $\delta{c_W}$:
\beqn
\label{cwCounterterm}
\amp^{(1)\delta{c_W}}_{\rm CM} &=& 2 \frac{\delta{c_W}}{c_W}
\vertmuCM(q_{W}) G^{\mu\nu}_{\CM}(q_{W}) \vertnuCM(q_{W}) = 
2\frac{\delta{c_W}}{c_W}\left( \vertmuZW G^{\mu\nu}_{\ZW} \vertnuZW +
\ord(\aem^2) \right)
\nonumber\\&=&
\left(\frac{\delta m_W^2}{m_W^2}-\frac{\delta m_Z^2}{m_Z^2}\right)\ampzZW+
\ord(\aem^3)
\nonumber\\&=& 
\left[2\frac{\delta{C_W}}{C_W} + \frac{ M_Z^2 i\Im{[\delta m_W^2]} - 
M_W^2 i \Im{[\delta m_Z^2]} }{M_Z^3 M_W}\frac{M_Z}{M_W}\right]\ampzZW+
\ord(\aem^3) 
\nonumber\\&=&
\left[2\frac{\delta{C_W}}{C_W} + i\frac{ M_W \Gamma_Z-M_Z \Gamma_W }{ M_Z^2}
\frac{1}{C_W}\right]\ampzZW +\ord(\aem^3) 
\nonumber \\&=&
\amp^{(1)\delta{C_W}}_{\rm ZW}+
i\frac{ M_W \Gamma_Z - M_Z \Gamma_W }{2 M_Z^2}\frac{\partial}{\partial C_W} 
\ampzZW +\ord(\aem^3)\,,
\eeqn
where the last equality assumed $\ampzZW\propto C_W^2$, so that
$\frac{1}{C_W}\ampzZW=\frac{1}{2}\frac{\partial}{\partial C_W}\ampzZW$. 
Note that the number of insertions of the counterterm $\delta {C_W}$ is 
always equal to the power of $C_W$ factorised in $\ampzZW$, hence rendering 
the above observation general.
By using eqs.~(\ref{RegularLoops}), (\ref{LoopLevelDY}), 
(\ref{sigmaContribs}), and~(\ref{cwCounterterm}), we can now explicitly 
verify that:
\beqn
&&\ampzCM-\ampzZW+\amp^{(1)}_{\rm CM}-\amp^{(1)}_{\rm ZW}=
\nonumber\\&&\phantom{aaaa}
\ampzCM-\ampzZW
+\Big(\amp^{(1){\rm reg}}_{\rm CM}-\amp^{(1),{\rm reg}}_{\rm ZW}\Big) 
+\Big(\amp^{(1)\delta{m_W^2}}_{\rm CM}-\amp^{(1)\delta{M_W^2}}_{\rm ZW}\Big) 
\nonumber\\&&\phantom{aaaa\ampzCM-\ampzZW}
+\Big(\amp^{(1)\delta{c_W}}_{\rm CM}-\amp^{(1)\delta{C_W}}_{\rm ZW}\Big)=
\nonumber\\&&\phantom{aaaa}
0+\ord(\aem^3)\,.
\label{analyticalCMLoopTest}
\eeqn
We can now proceed analogously to what has been done in eq.~(\ref{anCMTtest}).
Namely, one moves the two ZW amplitudes from the l.h.s.~to the r.h.s.~of
eq.~(\ref{analyticalCMLoopTest}), and then computes the absolute value
squared of both sides. In this way, one shows that the linear combination
of matrix elements that appears in the numerator of $\delta^{(1)}(\lambda)$
is of $\ord(\aem^4)$, thus proving that $\Delta^{(1)}$ is a finite number.

We conclude this section by presenting the numerical results of the
tests advocated here for a couple of representative processes. Such tests
can be run directly from the interactive \aNLOs\ interface by issuing
the command:

\vskip 0.25truecm
\noindent
~~\prompt\ {\tt check cms <process\_definition> <options>}

\vskip 0.25truecm
\noindent
where {\tt process\_definition} follows the usual~\cite{Alwall:2014hca}
tree-level or loop-level \aNLOs\ syntax (depending on whether one wants to 
evaluate the limit on the r.h.s.~of eq.~(\ref{D0def}) or that of 
eq.~(\ref{D1def})); further options can be added\footnote{The exhaustive 
list of all available options can be found at
\url{http://cp3.irmp.ucl.ac.be/projects/madgraph/wiki/ComplexMassScheme}.}. 
An explicit example of this is:

\vskip 0.25truecm
\noindent
~~\prompt\ {\tt check cms u d\~{} > e+ ve a [virt=QED] --tweak=alltweaks}

\vskip 0.25truecm
\noindent
In order to be definite, we limit ourselves to considering here the
one-loop test of eq.~(\ref{D1def}), which is obviously more involved
than its tree-level counterpart of eq.~(\ref{D0def}). We do so for
the two processes \mbox{$u\bar{d}\to e^+\nu_e\gamma$} and
\mbox{$gg\to e^+\nu_e e^-\bar{\nu}_e b\bb$}.

As was already said, the $\Delta^{(i)}$'s of eqs.~(\ref{D0def}) 
and~(\ref{D1def}) are computed by evaluating the arguments of the limits 
on the r.h.s.~of those equations (denoted there by $\delta^{(0)}(\lambda)$
and $\delta^{(1)}(\lambda)$, respectively) for increasingly smaller 
values of the scaling parameters $\lambda$. The results are then plotted 
for each value of $\lambda$, as is done here in fig.~\ref{fig:CMcheckFigs};
in this way, one can easily see the behaviour of the relevant matrix element
combination for \mbox{$\lambda\to 0$}.

\begin{figure}[t]
\begin{center}
\includegraphics[trim=75 240 45 70,clip,scale=0.43]{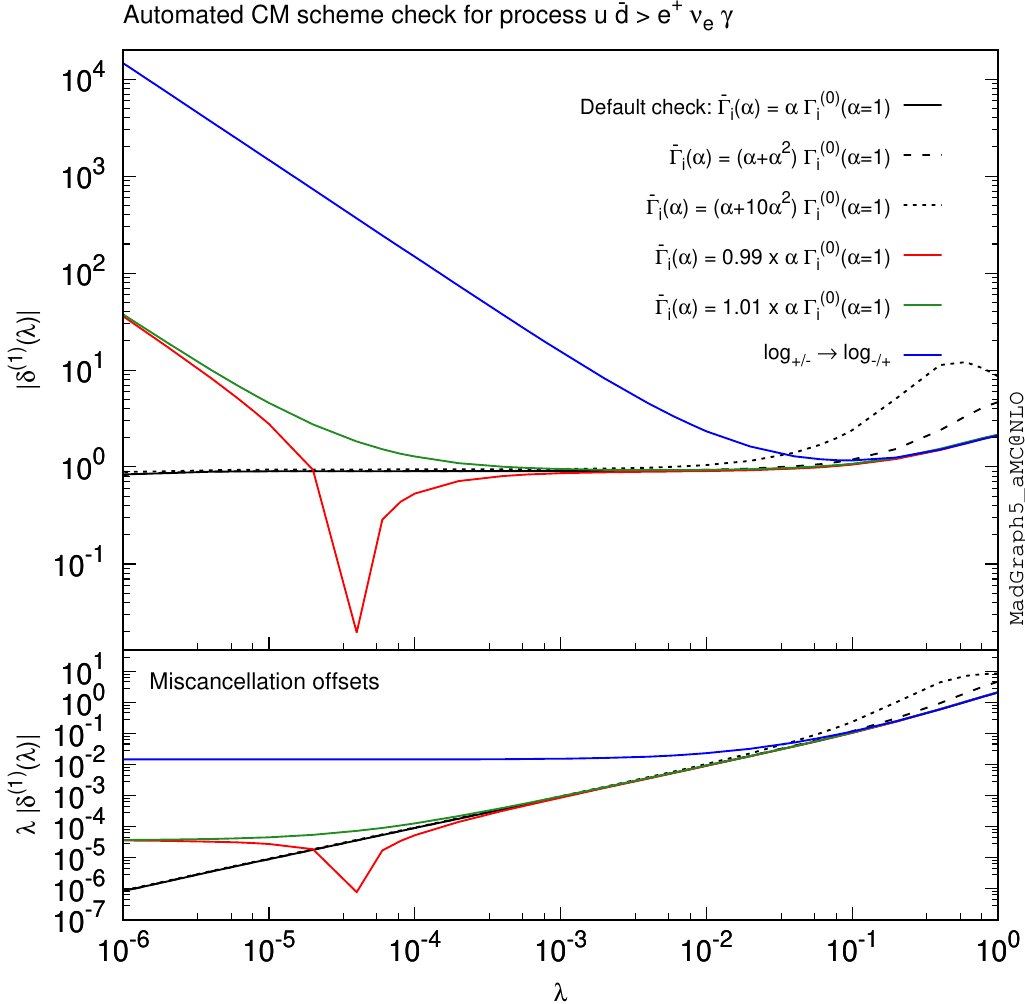}
\includegraphics[trim=75 240 45 70,clip,scale=0.43]{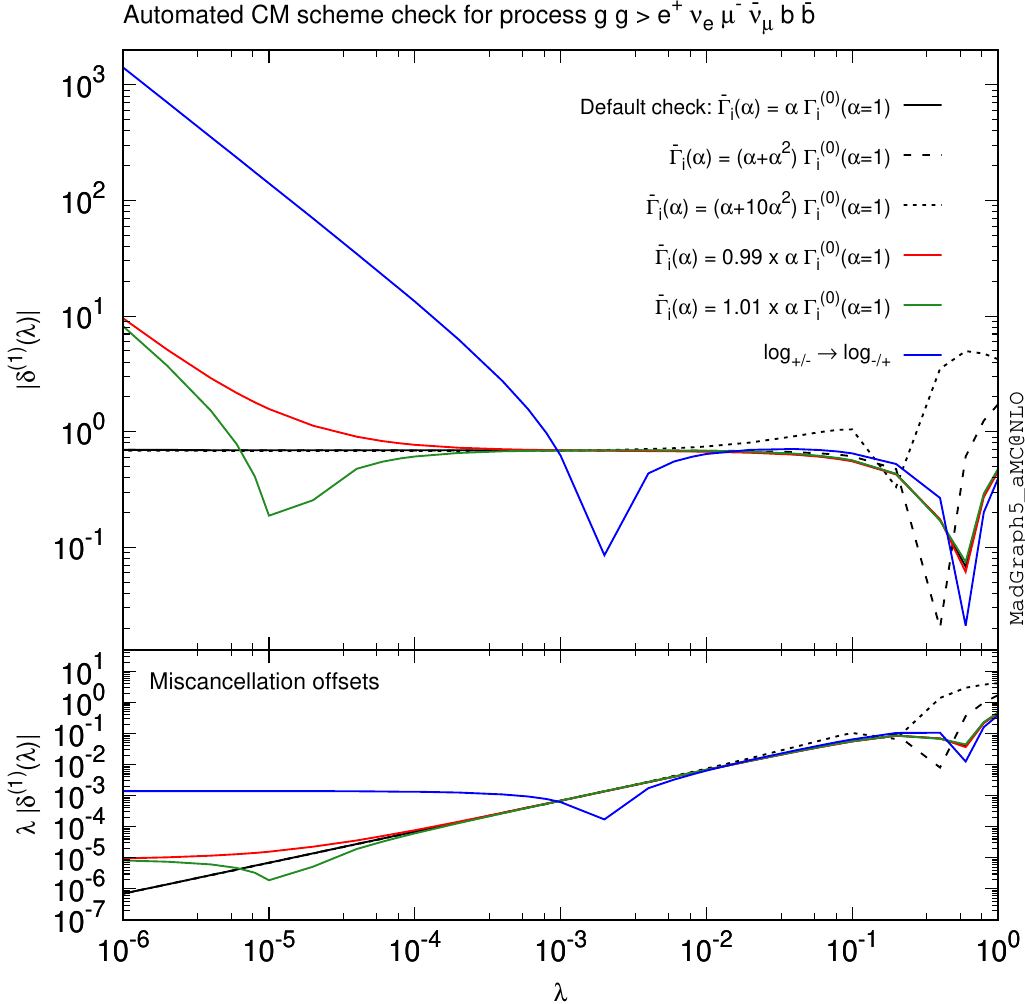}
\end{center}
\caption{\label{fig:CMcheckFigs} 
Tests of the CM-scheme implementation performed by \aNLOs, for the
$u\bar{d}\to e^+\nu_e\gamma$ (left panel) and
$gg\to e^+\nu_e e^-\bar{\nu}_e b\bb$ (right panel) processes,
with different choices for the widths and the Riemann sheet in the
logarithmic terms of the UV mass counterterms. See the text for
details.
}
\end{figure}
The left and right panels of fig.~\ref{fig:CMcheckFigs} display the
results relevant to the \mbox{$u\bar{d}\to e^+\nu_e\gamma$} and
\mbox{$gg\to e^+\nu_e e^-\bar{\nu}_e b\bb$} processes. In the main
frame, $\abs{\delta^{(1)}(\lambda)}$ is shown, while the inset presents
$\lambda\abs{\delta^{(1)}(\lambda)}$, which must tend to zero in the case
of a successful test. Both quantities are computed in different scenarios,
each of which corresponds to one curve in the main frame and one curve
in the inset of fig.~\ref{fig:CMcheckFigs}. More in detail, the three
black curves are obtained by setting:
\beqn
\bar{\Gamma}_i(\aem)&=&\aem\,\Gamma_i^{(0)}(\aem=1)\,,
\label{Ga1}
\\
\bar{\Gamma}_i(\aem)&=&\left(\aem+\aem^2\right)\Gamma_i^{(0)}(\aem=1)\,,
\label{Ga2}
\\
\bar{\Gamma}_i(\aem)&=&\left(\aem+10\aem^2\right)\Gamma_i^{(0)}(\aem=1)\,,
\label{Ga3}
\eeqn
for the solid, long-dashed, and short-dashed patterns respectively.
Equations~(\ref{Ga1})--(\ref{Ga3}) understand that these settings for
the widths apply to {\em all} relevant unstable particles (i.e., 
the $W$ boson for \mbox{$u\bar{d}\to e^+\nu_e\gamma$}, and the top quark, the 
$W$ boson and the $Z$ boson for \mbox{$gg\to e^+\nu_e e^-\bar{\nu}_e b\bb$} --
we point out that in the latter process all resonant and non-resonant 
contribution are included). The widths of eqs.~(\ref{Ga1})--(\ref{Ga3}) 
obey eq.~(\ref{CMGalim}); fig.~\ref{fig:CMcheckFigs} clearly shows that
the three choices differ in the large $\lambda$ region, but converge to
the same value when $\lambda\to 0$, and ultimately lead to a successful test.
Conversely, by setting:
\beqn
\bar{\Gamma}_i(\aem)&=&0.99\,\aem\,\Gamma_i^{(0)}(\aem=1)\,,
\label{Ga4}
\\
\bar{\Gamma}_i(\aem)&=&1.01\,\aem\,\Gamma_i^{(0)}(\aem=1)\,,
\label{Ga5}
\eeqn 
that correspond to the solid red and green curves, respectively,
the tests do fail, with $\delta^{(1)}(\lambda)$ departing from a 
constant behaviour for $\lambda\lesssim 10^{-3}$. We point out that 
this occurs for a mere $1$\% deviation from eq.~(\ref{CMGalim}), and 
demonstrates the high numerical sensitivity of the present checks.

Finally, the blue solid curves in fig.~\ref{fig:CMcheckFigs}, labelled
there as \mbox{$\log_{\pm}\to\log_{\mp}$}, correspond to an alteration of 
the CM-scheme implementation that consists in changing the selected
Riemann sheet (see eq.~(\ref{eq:logdef})) in the logarithmic part of the
UV mass counterterms of unstable particles. This leads to a violation 
of the consistency relation of eq.~(\ref{RiemannSheetDefCheck}),
which results in mis-cancellations much more severe than those induced by 
a small modification of the LO term in the functional form of the widths. 
This provides further evidence that any inconsistency in the CM-scheme
implementation is very unlikely to be undetected when the present
numerical validation procedure is carried out\footnote{We remark that
the plots of fig.~\ref{fig:CMcheckFigs} differ from those generated 
automatically by \aNLOs\ only in their layout, improved here for the 
sake of clarity. In order to get the curves that correspond to the 
incorrect choice of Riemann sheet for the logarithms and to the various
functional forms of the widths, the option {\tt --tweak=alltweaks} must be
specified. The $g g \rightarrow e^+ \nu_e \mu^- \bar{\nu_\mu} b \bar{b}$
process also necessitates to force quadruple precision (and a single
helicity for increased speed) with {\tt --CTModeRun=4 --helicity=1}.}.
Having said this, there is a subtle point that is worth making. Namely,
that by verifying that eq.~(\ref{RiemannSheetDefCheck}) is fulfilled
we establish that the appropriate Riemann sheet is used in the limit in 
which {\em all} widths vanish. This is not necessarily the same sheet
as that relevant to the physical (i.e.~with non-zero widths) configuration,
be it either because (in the language of sect.~\ref{AnalyticContinuationUVCT})
the trajectory is ``long'' (i.e.~$\bGa\ll\bM$ is not fulfilled), or because
its $\bga=0$ endpoint is not close to an OS-like configuration 
(i.e.~$\bGa_i\ll\bM_i$ is not fulfilled for some $i$). This does not 
happen in the SM, but there is no reason that prevents it from happening
in an arbitrary model, where thus we do not expect the test outlined in
this appendix to be particularly effective in detecting an incorrect
analytical continuation.

In conclusion, we have successfully run the automated numerical test 
discussed in this section for a variety of processes, in both the $\oG_\mu$ 
and the $\aem(m_Z)$ CM schemes, making sure that those collectively
involved all possible SM unstable particle. Together with the customary
IR-pole cancellation check, we have thus validated the \aNLOs\ implementation 
of the CM scheme in the SM.

\subsection{On the numerical study of trajectories and contour 
integration\label{sec:trj}}
In this section, we briefly sketch two numerical approaches to
the study of the trajectory ${\cal T}$ of eq.~(\ref{trjdef}). In
particular, our final goal is that of determining the difference
\mbox{$n_{+-}-n_{-+}$} that appears in eq.~(\ref{kvsnpm}).
We shall also briefly comment on the corresponding contour integration.

We start by observing that, irrespective of the winding number of
${\cal T}$, the real functions $\Re T(\bga)$ and $\Im T(\bga)$ of
the real variable \mbox{$\bga\in [0,\bGa]$} are single-valued.
Let:
\beq
0<\zeta_1<\zeta_2<\ldots <\zeta_n<\bGa\,,
\label{zetas}
\eeq
be the $n$ zeros of $\Im T(\bga)$ in the interval $(0,\bGa)$:
\beq
\Im T(\zeta_i)=0\,,\;\;\;\;\;\;\;\;1\le i\le n\,.
\eeq
Define:
\beqn
\sigma_i&=&\half\left(\zeta_{i+1}+\zeta_i\right)\,,
\;\;\;\;\;\;\;\;1\le i\le n-1\,,
\\
\sigma_0&=&0\,,
\\
\sigma_n&=&\bGa\,,
\label{sign}
\eeqn
so that:
\beqn
&&\sigma_{i-1}<\zeta_i<\sigma_i\,,
\\
&&\stepf\left(-\Im T(\sigma_{i-1})\;\Im T(\sigma_i) \right)=1\,,
\;\;\;\;\;\;\;\;1\le i\le n\,.
\label{oppsgn}
\eeqn
Then\footnote{We remind the reader that we work with the branch cut of
the logarithm on the negative real axis.}:
\beq
n_{+-}-n_{-+}=\sum_{i=1}^n\stepf(-\Re T(\zeta_i))
\Big[\stepf(\Im T(\sigma_{i-1}))-\stepf(-\Im T(\sigma_{i-1}))\Big]\,.
\label{npmvssig}
\eeq
Since eq.~(\ref{npmvssig}) gives the desired result in a closed form,
the only problem which remains to be solved is that of the determination
of the $\zeta_i$'s. In order to do that numerically, we proceed according
to the following steps:
\begin{itemize}
\item[0.] Define:
\beq
a=0\,,\;\;\;\;\;\;\;\;b=\bGa\,.
\eeq
Denote by ${\cal Z}=\{\}$ the set (presently empty) that will
contain the zeros $\zeta_i$'s, and by $I_{a,b}$ the interval:
\beq
I_{a,b}=[a,b]\,.
\eeq
\item[1.] Evaluate:
\beq
S_{ab}=\stepf\left(-\Im T(a)\;\Im T(b) \right)\,.
\label{Sab}
\eeq
If $S_{ab}=1$ go to step 2, otherwise go to step 3.
\item[2.]
Find numerically the ``first'' zero of $\Im T(\bga)$ in $[a,b]$ by using
a numerical routine\footnote{Different routines might have different
ideas about which zero is the first (the only one they return), in the case 
of multiple zeros. In particular, one must not assume the first zero to be
either the leftmost or the rightmost one in the given range. However, the 
final result of the procedure we propose  here does not depend on these 
differences.} (e.g.~{\tt RZEROX} of CERNLIB). Denote it by $\bar{\zeta}$, 
set \mbox{${\cal Z}={\cal Z}\cup\{\bar{\zeta}\}$}, and go to step 2a.
\begin{itemize}
\item[2a.] For a given number $N_\delta\gg 1$, set
\beq
\delta=\frac{b-a}{N_\delta}\,,
\eeq
and go to the next step.
\item[2b.] If
\beq
\bar{\zeta}-\delta<a\;\;\;\;\;\;{\rm or}\;\;\;\;\;\;
\bar{\zeta}+\delta>b\,,
\eeq
increase $N_\delta$ and return to step 2a. Otherwise, go to
the next step.
\item[2c.] If
\beq
\stepf\left(-\Im T(\bar{\zeta}-\delta)\;\Im T(\bar{\zeta}+\delta) \right)=0\,,
\eeq
increase $N_\delta$ and return to step 2a. Otherwise, go to
the next step.
\item[2d.] Repeat the procedure starting from step 1, by replacing
$I_{a,b}$ there with both $I_{a,\bar{\zeta}-\delta}$ and 
$I_{\bar{\zeta}+\delta,b}$. The interval
$I_{\bar{\zeta}-\delta,\bar{\zeta}+\delta}$ plays no further role.
\end{itemize}
\item[3.] Set:
\beq
c=\frac{a+b}{2}\,,
\eeq
and repeat the procedure starting from step 1, by replacing
$I_{a,b}$ there with both $I_{a,c}$ and $I_{c,b}$.
\end{itemize}
This is therefore an iterative procedure, whose core task is that of
finding one zero of $\Im T(\bga)$ in the interval $[a,b]$. The existence
of such a zero is guaranteed if the signs of $\Im T(a)$ and $\Im T(b)$ 
are opposite; this implies $S_{ab}=1$. When this is the case (step 2),
one of the zeros ($\bar{\zeta}$) is found, a small interval around it
is constructed ($I_{\bar{\zeta}-\delta,\bar{\zeta}+\delta}$), and 
the procedure repeated for each of the two intervals whose union is the 
complement of $I_{\bar{\zeta}-\delta,\bar{\zeta}+\delta}$ in $[a,b]$.
Conversely, if the sign of $\Im T(a)$ and $\Im T(b)$ are identical,
there could be either no zeros or an even number of zeros of $\Im T(\bga)$
in [a,b]. To search for them, this interval is bisected (step 3),
and the procedure repeated for the two resulting intervals.

When all of the zeros of $\Im T(\bga)$ are found, it is clear that there
will be an infinite loop consisting of a step 1--step 3 cycle relevant
to increasingly small intervals. One can avoid this by either imposing
an upper limit on the number of iterations, or a lower limit on the width
$b-a$ of the intervals considered. The latter option, being directly
associated with the geometry of the trajectory, seems to be more appealing.

Although numerical routines that find the zeros of a real function
are fast and efficient, it is desirable to have a less rigorous, but
much quicker, alternative to the procedure outlined so far. One possibility
stems from approximating the imaginary part of the trajectory with a sequence 
of segments. In other words, one introduces:
\beq
\sigma_i=i\,\frac{\bGa}{N_\sigma}\,,
\;\;\;\;\;\;\;\;0\le i\le N_\sigma\,,
\label{sidef}
\eeq
with $N_\sigma$ a given integer number. The idea is that of replacing
$\Im T(\bga)$ for \mbox{$\sigma_{i-1}\le\bga\le\sigma_i$} with a straight 
line that connects the two points $\Im T(\sigma_{i-1})$ and 
$\Im T(\sigma_i)$, i.e.~with:
\beq
\Im \widetilde{T}(\bga)\equiv
\Im T(\sigma_{i-1})+
\frac{\Im T(\sigma_i)-\Im T(\sigma_{i-1})}{\sigma_i-\sigma_{i-1}}\,
\Big(\bga-\sigma_{i-1}\Big)\,.
\label{Ttil}
\eeq
One then finds the zero of $\Im \widetilde{T}(\bga)$:
\beq
\Im \widetilde{T}(\zeta_i)=0\;\;\;\;\Longrightarrow\;\;\;\;
\zeta_i=\frac{\sigma_{i-1}\Im T(\sigma_i)-\sigma_i\Im T(\sigma_{i-1})}
{\Im T(\sigma_i)-\Im T(\sigma_{i-1})}\,.
\label{zetasol}
\eeq
Note that for this solution to be acceptable, we must have
\mbox{$\sigma_{i-1}<\zeta_i<\sigma_i$}, which happens when
$\Im T(\sigma_i)$ and $\Im T(\sigma_{i-1})$ have opposite signs.
This is nothing but eq.~(\ref{oppsgn}) (for a given $i$). In fact,
as the notation used here suggests, the roles of the $\sigma_i$
and $\zeta_i$ quantities introduced in eqs.~(\ref{sidef}) 
and~(\ref{zetasol}) are those of an approximation
to their counterparts in eqs.~(\ref{zetas})--(\ref{npmvssig}).
In fact, for the present case we can simply re-use eq.~(\ref{npmvssig}),
taking care of enforcing eq.~(\ref{oppsgn}) explicitly:
\beq
n_{+-}-n_{-+}=\sum_{i=1}^n\stepf(-\Re T(\zeta_i))
\stepf\left(-\Im T(\sigma_{i-1})\;\Im T(\sigma_i) \right)
\Big[\stepf(\Im T(\sigma_{i-1}))-\stepf(-\Im T(\sigma_{i-1}))\Big]\,.
\label{npmvssig2}
\eeq
Thus, this equation can be used both with the $\sigma_i$'s and $\zeta_i$'s
of eqs.~(\ref{zetas})--(\ref{sign}), or with those of eqs.~(\ref{sidef})
and~(\ref{zetasol}) by setting $n=N_\sigma$.

This procedure gives a correct result if the partition of the 
range $[0,\bGa]$ achieved by eq.~(\ref{sidef}) is sufficiently
fine-grained to capture the behaviour of $\Im T(\bga)$. This is
obviously the case for $N_\sigma\to\infty$, which suggests that
a self-diagnostic test (still not fully watertight) is that of
repeating the computation of eq.~(\ref{npmvssig2}) by increasing
$N_\sigma$ -- the same result must be obtained.
We point out that this constitutes an automated test. In a case-by-case
situation, one can always check {\em visually} that the discretisation of 
the trajectory according to eq.~(\ref{Ttil}) represents a continuous
deformation of the original curve that does not cross the origin.
Indeed, this is what has been done with the trajectories considered 
in sect.~\ref{AnalyticContinuationUVCT}, which have been dealt with 
by setting $N_\sigma=100$.

We point out that eq.~(\ref{npmvssig2}) is equivalent to the
straight-trajectory method when $N_\sigma=1$. Because of the leftmost
property in eq.~(\ref{DeltaArgprop}), a difference between the results
of eq.~(\ref{npmvssig2}) with $N_\sigma=1$ and with a sufficiently
large $N_\sigma$ implies the failure of the straight-trajectory approach,
which can only be due to the fact that the given trajectory cannot be
continuously deformed into a straight line without crossing the origin.

We now turn to discussing the contour integration that has led us 
to the results shown in fig.~\ref{fig:B0res}. In sect.~6.6 
of ref.~\cite{Passarino:2010qk}, the authors have proposed 
a strategy for integrating $I=\int_0^1 dx \log^{-}(V(x))$ with 
$V(x)=a x^2 + b x + c$, where the polynomial coefficients $a$, $b$, 
and $c$ can assume arbitrary complex values. This allows one to directly
evaluate the most general bubble function $B_0(p^2,\mu_1^2, \mu_2^2)$ 
with all arguments complex (under certain restrictive conditions -- see
eq.~(81) of that paper, and eq.~(37) there for the definition of $\log^{-}()$;
note that the latter differs from the $\log_{-}()$ function of 
eq.~(\ref{logshort})). In the complex plane, the branch cut of $\log^{-}()$ 
relevant to the integral $I$ is the set of all points $\bar{x}$ 
that satisfy the two conditions $\Re V(\bar{x})=0$ and $\Im V(\bar{x})>0$. 
The former condition results in two branches of an hyperbola, and the 
latter condition selects a subset of points (one of which possibly empty)
in each of these branches; we denote such subsets by $H_1$ and $H_2$.
If either (or both) $H_1$ or (and) $H_2$ crosses the real axis in the 
range $[0,1]$, then $I$ must be computed by deforming the $x\in [0,1]$ 
range into a contour ${\cal C}$ that lives in ${\mathbb C}$. The contour 
must be such that it crosses neither $H_1$ nor $H_2$, 
i.e.~${\cal C}\cap H_1=\emptyset$ and ${\cal C}\cap H_2=\emptyset$.
The authors of ref.~\cite{Passarino:2010qk} have proposed a general 
parametrisation for ${\cal C}$, given in terms of the coefficients $a$, 
$b$, and $c$. We have found that this parametrisation works as expected 
if both of the sets $H_i\cap [0,1]$, $i=1,2$ contain {\em at most one} 
point. Conversely, if for either $i=1$ or $i=2$ the set $H_i\cap [0,1]$ 
contains two points, then ${\cal C}\cap H_i\ne\emptyset$, and ${\cal C}$
is therefore not suited to numerical integration. One actual example of 
this situation is the case of configuration E of table~\ref{tab:trj}. We 
have addressed this issue by using different parameters w.r.t.~those that 
emerge from the proposal of ref.~\cite{Passarino:2010qk}\footnote{In 
the notation of eq.~(91) of ref.~\cite{Passarino:2010qk}, the chosen 
parameters for dealing with benchmark point E are: $\alpha_1=0$, 
$\alpha_2=0$, $\beta_1=-0.05$, $\beta_2=0$, and $\alpha_c=0.7$.}, 
while keeping the same functional parametrisation of ${\cal C}$ suggested
there. With these modified parameters we have obtained the results 
labelled as ``revised contour'' in sect.~\ref{AnalyticContinuationUVCT}.

\subsection{\SMWidth: an SM decay-width calculator at the NLO QCD+EW 
accuracy\label{sec:smwidth}}
As was documented in sect.~\ref{sec:CMtest}, in off-shell regions 
widths can be computed at the LO without spoiling the overall NLO
accuracy of the calculation. However, this is not true in general,
and NLO-accurate widths must be used. In the SM, the widths of the 
top quark~\cite{Jezabek:1988iv,Denner:1990ns,Czarnecki:1990kv,Li:1990qf,
Liu:1990py,Eilam:1991iz,Jezabek:1993wk,Basso:2015gca}, 
$W$ boson~\cite{Marciano:1974vg,Albert:1979ix,Inoue:1980ky,
Chang:1981qq,Consoli:1983yn,Bardin:1986fi,
Denner:1990tx}, $Z$ boson~\cite{Dine:1979qh,Celmaster:1979xr,Chang:1981qq, 
Consoli:1983yn,Czarnecki:1996ei}, and Higgs boson~\cite{Djouadi:1997yw}
have been known for a long time at the NLO accuracy in QCD and EW.
In spite of this, no single public tool provides one with all of these decay
widths at such an accuracy. We have amended this, by writing a 
self-contained package, dubbed \SMWidth, which can be used either 
standalone or within the \aNLOs\ framework. \SMWidth\ computes the
top, $W$, and $Z$ total widths\footnote{Widths relevant to specific
decay channels can be computed only by means of calls to functions 
which are not part of the standard user interface.} from 
first principles, and calls \HDecay~\cite{Djouadi:1997yw,Djouadi:2018xqq}
to obtain the Higgs width.
In what follows, we briefly describe its workings, with the standalone
usage presented in sect.~\ref{sec:SMWguide}, and that within \aNLOs\ in
sect.~\ref{sec:SMWaMC}. Sample results are given in sect.~\ref{sec:SMWres}.
Before going into that, we introduce some general features of the code.

\SMWidth\ is written in Fortran90, and is included in the \aNLOs\
distribution. It implements the analytically-integrated decay amplitudes 
of the top quark, $W$ boson, and $Z$ boson, by including NLO QCD and EW 
corrections. Such amplitudes have been generated by 
\FeynArts~\cite{Hahn:2000kx}, and the relevant loop and phase-space 
integrations have been carried out by means of an in-house {\tt Mathematica} 
module, whose output has been converted into the Fortran90 code; in 
this way, only elementary numerical 
operations\footnote{\OneLoop~\cite{vanHameren:2010cp} is called
on-the-fly as a library that returns the value of the $C_0$ function,
necessary in the computation of EW corrections to the top and $W$ widths.}
are performed by \SMWidth, and the program has thus a negligible CPU load.
The top-decay width is computed according to the following formula:
\beqn
\Gamma_t^{\rm NLO}&\equiv&\sum_{f_1\bar{f}_2\in D_W}
\Gamma_{\CM}^{\rm NLO}
\big(t\rightarrow b+(W^*\rightarrow f_1+\bar{f_2})\big)
\nonumber
\\*&=&
\Gamma_{\ZW}^{\rm NLO}\big(t\rightarrow b+W\big)
\nonumber
\\*&&\phantom{aaa}
+\!\!\sum_{f_1\bar{f}_2\in D_W}\Delta\Gamma_{\rm FW}
\big(t\rightarrow b+(W^*\rightarrow f_1+\bar{f_2})\big)
+\ord(\as^n\aem^m)\,,
\label{topdec}
\eeqn
with $n+m=3$, $n\ge 0$, $m\ge 1$, and:
\beqn
\Delta\Gamma_{\rm FW}
\big(t\rightarrow b+(W^*\rightarrow f_1+\bar{f_2})\big)&=&
\Gamma^{(0)}_{\rm CM}
\big(t\rightarrow b+(W^*\rightarrow f_1+\bar{f_2})\big)
\nonumber\\*&-&
\Gamma_{\ZW}^{(0)}\big(t\rightarrow b+W\big)
{\rm Br}\big(W\rightarrow f_1+\bar{f_2}\big)\,.
\label{eq:finitewidth}
\eeqn
As was done in appendix~\ref{sec:CMtest}, the subscripts \CM\ and \ZW\
denote CM-scheme and zero-width-setup results\footnote{\label{ft:SMW}
\SMWidth\ uses OS-type renormalisation conditions and inputs. The computation 
of the first term on the r.h.s.~of eq.~(\ref{eq:finitewidth}) is then 
performed by complexifying both the $W$ and $Z$ masses according to 
eq.~(\ref{complexMass}) (by using the internally computed LO values of 
$\Gamma_W$ and $\Gamma_Z$) and the relevant coupling factors. Note that
this is done with $\bM_i=M_i$ and $\bGa_i=\Gamma_i$, which is consistent
with the fact any difference between CM- and OS-type input parameters
would be beyond accuracy in eq.~(\ref{eq:finitewidth}), owing to
eqs.~(\ref{MvsMbar}) and~(\ref{CMwdefE}).}; for the definition of 
the latter, see eq.~(\ref{ZWdefinition}). By $W$ and $W^*$ we have
denoted a final-state on-shell $W$ boson, and its intermediate off-shell
counterpart, respectively. $D_W$ denotes the set of the fermion pairs
into which a $W$ boson can decay. The superscript ``NLO'' that appears
on the r.h.s.~of eq.~(\ref{topdec}) imply that the corresponding
process includes one-loop and real-emission corrections stemming from
QCD or EW interaction vertices. The superscripts ``$(0)$'' in
eq.~(\ref{eq:finitewidth}) indicate that the two widths are computed
at the tree level. Equation~(\ref{topdec}) can be easily generalised to 
take into account the decays of the top into down-type, non-bottom quarks.

As far as the Higgs boson is concerned, \SMWidth\ simply acts as a calling 
interface to \HDecay, which is embedded in our package. The user is thus 
expected to also cite the relevant \HDecay\ papers~\cite{Djouadi:1997yw,
Djouadi:2018xqq} in that case.

\subsubsection{Standalone usage\label{sec:SMWguide}}
We now turn to describing the standalone usage of \SMWidth. The package 
contains a module, {\tt test.f90}, which acts as the driver when working 
in standalone mode. The code is compiled with the shell command:
\cCode{}
> make -f makefile_test
\end{lstlisting}
We assume a {\tt gfortran} compiler with the same characteristics
as that employed when using \aNLOs. The test program embedded in
{\tt test.f90} can be run by executing:
\cCode{}
> ./test
\end{lstlisting}
There are two input files, one that sets the EW renormalisation scheme 
to be adopted, and one that sets the physics parameters\footnote{While
such parameters are fully under the user's control, one must not
vastly depart from the measured SM values, lest \SMWidth\ give incorrect
results. In particular, no new decay channel must open.}.
The former file, named {\tt scheme.dat}, contains a single integer,
equal to 1(2) for the $\aem(m_Z)$ ($\oG_\mu$) scheme. In the standalone
mode, the use of this file can be bypassed by setting the global variable
{\tt Decay\_scheme} equal to either 1 or 2.
As far as the physics inputs are concerned, they must be written
in a file {\tt phyinputs.dat}, which the Fortran subroutine 
{\tt ReadParamCard} is tasked to read as follows: 
\cCode{}
CALL ReadParamCard('phyinputs.dat')
\end{lstlisting}
As is suggested by the call above, the name {\tt phyinputs.dat} can
be freely changed by the user; what matters is that it contains the basic 
physics input parameters, namely the masses of the top quark and of the 
$W$, $Z$, and Higgs bosons, and the value of either $\aem(m_Z)^{-1}$ 
or $G^{(G_\mu)}_\mu$, depending on which EW scheme is chosen. We
assume \mbox{$\as(M_Z^\BW)=0.1184$} (with $M_Z^\BW=91.1876~\gev$,
see eq.~(\ref{ZmGBW})), and obtain the value of $\as(M_t)$, $\as(M_Z)$,
 and $\as(M_W)$ employed in the calculation of the NLO QCD corrections to 
the top, $Z$, and $W$ widths by means of a two-loop evolution. Two templates,
relevant to the two valid EW-scheme choices, are included in the package, 
{\tt param\_card\_MZ.dat} and {\tt param\_card\_Gmu.dat}. As these
names suggest, they are derived from the {\tt param\_card.dat} file
of \aNLOs.

The $W$-decay width can be computed as follows:
\cCode{}
width=SMWWidth(qcdord,qedord,finitemass)
\end{lstlisting}
The settings of {\tt qcdord} and {\tt qedord} determine whether QCD
and EW corrections, respectively, are included -- valid inputs are 1
to include the corrections, and 0 to exclude them. By setting the
entry {\tt finitemass} equal to {\tt .true.} the effects are included
due to the non-zero bottom, charm, tau, and muon masses. Otherwise,
when {\tt finitemass=.false.} all of these fermions are treated as massless.

In a manner fully analogous to the case of the $W$, the $Z$ decay width
is computed as follows:
\cCode{}
width=SMZWidth(qcdord,qedord,Wrad,finitemass)
\end{lstlisting}
The entries are the same as those relevant to the $W$ decays, and
have the same meanings. On top of those, the parameter {\tt Wrad} can be 
set equal to {\tt .true.} in order to include in the computation the 
contributions due to the channels $Z\to W^{\pm}+f_1+\bar{f}_2$ (where
the $W$ is on-shell and the fermions are treated as massless, regardless
of the value of {\tt finitemass}). These contributions are typically very 
small, owing to phase-space suppression.

The top-quark decay width can be calculated as follows:
\cCode{}
width=SMtWidth(qcdord,qedord,finitemass,wwidth)
\end{lstlisting}
The entries {\tt qcdord} and {\tt qedord} have the same meanings as
before. {\tt finitemass} is used here in order to include (when set
equal to {\tt .true.}) non-zero bottom mass effects -- the bottom
is regarded as massless when {\tt finitemass=.false.}. The other quarks
and the leptons are always treated as massless. Finally, by setting 
{\tt wwidth=.true.} one is able to include the contribution
due to a non-zero $W$ width, which is an essential ingredient in
order to attain NLO EW accuracy for a process that features intermediate 
top quarks -- this corresponds to the contribution of 
eq.~(\ref{eq:finitewidth}).

Finally, the width of Higgs boson is computed as follows:
\cCode{}
width=SMHWidth(idummy)
\end{lstlisting}
with {\tt idummy} is an INTEGER-type dummy argument. As was already
said, the function {\tt SMHWidth} simply acts as an interface to
\HDecay, with the input parameters given in {\tt param\_card.dat}
written onto the \HDecay\ input file {\tt hdecay.in}.

\subsubsection{Usage in \aNLOs\label{sec:SMWaMC}}
The usage of \SMWidth\ from within \aNLOs\ can proceed either in an
interactive manner, through the command line, or implicitly, through
suitable settings in an input file. In the former mode, the
syntax is as follows:

\vskip 0.25truecm
\noindent
~~\prompt\ {\tt compute\_widths <particle(s)> --nlo}

\vskip 0.25truecm
\noindent
where {\tt <particle(s)>} is the list of the labels (according to
the \aNLOs\ syntax) for all of the particles whose widths one wants
to compute. For example:

\vskip 0.25truecm
\noindent
~~\prompt\ {\tt compute\_widths w+ z h t --nlo}

\vskip 0.25truecm
\noindent
The option {\tt --nlo} specifies that the widths have to be computed by 
including both NLO QCD and EW corrections. In terms of the calls to the 
internal \SMWidth\ routines described in app.~\ref{sec:SMWguide}, this 
corresponds to {\tt qcdord}$=\!\!\!1$ and {\tt qedord}$=\!\!\!1$.
In the implicit mode, the physics-input file {\em must} be named
{\tt param\_card.dat}. The correct template for the latter is chosen 
automatically by \aNLOs\ upon loading the physics model (see below).
In turn, its contents (in particular, the presence of the numerical value of 
either $\aem(m_Z)$ or $G^{(G_\mu)}_\mu$) tell the code which EW scheme must
be adopted; thus, {\tt scheme.dat} is irrelevant in this mode of operation.
Finally, one must use the keyword {\tt Auto@NLO} in the decay 
block of {\tt param\_card.dat}, as per the following example: 

\begin{minipage}{\textwidth}
\fontsize{10pt}{10pt}\selectfont
\begin{verbatim} 
###################################
## INFORMATION FOR DECAY
###################################
DECAY   6 Auto@NLO # WT 
DECAY  23 Auto@NLO # WZ 
DECAY  24 Auto@NLO # WW 
DECAY  25 Auto@NLO # WH
\end{verbatim}  
\end{minipage}

\vskip 0.5truecm
\noindent
\aNLOs\ will then compute the widths for the top quark and the
heavy bosons at the NLO QCD+EW accuracy, by calling \SMWidth\ internally. 
This option is only available after importing one of the NLO-EW-compatible
\UFO\ models available in \aNLOs, namely either {\tt loop\_qcd\_qed\_sm} 
(for the $\aem(m_Z)$ EW scheme) or {\tt loop\_qcd\_qed\_sm\_Gmu} (for the
$\oG_\mu$ EW scheme). For consistency with these models, light-fermion
and $b$-mass effects are always neglected; the contributions of the channels 
$Z\to W^{\pm}+f_1+\bar{f}_2$ to the $Z$ width are also ignored. Conversely,
finite-$W$-width effects in top decays are included. In terms of the calls
to the internal \SMWidth\ routines described in app.~\ref{sec:SMWguide},
these settings correspond to {\tt finitemass=.false.}, {\tt Wrad=.false.}, 
and {\tt wwidth=.true.}.

\subsubsection{Illustrative numerical results\label{sec:SMWres}}
We conclude this section by reporting sample results obtained 
with \SMWidth. As was already said, in the case of the Higgs boson
they are obtained by an implicit call to \HDecay.

The values of the input parameters in the $\aem(m_Z)$ scheme are
given in table~\ref{tab:Wmzinp}. The light-fermion masses
have been set equal to the values reported in table~\ref{tab:lepmass}.
The renormalisation scale has been set equal to the mass of the 
mother particle, and a two-loop running has been used for $\as$.
\begin{table}[b]
\begin{center}
\begin{tabular}{cl|cl}\toprule
Parameter & value & Parameter & value
\\\midrule
$\aem(m_Z)^{-1}$ & \texttt{128.930} & $M_H$ & \texttt{125.0} 
\\
$M_t$ & \texttt{173.3}  & $y_{t}$ & \texttt{173.3}
\\
$M_W$ & \texttt{80.419}  & $V_{ij}$ & \texttt{$\delta_{ij}$} 
\\
$M_Z$ & \texttt{91.188}  & $\as(M_Z^\BW)$ & $0.1184$
\\\bottomrule
\end{tabular}
\caption{\label{tab:Wmzinp}
Inputs to \SMWidth\ in the $\aem(m_Z)$ scheme, with $M_Z^\BW=91.1876~\gev$;
see also footnote~\ref{ft:SMW}. The resulting value of the Fermi constant 
is $G^{(G_\mu)}_\mu=1.19875\cdot 10^{-5}$.
}
\end{center}
\end{table}
\begin{table}
\begin{center}
\begin{tabular}{cl|cl}\toprule
Parameter & value & Parameter & value
\\\midrule
$m_b$ & \texttt{4.49} & $m_c$ & \texttt{1.42}
\\
$m_{\tau}$ & \texttt{1.77684}    & $m_{\mu}$ & \texttt{0.105658367}
\\\bottomrule
\end{tabular}
\caption{\label{tab:lepmass}
Light-fermion masses given in inputs to \SMWidth.
}
\end{center}
\end{table}
The corresponding results are reported in table~\ref{widthMZ}, where we 
also show a breakdown into the various types of contributions to the
top quark, $W$, and $Z$ widths according to the following definitions:
\beqn
\Gamma_W&=&\Gamma^{\rm LO}_W
\left(1+\delta_{\as}+\delta_{\aem}+\delta_{m_f}\right)\,,
\nonumber\\
\Gamma_Z&=&\Gamma^{\rm LO}_Z
\left(1+\delta_{\as}+\delta_{\aem}+\delta_{m_f}\right)\,,
\nonumber\\
\Gamma_t&=&\Gamma^{\rm LO}_t
\left(1+\delta_{\as}+\delta_{\aem}+\delta_{m_f}+\delta_{\Gamma_W}\right)\,,
\eeqn
with $\delta_i$ the fractional corrections due to NLO QCD effects ($i=\as$),
NLO EW effects ($i=\aem$), finite fermion-mass effects ($i=m_f$, included 
only at the LO), and finite $W$-width effects ($i=\Gamma_W$ -- thus, 
\mbox{$\Gamma^{\rm LO}_t\delta_{\Gamma_W}$} is equal to the r.h.s.~of
eq.~(\ref{eq:finitewidth})).
\begin{table}
\begin{center}
\begin{tabular}{c||cccccc}
 ~ & $\Gamma$~[GeV] & $\Gamma^{\rm LO}$~[GeV] & $\delta_{\alpha_s}$ (\%) & 
     $\delta_{\alpha}$ (\%) & $\delta_{m_f}$ (\%) & $\delta_{\Gamma_W}$ (\%) \\
\hline
   $W^{\pm}$ & $2.08414$ & $2.10490$ &  $2.55$ & $-3.51$ & $-0.0238$ & - \\
   $Z$ & $2.48789$ & $2.51376$ & $2.61$ & $-3.60$ & $-0.0374$ & - \\
  $t$ & $1.36358$ & $1.54624$ &  $-8.58$ & $-1.41$ & $-0.239$ & $-1.58$ \\
  $H$ &  $4.187\times 10^{-3}$ & -  & - & - & - & - \\
\end{tabular}
\end{center}
\caption{\label{widthMZ}
SM-particle widths computed by \SMWidth\ in the $\aem(m_Z)$ scheme. 
}
\end{table}

\begin{table}
\begin{center}
\begin{tabular}{c||cccccc}
 ~ & $\Gamma$~[GeV] & $\Gamma^{\rm LO}$~[GeV] & $\delta_{\alpha_s}$ (\%) & 
  $\delta_{\alpha}$ (\%) & $\delta_{m_f}$ (\%) & $\delta_{\Gamma_W}$ (\%) \\
\hline
   $W^{\pm}$ & $2.09241$ & $2.04808$ &  $2.55$ & $-0.364$ & $-0.0238$ & - \\
   $Z$ & $2.49785$ & $2.44591$ & $2.61$ & $-0.444$ & $-0.0374$ & - \\
  $t$ & $1.37398$ & $1.50450$	&  $-8.58$ & $1.68$ & $-0.239$ & $-1.54$ \\
  $H$ & $4.075\times 10^{-3}$ & - & - & - & - & -\\
\end{tabular}
\end{center}
\caption{\label{widthGmu}
As in table~\ref{widthMZ}, for the $\oG_\mu$ scheme.}
\end{table}
In the $\oG_\mu$ scheme, the same parameters as in tables~\ref{tab:Wmzinp}
and~\ref{tab:lepmass} are employed, with the exception of $\aem(m_Z)$, 
which is replaced by $G^{(G_\mu)}_\mu$. This in turn leads to a different 
value for the QED coupling constant, which we denote by $\aem_{\oG_\mu}$. 
The situation is summarised as follows:
\bqa
G^{(G_\mu)}_\mu=1.16639\cdot 10^{-5}\;\;\;\;
\longrightarrow\;\;\;\; 
\aem_{\oG_\mu}^{-1}=132.23\,.
\eqa
The corresponding results are presented in table~\ref{widthGmu}. 

We have checked numerically that the contribution of the channels 
$Z\to W^{\pm}f_1\bar{f}_2$ to the total $Z$ boson width is negligible
owing to phase-space suppression, in spite of being formally
of $\mathcal{O}(\alpha^2)$. As far as the top decay is concerned,
we see that $\delta_{\alpha}$ and $\delta_{\Gamma_W}$ give comparable
contributions (in absolute value), which thus must be both taken into 
account in order to have a sensible result at the NLO EW accuracy. 
Conversely, the impact of $\delta_{m_f}$ is significantly smaller
(for the top) or negligible (for the $W$ and the $Z$), and compatible 
with being thought as due to power corrections $(m_f/M_{t,W,Z})^2$.
Where possible, the results of \SMWidth\ have been cross-checked with 
those available in the literature~\cite{Jezabek:1993wk,Denner:1991kt}.
We also point out that \aNLOs\ is itself capable of computing the $W$,
$Z$, and top-quark total widths with the same accuracy as \SMWidth\
(obviously, with running times longer than those of the latter);
we have verified that the predictions of the two codes are in excellent
agreement with each other.

\phantomsection
\addcontentsline{toc}{section}{References}
\bibliographystyle{JHEP}
\bibliography{EWauto}

\providecommand{\href}[2]{#2}\begingroup\raggedright\begin{thebibliography}{10%
0}

\bibitem{Bendavid:2018nar}
J.~Bendavid et~al., \emph{{Les Houches 2017: Physics at TeV Colliders Standard
  Model Working Group Report}},  2018.
\newblock \href{http://arxiv.org/abs/1803.07977}{{\tt 1803.07977}}.

\bibitem{Kuroda:1990wn}
M.~Kuroda, G.~Moultaka and D.~Schildknecht, \emph{{Direct one loop
  renormalization of SU(2)-L x U(1)-Y four fermion processes and running
  coupling constants}},
  \href{http://dx.doi.org/10.1016/0550-3213(91)90252-S}{\emph{Nucl. Phys.} {\bf
  B350} (1991) 25--72}.

\bibitem{Degrassi:1992ue}
G.~Degrassi and A.~Sirlin, \emph{{Gauge invariant selfenergies and vertex parts
  of the Standard Model in the pinch technique framework}},
  \href{http://dx.doi.org/10.1103/PhysRevD.46.3104}{\emph{Phys. Rev.} {\bf D46}
  (1992) 3104--3116}.

\bibitem{Ciafaloni:1998xg}
P.~Ciafaloni and D.~Comelli, \emph{{Sudakov enhancement of electroweak
  corrections}},
  \href{http://dx.doi.org/10.1016/S0370-2693(98)01541-X}{\emph{Phys. Lett.}
  {\bf B446} (1999) 278--284}, [\href{http://arxiv.org/abs/hep-ph/9809321}{{\tt
  hep-ph/9809321}}].

\bibitem{Ciafaloni:2000df}
M.~Ciafaloni, P.~Ciafaloni and D.~Comelli, \emph{{Bloch-Nordsieck violating
  electroweak corrections to inclusive TeV scale hard processes}},
  \href{http://dx.doi.org/10.1103/PhysRevLett.84.4810}{\emph{Phys. Rev. Lett.}
  {\bf 84} (2000) 4810--4813}, [\href{http://arxiv.org/abs/hep-ph/0001142}{{\tt
  hep-ph/0001142}}].

\bibitem{Denner:2000jv}
A.~Denner and S.~Pozzorini, \emph{{One loop leading logarithms in electroweak
  radiative corrections. 1. Results}},
  \href{http://dx.doi.org/10.1007/s100520100551}{\emph{Eur. Phys. J.} {\bf C18}
  (2001) 461--480}, [\href{http://arxiv.org/abs/hep-ph/0010201}{{\tt
  hep-ph/0010201}}].

\bibitem{Denner:2001gw}
A.~Denner and S.~Pozzorini, \emph{{One loop leading logarithms in electroweak
  radiative corrections. 2. Factorization of collinear singularities}},
  \href{http://dx.doi.org/10.1007/s100520100721}{\emph{Eur. Phys. J.} {\bf C21}
  (2001) 63--79}, [\href{http://arxiv.org/abs/hep-ph/0104127}{{\tt
  hep-ph/0104127}}].

\bibitem{Actis:2016mpe}
S.~Actis, A.~Denner, L.~Hofer, J.-N. Lang, A.~Scharf and S.~Uccirati,
  \emph{{RECOLA: REcursive Computation of One-Loop Amplitudes}},
  \href{http://dx.doi.org/10.1016/j.cpc.2017.01.004}{\emph{Comput. Phys.
  Commun.} {\bf 214} (2017) 140--173},
  [\href{http://arxiv.org/abs/1605.01090}{{\tt 1605.01090}}].

\bibitem{Actis:2012qn}
S.~Actis, A.~Denner, L.~Hofer, A.~Scharf and S.~Uccirati, \emph{{Recursive
  generation of one-loop amplitudes in the Standard Model}},
  \href{http://dx.doi.org/10.1007/JHEP04(2013)037}{\emph{JHEP} {\bf 04} (2013)
  037}, [\href{http://arxiv.org/abs/1211.6316}{{\tt 1211.6316}}].

\bibitem{Gleisberg:2008ta}
T.~Gleisberg, S.~Hoeche, F.~Krauss, M.~Schonherr, S.~Schumann, F.~Siegert
  et~al., \emph{{Event generation with SHERPA 1.1}},
  \href{http://dx.doi.org/10.1088/1126-6708/2009/02/007}{\emph{JHEP} {\bf 02}
  (2009) 007}, [\href{http://arxiv.org/abs/0811.4622}{{\tt 0811.4622}}].

\bibitem{Schonherr:2017qcj}
M.~Schonherr, \emph{{An automated subtraction of NLO EW infrared divergences}},
  \href{http://dx.doi.org/10.1140/epjc/s10052-018-5600-z}{\emph{Eur. Phys. J.}
  {\bf C78} (2018) 119}, [\href{http://arxiv.org/abs/1712.07975}{{\tt
  1712.07975}}].

\bibitem{Cascioli:2011va}
F.~Cascioli, P.~Maierhofer and S.~Pozzorini, \emph{{Scattering Amplitudes with
  Open Loops}},
  \href{http://dx.doi.org/10.1103/PhysRevLett.108.111601}{\emph{Phys. Rev.
  Lett.} {\bf 108} (2012) 111601}, [\href{http://arxiv.org/abs/1111.5206}{{\tt
  1111.5206}}].

\bibitem{Cullen:2011ac}
G.~Cullen, N.~Greiner, G.~Heinrich, G.~Luisoni, P.~Mastrolia, G.~Ossola et~al.,
  \emph{{Automated One-Loop Calculations with GoSam}},
  \href{http://dx.doi.org/10.1140/epjc/s10052-012-1889-1}{\emph{Eur. Phys. J.}
  {\bf C72} (2012) 1889}, [\href{http://arxiv.org/abs/1111.2034}{{\tt
  1111.2034}}].

\bibitem{Cullen:2014yla}
G.~Cullen et~al., \emph{{G$\scriptsize{O}$S$\scriptsize{AM}$-2.0: a tool for
  automated one-loop calculations within the Standard Model and beyond}},
  \href{http://dx.doi.org/10.1140/epjc/s10052-014-3001-5}{\emph{Eur. Phys. J.}
  {\bf C74} (2014) 3001}, [\href{http://arxiv.org/abs/1404.7096}{{\tt
  1404.7096}}].

\bibitem{Frederix:2008hu}
R.~Frederix, T.~Gehrmann and N.~Greiner, \emph{{Automation of the Dipole
  Subtraction Method in MadGraph/MadEvent}},
  \href{http://dx.doi.org/10.1088/1126-6708/2008/09/122}{\emph{JHEP} {\bf 09}
  (2008) 122}, [\href{http://arxiv.org/abs/0808.2128}{{\tt 0808.2128}}].

\bibitem{Gehrmann:2010ry}
T.~Gehrmann and N.~Greiner, \emph{{Photon Radiation with MadDipole}},
  \href{http://dx.doi.org/10.1007/JHEP12(2010)050}{\emph{JHEP} {\bf 12} (2010)
  050}, [\href{http://arxiv.org/abs/1011.0321}{{\tt 1011.0321}}].

\bibitem{Alwall:2014hca}
J.~Alwall, R.~Frederix, S.~Frixione, V.~Hirschi, F.~Maltoni, O.~Mattelaer
  et~al., \emph{{The automated computation of tree-level and next-to-leading
  order differential cross sections, and their matching to parton shower
  simulations}}, \href{http://dx.doi.org/10.1007/JHEP07(2014)079}{\emph{JHEP}
  {\bf 07} (2014) 079}, [\href{http://arxiv.org/abs/1405.0301}{{\tt
  1405.0301}}].

\bibitem{Denner:2014ina}
A.~Denner, L.~Hofer, A.~Scharf and S.~Uccirati, \emph{{Electroweak corrections
  to lepton pair production in association with two hard jets at the LHC}},
  \href{http://dx.doi.org/10.1007/JHEP01(2015)094}{\emph{JHEP} {\bf 01} (2015)
  094}, [\href{http://arxiv.org/abs/1411.0916}{{\tt 1411.0916}}].

\bibitem{Denner:2014wka}
A.~Denner, R.~Feger and A.~Scharf, \emph{{Irreducible background and
  interference effects for Higgs-boson production in association with a
  top-quark pair}},
  \href{http://dx.doi.org/10.1007/JHEP04(2015)008}{\emph{JHEP} {\bf 04} (2015)
  008}, [\href{http://arxiv.org/abs/1412.5290}{{\tt 1412.5290}}].

\bibitem{Denner:2015yca}
A.~Denner and R.~Feger, \emph{{NLO QCD corrections to off-shell top-antitop
  production with leptonic decays in association with a Higgs boson at the
  LHC}}, \href{http://dx.doi.org/10.1007/JHEP11(2015)209}{\emph{JHEP} {\bf 11}
  (2015) 209}, [\href{http://arxiv.org/abs/1506.07448}{{\tt 1506.07448}}].

\bibitem{Kallweit:2014xda}
S.~Kallweit, J.~M. Lindert, P.~Maierhofer, S.~Pozzorini and M.~Schonherr,
  \emph{{NLO electroweak automation and precise predictions for W+multijet
  production at the LHC}},
  \href{http://dx.doi.org/10.1007/JHEP04(2015)012}{\emph{JHEP} {\bf 04} (2015)
  012}, [\href{http://arxiv.org/abs/1412.5157}{{\tt 1412.5157}}].

\bibitem{Frixione:2014qaa}
S.~Frixione, V.~Hirschi, D.~Pagani, H.~S. Shao and M.~Zaro, \emph{{Weak
  corrections to Higgs hadroproduction in association with a top-quark pair}},
  \href{http://dx.doi.org/10.1007/JHEP09(2014)065}{\emph{JHEP} {\bf 09} (2014)
  065}, [\href{http://arxiv.org/abs/1407.0823}{{\tt 1407.0823}}].

\bibitem{Chiesa:2015mya}
M.~Chiesa, N.~Greiner and F.~Tramontano, \emph{{Automation of electroweak
  corrections for LHC processes}},
  \href{http://dx.doi.org/10.1088/0954-3899/43/1/013002}{\emph{J. Phys.} {\bf
  G43} (2016) 013002}, [\href{http://arxiv.org/abs/1507.08579}{{\tt
  1507.08579}}].

\bibitem{Kallweit:2015dum}
S.~Kallweit, J.~M. Lindert, P.~Maierhofer, S.~Pozzorini and M.~Schonherr,
  \emph{{NLO QCD+EW predictions for V + jets including off-shell vector-boson
  decays and multijet merging}},
  \href{http://dx.doi.org/10.1007/JHEP04(2016)021}{\emph{JHEP} {\bf 04} (2016)
  021}, [\href{http://arxiv.org/abs/1511.08692}{{\tt 1511.08692}}].

\bibitem{Frixione:2015zaa}
S.~Frixione, V.~Hirschi, D.~Pagani, H.~S. Shao and M.~Zaro, \emph{{Electroweak
  and QCD corrections to top-pair hadroproduction in association with heavy
  bosons}}, \href{http://dx.doi.org/10.1007/JHEP06(2015)184}{\emph{JHEP} {\bf
  06} (2015) 184}, [\href{http://arxiv.org/abs/1504.03446}{{\tt 1504.03446}}].

\bibitem{Biedermann:2016guo}
B.~Biedermann, M.~Billoni, A.~Denner, S.~Dittmaier, L.~Hofer, B.~Jaeger et~al.,
  \emph{{Next-to-leading-order electroweak corrections to $pp \to W^+W^-\to$ 4
  leptons at the LHC}},
  \href{http://dx.doi.org/10.1007/JHEP06(2016)065}{\emph{JHEP} {\bf 06} (2016)
  065}, [\href{http://arxiv.org/abs/1605.03419}{{\tt 1605.03419}}].

\bibitem{Biedermann:2016yvs}
B.~Biedermann, A.~Denner, S.~Dittmaier, L.~Hofer and B.~Jaeger,
  \emph{{Electroweak corrections to $pp \to \mu^+\mu^-e^+e^- + X$ at the LHC: a
  Higgs background study}},
  \href{http://dx.doi.org/10.1103/PhysRevLett.116.161803}{\emph{Phys. Rev.
  Lett.} {\bf 116} (2016) 161803}, [\href{http://arxiv.org/abs/1601.07787}{{\tt
  1601.07787}}].

\bibitem{Denner:2016jyo}
A.~Denner and M.~Pellen, \emph{{NLO electroweak corrections to off-shell
  top-antitop production with leptonic decays at the LHC}},
  \href{http://dx.doi.org/10.1007/JHEP08(2016)155}{\emph{JHEP} {\bf 08} (2016)
  155}, [\href{http://arxiv.org/abs/1607.05571}{{\tt 1607.05571}}].

\bibitem{Biedermann:2016yds}
B.~Biedermann, A.~Denner and M.~Pellen, \emph{{Large electroweak corrections to
  vector-boson scattering at the Large Hadron Collider}},
  \href{http://dx.doi.org/10.1103/PhysRevLett.118.261801}{\emph{Phys. Rev.
  Lett.} {\bf 118} (2017) 261801}, [\href{http://arxiv.org/abs/1611.02951}{{\tt
  1611.02951}}].

\bibitem{Biedermann:2016lvg}
B.~Biedermann, A.~Denner, S.~Dittmaier, L.~Hofer and B.~Jager,
  \emph{{Next-to-leading-order electroweak corrections to the production of
  four charged leptons at the LHC}},
  \href{http://dx.doi.org/10.1007/JHEP01(2017)033}{\emph{JHEP} {\bf 01} (2017)
  033}, [\href{http://arxiv.org/abs/1611.05338}{{\tt 1611.05338}}].

\bibitem{Denner:2016wet}
A.~Denner, J.-N. Lang, M.~Pellen and S.~Uccirati, \emph{{Higgs production in
  association with off-shell top-antitop pairs at NLO EW and QCD at the LHC}},
  \href{http://dx.doi.org/10.1007/JHEP02(2017)053}{\emph{JHEP} {\bf 02} (2017)
  053}, [\href{http://arxiv.org/abs/1612.07138}{{\tt 1612.07138}}].

\bibitem{Frederix:2016ost}
R.~Frederix, S.~Frixione, V.~Hirschi, D.~Pagani, H.-S. Shao and M.~Zaro,
  \emph{{The complete NLO corrections to dijet hadroproduction}},
  \href{http://dx.doi.org/10.1007/JHEP04(2017)076}{\emph{JHEP} {\bf 04} (2017)
  076}, [\href{http://arxiv.org/abs/1612.06548}{{\tt 1612.06548}}].

\bibitem{Pagani:2016caq}
D.~Pagani, I.~Tsinikos and M.~Zaro, \emph{{The impact of the photon PDF and
  electroweak corrections on $t \bar{t}$ distributions}},
  \href{http://dx.doi.org/10.1140/epjc/s10052-016-4318-z}{\emph{Eur. Phys. J.}
  {\bf C76} (2016) 479}, [\href{http://arxiv.org/abs/1606.01915}{{\tt
  1606.01915}}].

\bibitem{Biedermann:2017yoi}
B.~Biedermann, S.~Brauer, A.~Denner, M.~Pellen, S.~Schumann and J.~M. Thompson,
  \emph{{Automation of NLO QCD and EW corrections with Sherpa and Recola}},
  \href{http://dx.doi.org/10.1140/epjc/s10052-017-5054-8}{\emph{Eur. Phys. J.}
  {\bf C77} (2017) 492}, [\href{http://arxiv.org/abs/1704.05783}{{\tt
  1704.05783}}].

\bibitem{Kallweit:2017khh}
S.~Kallweit, J.~M. Lindert, S.~Pozzorini and M.~Schonherr, \emph{{NLO QCD+EW
  predictions for $2\ell2\nu$ diboson signatures at the LHC}},
  \href{http://arxiv.org/abs/1705.00598}{{\tt 1705.00598}}.

\bibitem{Biedermann:2017bss}
B.~Biedermann, A.~Denner and M.~Pellen, \emph{{Complete NLO corrections to
  ${\rm W}^+{\rm W}^+$ scattering and its irreducible background at the LHC}},
  \href{http://arxiv.org/abs/1708.00268}{{\tt 1708.00268}}.

\bibitem{Biedermann:2017oae}
B.~Biedermann, A.~Denner and L.~Hofer, \emph{{Next-to-leading-order electroweak
  corrections to the production of three charged leptons plus missing energy at
  the LHC}},  \href{http://arxiv.org/abs/1708.06938}{{\tt 1708.06938}}.

\bibitem{Chiesa:2017gqx}
M.~Chiesa, N.~Greiner, M.~Schoenherr and F.~Tramontano, \emph{{Electroweak
  corrections to diphoton plus jets}},
  \href{http://arxiv.org/abs/1706.09022}{{\tt 1706.09022}}.

\bibitem{Czakon:2017wor}
M.~Czakon, D.~Heymes, A.~Mitov, D.~Pagani, I.~Tsinikos and M.~Zaro,
  \emph{{Top-pair production at the LHC through NNLO QCD and NLO EW}},
  \href{http://arxiv.org/abs/1705.04105}{{\tt 1705.04105}}.

\bibitem{Frederix:2017wme}
R.~Frederix, D.~Pagani and M.~Zaro, \emph{{Large NLO corrections in
  $t\bar{t}W^{\pm}$ and $t\bar{t}t\bar{t}$ hadroproduction from supposedly
  subleading EW contributions}},
  \href{http://dx.doi.org/10.1007/JHEP02(2018)031}{\emph{JHEP} {\bf 02} (2018)
  031}, [\href{http://arxiv.org/abs/1711.02116}{{\tt 1711.02116}}].

\bibitem{Gutschow:2018tuk}
C.~Gtschow, J.~M. Lindert and M.~Schonherr, \emph{{Multi-jet merged top-pair
  production including electroweak corrections}},
  \href{http://arxiv.org/abs/1803.00950}{{\tt 1803.00950}}.

\bibitem{Czakon:2017lgo}
M.~Czakon, D.~Heymes, A.~Mitov, D.~Pagani, I.~Tsinikos and M.~Zaro, \emph{{The
  top-quark charge asymmetry at the LHC and Tevatron through NNLO QCD and NLO
  EW}},  \href{http://arxiv.org/abs/1711.03945}{{\tt 1711.03945}}.

\bibitem{Maltoni:2017ims}
F.~Maltoni, D.~Pagani, A.~Shivaji and X.~Zhao, \emph{{Trilinear Higgs coupling
  determination via single-Higgs differential measurements at the LHC}},
  \href{http://dx.doi.org/10.1140/epjc/s10052-017-5410-8}{\emph{Eur. Phys. J.}
  {\bf C77} (2017) 887}, [\href{http://arxiv.org/abs/1709.08649}{{\tt
  1709.08649}}].

\bibitem{Plehn:2015cta}
M.~L. Mangano, T.~Plehn, P.~Reimitz, T.~Schell and H.-S. Shao, \emph{{Measuring
  the Top Yukawa Coupling at 100 TeV}},
  \href{http://dx.doi.org/10.1088/0954-3899/43/3/035001}{\emph{J. Phys.} {\bf
  G43} (2016) 035001}, [\href{http://arxiv.org/abs/1507.08169}{{\tt
  1507.08169}}].

\bibitem{Hirschi:2015iia}
V.~Hirschi and O.~Mattelaer, \emph{{Automated event generation for loop-induced
  processes}}, \href{http://dx.doi.org/10.1007/JHEP10(2015)146}{\emph{JHEP}
  {\bf 10} (2015) 146}, [\href{http://arxiv.org/abs/1507.00020}{{\tt
  1507.00020}}].

\bibitem{Frixione:1995ms}
S.~Frixione, Z.~Kunszt and A.~Signer, \emph{{Three jet cross-sections to
  next-to-leading order}},
  \href{http://dx.doi.org/10.1016/0550-3213(96)00110-1}{\emph{Nucl. Phys.} {\bf
  B467} (1996) 399--442}, [\href{http://arxiv.org/abs/hep-ph/9512328}{{\tt
  hep-ph/9512328}}].

\bibitem{Frixione:1997np}
S.~Frixione, \emph{{A General approach to jet cross-sections in QCD}},
  \href{http://dx.doi.org/10.1016/S0550-3213(97)00574-9}{\emph{Nucl. Phys.}
  {\bf B507} (1997) 295--314}, [\href{http://arxiv.org/abs/hep-ph/9706545}{{\tt
  hep-ph/9706545}}].

\bibitem{Denner:1999gp}
A.~Denner, S.~Dittmaier, M.~Roth and D.~Wackeroth, \emph{{Predictions for all
  processes e+ e- $\to$ 4 fermions + gamma}},
  \href{http://dx.doi.org/10.1016/S0550-3213(99)00437-X}{\emph{Nucl. Phys.}
  {\bf B560} (1999) 33--65}, [\href{http://arxiv.org/abs/hep-ph/9904472}{{\tt
  hep-ph/9904472}}].

\bibitem{Denner:2005fg}
A.~Denner, S.~Dittmaier, M.~Roth and L.~H. Wieders, \emph{{Electroweak
  corrections to charged-current e+ e- $\to$ 4 fermion processes: Technical
  details and further results}},
  \href{http://dx.doi.org/10.1016/j.nuclphysb.2011.09.001,
  10.1016/j.nuclphysb.2005.06.033}{\emph{Nucl. Phys.} {\bf B724} (2005)
  247--294}, [\href{http://arxiv.org/abs/hep-ph/0505042}{{\tt
  hep-ph/0505042}}].

\bibitem{Frederix:2009yq}
R.~Frederix, S.~Frixione, F.~Maltoni and T.~Stelzer, \emph{{Automation of
  next-to-leading order computations in QCD: The FKS subtraction}},
  \href{http://dx.doi.org/10.1088/1126-6708/2009/10/003}{\emph{JHEP} {\bf 10}
  (2009) 003}, [\href{http://arxiv.org/abs/0908.4272}{{\tt 0908.4272}}].

\bibitem{Frederix:2016rdc}
R.~Frederix, S.~Frixione, A.~S. Papanastasiou, S.~Prestel and P.~Torrielli,
  \emph{{Off-shell single-top production at NLO matched to parton showers}},
  \href{http://dx.doi.org/10.1007/JHEP06(2016)027}{\emph{JHEP} {\bf 06} (2016)
  027}, [\href{http://arxiv.org/abs/1603.01178}{{\tt 1603.01178}}].

\bibitem{Ossola:2006us}
G.~Ossola, C.~G. Papadopoulos and R.~Pittau, \emph{{Reducing full one-loop
  amplitudes to scalar integrals at the integrand level}},
  \href{http://dx.doi.org/10.1016/j.nuclphysb.2006.11.012}{\emph{Nucl. Phys.}
  {\bf B763} (2007) 147--169}, [\href{http://arxiv.org/abs/hep-ph/0609007}{{\tt
  hep-ph/0609007}}].

\bibitem{Mastrolia:2012bu}
P.~Mastrolia, E.~Mirabella and T.~Peraro, \emph{{Integrand reduction of
  one-loop scattering amplitudes through Laurent series expansion}},
  \href{http://dx.doi.org/10.1007/JHEP11(2012)128,
  10.1007/JHEP06(2012)095}{\emph{JHEP} {\bf 06} (2012) 095},
  [\href{http://arxiv.org/abs/1203.0291}{{\tt 1203.0291}}].

\bibitem{Passarino:1978jh}
G.~Passarino and M.~Veltman, \emph{{One Loop Corrections for $e^+e^-$
  Annihilation Into $\mu^+ \mu^-$ in the Weinberg Model}},
  \href{http://dx.doi.org/10.1016/0550-3213(79)90234-7}{\emph{Nucl.Phys.} {\bf
  B160} (1979) 151}.

\bibitem{Davydychev:1991va}
A.~I. Davydychev, \emph{{A Simple formula for reducing Feynman diagrams to
  scalar integrals}},
  \href{http://dx.doi.org/10.1016/0370-2693(91)91715-8}{\emph{Phys.Lett.} {\bf
  B263} (1991) 107--111}.

\bibitem{Denner:2005nn}
A.~Denner and S.~Dittmaier, \emph{{Reduction schemes for one-loop tensor
  integrals}},
  \href{http://dx.doi.org/10.1016/j.nuclphysb.2005.11.007}{\emph{Nucl. Phys.}
  {\bf B734} (2006) 62--115}, [\href{http://arxiv.org/abs/hep-ph/0509141}{{\tt
  hep-ph/0509141}}].

\bibitem{Hirschi:2011pa}
V.~Hirschi, R.~Frederix, S.~Frixione, M.~V. Garzelli, F.~Maltoni et~al.,
  \emph{{Automation of one-loop QCD corrections}},
  \href{http://dx.doi.org/10.1007/JHEP05(2011)044}{\emph{JHEP} {\bf 1105}
  (2011) 044}, [\href{http://arxiv.org/abs/1103.0621}{{\tt 1103.0621}}].

\bibitem{Ossola:2007ax}
G.~Ossola, C.~G. Papadopoulos and R.~Pittau, \emph{{CutTools: A Program
  implementing the OPP reduction method to compute one-loop amplitudes}},
  \href{http://dx.doi.org/10.1088/1126-6708/2008/03/042}{\emph{JHEP} {\bf 03}
  (2008) 042}, [\href{http://arxiv.org/abs/0711.3596}{{\tt 0711.3596}}].

\bibitem{Peraro:2014cba}
T.~Peraro, \emph{{Ninja: Automated Integrand Reduction via Laurent Expansion
  for One-Loop Amplitudes}},
  \href{http://dx.doi.org/10.1016/j.cpc.2014.06.017}{\emph{Comput. Phys.
  Commun.} {\bf 185} (2014) 2771--2797},
  [\href{http://arxiv.org/abs/1403.1229}{{\tt 1403.1229}}].

\bibitem{Hirschi:2016mdz}
V.~Hirschi and T.~Peraro, \emph{{Tensor integrand reduction via Laurent
  expansion}}, \href{http://dx.doi.org/10.1007/JHEP06(2016)060}{\emph{JHEP}
  {\bf 06} (2016) 060}, [\href{http://arxiv.org/abs/1604.01363}{{\tt
  1604.01363}}].

\bibitem{ShaoIREGI}
H.-S. Shao, \emph{Iregi user manual, unpublished}, .

\bibitem{Denner:2016kdg}
A.~Denner, S.~Dittmaier and L.~Hofer, \emph{{Collier: a fortran-based Complex
  One-Loop LIbrary in Extended Regularizations}},
  \href{http://dx.doi.org/10.1016/j.cpc.2016.10.013}{\emph{Comput. Phys.
  Commun.} {\bf 212} (2017) 220--238},
  [\href{http://arxiv.org/abs/1604.06792}{{\tt 1604.06792}}].

\bibitem{Frixione:2002ik}
S.~Frixione and B.~R. Webber, \emph{{Matching NLO QCD computations and parton
  shower simulations}},
  \href{http://dx.doi.org/10.1088/1126-6708/2002/06/029}{\emph{JHEP} {\bf 06}
  (2002) 029}, [\href{http://arxiv.org/abs/hep-ph/0204244}{{\tt
  hep-ph/0204244}}].

\bibitem{Christensen:2008py}
N.~D. Christensen and C.~Duhr, \emph{{FeynRules - Feynman rules made easy}},
  \href{http://dx.doi.org/10.1016/j.cpc.2009.02.018}{\emph{Comput.Phys.Commun.}
  {\bf 180} (2009) 1614--1641}, [\href{http://arxiv.org/abs/0806.4194}{{\tt
  0806.4194}}].

\bibitem{Christensen:2009jx}
N.~D. Christensen, P.~de~Aquino, C.~Degrande, C.~Duhr, B.~Fuks et~al., \emph{{A
  Comprehensive approach to new physics simulations}},
  \href{http://dx.doi.org/10.1140/epjc/s10052-011-1541-5}{\emph{Eur.Phys.J.}
  {\bf C71} (2011) 1541}, [\href{http://arxiv.org/abs/0906.2474}{{\tt
  0906.2474}}].

\bibitem{Christensen:2010wz}
N.~D. Christensen, C.~Duhr, B.~Fuks, J.~Reuter and C.~Speckner,
  \emph{{Introducing an interface between WHIZARD and FeynRules}},
  \href{http://dx.doi.org/10.1140/epjc/s10052-012-1990-5}{\emph{Eur.Phys.J.}
  {\bf C72} (2012) 1990}, [\href{http://arxiv.org/abs/1010.3251}{{\tt
  1010.3251}}].

\bibitem{Duhr:2011se}
C.~Duhr and B.~Fuks, \emph{{A superspace module for the FeynRules package}},
  \href{http://dx.doi.org/10.1016/j.cpc.2011.06.009}{\emph{Comput.Phys.Commun.}
  {\bf 182} (2011) 2404--2426}, [\href{http://arxiv.org/abs/1102.4191}{{\tt
  1102.4191}}].

\bibitem{Alloul:2013bka}
A.~Alloul, N.~D. Christensen, C.~Degrande, C.~Duhr and B.~Fuks,
  \emph{{FeynRules 2.0 - A complete toolbox for tree-level phenomenology}},
  \href{http://arxiv.org/abs/1310.1921}{{\tt 1310.1921}}.

\bibitem{Alloul:2013fw}
A.~Alloul, J.~D'Hondt, K.~De~Causmaecker, B.~Fuks and M.~Rausch~de Traubenberg,
  \emph{{Automated mass spectrum generation for new physics}},
  \href{http://dx.doi.org/10.1140/epjc/s10052-013-2325-x}{\emph{Eur.Phys.J.}
  {\bf C73} (2013) 2325}, [\href{http://arxiv.org/abs/1301.5932}{{\tt
  1301.5932}}].

\bibitem{Degrande:2014vpa}
C.~Degrande, \emph{{Automatic evaluation of UV and R2 terms for beyond the
  Standard Model Lagrangians: a proof-of-principle}},
  \href{http://dx.doi.org/10.1016/j.cpc.2015.08.015}{\emph{Comput. Phys.
  Commun.} {\bf 197} (2015) 239--262},
  [\href{http://arxiv.org/abs/1406.3030}{{\tt 1406.3030}}].

\bibitem{Harland-Lang:2016lhw}
L.~A. Harland-Lang, V.~A. Khoze and M.~G. Ryskin, \emph{{Sudakov effects in
  photon-initiated processes}},
  \href{http://dx.doi.org/10.1016/j.physletb.2016.08.004}{\emph{Phys. Lett.}
  {\bf B761} (2016) 20--24}, [\href{http://arxiv.org/abs/1605.04935}{{\tt
  1605.04935}}].

\bibitem{Mele:1990cw}
B.~Mele and P.~Nason, \emph{{The Fragmentation function for heavy quarks in
  QCD}}, \href{http://dx.doi.org/10.1016/0550-3213(91)90597-Q,
  10.1016/j.nuclphysb.2017.05.005}{\emph{Nucl. Phys.} {\bf B361} (1991)
  626--644}.

\bibitem{Sjostrand:2014zea}
T.~Sj{\"o}strand, S.~Ask, J.~R. Christiansen, R.~Corke, N.~Desai, P.~Ilten
  et~al., \emph{{An Introduction to PYTHIA 8.2}},
  \href{http://dx.doi.org/10.1016/j.cpc.2015.01.024}{\emph{Comput. Phys.
  Commun.} {\bf 191} (2015) 159--177},
  [\href{http://arxiv.org/abs/1410.3012}{{\tt 1410.3012}}].

\bibitem{Bellm:2015jjp}
J.~Bellm et~al., \emph{{Herwig 7.0/Herwig++ 3.0 release note}},
  \href{http://dx.doi.org/10.1140/epjc/s10052-016-4018-8}{\emph{Eur. Phys. J.}
  {\bf C76} (2016) 196}, [\href{http://arxiv.org/abs/1512.01178}{{\tt
  1512.01178}}].

\bibitem{Frixione:2011kh}
S.~Frixione, \emph{{Colourful FKS subtraction}},
  \href{http://dx.doi.org/10.1007/JHEP09(2011)091}{\emph{JHEP} {\bf 09} (2011)
  091}, [\href{http://arxiv.org/abs/1106.0155}{{\tt 1106.0155}}].

\bibitem{Frederix:2011ss}
R.~Frederix, S.~Frixione, V.~Hirschi, F.~Maltoni, R.~Pittau and P.~Torrielli,
  \emph{{Four-lepton production at hadron colliders: aMC@NLO predictions with
  theoretical uncertainties}},
  \href{http://dx.doi.org/10.1007/JHEP02(2012)099}{\emph{JHEP} {\bf 02} (2012)
  099}, [\href{http://arxiv.org/abs/1110.4738}{{\tt 1110.4738}}].

\bibitem{Ellis:1979sj}
R.~K. Ellis, M.~A. Furman, H.~E. Haber and I.~Hinchliffe, \emph{{Large
  Corrections to High p(T) Hadron-Hadron Scattering in QCD}},
  \href{http://dx.doi.org/10.1016/0550-3213(80)90010-3}{\emph{Nucl. Phys.} {\bf
  B173} (1980) 397--421}.

\bibitem{Aversa:1988vb}
F.~Aversa, P.~Chiappetta, M.~Greco and J.~P. Guillet, \emph{{QCD Corrections to
  Parton-Parton Scattering Processes}},
  \href{http://dx.doi.org/10.1016/0550-3213(89)90288-5}{\emph{Nucl. Phys.} {\bf
  B327} (1989) 105}.

\bibitem{Stuart:1991xk}
R.~G. Stuart, \emph{{Gauge invariance, analyticity and physical observables at
  the Z0 resonance}},
  \href{http://dx.doi.org/10.1016/0370-2693(91)90653-8}{\emph{Phys. Lett.} {\bf
  B262} (1991) 113--119}.

\bibitem{Stuart:1991cc}
R.~G. Stuart, \emph{{General renormalization of the gauge invariant
  perturbation expansion near the Z0 resonance}},
  \href{http://dx.doi.org/10.1016/0370-2693(91)91842-J}{\emph{Phys. Lett.} {\bf
  B272} (1991) 353--358}.

\bibitem{Sirlin:1991rt}
A.~Sirlin, \emph{{Observations concerning mass renormalization in the
  electroweak theory}},
  \href{http://dx.doi.org/10.1016/0370-2693(91)91254-S}{\emph{Phys. Lett.} {\bf
  B267} (1991) 240--242}.

\bibitem{Sirlin:1991fd}
A.~Sirlin, \emph{{Theoretical considerations concerning the Z0 mass}},
  \href{http://dx.doi.org/10.1103/PhysRevLett.67.2127}{\emph{Phys. Rev. Lett.}
  {\bf 67} (1991) 2127--2130}.

\bibitem{Sterman:1994ce}
G.~F. Sterman, \emph{{An Introduction to quantum field theory}}.
\newblock Cambridge University Press, 1993.

\bibitem{Aeppli:1993cb}
A.~Aeppli, F.~Cuypers and G.~J. van Oldenborgh, \emph{{O(Gamma) corrections to
  W pair production in e+ e- and gamma gamma collisions}},
  \href{http://dx.doi.org/10.1016/0370-2693(93)91259-P}{\emph{Phys. Lett.} {\bf
  B314} (1993) 413--420}, [\href{http://arxiv.org/abs/hep-ph/9303236}{{\tt
  hep-ph/9303236}}].

\bibitem{Aeppli:1993rs}
A.~Aeppli, G.~J. van Oldenborgh and D.~Wyler, \emph{{Unstable particles in one
  loop calculations}},
  \href{http://dx.doi.org/10.1016/0550-3213(94)90195-3}{\emph{Nucl. Phys.} {\bf
  B428} (1994) 126--146}, [\href{http://arxiv.org/abs/hep-ph/9312212}{{\tt
  hep-ph/9312212}}].

\bibitem{Chapovsky:2001zt}
A.~P. Chapovsky, V.~A. Khoze, A.~Signer and W.~J. Stirling,
  \emph{{Nonfactorizable corrections and effective field theories}},
  \href{http://dx.doi.org/10.1016/S0550-3213(01)00577-6}{\emph{Nucl. Phys.}
  {\bf B621} (2002) 257--302}, [\href{http://arxiv.org/abs/hep-ph/0108190}{{\tt
  hep-ph/0108190}}].

\bibitem{Beneke:2003xh}
M.~Beneke, A.~P. Chapovsky, A.~Signer and G.~Zanderighi, \emph{{Effective
  theory approach to unstable particle production}},
  \href{http://dx.doi.org/10.1103/PhysRevLett.93.011602}{\emph{Phys. Rev.
  Lett.} {\bf 93} (2004) 011602},
  [\href{http://arxiv.org/abs/hep-ph/0312331}{{\tt hep-ph/0312331}}].

\bibitem{Beneke:2004km}
M.~Beneke, A.~P. Chapovsky, A.~Signer and G.~Zanderighi, \emph{{Effective
  theory calculation of resonant high-energy scattering}},
  \href{http://dx.doi.org/10.1016/j.nuclphysb.2004.03.016}{\emph{Nucl. Phys.}
  {\bf B686} (2004) 205--247}, [\href{http://arxiv.org/abs/hep-ph/0401002}{{\tt
  hep-ph/0401002}}].

\bibitem{Denner:1991kt}
A.~Denner, \emph{{Techniques for calculation of electroweak radiative
  corrections at the one loop level and results for W physics at LEP-200}},
  \href{http://dx.doi.org/10.1002/prop.2190410402}{\emph{Fortsch. Phys.} {\bf
  41} (1993) 307--420}, [\href{http://arxiv.org/abs/0709.1075}{{\tt
  0709.1075}}].

\bibitem{Bauer:2012gn}
T.~Bauer, J.~Gegelia, G.~Japaridze and S.~Scherer, \emph{{Complex-mass scheme
  and perturbative unitarity}},
  \href{http://dx.doi.org/10.1142/S0217751X12501783}{\emph{Int. J. Mod. Phys.}
  {\bf A27} (2012) 1250178}, [\href{http://arxiv.org/abs/1211.1684}{{\tt
  1211.1684}}].

\bibitem{Denner:2014zga}
A.~Denner and J.-N. Lang, \emph{{The Complex-Mass Scheme and Unitarity in
  perturbative Quantum Field Theory}},
  \href{http://dx.doi.org/10.1140/epjc/s10052-015-3579-2}{\emph{Eur. Phys. J.}
  {\bf C75} (2015) 377}, [\href{http://arxiv.org/abs/1406.6280}{{\tt
  1406.6280}}].

\bibitem{tHooft:1978jhc}
G.~'t~Hooft and M.~J.~G. Veltman, \emph{{Scalar One Loop Integrals}},
  \href{http://dx.doi.org/10.1016/0550-3213(79)90605-9}{\emph{Nucl. Phys.} {\bf
  B153} (1979) 365--401}.

\bibitem{Ellis:2007qk}
R.~K. Ellis and G.~Zanderighi, \emph{{Scalar one-loop integrals for QCD}},
  \href{http://dx.doi.org/10.1088/1126-6708/2008/02/002}{\emph{JHEP} {\bf 0802}
  (2008) 002}, [\href{http://arxiv.org/abs/0712.1851}{{\tt 0712.1851}}].

\bibitem{Passarino:2010qk}
G.~Passarino, C.~Sturm and S.~Uccirati, \emph{{Higgs Pseudo-Observables, Second
  Riemann Sheet and All That}},
  \href{http://dx.doi.org/10.1016/j.nuclphysb.2010.03.013}{\emph{Nucl. Phys.}
  {\bf B834} (2010) 77--115}, [\href{http://arxiv.org/abs/1001.3360}{{\tt
  1001.3360}}].

\bibitem{Veltman:1963th}
M.~J.~G. Veltman, \emph{{Unitarity and causality in a renormalizable field
  theory with unstable particles}},
  \href{http://dx.doi.org/10.1016/S0031-8914(63)80277-3}{\emph{Physica} {\bf
  29} (1963) 186--207}.

\bibitem{Ciafaloni:2006qu}
P.~Ciafaloni and D.~Comelli, \emph{{The Importance of weak bosons emission at
  LHC}}, \href{http://dx.doi.org/10.1088/1126-6708/2006/09/055}{\emph{JHEP}
  {\bf 09} (2006) 055}, [\href{http://arxiv.org/abs/hep-ph/0604070}{{\tt
  hep-ph/0604070}}].

\bibitem{Baur:2006sn}
U.~Baur, \emph{{Weak Boson Emission in Hadron Collider Processes}},
  \href{http://dx.doi.org/10.1103/PhysRevD.75.013005}{\emph{Phys. Rev.} {\bf
  D75} (2007) 013005}, [\href{http://arxiv.org/abs/hep-ph/0611241}{{\tt
  hep-ph/0611241}}].

\bibitem{Bell:2010gi}
G.~Bell, J.~H. Kuhn and J.~Rittinger, \emph{{Electroweak Sudakov Logarithms and
  Real Gauge-Boson Radiation in the TeV Region}},
  \href{http://dx.doi.org/10.1140/epjc/s10052-010-1489-x}{\emph{Eur. Phys. J.}
  {\bf C70} (2010) 659--671}, [\href{http://arxiv.org/abs/1004.4117}{{\tt
  1004.4117}}].

\bibitem{Stirling:2012ak}
W.~J. Stirling and E.~Vryonidou, \emph{{Electroweak corrections and
  Bloch-Nordsieck violations in 2-to-2 processes at the LHC}},
  \href{http://dx.doi.org/10.1007/JHEP04(2013)155}{\emph{JHEP} {\bf 04} (2013)
  155}, [\href{http://arxiv.org/abs/1212.6537}{{\tt 1212.6537}}].

\bibitem{Chiesa:2013yma}
M.~Chiesa, G.~Montagna, L.~Barze, M.~Moretti, O.~Nicrosini, F.~Piccinini
  et~al., \emph{{Electroweak Sudakov Corrections to New Physics Searches at the
  LHC}}, \href{http://dx.doi.org/10.1103/PhysRevLett.111.121801}{\emph{Phys.
  Rev. Lett.} {\bf 111} (2013) 121801},
  [\href{http://arxiv.org/abs/1305.6837}{{\tt 1305.6837}}].

\bibitem{Christiansen:2014kba}
J.~R. Christiansen and T.~Sj{\"o}strand, \emph{{Weak Gauge Boson Radiation in
  Parton Showers}},
  \href{http://dx.doi.org/10.1007/JHEP04(2014)115}{\emph{JHEP} {\bf 04} (2014)
  115}, [\href{http://arxiv.org/abs/1401.5238}{{\tt 1401.5238}}].

\bibitem{Krauss:2014yaa}
F.~Krauss, P.~Petrov, M.~Schoenherr and M.~Spannowsky, \emph{{Measuring
  collinear W emissions inside jets}},
  \href{http://dx.doi.org/10.1103/PhysRevD.89.114006}{\emph{Phys. Rev.} {\bf
  D89} (2014) 114006}, [\href{http://arxiv.org/abs/1403.4788}{{\tt
  1403.4788}}].

\bibitem{Bauer:2016kkv}
C.~W. Bauer and N.~Ferland, \emph{{Resummation of electroweak Sudakov
  logarithms for real radiation}},
  \href{http://dx.doi.org/10.1007/JHEP09(2016)025}{\emph{JHEP} {\bf 09} (2016)
  025}, [\href{http://arxiv.org/abs/1601.07190}{{\tt 1601.07190}}].

\bibitem{Degrande:2011ua}
C.~Degrande, C.~Duhr, B.~Fuks, D.~Grellscheid, O.~Mattelaer et~al., \emph{{UFO
  - The Universal FeynRules Output}},
  \href{http://dx.doi.org/10.1016/j.cpc.2012.01.022}{\emph{Comput.Phys.Commun.}
  {\bf 183} (2012) 1201--1214}, [\href{http://arxiv.org/abs/1108.2040}{{\tt
  1108.2040}}].

\bibitem{Patrignani:2016xqp}
{\scshape Particle Data Group} collaboration, C.~Patrignani et~al.,
  \emph{{Review of Particle Physics}},
  \href{http://dx.doi.org/10.1088/1674-1137/40/10/100001}{\emph{Chin. Phys.}
  {\bf C40} (2016) 100001}.

\bibitem{Kniehl:1998fn}
B.~A. Kniehl and A.~Sirlin, \emph{{Differences between the pole and on-shell
  masses and widths of the Higgs boson}},
  \href{http://dx.doi.org/10.1103/PhysRevLett.81.1373}{\emph{Phys. Rev. Lett.}
  {\bf 81} (1998) 1373--1376}, [\href{http://arxiv.org/abs/hep-ph/9805390}{{\tt
  hep-ph/9805390}}].

\bibitem{Kniehl:2002wn}
B.~A. Kniehl, C.~P. Palisoc and A.~Sirlin, \emph{{Elimination of threshold
  singularities in the relation between on shell and pole widths}},
  \href{http://dx.doi.org/10.1103/PhysRevD.66.057902}{\emph{Phys. Rev.} {\bf
  D66} (2002) 057902}, [\href{http://arxiv.org/abs/hep-ph/0205304}{{\tt
  hep-ph/0205304}}].

\bibitem{Grassi:2001bz}
P.~A. Grassi, B.~A. Kniehl and A.~Sirlin, \emph{{Width and partial widths of
  unstable particles in the light of the Nielsen identities}},
  \href{http://dx.doi.org/10.1103/PhysRevD.65.085001}{\emph{Phys. Rev.} {\bf
  D65} (2002) 085001}, [\href{http://arxiv.org/abs/hep-ph/0109228}{{\tt
  hep-ph/0109228}}].

\bibitem{Butterworth:2015oua}
J.~Butterworth et~al., \emph{{PDF4LHC recommendations for LHC Run II}},
  \href{http://dx.doi.org/10.1088/0954-3899/43/2/023001}{\emph{J. Phys.} {\bf
  G43} (2016) 023001}, [\href{http://arxiv.org/abs/1510.03865}{{\tt
  1510.03865}}].

\bibitem{Manohar:2016nzj}
A.~Manohar, P.~Nason, G.~P. Salam and G.~Zanderighi, \emph{{How bright is the
  proton? A precise determination of the photon parton distribution function}},
  \href{http://dx.doi.org/10.1103/PhysRevLett.117.242002}{\emph{Phys. Rev.
  Lett.} {\bf 117} (2016) 242002}, [\href{http://arxiv.org/abs/1607.04266}{{\tt
  1607.04266}}].

\bibitem{Buckley:2014ana}
A.~Buckley, J.~Ferrando, S.~Lloyd, K.~Nordstrom, B.~Page, M.~Rafenacht et~al.,
  \emph{{LHAPDF6: parton density access in the LHC precision era}},
  \href{http://dx.doi.org/10.1140/epjc/s10052-015-3318-8}{\emph{Eur. Phys. J.}
  {\bf C75} (2015) 132}, [\href{http://arxiv.org/abs/1412.7420}{{\tt
  1412.7420}}].

\bibitem{Cacciari:2008gp}
M.~Cacciari, G.~P. Salam and G.~Soyez, \emph{{The Anti-k(t) jet clustering
  algorithm}},
  \href{http://dx.doi.org/10.1088/1126-6708/2008/04/063}{\emph{JHEP} {\bf 04}
  (2008) 063}, [\href{http://arxiv.org/abs/0802.1189}{{\tt 0802.1189}}].

\bibitem{Cacciari:2011ma}
M.~Cacciari, G.~P. Salam and G.~Soyez, \emph{{FastJet User Manual}},
  \href{http://dx.doi.org/10.1140/epjc/s10052-012-1896-2}{\emph{Eur. Phys. J.}
  {\bf C72} (2012) 1896}, [\href{http://arxiv.org/abs/1111.6097}{{\tt
  1111.6097}}].

\bibitem{vanHameren:2010cp}
A.~van Hameren, \emph{{OneLOop: For the evaluation of one-loop scalar
  functions}}, \href{http://dx.doi.org/10.1016/j.cpc.2011.06.011}{\emph{Comput.
  Phys. Commun.} {\bf 182} (2011) 2427--2438},
  [\href{http://arxiv.org/abs/1007.4716}{{\tt 1007.4716}}].

\bibitem{Dittmaier:2001ay}
S.~Dittmaier and M.~Kramer, 1, \emph{{Electroweak radiative corrections to W
  boson production at hadron colliders}},
  \href{http://dx.doi.org/10.1103/PhysRevD.65.073007}{\emph{Phys. Rev.} {\bf
  D65} (2002) 073007}, [\href{http://arxiv.org/abs/hep-ph/0109062}{{\tt
  hep-ph/0109062}}].

\bibitem{Kuhn:2007cv}
J.~H. Kuhn, A.~Kulesza, S.~Pozzorini and M.~Schulze, \emph{{Electroweak
  corrections to hadronic production of W bosons at large transverse momenta}},
  \href{http://dx.doi.org/10.1016/j.nuclphysb.2007.12.029}{\emph{Nucl. Phys.}
  {\bf B797} (2008) 27--77}, [\href{http://arxiv.org/abs/0708.0476}{{\tt
  0708.0476}}].

\bibitem{Hollik:2007sq}
W.~Hollik, T.~Kasprzik and B.~A. Kniehl, \emph{{Electroweak corrections to
  W-boson hadroproduction at finite transverse momentum}},
  \href{http://dx.doi.org/10.1016/j.nuclphysb.2007.09.013}{\emph{Nucl. Phys.}
  {\bf B790} (2008) 138--159}, [\href{http://arxiv.org/abs/0707.2553}{{\tt
  0707.2553}}].

\bibitem{Denner:2009gj}
A.~Denner, S.~Dittmaier, T.~Kasprzik and A.~Muck, \emph{{Electroweak
  corrections to W + jet hadroproduction including leptonic W-boson decays}},
  \href{http://dx.doi.org/10.1088/1126-6708/2009/08/075}{\emph{JHEP} {\bf 08}
  (2009) 075}, [\href{http://arxiv.org/abs/0906.1656}{{\tt 0906.1656}}].

\bibitem{Denner:2012ts}
A.~Denner, S.~Dittmaier, T.~Kasprzik and A.~Muck, \emph{{Electroweak
  corrections to monojet production at the LHC}},
  \href{http://dx.doi.org/10.1140/epjc/s10052-013-2297-x}{\emph{Eur. Phys. J.}
  {\bf C73} (2013) 2297}, [\href{http://arxiv.org/abs/1211.5078}{{\tt
  1211.5078}}].

\bibitem{Badger:2016bpw}
J.~R. Andersen et~al., \emph{{Les Houches 2015: Physics at TeV Colliders
  Standard Model Working Group Report}},  in \emph{{9th Les Houches Workshop on
  Physics at TeV Colliders (PhysTeV 2015) Les Houches, France, June 1-19,
  2015}}, 2016.
\newblock \href{http://arxiv.org/abs/1605.04692}{{\tt 1605.04692}}.

\bibitem{Frixione:1992pj}
S.~Frixione, P.~Nason and G.~Ridolfi, \emph{{Strong corrections to W Z
  production at hadron colliders}},
  \href{http://dx.doi.org/10.1016/0550-3213(92)90668-2}{\emph{Nucl. Phys.} {\bf
  B383} (1992) 3--44}.

\bibitem{Butterworth:2008iy}
J.~M. Butterworth, A.~R. Davison, M.~Rubin and G.~P. Salam, \emph{{Jet
  substructure as a new Higgs search channel at the LHC}},
  \href{http://dx.doi.org/10.1103/PhysRevLett.100.242001}{\emph{Phys. Rev.
  Lett.} {\bf 100} (2008) 242001}, [\href{http://arxiv.org/abs/0802.2470}{{\tt
  0802.2470}}].

\bibitem{Bauer:2009km}
C.~W. Bauer and B.~O. Lange, \emph{{Scale setting and resummation of logarithms
  in pp $\to$ V + jets}},  \href{http://arxiv.org/abs/0905.4739}{{\tt
  0905.4739}}.

\bibitem{Rubin:2010xp}
M.~Rubin, G.~P. Salam and S.~Sapeta, \emph{{Giant QCD K-factors beyond NLO}},
  \href{http://dx.doi.org/10.1007/JHEP09(2010)084}{\emph{JHEP} {\bf 09} (2010)
  084}, [\href{http://arxiv.org/abs/1006.2144}{{\tt 1006.2144}}].

\bibitem{Baur:2001ze}
U.~Baur, O.~Brein, W.~Hollik, C.~Schappacher and D.~Wackeroth,
  \emph{{Electroweak radiative corrections to neutral current Drell-Yan
  processes at hadron colliders}},
  \href{http://dx.doi.org/10.1103/PhysRevD.65.033007}{\emph{Phys. Rev.} {\bf
  D65} (2002) 033007}, [\href{http://arxiv.org/abs/hep-ph/0108274}{{\tt
  hep-ph/0108274}}].

\bibitem{Dittmaier:2009cr}
S.~Dittmaier and M.~Huber, \emph{{Radiative corrections to the neutral-current
  Drell-Yan process in the Standard Model and its minimal supersymmetric
  extension}}, \href{http://dx.doi.org/10.1007/JHEP01(2010)060}{\emph{JHEP}
  {\bf 01} (2010) 060}, [\href{http://arxiv.org/abs/0911.2329}{{\tt
  0911.2329}}].

\bibitem{Kuhn:2005az}
J.~H. Kuhn, A.~Kulesza, S.~Pozzorini and M.~Schulze, \emph{{One-loop weak
  corrections to hadronic production of Z bosons at large transverse momenta}},
  \href{http://dx.doi.org/10.1016/j.nuclphysb.2005.08.019}{\emph{Nucl. Phys.}
  {\bf B727} (2005) 368--394}, [\href{http://arxiv.org/abs/hep-ph/0507178}{{\tt
  hep-ph/0507178}}].

\bibitem{Denner:2011vu}
A.~Denner, S.~Dittmaier, T.~Kasprzik and A.~Muck, \emph{{Electroweak
  corrections to dilepton + jet production at hadron colliders}},
  \href{http://dx.doi.org/10.1007/JHEP06(2011)069}{\emph{JHEP} {\bf 06} (2011)
  069}, [\href{http://arxiv.org/abs/1103.0914}{{\tt 1103.0914}}].

\bibitem{Hollik:2015pja}
W.~Hollik, B.~A. Kniehl, E.~S. Scherbakova and O.~L. Veretin,
  \emph{{Electroweak corrections to $Z$-boson hadroproduction at finite
  transverse momentum}},
  \href{http://dx.doi.org/10.1016/j.nuclphysb.2015.09.018}{\emph{Nucl. Phys.}
  {\bf B900} (2015) 576--602}, [\href{http://arxiv.org/abs/1504.07574}{{\tt
  1504.07574}}].

\bibitem{Gieseke:2014gka}
S.~Gieseke, T.~Kasprzik and J.~H. Kuhn, \emph{{Vector-boson pair production and
  electroweak corrections in HERWIG++}},
  \href{http://dx.doi.org/10.1140/epjc/s10052-014-2988-y}{\emph{Eur. Phys. J.}
  {\bf C74} (2014) 2988}, [\href{http://arxiv.org/abs/1401.3964}{{\tt
  1401.3964}}].

\bibitem{Billoni:2013aba}
M.~Billoni, S.~Dittmaier, B.~Jaeger and C.~Speckner, \emph{{Next-to-leading
  order electroweak corrections to $pp\to W^+W^- \to 4$ leptons at the LHC in
  double-pole approximation}},
  \href{http://dx.doi.org/10.1007/JHEP12(2013)043}{\emph{JHEP} {\bf 12} (2013)
  043}, [\href{http://arxiv.org/abs/1310.1564}{{\tt 1310.1564}}].

\bibitem{Frixione:1993yp}
S.~Frixione, \emph{{A Next-to-leading order calculation of the cross-section
  for the production of W+ W- pairs in hadronic collisions}},
  \href{http://dx.doi.org/10.1016/0550-3213(93)90435-R}{\emph{Nucl. Phys.} {\bf
  B410} (1993) 280--324}.

\bibitem{Baglio:2013toa}
J.~Baglio, L.~D. Ninh and M.~M. Weber, \emph{{Massive gauge boson pair
  production at the LHC: a next-to-leading order story}},
  \href{http://dx.doi.org/10.1103/PhysRevD.94.099902,
  10.1103/PhysRevD.88.113005}{\emph{Phys. Rev.} {\bf D88} (2013) 113005},
  [\href{http://arxiv.org/abs/1307.4331}{{\tt 1307.4331}}].

\bibitem{Gehrmann:2014fva}
T.~Gehrmann, M.~Grazzini, S.~Kallweit, P.~Maierhofer, A.~von Manteuffel,
  S.~Pozzorini et~al., \emph{{$W^+W^-$ Production at Hadron Colliders in Next
  to Next to Leading Order QCD}},
  \href{http://dx.doi.org/10.1103/PhysRevLett.113.212001}{\emph{Phys. Rev.
  Lett.} {\bf 113} (2014) 212001}, [\href{http://arxiv.org/abs/1408.5243}{{\tt
  1408.5243}}].

\bibitem{Ciccolini:2003jy}
M.~L. Ciccolini, S.~Dittmaier and M.~Kramer, \emph{{Electroweak radiative
  corrections to associated WH and ZH production at hadron colliders}},
  \href{http://dx.doi.org/10.1103/PhysRevD.68.073003}{\emph{Phys. Rev.} {\bf
  D68} (2003) 073003}, [\href{http://arxiv.org/abs/hep-ph/0306234}{{\tt
  hep-ph/0306234}}].

\bibitem{Denner:2011id}
A.~Denner, S.~Dittmaier, S.~Kallweit and A.~Muck, \emph{{Electroweak
  corrections to Higgs-strahlung off W/Z bosons at the Tevatron and the LHC
  with HAWK}}, \href{http://dx.doi.org/10.1007/JHEP03(2012)075}{\emph{JHEP}
  {\bf 03} (2012) 075}, [\href{http://arxiv.org/abs/1112.5142}{{\tt
  1112.5142}}].

\bibitem{Granata:2017iod}
F.~Granata, J.~M. Lindert, C.~Oleari and S.~Pozzorini, \emph{{NLO QCD+EW
  predictions for HV and HV+jet production including parton-shower effects}},
  \href{http://arxiv.org/abs/1706.03522}{{\tt 1706.03522}}.

\bibitem{Ciccolini:2007jr}
M.~Ciccolini, A.~Denner and S.~Dittmaier, \emph{{Strong and electroweak
  corrections to the production of Higgs + 2jets via weak interactions at the
  LHC}}, \href{http://dx.doi.org/10.1103/PhysRevLett.99.161803}{\emph{Phys.
  Rev. Lett.} {\bf 99} (2007) 161803},
  [\href{http://arxiv.org/abs/0707.0381}{{\tt 0707.0381}}].

\bibitem{Ciccolini:2007ec}
M.~Ciccolini, A.~Denner and S.~Dittmaier, \emph{{Electroweak and QCD
  corrections to Higgs production via vector-boson fusion at the LHC}},
  \href{http://dx.doi.org/10.1103/PhysRevD.77.013002}{\emph{Phys. Rev.} {\bf
  D77} (2008) 013002}, [\href{http://arxiv.org/abs/0710.4749}{{\tt
  0710.4749}}].

\bibitem{Nhung:2013tfu}
D.~T. Nhung, L.~D. Ninh and M.~M. Weber, \emph{{NLO $W W Z$ production at the
  LHC}},  in \emph{{Proceedings, 9th Rencontres du Vietnam: Windows on the
  Universe: Quy Nhon, Vietnam, August 11-17, 2013}}, pp.~219--222, 2013.
\newblock \href{http://arxiv.org/abs/1310.6159}{{\tt 1310.6159}}.

\bibitem{Yong-Bai:2015xna}
S.~Yong-Bai, Z.~Ren-You, M.~Wen-Gan, L.~Xiao-Zhou, Z.~Yu and G.~Lei, \emph{{NLO
  QCD + NLO EW corrections to $WZZ$ productions with leptonic decays at the
  LHC}}, \href{http://dx.doi.org/10.1007/JHEP10(2016)156,
  10.1007/JHEP10(2015)186}{\emph{JHEP} {\bf 10} (2015) 186},
  [\href{http://arxiv.org/abs/1507.03693}{{\tt 1507.03693}}].

\bibitem{Hong:2016aek}
W.~Hong, Z.~Ren-You, M.~Wen-Gan, G.~Lei, L.~Xiao-Zhou and W.~Shao-Ming,
  \emph{{NLO QCD + EW corrections to ZZZ production with subsequent leptonic
  decays at the LHC}},
  \href{http://dx.doi.org/10.1088/0954-3899/43/11/115001}{\emph{J. Phys.} {\bf
  G43} (2016) 115001}, [\href{http://arxiv.org/abs/1610.05876}{{\tt
  1610.05876}}].

\bibitem{Yong-Bai:2016sal}
Y.-B. Shen, R.-Y. Zhang, W.-G. Ma, X.-Z. Li and L.~Guo, \emph{{NLO QCD and
  electroweak corrections to WWW production at the LHC}},
  \href{http://dx.doi.org/10.1103/PhysRevD.95.073005}{\emph{Phys. Rev.} {\bf
  D95} (2017) 073005}, [\href{http://arxiv.org/abs/1605.00554}{{\tt
  1605.00554}}].

\bibitem{Dittmaier:2017bnh}
S.~Dittmaier, A.~Huss and G.~Knippen, \emph{{Next-to-leading-order QCD and
  electroweak corrections to WWW production at proton-proton colliders}},
  \href{http://arxiv.org/abs/1705.03722}{{\tt 1705.03722}}.

\bibitem{Frixione:2008yi}
S.~Frixione, E.~Laenen, P.~Motylinski, B.~R. Webber and C.~D. White,
  \emph{{Single-top hadroproduction in association with a W boson}},
  \href{http://dx.doi.org/10.1088/1126-6708/2008/07/029}{\emph{JHEP} {\bf 07}
  (2008) 029}, [\href{http://arxiv.org/abs/0805.3067}{{\tt 0805.3067}}].

\bibitem{White:2009yt}
C.~D. White, S.~Frixione, E.~Laenen and F.~Maltoni, \emph{{Isolating Wt
  production at the LHC}},
  \href{http://dx.doi.org/10.1088/1126-6708/2009/11/074}{\emph{JHEP} {\bf 11}
  (2009) 074}, [\href{http://arxiv.org/abs/0908.0631}{{\tt 0908.0631}}].

\bibitem{Weydert:2009vr}
C.~Weydert, S.~Frixione, M.~Herquet, M.~Klasen, E.~Laenen, T.~Plehn et~al.,
  \emph{{Charged Higgs boson production in association with a top quark in
  MC@NLO}}, \href{http://dx.doi.org/10.1140/epjc/s10052-010-1320-8}{\emph{Eur.
  Phys. J.} {\bf C67} (2010) 617--636},
  [\href{http://arxiv.org/abs/0912.3430}{{\tt 0912.3430}}].

\bibitem{Re:2010bp}
E.~Re, \emph{{Single-top Wt-channel production matched with parton showers
  using the POWHEG method}},
  \href{http://dx.doi.org/10.1140/epjc/s10052-011-1547-z}{\emph{Eur. Phys. J.}
  {\bf C71} (2011) 1547}, [\href{http://arxiv.org/abs/1009.2450}{{\tt
  1009.2450}}].

\bibitem{Binoth:2011xi}
T.~Binoth, D.~Goncalves~Netto, D.~Lopez-Val, K.~Mawatari, T.~Plehn and
  I.~Wigmore, \emph{{Automized Squark-Neutralino Production to Next-to-Leading
  Order}}, \href{http://dx.doi.org/10.1103/PhysRevD.84.075005}{\emph{Phys.
  Rev.} {\bf D84} (2011) 075005}, [\href{http://arxiv.org/abs/1108.1250}{{\tt
  1108.1250}}].

\bibitem{GoncalvesNetto:2012yt}
D.~Goncalves-Netto, D.~Lopez-Val, K.~Mawatari, T.~Plehn and I.~Wigmore,
  \emph{{Automated Squark and Gluino Production to Next-to-Leading Order}},
  \href{http://dx.doi.org/10.1103/PhysRevD.87.014002}{\emph{Phys. Rev.} {\bf
  D87} (2013) 014002}, [\href{http://arxiv.org/abs/1211.0286}{{\tt
  1211.0286}}].

\bibitem{Gavin:2013kga}
R.~Gavin, C.~Hangst, M.~Kr{\"a}mer, M.~M{\"u}hlleitner, M.~Pellen, E.~Popenda
  et~al., \emph{{Matching Squark Pair Production at NLO with Parton Showers}},
  \href{http://dx.doi.org/10.1007/JHEP10(2013)187}{\emph{JHEP} {\bf 10} (2013)
  187}, [\href{http://arxiv.org/abs/1305.4061}{{\tt 1305.4061}}].

\bibitem{Gavin:2014yga}
R.~Gavin, C.~Hangst, M.~Kr{\"a}mer, M.~M{\"u}hlleitner, M.~Pellen, E.~Popenda
  et~al., \emph{{Squark Production and Decay matched with Parton Showers at
  NLO}}, \href{http://dx.doi.org/10.1140/epjc/s10052-014-3243-2}{\emph{Eur.
  Phys. J.} {\bf C75} (2015) 29}, [\href{http://arxiv.org/abs/1407.7971}{{\tt
  1407.7971}}].

\bibitem{Demartin:2016axk}
F.~Demartin, B.~Maier, F.~Maltoni, K.~Mawatari and M.~Zaro, \emph{{tWH
  associated production at the LHC}},
  \href{http://dx.doi.org/10.1140/epjc/s10052-017-4601-7}{\emph{Eur. Phys. J.}
  {\bf C77} (2017) 34}, [\href{http://arxiv.org/abs/1607.05862}{{\tt
  1607.05862}}].

\bibitem{Tait:1999cf}
T.~M.~P. Tait, \emph{{The $t W^{-}$ mode of single top production}},
  \href{http://dx.doi.org/10.1103/PhysRevD.61.034001}{\emph{Phys. Rev.} {\bf
  D61} (1999) 034001}, [\href{http://arxiv.org/abs/hep-ph/9909352}{{\tt
  hep-ph/9909352}}].

\bibitem{Zhu:2001hw}
S.~Zhu, \emph{{Next-to-leading order QCD corrections to $bg\to tW^-$ at CERN
  large hadron collider}},
  \href{http://dx.doi.org/10.1016/S0370-2693(02)01952-4,
  10.1016/S0370-2693(01)01404-6}{\emph{Phys. Lett.} {\bf B524} (2002)
  283--288}, [\href{http://arxiv.org/abs/hep-ph/0109269}{{\tt
  hep-ph/0109269}}].

\bibitem{Cascioli:2013wga}
F.~Cascioli, S.~Kallweit, P.~Maierhofer and S.~Pozzorini, \emph{{A unified NLO
  description of top-pair and associated Wt production}},
  \href{http://dx.doi.org/10.1140/epjc/s10052-014-2783-9}{\emph{Eur. Phys. J.}
  {\bf C74} (2014) 2783}, [\href{http://arxiv.org/abs/1312.0546}{{\tt
  1312.0546}}].

\bibitem{Beenakker:1996ch}
W.~Beenakker, R.~Hopker, M.~Spira and P.~M. Zerwas, \emph{{Squark and gluino
  production at hadron colliders}},
  \href{http://dx.doi.org/10.1016/S0550-3213(97)80027-2}{\emph{Nucl. Phys.}
  {\bf B492} (1997) 51--103}, [\href{http://arxiv.org/abs/hep-ph/9610490}{{\tt
  hep-ph/9610490}}].

\bibitem{Berger:2003sm}
E.~L. Berger, T.~Han, J.~Jiang and T.~Plehn, \emph{{Associated production of a
  top quark and a charged Higgs boson}},
  \href{http://dx.doi.org/10.1103/PhysRevD.71.115012}{\emph{Phys. Rev.} {\bf
  D71} (2005) 115012}, [\href{http://arxiv.org/abs/hep-ph/0312286}{{\tt
  hep-ph/0312286}}].

\bibitem{Dao:2010nu}
T.~N. Dao, W.~Hollik and D.~N. Le, \emph{{$W^\mp H^\pm$ production and CP
  asymmetry at the LHC}}, \href{http://dx.doi.org/10.1103/PhysRevD.83.079903,
  10.1103/PhysRevD.83.075003}{\emph{Phys. Rev.} {\bf D83} (2011) 075003},
  [\href{http://arxiv.org/abs/1011.4820}{{\tt 1011.4820}}].

\bibitem{Hollik:2012rc}
W.~Hollik, J.~M. Lindert and D.~Pagani, \emph{{NLO corrections to squark-squark
  production and decay at the LHC}},
  \href{http://dx.doi.org/10.1007/JHEP03(2013)139}{\emph{JHEP} {\bf 03} (2013)
  139}, [\href{http://arxiv.org/abs/1207.1071}{{\tt 1207.1071}}].

\bibitem{Yu:2014cka}
Y.~Zhang, W.-G. Ma, R.-Y. Zhang, C.~Chen and L.~Guo, \emph{{QCD NLO and EW NLO
  corrections to $t\bar{t}H$ production with top quark decays at hadron
  collider}},
  \href{http://dx.doi.org/10.1016/j.physletb.2014.09.022}{\emph{Phys. Lett.}
  {\bf B738} (2014) 1--5}, [\href{http://arxiv.org/abs/1407.1110}{{\tt
  1407.1110}}].

\bibitem{Dittmaier:2012kx}
S.~Dittmaier, A.~Huss and C.~Speckner, \emph{{Weak radiative corrections to
  dijet production at hadron colliders}},
  \href{http://dx.doi.org/10.1007/JHEP11(2012)095}{\emph{JHEP} {\bf 11} (2012)
  095}, [\href{http://arxiv.org/abs/1210.0438}{{\tt 1210.0438}}].

\bibitem{Beccaria:2006ir}
M.~Beccaria, G.~Macorini, F.~M. Renard and C.~Verzegnassi, \emph{{Single top
  production in the t-channel at LHC: A Realistic test of electroweak models}},
  \href{http://dx.doi.org/10.1103/PhysRevD.74.013008}{\emph{Phys. Rev.} {\bf
  D74} (2006) 013008}, [\href{http://arxiv.org/abs/hep-ph/0605108}{{\tt
  hep-ph/0605108}}].

\bibitem{Mirabella:2008gj}
E.~Mirabella, \emph{{Electroweak Corrections to t-channel single top production
  at the LHC}}, \href{http://dx.doi.org/10.1393/ncb/i2008-10669-4}{\emph{Nuovo
  Cim.} {\bf B123} (2008) 1111--1117},
  [\href{http://arxiv.org/abs/0811.2051}{{\tt 0811.2051}}].

\bibitem{Bardin:2010mz}
D.~Bardin, S.~Bondarenko, L.~Kalinovskaya, V.~Kolesnikov and W.~von Schlippe,
  \emph{{Electroweak Radiative Corrections to Single-top Production}},
  \href{http://dx.doi.org/10.1140/epjc/s10052-010-1533-x}{\emph{Eur. Phys. J.}
  {\bf C71} (2011) 1533}, [\href{http://arxiv.org/abs/1008.1859}{{\tt
  1008.1859}}].

\bibitem{Dror:2015nkp}
J.~A. Dror, M.~Farina, E.~Salvioni and J.~Serra, \emph{{Strong tW Scattering at
  the LHC}}, \href{http://dx.doi.org/10.1007/JHEP01(2016)071}{\emph{JHEP} {\bf
  01} (2016) 071}, [\href{http://arxiv.org/abs/1511.03674}{{\tt 1511.03674}}].

\bibitem{Degrande:2016hyf}
C.~Degrande, R.~Frederix, V.~Hirschi, M.~Ubiali, M.~Wiesemann and M.~Zaro,
  \emph{{Accurate predictions for charged Higgs production: Closing the
  $m_{H^{\pm}}\sim m_t$ window}},
  \href{http://dx.doi.org/10.1016/j.physletb.2017.06.037}{\emph{Phys. Lett.}
  {\bf B772} (2017) 87--92}, [\href{http://arxiv.org/abs/1607.05291}{{\tt
  1607.05291}}].

\bibitem{Alwall:2011uj}
J.~Alwall, M.~Herquet, F.~Maltoni, O.~Mattelaer and T.~Stelzer, \emph{{MadGraph
  5 : Going Beyond}},
  \href{http://dx.doi.org/10.1007/JHEP06(2011)128}{\emph{JHEP} {\bf 1106}
  (2011) 128}, [\href{http://arxiv.org/abs/1106.0522}{{\tt 1106.0522}}].

\bibitem{Bertone:2015lqa}
V.~Bertone, S.~Carrazza, D.~Pagani and M.~Zaro, \emph{{On the Impact of Lepton
  PDFs}}, \href{http://dx.doi.org/10.1007/JHEP11(2015)194}{\emph{JHEP} {\bf 11}
  (2015) 194}, [\href{http://arxiv.org/abs/1508.07002}{{\tt 1508.07002}}].

\bibitem{Bauer:2017isx}
C.~W. Bauer, N.~Ferland and B.~R. Webber, \emph{{Standard Model Parton
  Distributions at Very High Energies}},
  \href{http://dx.doi.org/10.1007/JHEP08(2017)036}{\emph{JHEP} {\bf 08} (2017)
  036}, [\href{http://arxiv.org/abs/1703.08562}{{\tt 1703.08562}}].

\bibitem{Jezabek:1988iv}
M.~Jezabek and J.~H. Kuhn, \emph{{QCD Corrections to Semileptonic Decays of
  Heavy Quarks}},
  \href{http://dx.doi.org/10.1016/0550-3213(89)90108-9}{\emph{Nucl. Phys.} {\bf
  B314} (1989) 1--6}.

\bibitem{Denner:1990ns}
A.~Denner and T.~Sack, \emph{{The Top width}},
  \href{http://dx.doi.org/10.1016/0550-3213(91)90530-B}{\emph{Nucl. Phys.} {\bf
  B358} (1991) 46--58}.

\bibitem{Czarnecki:1990kv}
A.~Czarnecki, \emph{{QCD corrections to the decay $t\to Wb$ in dimensional
  regularization}},
  \href{http://dx.doi.org/10.1016/0370-2693(90)90571-M}{\emph{Phys. Lett.} {\bf
  B252} (1990) 467--470}.

\bibitem{Li:1990qf}
C.~S. Li, R.~J. Oakes and T.~C. Yuan, \emph{{QCD corrections to $t \to W^{+}
  b$}}, \href{http://dx.doi.org/10.1103/PhysRevD.43.3759}{\emph{Phys. Rev.}
  {\bf D43} (1991) 3759--3762}.

\bibitem{Liu:1990py}
J.-a. Liu and Y.-P. Yao, \emph{{One loop radiative corrections to a heavy top
  decay in the standard model}},
  \href{http://dx.doi.org/10.1142/S0217751X91002331}{\emph{Int. J. Mod. Phys.}
  {\bf A6} (1991) 4925--4948}.

\bibitem{Eilam:1991iz}
G.~Eilam, R.~R. Mendel, R.~Migneron and A.~Soni, \emph{{Radiative corrections
  to top quark decay}},
  \href{http://dx.doi.org/10.1103/PhysRevLett.66.3105}{\emph{Phys. Rev. Lett.}
  {\bf 66} (1991) 3105--3108}.

\bibitem{Jezabek:1993wk}
M.~Jezabek and J.~H. Kuhn, \emph{{The Top width: Theoretical update}},
  \href{http://dx.doi.org/10.1103/PhysRevD.49.4970,
  10.1103/PhysRevD.48.R1910}{\emph{Phys. Rev.} {\bf D48} (1993) R1910--R1913},
  [\href{http://arxiv.org/abs/hep-ph/9302295}{{\tt hep-ph/9302295}}].

\bibitem{Basso:2015gca}
L.~Basso, S.~Dittmaier, A.~Huss and L.~Oggero, \emph{{Techniques for the
  treatment of IR divergences in decay processes at NLO and application to the
  top-quark decay}},
  \href{http://dx.doi.org/10.1140/epjc/s10052-016-3878-2}{\emph{Eur. Phys. J.}
  {\bf C76} (2016) 56}, [\href{http://arxiv.org/abs/1507.04676}{{\tt
  1507.04676}}].

\bibitem{Marciano:1974vg}
W.~J. Marciano and A.~Sirlin, \emph{{Deviations from electron-muon universality
  in the leptonic decays of the intermediate bosons}},
  \href{http://dx.doi.org/10.1103/PhysRevD.8.3612}{\emph{Phys. Rev.} {\bf D8}
  (1973) 3612--3615}.

\bibitem{Albert:1979ix}
D.~Albert, W.~J. Marciano, D.~Wyler and Z.~Parsa, \emph{{Decays of Intermediate
  Vector Bosons, Radiative Corrections and QCD Jets}},
  \href{http://dx.doi.org/10.1016/0550-3213(80)90208-4}{\emph{Nucl. Phys.} {\bf
  B166} (1980) 460--492}.

\bibitem{Inoue:1980ky}
K.~Inoue, A.~Kakuto, H.~Komatsu and S.~Takeshita, \emph{{Radiative Corrections
  for $W \to e$ Anti-neutrino Decay in the {Weinberg-Salam} Model With
  Arbitrary Number of Generations}},
  \href{http://dx.doi.org/10.1143/PTP.64.1008}{\emph{Prog. Theor. Phys.} {\bf
  64} (1980) 1008}.

\bibitem{Chang:1981qq}
T.~H. Chang, K.~J.~F. Gaemers and W.~L. van Neerven, \emph{{QCD Corrections to
  the Mass and Width of the Intermediate Vector Bosons}},
  \href{http://dx.doi.org/10.1016/0550-3213(82)90407-2}{\emph{Nucl. Phys.} {\bf
  B202} (1982) 407--436}.

\bibitem{Consoli:1983yn}
M.~Consoli, S.~Lo~Presti and L.~Maiani, \emph{{Higher Order Effects and the
  Vector Boson Physical Parameters}},
  \href{http://dx.doi.org/10.1016/0550-3213(83)90066-4}{\emph{Nucl. Phys.} {\bf
  B223} (1983) 474--500}.

\bibitem{Bardin:1986fi}
D.~{\relax Yu}. Bardin, S.~Riemann and T.~Riemann, \emph{{Electroweak One Loop
  Corrections to the Decay of the Charged Vector Boson}},
  \href{http://dx.doi.org/10.1007/BF01441360}{\emph{Z. Phys.} {\bf C32} (1986)
  121--125}.

\bibitem{Denner:1990tx}
A.~Denner and T.~Sack, \emph{{The W boson width}},
  \href{http://dx.doi.org/10.1007/BF01560267}{\emph{Z. Phys.} {\bf C46} (1990)
  653--663}.

\bibitem{Dine:1979qh}
M.~Dine and J.~R. Sapirstein, \emph{{Higher Order QCD Corrections in e+ e-
  Annihilation}},
  \href{http://dx.doi.org/10.1103/PhysRevLett.43.668}{\emph{Phys. Rev. Lett.}
  {\bf 43} (1979) 668}.

\bibitem{Celmaster:1979xr}
W.~Celmaster and R.~J. Gonsalves, \emph{{An Analytic Calculation of Higher
  Order Quantum Chromodynamic Corrections in e+ e- Annihilation}},
  \href{http://dx.doi.org/10.1103/PhysRevLett.44.560}{\emph{Phys. Rev. Lett.}
  {\bf 44} (1980) 560}.

\bibitem{Czarnecki:1996ei}
A.~Czarnecki and J.~H. Kuhn, \emph{{Nonfactorizable QCD and electroweak
  corrections to the hadronic Z boson decay rate}},
  \href{http://dx.doi.org/10.1103/PhysRevLett.77.3955}{\emph{Phys. Rev. Lett.}
  {\bf 77} (1996) 3955--3958}, [\href{http://arxiv.org/abs/hep-ph/9608366}{{\tt
  hep-ph/9608366}}].

\bibitem{Djouadi:1997yw}
A.~Djouadi, J.~Kalinowski and M.~Spira, \emph{{HDECAY: A Program for Higgs
  boson decays in the standard model and its supersymmetric extension}},
  \href{http://dx.doi.org/10.1016/S0010-4655(97)00123-9}{\emph{Comput. Phys.
  Commun.} {\bf 108} (1998) 56--74},
  [\href{http://arxiv.org/abs/hep-ph/9704448}{{\tt hep-ph/9704448}}].

\bibitem{Djouadi:2018xqq}
A.~Djouadi, J.~Kalinowski, M.~Muehlleitner and M.~Spira, \emph{{HDECAY:
  Twenty$_{++}$ Years After}},  \href{http://arxiv.org/abs/1801.09506}{{\tt
  1801.09506}}.

\bibitem{Hahn:2000kx}
T.~Hahn, \emph{{Generating Feynman diagrams and amplitudes with FeynArts 3}},
  \href{http://dx.doi.org/10.1016/S0010-4655(01)00290-9}{\emph{Comput. Phys.
  Commun.} {\bf 140} (2001) 418--431},
  [\href{http://arxiv.org/abs/hep-ph/0012260}{{\tt hep-ph/0012260}}].

\end{thebibliography}\endgroup

\end{document}